\documentclass[12pt]{report}
\usepackage{epsfig}

\newif\ifdimspec

\def\checkdim#1{\ifx#1\end \let\next=\relax
  \else \ifcat#1a \dimspectrue \fi \let\next=\checkdim\fi \next}

\newcommand{\ee}{e^+e^-}
\newcommand{\qq}{q\bar q}
\renewcommand{\bar}[1]{\overline{#1}}
\newcommand{\bold}[1]{\setbox0=\hbox{$#1$}
              \kern-.025em\copy0\kern-\wd0
              \kern.05em\copy0\kern-\wd0
              \kern-.025em\raise.0433em\box0 }
\newcommand{\etal}{{\em et al.}}

\newcommand{\eg}{{\em e.g.}}

\newcommand{\lsim}{\buildrel < \over {_\sim}}

\renewcommand{\arraystretch}{1.3}

\begin{document}
\pagenumbering{roman}
\thispagestyle{empty}

\begin{flushright}
BNL 52--502\\
FERMILAB--PUB--96/112\\
LBNL--PUB--5425\\
SLAC--Report--485\\
UCRL--ID--124160\\
UC--414\\
June, 1996
\end{flushright}
\vspace{2cm}

\begin{center}
{\Huge Physics and Technology\\
of the Next Linear Collider:\\ 
\mbox{ }\\ 
A Report Submitted to Snowmass '96\\
by the NLC Zeroth-Order Design Group\\
and the NLC Physics Working Group\\}
\end {center}
\vspace{10pt}
\enlargethispage*{20pt}

\newpage

\begin{center}
{\Huge \bf The NLC Accelerator Design Group\\
and\\
The NLC Physics Working Group\\}
\end {center}
\vskip10pt
\enlargethispage*{20pt}

\medskip
 
\medskip\noindent
S. Kuhlman;\\
{\sl Argonne National Laboratory, Chicago, Illinois, USA}

\medskip\noindent
P. Minkowski;\\
{\sl University of Bern, Bern, Switzerland}

\medskip\noindent
W. Marciano,
F.~Paige;\\ 
{\sl Brookhaven National Laboratory, Upton, Long Island, New York, USA}

\medskip\noindent
V.~Telnov;\\
{\sl Budker Institute for Nuclear Physics, Novosibirsk, Russia}

\medskip\noindent
J. F. Gunion, 
T. Han,
S.~Lidia;\\
{\sl University of California, Davis, California, USA}

\medskip\noindent
J.~Rosenzweig;\\
{\sl University of California, Los Angeles, California,
USA}

\medskip\noindent
J. Wudka;\\
{\sl University of California, Riverside, California, USA}

\medskip\noindent
N.~M.~Kroll;\\
{\sl University of California, San Diego, California, USA}

\medskip\noindent
D.~A. Bauer, 
H.~Nelson;\\
{\sl University of California, Santa Barbara, California, USA}

\medskip\noindent
H.~Haber, 
C.~Heusch, 
B.~Schumm;\\
{\sl University of California, Santa Cruz, California, USA}

\medskip\noindent
M. Gintner, 
S. Godfrey, 
P. Kalyniak;\\
{\sl Carleton University, Ottawa, Canada}

\medskip\noindent
L.~Rinolfi;\\
{\sl CERN, 
Geneva, Switzerland}

\medskip\noindent
A. Barker, 
M. Danielson, 
S. Fahey, 
M. Goluboff, 
U. Nauenberg, 
D. L. Wagner;\\
{\sl University of Colorado, Boulder, Colorado, USA}

\medskip\noindent
P. C. Rowson;\\
{\sl Columbia University, New York, New York, USA}

\medskip\noindent
J.~A.~Holt,
K. Maeshima, 
R. Raja;\\
{\sl Fermilab National Laboratory, Batavia, Illinois, USA}
% ...

\medskip\noindent
B.~F.~L.~Ward, 
A. Weidemann;\\
{\sl University of Tennessee, Knoxville, Tennessee, USA}

\medskip\noindent
L.~Sawyer, 
A.~White;\\
{\sl University of Texas at Arlington, Arlington, Texas, USA}

\nopagebreak
\medskip\noindent
C.~Baltay, 
S.~Manly;\\
{\sl Yale University, New Haven, Connecticut, USA}

\newpage

\section*{\it PREFACE}
\addcontentsline{toc}{chapter}{Preface}

We present the prospects for the next generation of high-energy physics
experiments with electron-positron colliding beams.  This  report
summarizes the current status of the design and technological basis of
a linear collider of center-of-mass energy 0.5--1.5 TeV, and the
opportunities for high-energy physics experiments that this machine is
expected to open.

Over the past two decades, particle physics experiments have made an
increasingly precise confirmation of the ``Standard Model'' of strong,
weak, and electromagnetic interactions.  High-energy physicists feel
confident that the basic structure of these once-mysterious
interactions of elementary particles is now well understood.  But the
verification of this model has brought with it the realization that
there is a missing piece to the story: although the structure of the
weak interactions is based on a symmetry principle, we observe that
symmetry to be broken, by an agent that we do not yet know.  This
agent, whatever its source, must provide new physical phenomena  at the
TeV energy scale.

The Large Hadron Collider (LHC) in Europe offers an entry into this
energy regime with significant opportunity for discovery of new
phenomena.  An electron-positron collider at this next step in energy,
the Next Linear Collider (NLC), will provide a complementary program of
experiments with unique opportunities for both  discovery and 
precision measurement. To understand the nature of the new phenomena at
the TeV scale, to see how they fit together with the known particles
and interactions into a grander picture, both of these facilities will
be required.

In particular, electron-positron colliders offer specific features that
are essential to understand the nature of these new interactions
whatever their source.  They allow precise and detailed studies of the
two known particles that couple most strongly to these interactions,
the $W$ boson and the top quark.  They provide a clean environment for
the discovery of new particles whatever their nature, and they provide
special tools, such as the use of electron beam polarization, to
dissect the couplings of those particles.

All of this would be merely theoretical if the next-generation linear
collider could not be realized.  But, in the past few years, the
technology of the linear collider has come of age.  The experience
gained from the operation of the Stanford Linear Collider (SLC) has
provided a firm foundation to the design choices for
the NLC. The fundamental new technologies needed to
construct the NLC  have been demonstrated experimentally. Microwave
power sources have exceeded requirements for the initial stage of the
NLC, and critical tests assure us that this technology can be expected
to drive beams to a center-of-mass energies of a TeV or more. Essential
demonstrations of prototype collider subsystems have either taken place
or are now underway:  the  Final Focus Test Beam has already operated
successfully; a linear accelerator and a damping
ring will be operated within the next year.  A detailed feasibility
study, the ``Zeroth-Order Design Report'' (ZDR), has shown that these
components can be integrated into a complete machine design.

 \nopagebreak
The Next Linear Collider can be constructed, and it will play an
essential role in our understanding of physics at the TeV energy scale.

\newpage
$$ $$
\newpage

\newpage
\tableofcontents
\newpage

\chapter{The Next Linear Collider}
\pagenumbering{arabic}
\vspace*{-1cm}

\section{Goals for the Next Linear Collider}

\begin{figure}[htb]
\leavevmode
\centerline{\epsfbox{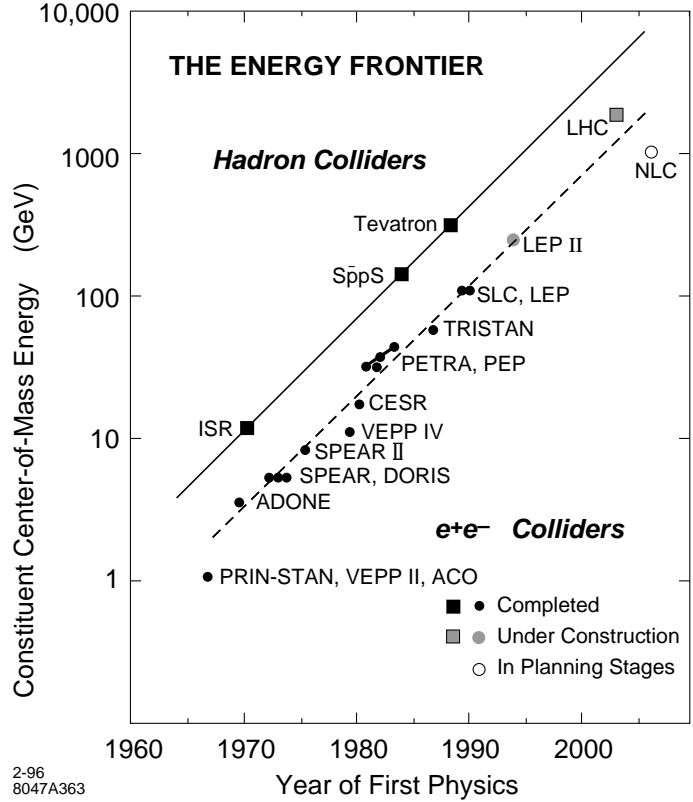}}
\centerline{\parbox{5in}{\caption[*]
{The energy frontier of particle physics. The effective
constituent energy of existing and planned colliders and the year of
first physics results from each.}
\label{fig:energy}}}
\end{figure}

For the past 25 years accelerator facilities with colliding beams have
been the forefront instruments used to study elementary particle
physics at high energies (Fig.~\ref{fig:energy}). Both hadron-hadron
and electron-positron colliders have been used to make important
observations and discoveries. Direct observations of the $W^\pm$ and
$Z^0$ bosons at CERN and investigations of the top quark at Fermilab
are examples of physics done at hadron colliders. Electron-positron
colliders provide well-controlled and well-understood experimental
environments in which new phenomena stand out and precise measurements
can be made. The discoveries of the charm quark and $\tau$ lepton at
SPEAR, discovery of the gluon and establishment of QCD at PETRA and
PEP, and precision exploration of electroweak phenomena at the SLC and
LEP are highlights of the results produced by experiments at
electron-positron colliders.

The ability to study nature with these two different kinds of
instruments has proven essential to the advancement of our
understanding of particle physics.  This will remain true as we seek
answers to questions posed at the TeV energy scale:

\begin{itemize}
\item 
What is the top quark, and what are its interactions?
\item 
Why is the symmetry of the electroweak interaction broken, and what is
the origin of mass?
\item 
Do Higgs particles exist? If so, how many, and what are their
structures and interactions?
\item 
Is the world supersymmetric, and if so, what is its structure, and is
this supersymmetry part of a larger unification of nature?
\item 
Are quarks, leptons, and gauge bosons fundamental particles, or are
they more complex?
\nopagebreak[1]
\item 
Are there other new particles or interactions, and what might nature
contain that we have not yet imagined?
\end{itemize}

The Large Hadron Collider (LHC) in Europe offers an entry into the TeV
energy regime with significant opportunity for discovery of new
phenomena.  The planned participation in the design, construction, and
utilization of this collider by nations around the world will make the
LHC the first truly global facility for the study of particle physics. 
This will be an exciting and important step in the continuing evolution
of our science.

The companion electron-positron collider at this next step in energy,
the Next Linear Collider (NLC), will provide a complementary program of
experiments with unique opportunities for both discovery and precision
measurement.  To understand the nature of physics at the TeV scale, to
see how the new phenomena we expect to find there fit together with the
known particles and interactions into a grander picture, both the LHC
and the NLC will be required.

Studies of physics goals and requirements for the next-generation
electron-positron collider began in 1987-88 in the United States
\cite{ahn88,sno88,sno90}, Europe \cite{lat87, des90}, and Japan
\cite{jlc89, jlc90}. These regional studies have evolved into a series
of internationally sponsored and organized workshops
\cite{fin91,haw93,jap95} that continue to build an important consensus
on the goals and specifications of the Next Linear Collider. This
document is both a part of this process, and input to deliberations by
the U.S. particle physics community that will take place this Summer at
Snowmass, Colorado. To prepare for Snowmass, a series of workshops was
held over the past year at locations throughout the United States.
Working groups were established at a first meeting in Estes Park,
Colorado to provide a framework for people to participate in the
discussions of various topics in physics and experimentation at linear
colliders.  These groups continued to meet at subsequent workshops held
at Fermilab, SLAC, and Brookhaven National Laboratory.  This document
contains a written summary from these workshops.

\begin{figure}[htb]
\leavevmode
\centerline{\epsfxsize=3.5in\epsfbox{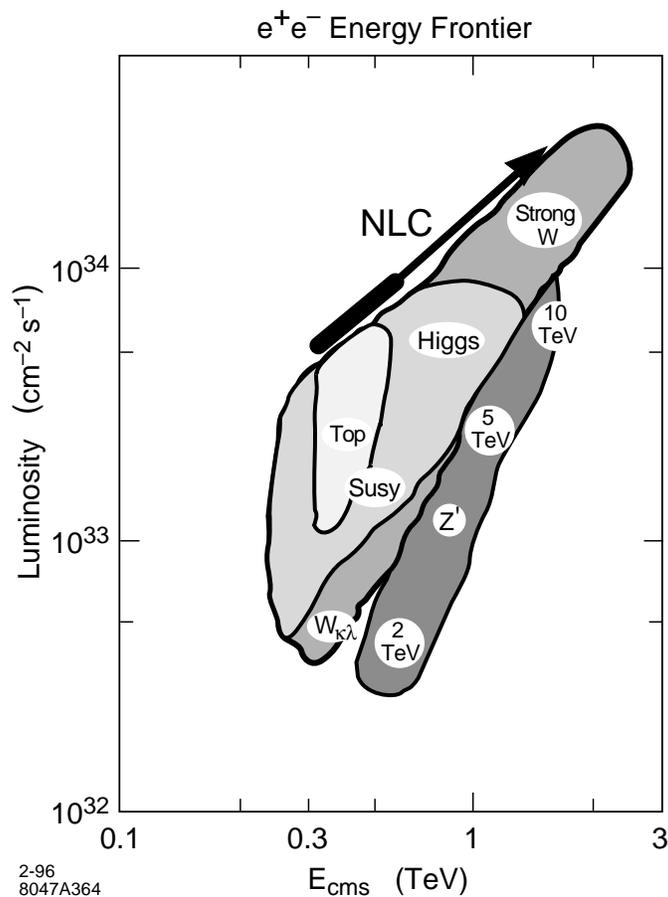}}
\centerline{\parbox{5in}
{\caption[*]{Physics goals for a TeV-scale $e^+e^-$ collider.}
\label{fig:goals}}}
\end{figure}

A picture has emerged of a high-performance collider able to explore a
broad range of center of mass energies from a few hundred GeV to a TeV
and beyond (Fig.~\ref{fig:goals}).  The goals of particle physics at
the TeV scale require luminosities of approximately
$10^{34}$cm$^{-2}$sec$^{-1}$, and reliable technologies that can
provide large integrated data samples. It is important that the beam
energy spread remain well controlled, and that backgrounds created by
lost particles and radiation from the beams be maintained at low
levels. This will assure that the clean experimental environment
historically offered by electron-positron colliders remains intact.
Beam polarization is an additional tool available at a linear collider
that provides new and revealing views of particle physics, and this too
is a requirement for any future collider.

In this first chapter we introduce the accelerator physics and
technologies of the Next Linear Collider, discuss design choices and
philosophies, and provide a brief status report on the R\&D program
that is being carried out in support of the NLC design effort.  The
second chapter of this document concentrates on the physics program of
the NLC{}.  The final chapter gives a more detailed overview of the
accelerator design.  A companion document, {\it A Zeroth-Order Design
Report for the Next Linear Collider}, that contains results from a
rather extensive feasibility study of the NLC, has also been prepared
\cite{newref}. This may be of further interest to readers.

\section{Accelerator Design Choices}
\label{sec:choice}

\begin{figure}[htbp]
\leavevmode
\centerline{\epsfysize=5in\epsfbox{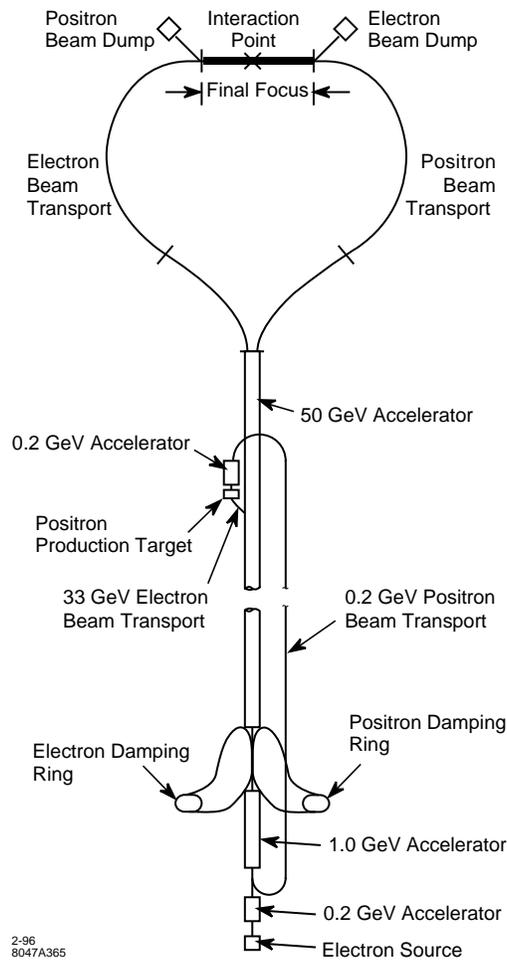}}
\centerline{\parbox{5in}{\caption[*]{The Stanford Linear Collider
(SLC).}
\label{fig:slc}}}
\end{figure}

\subsection{The Stanford Linear Collider}

\begin{figure}[htb]
\leavevmode
\centerline{\epsfxsize=4.25in\epsfbox{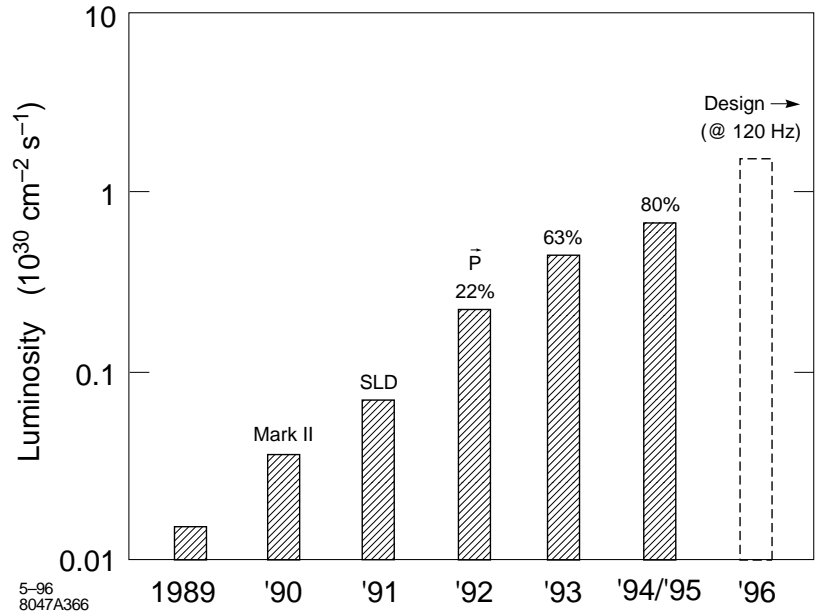}}
\centerline{\parbox{5in}
{\caption[*]{Performance of the SLC from early commissioning. 
Polarization of the electron beam is also shown.}
\label{perf}}}
\end{figure}

The Stanford Linear Collider (Fig.~\ref{fig:slc}) was conceived and
built to accomplish two goals: to study particle physics at the
100-GeV energy scale, and to develop the accelerator physics and
technology necessary for the realization of future high-energy
colliders. The SLC was completed in 1987 and provided a first look at
the physics of the $Z^0$ in 1989. In time, the luminosity provided by
this machine has grown steadily (Fig.~\ref{perf}), and has allowed
particle physicists to make unique and important studies of the $Z^0$
and its decays.

The design of the Next Linear Collider (NLC) presented in this document
is intimately connected with experiences gained at the SLC. Our choices
of technologies and philosophies of design have direct links to these
experiences and considerable overlap with them. Lessons have been
learned and techniques developed at the SLC that are relevant to the
design and implementation of every part and system of the NLC:

\begin{itemize}
\item Injectors\\
      $-$  Stabilized high-power electron sources\\
      $-$  Polarized electrons\\
      $-$  High-power targets and positron capture
\item Damping Rings\\
      $-$  Stabilized fast (50 ns) injection and extraction systems\\
      $-$  Sub-picosecond phase synchronization with linac rf systems
\item Linear Accelerator\\
      $-$  Beam Acceleration\\
      \hspace*{2em} Management of large rf systems\\
      \hspace*{2em} Rf phase control\\
      \hspace*{2em} ``Time-slot'' compensation\\
      \hspace*{2em} Short-range longitudinal wake compensation\\
      \hspace*{2em} Multibunch beam loading compensation\\
      $-$  Emittance Preservation\\
      \hspace*{2em} Beam-based alignment\\
      \hspace*{2em} LEM---lattice/energy matching\\
      \hspace*{2em} BNS damping \\
      \hspace*{2em} Coherent wakefield cancellation\\
      \hspace*{2em} Dispersion-free steering
\item Final Focus Systems\\
      $-$  Second-order chromatic optics and tuning\\
      $-$  Precision diagnostics\\
      $-$  Beam-beam control and tuning   
\item Experimentation\\
      $-$  Theory and modeling of backgrounds\\
      $-$  Vulnerability of detector technologies\\
      $-$  Collimation---theory and implementation
\item Systems Performance and Operation\\
      $-$  Precision instrumentation---BPMs and wirescanners\\
      $-$  Feedback theory and implementation\\
      $-$  Importance of on-line modeling and analysis\\
      $-$  Automated diagnostics and tuning \\
      $-$  Mechanical stabilization of supports and components\\
      $-$  Thermal stabilization of supports and components\\
      $-$  Reliability\\
      $-$  History monitoring (from seconds to years)
\end{itemize}

\subsection{Future Linear Colliders}
\label{sub:fut} 

\begin{figure}[htb]
\leavevmode
\centerline{\epsfbox{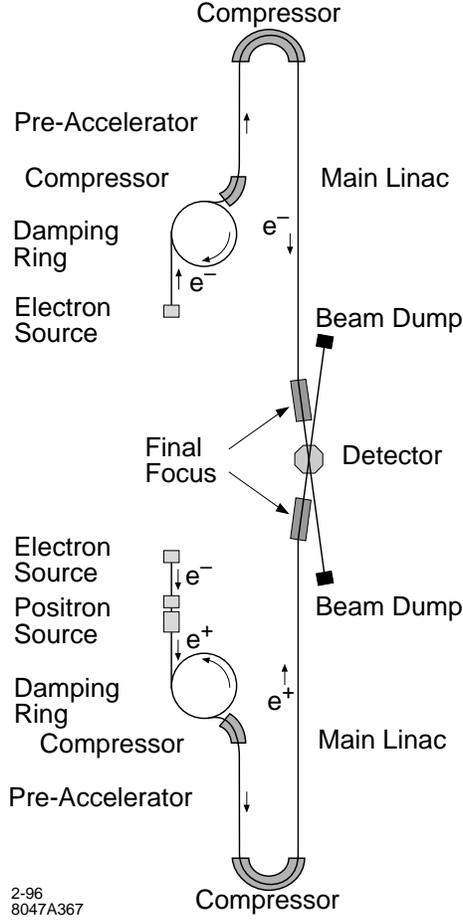}}
\caption[*]{Schematic of a TeV-scale linear collider.}
\label{fig:layout}
\end{figure}

The basic components of any linear collider are those already
incorporated into the SLC; a generic collider complex is
diagrammed in Fig.~\ref{fig:layout}. 
 The energy of such a future collider must be
ten times that of the SLC, and a TeV-scale collider must be
able to deliver luminosities that are several orders of magnitude
greater than those achieved at the SLC. Trains of bunches of electrons
and positrons are created, condensed in damping rings, accelerated to
high energy, focused to small spots, and collided to produce a
luminosity given by
\begin{equation}
   L = \frac{n N^2 H f}{ 4\pi \sigma_x^{\ast}\sigma_y^{\ast}}
   \quad ,
\label{eq:bright}
\end{equation}
where
\begin{center}
\begin{tabular}{rl}
       $n$\  =  &  number of bunches per train,  \\
       $N$\  =  &   number of particles per bunch,  \\
       $H$\  =  &   beam pinch enhancement,  \\
       $f$\  =  &  machine repetition rate, \\
\end{tabular}
\end{center}
and $\sigma_x^{\ast}$ and $\sigma_y^{\ast}$ are the horizontal and
vertical beam dimensions at the collision point.
Equation~\ref{eq:bright} can be written as
\begin{equation}
    L = \frac{1}{4 \pi E}\ 
        \frac{NH}{\sigma_x^{\ast}}\ 
        \frac{P}{\sigma_y^{\ast}}
   \quad ,
\label{eq:zzz}
\end{equation}
where $P$ is the average power in each beam. The factor
$N/\sigma_x^{\ast}$  determines the number of beamstrahlung photons
emitted during the beam-beam interaction, and since these photons will
alter the effective spread in beam collision energies and can create
backgrounds in experimental detectors, this factor is highly
constrained. It is mainly the last ratio, $P/\sigma^*_y$, that can be
addressed by accelerator technology; high luminosity corresponds to
high beam power and/or small beam spots. These two parameters pose
different, and in many cases contrary, challenges to the accelerator
physicist, and several technologies that represent differing degrees of
compromise between beam power and spot size are being developed.
Table~\ref{tab:param} summarizes the initial stage of the mainstream
design choices.

\begin{table}[ht]
\centering
\caption[*]{Linear collider design parameters (E$_{\rm cms}$ = 500 GeV).}
\label{tab:param}
\bigskip
\begin{tabular}{|l|ccc|ccc|} \hline \hline
   &  Frequency   & Gradient &  Total Length &  Beam Power &
    $\sigma_y$  &   Luminosity  \\
   &  (GHz)  &   (MV/m)  & (km)  &  (MW) &  (nm) &
           ($10^{33}\,$cm$^{-2}\,$s$^{-1}$)   \\
\hline
SuperC  &   1.3  &  25  &   30  &   8.2  &   19   &   6   \\
S-Band  &   3.0  &  21  &   30  &   7.3  &   15   &   5   \\
X-Band  &   11.4 &  50  &   16  &   4.8  &   5.5  &   6   \\
2-Beam  &   30.0 &  80  &    9  &   2.7  &   7.5  &   5   \\
\hline\hline
\end{tabular}
\end{table}

Each of the technologies in Table~\ref{tab:param} is being
pursued by physicists and engineers at laboratories around the globe.
This strong international effort is remarkably well coordinated
through collaborations that together provide a
set of test facilities to address each of the important aspects of the
collider design and implementation.
A summary of the facilities presently in operation or under
construction is given in Table~\ref{tab:test}.

\begin{table}[ht]
\centering
\caption[*]{Linear collider test facilities around the world.}
\label{tab:test}
\bigskip
\begin{tabular}{|lllc|} \hline \hline
\multicolumn{1}{|c}{Facility}  &
\multicolumn{1}{c}{Location}  &
\multicolumn{1}{c}{Goal}  &
 Operations    \\
\hline
 SLC   &   SLAC  &   Prototype Collider  &   1988   \\
 ATF   &   KEK   &   Injector and Damping Ring   &   1995    \\
 TTF   &   DESY  &   SuperC Linac        &   1997         \\
 SBTF  &   DESY  &   S-band Linac        &   1996         \\
 NLCTA &   SLAC  &   X-band Linac        &   1996         \\
 CTF   &   CERN  &   2-Beam Linac        &   1996         \\
 FFTB  &   SLAC  &   Final Focus/IR      &   1994         \\
\hline \hline
\end{tabular}
\end{table}
\clearpage

\section{The Next Linear Collider }
\label{sec:next} 

\subsection{Technology Choice and Design Philosophy}
\label{sub:tech} 

\begin{figure}[htb]
\leavevmode
\centerline{\epsfbox{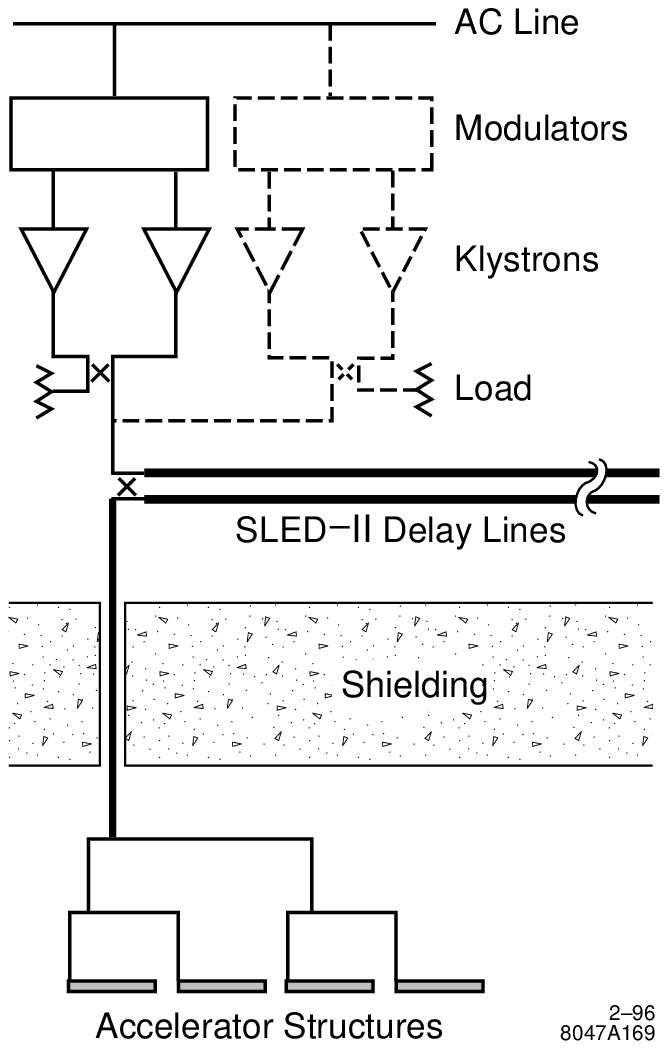}}
\centerline{\parbox{5in}{\caption[*]{Normal-conducting rf
system module in NLC main linacs. The dashed elements 
are expected to be necessary to
reach 1 TeV cms energy.}
\label{fig:norm}}}
\end{figure}

The goal to reach 1 to 1.5-TeV cms energy with luminosities of
$10^{34}\mbox{\,cm}^{-2}\,$s$^{-1}$ or more and our experiences with
the SLC, guide our choice of technologies for the NLC. We believe that
the most natural match to these design goals is made with
normal-conducting X-band (11.424 GHz) microwave components patterned
after the S-band technology used in the SLC. A schematic of a section
of the rf system of the NLC is shown in Fig.~\ref{fig:norm}. Our
choice of technology has required the development of new advanced rf klystrons
and pulse-compression systems, but provides confidence that
accelerating gradients of 50--100 MV/m  can be achieved and used in the
implementation of the collider. The technical risk of building a
collider with new X-band technologies is perhaps greater than simply
building a larger SLC at S-Band, but the goal to reach 1--1.5 TeV is
substantially more assured, and capital costs to reach these energies
will be lower.

The NLC is designed with nominal cms energy of 1 TeV. It is envisaged
to be built with an initial rf system able to drive the beams to
0.5-TeV cms energy, but with all infrastructure and beam lines able to
support 1 TeV. The rf system design incorporates the ability to
replace and add modulators and klystrons without access to the
accelerator beam line (dashed lines in Fig.~\ref{fig:norm}), so an
unobtrusive, smooth, and adiabatic transition from 0.5 TeV to 1 TeV
cms energy can be made with modest and expected improvements in X-band
technology. This allows the collider to begin operation with the
greatest of margins in cost and performance, and provides an excellent
match to the anticipated physics goals at the energy frontier
(Fig.~\ref{fig:goals}). Our philosophy is akin to that taken previously
in the construction of the SLAC linac which provided a 17-GeV electron
beam at its inauguration, was improved to 35 GeV, and with continued
advances in S-band technology, now provides 50-GeV electrons and
positrons for the SLC.

The NLC design also incorporates multiple paths to further upgrade the
cms energy to 1.5 TeV. The ``trombone'' shape of the collider layout
would easily accommodate a straightforward albeit expensive increase in
the length of the main accelerators without requiring extensive
modification of the remainder of the complex. This final energy might
also be accomplished by development of new, more efficient, X-band
technologies; for example, gridded klystrons, cluster klystrons, or
relativistic two-beam klystrons.

The highest-level parameters of the NLC are listed in
Table~\ref{tab:pres}. At each of the nominal 0.5- and 1-TeV cms
energies, three sets of parameters define the operating plane of the
collider. The expected luminosity is constant over the operating plane,
but is achieved with differing combinations of beam current and spot
size. This  provides a {\it region} in parameter space where the collider can
be operated. Construction and operational tolerances for the various
subsystems of the collider are set by the most difficult 
portion of the operating
region. For example, the more difficult
parameters for the final focus are those of case (a) in Table 
\ref{tab:pres}, for which the
beam height is smallest. In contrast, preserving the emittance of the
beam in the linac is more difficult in case (c), in which the beam
charge is highest and the bunch length longest. This design philosophy
builds significant margin into the underlying parameters of the
collider.

\begin{table}[htp]
\centering
{\footnotesize\caption[*]{High-level parameters and operating region in
parameter space of the NLC.}
\label{tab:pres}
\bigskip
\begin{tabular}{|l||ccc|ccc|} \hline\hline
   &  NLC-Ia & NLC-Ib & NLC-Ic & NLC-IIa & NLC-IIb & NLC-IIc  \\
\hline
Nominal CMS Energy (TeV)  & \multicolumn{3}{c|}{0.5} & \multicolumn{3}{c|}{1.0} \\
Repition Rate (Hz)      & \multicolumn{3}{c|}{180}  & \multicolumn{3}{c|}{120} \\
Bunches Pulse     & \multicolumn{3}{c|}{90}  & \multicolumn{3}{c|}{90}  \\
Bunch Separation (ns)      & \multicolumn{3}{c|}{1.4} & \multicolumn{3}{c|}{1.4}  \\
Bunch Charge ($10^{10}$)  & 0.65 & 0.75 & 0.85 & 0.95 & 1.10 & 1.25 \\
Beam Power (MW)        & 4.2 & 4.8 & 5.5  & 6.8 & 7.9 & 9.0 \\
$\sigma_x$ at IP (nm) & 264  & 294   & 294  & 231  & 250  & 284  \\
$\sigma_y$ at IP (nm) & 5.1  & 6.3   & 7.8  & 4.4  & 5.1  & 6.5  \\
$\sigma_z$ at IP ($\mu$m)        & 100      & 125       & 150      & 125      & 150      & 150  \\
Pinch Enhancement H              & 1.4      & 1.4       & 1.5      & 1.4      & 1.4      & 1.5  \\
Beamstrahlung $\delta_E$ (\%)  & 3.5      & 3.2       & 3.5      & 12.6     & 12.6     & 12.1 \\
No. Photons per $e^-/e^+$       & 0.97     & 1.02      & 1.16     & 1.65     & 1.77     & 1.74 \\
Max.\ Beam Energy (GeV) & 267 & 250 & 232  & 529 & 500 & 468 \\
Luminosity ($10^{33}$)    & 5.8 & 5.5 & 6.0 & 10.2 & 11.0 & 10.6  \\
No. Klystrons      & \multicolumn{3}{c|}{4528} & \multicolumn{3}{c|}{9816} \\
Klystron Peak Power (MW) & \multicolumn{3}{c|}{50} & \multicolumn{3}{c|}{75}  \\
Pulse Compression Gain        & \multicolumn{3}{c|}{3.6} & \multicolumn{3}{c|}{3.6}  \\
Unloaded Gradient (MV/m)  & \multicolumn{3}{c|}{50} & \multicolumn{3}{c|}{85} \\
Total Linac Length (km)   & \multicolumn{3}{c|}{17.6} & \multicolumn{3}{c|}{19.1} \\
Beam Delivery Length (km)   & \multicolumn{3}{c|}{10.4} & \multicolumn{3}{c|}{10.4} \\
Total Site Length (km) & \multicolumn{3}{c|}{30.5} & \multicolumn{3}{c|}{30.5} \\
Total Linac AC Power (MW)      & \multicolumn{3}{c|}{120} & \multicolumn{3}{c|}{193}  \\
\hline \hline
\end{tabular}}
\end{table}

An important element in the design strategy of the NLC is the use of
the beam to measure and correct or compensate for errors in electrical
and mechanical parameters of the accelerator. These techniques, many in
extensive use at the SLC and FFTB, are able to achieve far greater
accuracy than is possible during fabrication and installation of
components. For example, the use of optical matching and beam-based
alignment algorithms considerably loosen tolerances required on magnet
strengths and positioning. These procedures require accurate
measurement of the properties of the beam and extensive online modeling
and control software. The existence of instrumentation suitable for
these purposes is an important aspect of the readiness of technologies
for the collider.

Additional performance overhead has been included in the designs of
most subsystems of the NLC{}. Errors that we anticipate will occur
during machine tuning operations have been taken into account. For
example, the injector systems are designed to provide 20\%\ more charge
than is indicated in Table~\ref{tab:pres}. Fabrication and alignment
tolerances for main linac structures are specified without assuming
benefit from certain global tuning methods such as coherent wakefield
cancellation. These are powerful techniques in routine practice at the
SLC, but our philosophy is to use them only to provide operational
margin. We also recognize that the beam-based tuning described above
cannot be done with perfect accuracy. For example, we have analyzed the
tuning procedure for the final focus and estimated a 30\%\ increase in
the spot size at the IP due to errors that we anticipate will occur in
measuring and correcting aberrations inherent in the optics. (This is
included in Table~\ref{tab:pres}.) This layered approach to
specification of collider performance is an important part of our
design philosophy.

\subsection{Status Report on Technologies for the NLC}
\label{sub:status} 

\begin{figure}[htb]
\leavevmode
\centerline{\epsfbox{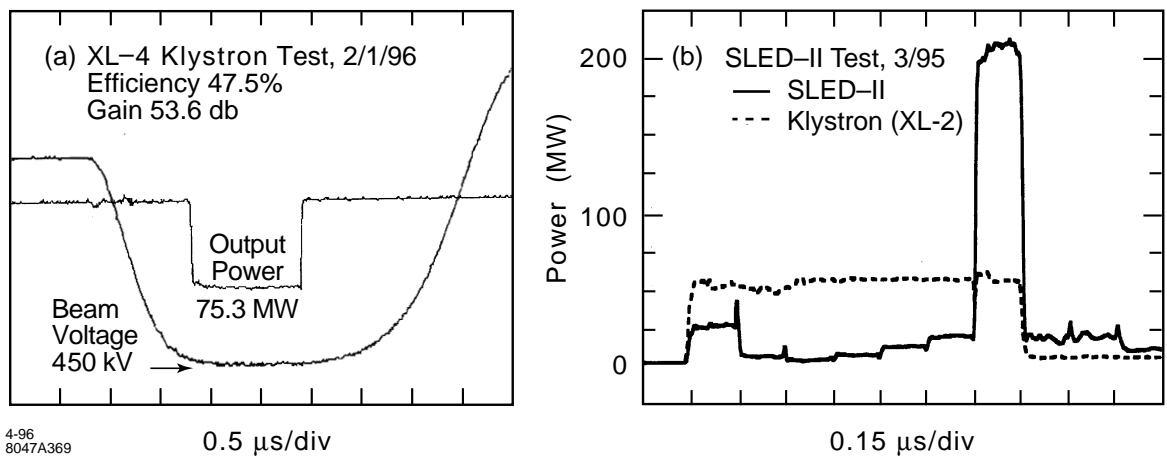}}
\centerline{\parbox{5in}
{\caption[*]{Results of tests of X-band rf components: (a) high-power
klystrons, and (b) pulse compression systems.}
\label{fig:result}}}
\end{figure}

Progress in development of X-band rf components has been impressive in
recent years. Prototype klystrons now produce 50-MW pulses, over 1.5
microseconds long, with performance characteristics that are correctly
modeled by computer codes.  The most recent prototype produces 75-MW
pulses, one microsecond long.  This exceeds the requirements of the
initial 0.5-TeV stage of the NLC, and indeed approaches the
requirements for 1-TeV cms energy. Tests of rf pulse-compression
transformers have exceeded most goals of the NLC, and high-power rf
windows and mode converters that allow high-efficiency transfer of
power between components have been successfully tested. Examples of
some of these results are shown in Fig.~\ref{fig:result}.

\begin{figure}[htb]
\leavevmode
\centerline{\epsfbox{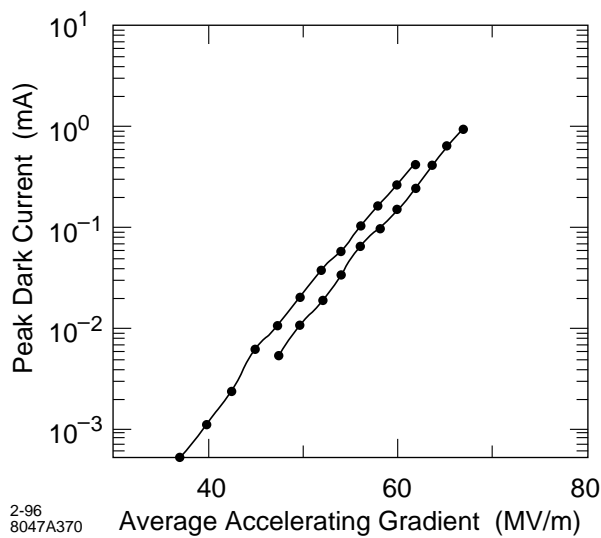}}
\centerline{\parbox{5in}
{\caption[*]{Processing of X-band accelerator structures to high 
gradient.}
\label{fig:proc}}}
\end{figure}

The voltage gradient that can be used in a particle accelerator can be
limited by the dark current created when electrons are drawn from the
surfaces of the accelerator structures and captured on the accelerating
rf wave. For a given rf frequency, there is a well-defined gradient
beyond which some electrons emitted at rest will be captured and
accelerated to relativistic velocities. This threshold gradient is
about 16 MV/m at S-band, and scales to 64 MV/m at X-band. These are not
actual limits to gradients that can be utilized in an accelerator since
much of the charge is swept aside by the focusing quadrupoles of the
machine lattice, but the dark current will grow rapidly above these
values, and may adversely affect the primary beam or interfere with
instrumentation needed for tuning. Gradients somewhat above the capture
threshold are likely to be useful in practice, but the operational limits
are not well known since no large-scale high-performance facility has
been operated significantly above capture threshold. Expected thresholds
of dark currents in S-band and X-band structures have been confirmed,
and it has been proven that (unloaded) gradients as large as 70 MV/m
can be used at X-band (Fig.~\ref{fig:proc}).

The electro-mechanical design of the structures of the main accelerator
must not only produce the desired gradient, but must also minimize
wakefields excited by the passage of the beam. The retarded
electromagnetic fields left by each particle can disrupt the
trajectories of particles that follow it through the accelerator. Many
techniques to control the effects of the short-distance, intrabunch
wakefields have been developed, tested, and put into use at the SLC. It
will be necessary to also control long-range wakefields at the NLC in
order to allow trains of closely spaced bunches to be accelerated on
each rf pulse.
  
\begin{figure}[htb]
\leavevmode
\centerline{\epsfbox{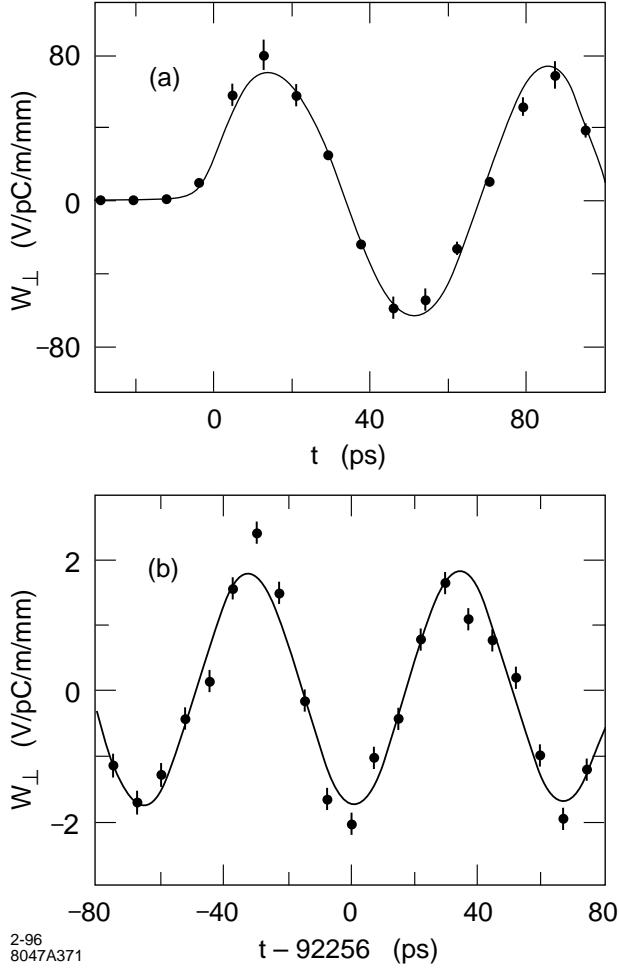}}
\centerline{\parbox{5in}
{\caption[*]{Measured and predicted transverse dipole wakefields 
in a 1.8-m-long X-band accelerator structure.}
\label{fig:meas}}}
\end{figure}

Structures in which wakefields are suppressed by careful tuning
of their response to the passage of the beam have been
developed, and tests have been performed at a facility (ASSET)
installed in the SLAC linac (Fig.~\ref{fig:meas}). Agreement with
theoretical expectations is excellent and lends confidence to the
design and manufacture of these structures. A more advanced design that
further mitigates the long-range wakefields by coupling deflecting rf
modes to external energy-absorbing materials has been completed, and a
prototype of this new structure is being readied for testing in ASSET
as well.

Work remains to be done on X-band rf technologies, but with prototype
components now in hand, tests of completely integrated systems have
begun. A fully engineered test accelerator is under construction at
SLAC that will allow optimization of rf systems and provide experience
with beam operations at X-band frequencies. This test accelerator will
be a 40-m long beam line containing six 1.8-m-long X-band structures
powered by 50--75 MW klystrons to an accelerating gradient of 50--85
MV/m. Commissioning of this facility has begun, and operations are
expected to be underway by the end of this year (Table~\ref{tab:test}).

The spot sizes that must be produced at the interaction point of the
NLC represent significant extrapolations from those achieved at the
SLC. It is important to demonstrate that it is possible to demagnify a
beam by the large factor needed in the NLC. An experiment has been
performed by the Final Focus Test Beam (FFTB) Collaboration to show that
such large demagnifications can be achieved. The FFTB is a prototype
beam line installed in a channel located at the end of the SLAC linac
at zero degrees extraction angle. The FFTB lattice is designed to
produce a focal point at which the beam height can be demagnified by a
factor of 380 to reduce the SLC beam ($\gamma \varepsilon_y = 2 \times
10^{-6}$ m-rad) to a size smaller than 100 nm. The demagnification
factor of the FFTB beam line is well in excess of that needed for the
NLC.

\begin{figure}[htb]
\leavevmode
\centerline{\epsfxsize=3.5in\epsfbox{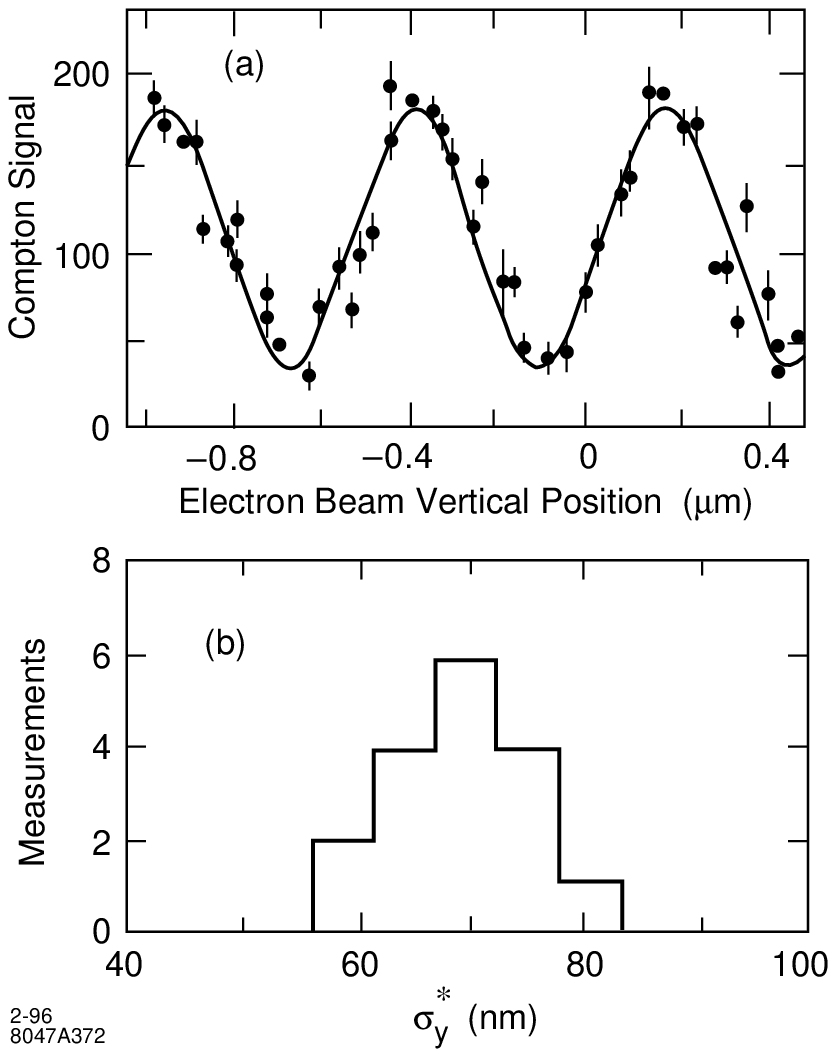}}
\centerline{\parbox{5in}
{\caption[*]{Measurement of 70-nm beam spots with a laser-Compton
beam size monitor in the FFTB. (a) The rate of Compton scatters from a
laser interference pattern used to determine the beam size, in this
case 73 nm. (b) Repeatability of spot measurement over periods of
several hours.}
\label{fig:spots}}}
\end{figure}

The FFTB optics are chromatically corrected to third order in the beam
energy spread. (The SLC is corrected to second order.) All magnetic
elements are mounted on precision stages that can be remotely
positioned with step size of about 0.3 micron, and beam-based
alignment procedures were developed that successfully place these
elements to within 5--15 microns of an ideal smooth trajectory. New
state-of-the-art instruments were developed and used to measure the
FFTB beam positions and spot sizes. Following a brief shake-down run in
August of 1993, data were taken with the FFTB during a three-week
period in April and May of 1994. Beam demagnifications of 320 and spot
sizes of 70 nm were controllably produced during this period.
Measurements of these beams are shown in Fig.~\ref{fig:spots}. The
design of the NLC final focus follows that of the FFTB, and the
experiences gained from the FFTB are incorporated into the tuning
strategies for the NLC.

Important advances have also been made in instrumentation required to
measure and control properties of the beams. The SLC control system has
evolved dramatically over the past years to include extensive online
modeling and automation of data analysis and tuning procedures.
Scheduled procedures use sets of wire scanners to make complete
measurements of the beam phase space, and provide recorded histories of
machine performance. Online data-analysis packages are able to
reconstruct fully coupled non-linear optical systems. Beam-based
feedback and feedforward loops are in routine operation in the SLC with over
100 loops providing control of beam trajectories and energies. Beam
position monitors have been developed for the FFTB that achieve
pulse-to-pulse resolutions of 1 micron, and new position monitors have
recently been installed that are able to measure beam motions of 100
nm. The FFTB focal-point spot monitors have demonstrated techniques to
measure beam sizes of {30--40 nm}, and extrapolation of these
techniques to sizes as small as 10 nm is expected to be successful.

\section{Outlook for the Next Linear Collider}

As the SLC has systematically increased its luminosity, the accelerator
physics and technologies of linear colliders have matured.  Experiences
and lessons learned from the task of making this first collider perform
as an instrument for particle physics studies make a firm
foundation on which to base the design and technology choices for the
next linear collider.  At the same time, essential demonstrations of
new collider technologies have either taken place or soon will be
underway. The experimental program with the FFTB is providing the
experience needed to evaluate limitations to designs of final focus and
interaction regions.  The ability to demagnify beams by the amount
required for the NLC has already been achieved.  Microwave power
sources have exceeded requirements for the initial stage of the NLC,
and critical tests assure us that this technology can be expected to
drive beams to center-of-mass energies of a TeV or more.  Fully
integrated test accelerators are presently under construction at CERN,
DESY, KEK, and SLAC that will soon provide answers to questions of
technical optimization and costs of the major components of a TeV-scale
collider.

Given the great international interest and commitment to the goals of a
TeV-scale high-performance $e^+e^-$ collider, it is certain that the
final design, construction, and utilization of such a collider will be
a global effort. It is important that the scientific community put into
place foundations for such a collaboration. The international character
of the linear collider project is already reflected in the
collaborations at work on the accelerator physics and technology of
linear colliders, and in the process of international discussion and
review of progress in the field \cite{ref:neww}. It is essential that
we continue to build on this base of understanding and cooperation, and
make certain that all involved in this enterprise are full parties in its
final realization.

\newpage

 $$ $$ 
 
 \newpage
  
\chapter{Physics Goals of the Next Linear Collider}
\section{Introduction}
 
During the past several decades, significant advances have been made in
elementary  particle physics. We now have a renormalizable quantum
field theory of strong and electroweak interactions, based on the
principle of local $SU(3)_c\times SU(2)_L\times U(1)_Y$ gauge
invariance.  That theory properly describes the interactions of all
known particles, incorporating the proven symmetries and successes of
quantum electrodynamics, the quark model, and low energy V--A theory. It
correctly predicted weak neutral currents, the now observed gluons and
weak gauge bosons, and the special properties of the heavy fermions
$\tau$, $b$, and $t$. Since it is a renormalizable theory, its
predictions can be tested at the quantum loop level by high precision
experiments. It has already confronted a wealth of data at the level of
1\% or better without any significant evidence of inconsistency.
Because of those impressive successes, the $SU(3)_c\times SU(2)_L\times
U(1)_Y$ theory has been given the title ``The Standard Model'', a
designation which establishes it as the paradigm against which future
experimental findings and new theoretical ideas must be compared.

The Standard Model cannot be  the final theory of Nature, but it does
represent the completion of a major stage toward the uncovering of that
theory. To make further progress, we must examine both the strengths
and failings of this model and direct experimental effort toward the
weakest points in its structure.

The Standard Model is based on the interactions of fermions and vector
gauge bosons. The fermions are grouped into three generations of
leptons and quarks which span an enormous mass range.  Their newest
member, the top quark, is exceptionally heavy.   Why is the top so
massive, or why are the other fermions so light?  This question
highlights the broader problem of why Nature chose to repeat the
fermion generations three times and endow quarks and leptons with their
observed pattern of masses and mixing. It is likely that future intense
scrutiny of the top quark's properties will provide new insights
\nopagebreak[1]
regarding this important problem.
 
The vector bosons of the Standard Model are grouped into eight massless
gluons of $SU(3)_c$ which mediate the strong interactions, plus the
$W^\pm$, $Z$, and $\gamma$ which  are  responsible for electroweak
interactions. The SU(3)$_c$ gauge theory, called quantum chromodynamics
(QCD), taken on its own, is an ideal theory. It has no arbitrary or
free parameters but can, in principle, explain all hadronic dynamics
including confinement, asymptotic freedom, proton structure, and baryon
and meson spectroscopy.  Confirming those properties and uncovering
additional more subtle features of QCD remains an important
experimental and theoretical challenge.
 
In contrast with QCD, the electroweak sector has many arbitrary
parameters. Most stem from the Higgs mechanism which is used to break
the SU(2)$_L\times{}$U(1)$_Y$ symmetry and endow particles with mass.
In the simplest realization of this symmetry breaking, one introduces a
scalar doublet $\phi$, the Higgs field, which obtains a vacuum
expectation value $v$. This assumption introduces into the theory an
electroweak mass scale $v\simeq250$ GeV. The masses of the $W$ and $Z$
bosons and the various quarks and leptons are proportional to $v$.
Their disparity reflects extreme differences in their couplings to the
scalar field $\phi$. It is true that this simple model with one Higgs
field can parametrize all electroweak masses, quark mixing, and even CP
violation.  But it does not provide insight into any of these
phenomena, or even into the basic fact that the electroweak gauge
symmetry is spontaneously broken. The many unanswered questions
associated with the Higgs field, or whatever more complicated structure
leads to the breaking of the electroweak symmetry, call for experiments
which thoroughly explore this sector.

An important testable prediction of the simple Higgs model is the
existence of a neutral spin-0 remnant particle  $H$, called the
Higgs scalar. Its mass depends on the self-coupling $\lambda$ of the
Higgs field through the relation 
\begin{equation} 
m_H=\sqrt{2\lambda} v \ ,
\label{higgsmass} 
\end{equation}
but it is unspecified as long as $\lambda$ is unknown. There is
an  experimental lower bound on $m_H$ of 65 GeV from direct searches at
LEP.  That search reach is expected to be extended up to about 90 GeV
at LEP II.  There is also an approximate upper bound on the Higgs mass
$m_H\lsim 800$  GeV from theoretical bounds on $\lambda$. For example,
perturbative partial wave unitarity in high energy scattering of
longitudinal $W$ bosons, $W_LW_L\rightarrow W_LW_L$ requires
$|\lambda|\lsim8\pi/5$. This gives a large window in which to search. 
However, there is a much stronger upper bound which comes from the
stronger assumption that the Higgs boson is a fundamental particle with
no nonperturbative  interactions up to the grand unification scale. 
This requires $m_H \lsim 200$ GeV; we will refer to a Higgs boson
satisfying this hypothesis as a `light Higgs'.

Even more daunting than the problem of finding the Higgs boson $H$ in
the context of the simple Higgs theory is the prospect that this
theory is inadequate to correctly describe the weak interaction scale. 
This simplest theory  has theoretical problems of self-consistency,
particularly when it is extrapolated to a unified theory at high
energies.  Also, the fact that its pattern of couplings must be
input without any explanation is a sign that this theory is only a
parametrization of electroweak symmetry breaking rather than being a
fundamental explanation of this phenomenon.  This state of affairs has
led to many speculations on the true symmetry breaking mechanism and,
from there, to interesting new physics possibilities beyond the
Standard Model with observable manifestations at high energy.
 
In order to build a theory in which electroweak symmetry is naturally
broken by the expectation value of a fundamental Higgs field, it is
necessary to incorporate supersymmetry (SUSY) at the weak interaction
scale.  That elegant boson-fermion symmetry allows a simple connection
to gauge or string theory unification and provides a logic for the
symmetry-breaking form of the Higgs potential.  Achieving these goals,
however, requires introducing novel partners for all Standard Model
particles. It also requires at least two Higgs doublets and thus
predicts five remnant scalars, $h$, $H$, $A$, $H^\pm$. The $h$  should
have a mass below about $150$ GeV and should be  most similar to the
Standard Model Higgs boson. Finding that particle and determining its
properties may be our first window to supersymmetry. If supersymmetry
does indeed appear below 1 TeV, there will be a wealth of supersymmetry
partner  spectroscopy waiting to be explored. Currently, supersymmetry
has no direct experimental support. However, there are two very
suggestive pieces of evidence that are in favor of this theory.  The
first is the values of the $SU(3)\times SU(2)\times U(1)$ coupling
constants. These coupling constants are in just the relation predicted
by a supersymmetric grand unified theory.  The second is the tendency
of the precision electroweak data to favor a light Higgs boson, which
is an indication that the mechanism of electroweak symmetry breaking
involves weakly-coupled fields.
 
Alternatively, one might imagine that there is no fundamental Higgs
field, and that the electroweak symmetry is broken dynamically by
fermion-antifermion condensation due to new strong forces at high
energy. Scenarios ranging from $t\bar t$ condensation to complex
extended technicolor models have been proposed. Their basic premise is
very appealing, but no  compelling model exists. Nevertheless, the
generic idea of new underlying strong dynamics gives rise to testable
consequences for anomalous top and gauge boson couplings and high
energy scattering behavior.

This issue of whether the mechanism of electroweak symmetry breaking is
weak-coupling or strong-coupling is the most important question in
elementary particle physics today.  The NLC should resolve it
definitively.  For the case in which this physics is weak-coupling, the
NLC should have a rich experimental program involving the detailed
study of Higgs bosons and supersymmetric particles.  The precise
spectrum and branching ratio determinations for these particles should
give information which, like the values of the strong and electroweak
coupling constants, can be extrapolated to the unification scale.  This
scenario offers the tantalizing possibility that experimental data
collected at the NLC would be directly relevant to supergravity and
superstring theories at very high energy.  On the other hand, if the
mechanism of electroweak symmetry breaking is strong-coupling, this
could imply a new spectroscopy at the TeV energy scale which the NLC
might access directly.
 
In addition to these two options which relate directly to the physics
of electroweak symmetry breaking, there are many other possibilities
for new physics at the TeV energy scale.  These include larger gauge
groups with additional $W^\prime$ and $Z^\prime$ gauge bosons, heavy
new fermions, and additional scalars. Many of these possibilities are
realized in specific models of electroweak symmetry breaking, so a
broad-based search for new phenomena is an essential part of the
experimental program devoted to this question. The most direct way to
uncover such new particles and their associated phenomena is to search
at very high energies above particle production threshold. Important
indirect evidence can also be inferred from precision studies of
Standard Model parameters such as $m_W$, $\sin^2\theta_W$, and the
couplings of heavy quarks and $W$ bosons to the $\gamma$ and  $Z^0$.
 
For the exploration of all of these possibilities, which defines the
next step in experimental high-energy physics, the Next Linear Collider
(NLC) will play an essential role.  We envisage this machine as an
$e^+e^-$ collider which operates initially at a center of mass energy
of about 500 GeV and can be upgraded to 1.5 TeV, providing a luminosity
corresponding to $10^4$ events per year for a process with  the point
cross section for QED pair production. This machine will employ
polarized electrons  and offers the possibility of $e^-e^-$, $e\gamma$,
and $\gamma\gamma$ collider options.  With such a facility, it is
possible to carry out crucial and unique  experiments across the whole
range of possibilities we have just described for the physics of the
weak interaction scale. In this report, we will summarize the
capabilities of the NLC to explore the physics of the weak interaction
scale across this broad front. A design for the NLC is presented in an
accompanying report \cite{ZDR}.

In Section 2 of this chapter, we will summarize the basic conclusions
of this report relevant to the physics studies, including the basic
accelerator parameters of energy and luminosity.  We will also describe
the basic assumptions on detector performance that we will use to
describe the physics capabilities of this machine. In Sections 11 and
12, after our discussion of the physics opportunities that the NLC will
provide, we will give a more detailed description of a detector design
and the constraints on the detector which come both from the physics
goals and from the accelerator.

One of the first physics goals of the NLC will be the detailed study of
the top quark at its threshold and just above.  We will explain in
Section 3  the special features of the $t\bar t$ threshold region which
make it a unique laboratory for the precision measurement of the top
mass and width, the QCD coupling of the top quark, and the possible
couplings to the Higgs boson and other new particles. We will also
describe how the NLC will make precision measurements of the couplings
of top to electroweak gauge bosons, couplings which might contain
signals of new strong interactions which connect top to the sector
responsible for electroweak symmetry breaking.

Whether the electroweak gauge symmetry is broken by fundamental Higgs
bosons or by new high-energy strong interactions, the NLC will bring
important contributions to the experimental study of this sector. First
of all, though the LHC and other facilities have the capability to find
a light Higgs boson in many decay channels, the NLC is the only planned
facility at which the existence of a light Higgs boson can be ruled out
in a model-independent way. If the light Higgs boson is indeed present,
we will show in Section 4 that the NLC will be able not only to
discover this particle but also to characterize many of its
interactions.  We  will show that the NLC has a unique capability to
determine the couplings of the Higgs boson to $Z$ and $W$, to heavy
quarks and leptons, and to photons.  These measurements dovetail nicely
with the expected measurement of the Higgs production cross section
from gluon fusion at the LHC to give the complete phenomenological
profile of this particle.

If the presence of a relatively light fundamental Higgs particle is
accompanied by the appearance of supersymmetry at the TeV scale, the
NLC can perform crucial experiments to characterize the new
supersymmetric particles.  We will show in Section 5 that the NLC can
detect the supersymmetric partners of $W$ and $Z$ over essentially the
complete range of parameters expected in the model. But, even more
importantly, the NLC can measure the masses and mixing angles of these
particles and, in so doing, determine the most important underlying
parameters of supersymmetry.  This determination of parameters will be
essential not only for the exploration of the physics of fermion
partners at $e^+e^-$ colliders, but also for the extraction of detailed
information about the underlying theory from the complementary
signatures of supersymmetry seen at hadron colliders.

If electroweak gauge symmetry is broken by new forces at high energy,
one can look for the signs of these forces in the couplings of $W$
bosons to the $\gamma$ and $Z$ and in the study of $WW$ scattering. We
will show in Section 6 that the NLC is an ideal machine for the study
of the gauge couplings of the $W$, capable of achieving parts per mil
precision on the $W$ form factors.  We will show in Section 7 that the
NLC at the high end of its energy range can achieve constraints on $WW$
scattering comparable to those of the LHC, in an environment with a
number of qualitative advantages. We will also show that the NLC also
offers  new windows into $WW$ interactions through the precision study
of $e^+e^-\rightarrow W^+W^-$ and through high-energy $t\bar t$
production.

Finally, these capabilities of the NLC to explore specific models of
electroweak symmetry breaking are  balanced by the ability of this
facility to perform broad searches for novel fermions, scalars, and
gauge bosons.  We will describe the abilities of the NLC to search for
exotic particles in $e^+e^-$ annihilation in Section 8.  In Section~9, we
will show how this broad capability is extended further by the
availability of $e^-e^-$, $e\gamma$, and $\gamma\gamma$ collisions.  In
Section 10, we will show that the NLC will also contribute to the
future program in the study of the strong interactions, in particular,
through the precision measurement of $\alpha_s$.

Section 13  will present our  conclusions.  We will review the unique
capabilities of the NLC and contrast its prospects with those of the
next generation of hadron colliders.

\newpage

\section{Standard Model Processes and Simulations}


We begin by describing the basic assumptions underlying our study of the 
physics capabilities of the NLC.  We will briefly discuss the expected 
energy and luminosity that the NLC will provide, the performance of the
detector that we expect to have available, our simulation methods, and 
the magnitudes of the most important standard model background  
processes.

\subsection{Accelerator and Detector}

The NLC is envisaged as the first full-scale $e^+e^-$ linear collider, a
machine designed from the beginning with the goal of high-luminosity
colliding beam physics and one which takes account of the lessons of
its prototype, the SLC.  The NLC will be designed for an initial energy
of 500 GeV in the center of mass, with  an upgrade path to 1.5 TeV. It
will provide a luminosity sufficient for a thorough experimental
program on $e^+e^-$ annihilation to standard and exotic particle pairs. It
will provide a highly polarized $e^-$ beam, and possibly also a
polarized positron beam. Our basic assumptions on luminosity as a
function of energy and on polarization are given in Table
\ref{Billstable}.  These assumptions are justified in the description
of the accelerator design given in Chapter 3 of this report and, at 
greater length, in \cite{ZDR}.

\begin{table}[ht]
\centering
\caption{Basic Parameters of the Next Linear Collider}
\bigskip 
\begin{tabular}{|lrcc|} \hline \hline
Energy (GeV)   & & Luminosity (cm$^{-2}$s$^{-1}$)& \\ 
&500 GeV           &  $5\times10^{33}$              & \\
&1000 GeV          &  $1\times10^{34}$              & \\
&1500 GeV          &  $1\times10^{34}$              & \\ 
Polarization:  &   &    &        \\
 &  & 80\% $e^-$,  0\% $e^+$          & Initial \\
 &   & 90\% $e^-$  65\% $e^+$         & Possible
                  \\ \hline \hline
\end{tabular}
\label{Billstable}
\end{table}

\begin{figure}[htbp]
\leavevmode
\centerline{\epsfxsize=3in\epsfbox{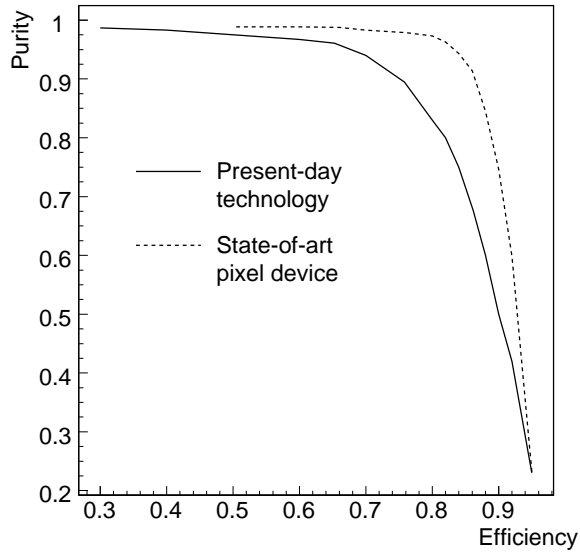} }
\centerline{\parbox{5in}{\caption[*]{Efficiency versus purity relation for
$b$-tagging with the NLC detector.}
\label{vtxcurve}}}
\end{figure}

The  NLC experiments can be carried out with a standard $4\pi$
multipurpose detector similar to those at LEP or SLC.  In our concept
of this detector, we include some innovations such as all-silicon
tracking to minimize the effect of machine-related backgrounds, but for
the most part the demands we have made on the detector are
straightforwardly met. The essential performance assumptions we have
made are shown in Table \ref{smrtable}.  Because of the small beam spot
sizes at a linear collider, which allows us to bring a CCD vertex
detector within 2 cm of the interaction point, the detector should have
excellent $b$-tagging capabilities.  The assumed curve of efficiency
versus purity for $b$-tagging is shown in Fig. \ref{vtxcurve}; the
performance required has already been demonstrated in the SLD vertex
detector.

\begin{table}[ht]
\centering
\caption{Summary of the detector parametrization used in the simulations.
Smeared quantities are denoted in the table by a subscript $s$.}
\label{smrtable}
\bigskip
   \begin{tabular}{|c|c|c|} \hline \hline
Particle & Energy & Momentum \\  \hline
Electrons & $\frac{\delta E}{E} = \frac{12\%}{\sqrt{E}} + 1.0\%$
 & $P_{s}^{2}=E_{s}^{2}-m_{e}^{2}$  \\
  & $E_{s}=E+\delta E$ & \\  \hline
Photons & $\frac{\delta E}{E} = \frac{12\%}{\sqrt{E}} + 1.0\%$ 
  & $P_{s}=E_{s}$ \\
  & $E_{s}=E+\delta E$ & \\ 
  & & \\ \hline
Neutral Hadrons & $\frac{\delta E}{E} = \frac{45\%}{\sqrt{E}} + 2.0\%$ 
 & $P_{s}^{2}=E_{s}^{2}-m_{\pi}^{2}$  \\
  & $E_{s}=E+\delta E$ & \\   \hline
Charged Hadrons &  $E_{s}^{2}=P_{s}^{2}+m_{\pi}^{2}$
 & $\frac{\delta P_{x,y}}{P_{x,y}^{2}}=0.0005 \oplus 
\frac{0.0015}{P_{x,y} \sqrt{P}(\sin\theta)^{2.5}}$ \\
 &  &  $\frac{\delta P_{z}}{P_{z}^{2}}= \frac{0.0015}{P_{x,y} \sqrt{P}
(\sin\theta)^{2.5}}$ \\
 & & $(P_{s})_{i}=P_{i}+\delta P_{i}$ \\
\hline \hline
\end{tabular}
\medskip
\end{table}

\subsection{Simulations}

In the studies presented here, the detector model has been used in
concert with a set of familiar and newly-written simulation programs.
In general, the background processes were generated by PYTHIA 5.7
\cite{pythia}, except for the background from $e^+e^-\rightarrow
W^+W^-$, which plays an especially important role at linear collider
energies. For this process we used a new generator \cite{barkmorioka}
which is  based on the formalism for this reaction presented by 
Hagiwara \etal~\cite{hpzh}. This generator computed the total amplitude
for $W^+W^-$ production and subsequent decay to four fermions,
retaining the full spin correlations through the process. It did make
the approximation of treating the $W$'s as on-shell particles, but it
properly treated the effects of initial state electron polarization,
beamstrahlung, collinear multi-photon initial state bremsstrahlung, and
a nonzero $W$ boson decay width. The same Monte Carlo program was used
in the studies of nonstandard $W$ physics reported in Sections 6 and 7.
Many of the other analyses used specialized generators at this level of
sophistication to simulate  the new physics processes.  These are
described in the various sections of this report. Except where it is
reported otherwise, the hadronization of partons and subsequent decays
were performed by JETSET 7.4 \cite{pythia}.

Four-vectors of stable particles emerging from the simulated reaction
were adjusted by a detector resolution smearing routine, which
implemented the parametrization summarized in Table~\ref{smrtable}. All
quantities were parametrized as a function of theta.  The smearing
assumed Gaussian errors and populated tails out to 3.5$\sigma$. The
parametrization assumed a dead cone about the beampipe of 150 mrad
($\cos\theta$ = 0.99). The neutral particle and charged particle
detection efficiencies were each taken to be 98\%. For neutral hadrons,
the momentum directions were Gaussian smeared in a cone about the
original direction with a half-width of 15 mrad to simulate finite
directional resolution.  For photons and electrons, the directions were
smeared by a cone of half-width 10 mrad.

\subsection{Standard Model Processes at the NLC}

Standard model processes, in addition to being interesting in their own
right, are the background to searches for new physics at the NLC. Many
of the standard model reactions at the NLC are familiar at lower
energies and need only be extrapolated to higher energies. However, new
processes, such as the pair production of gauge bosons, emerge as
dominant reactions.

\begin{figure}[htb]
\leavevmode
\centerline{\epsfig{file=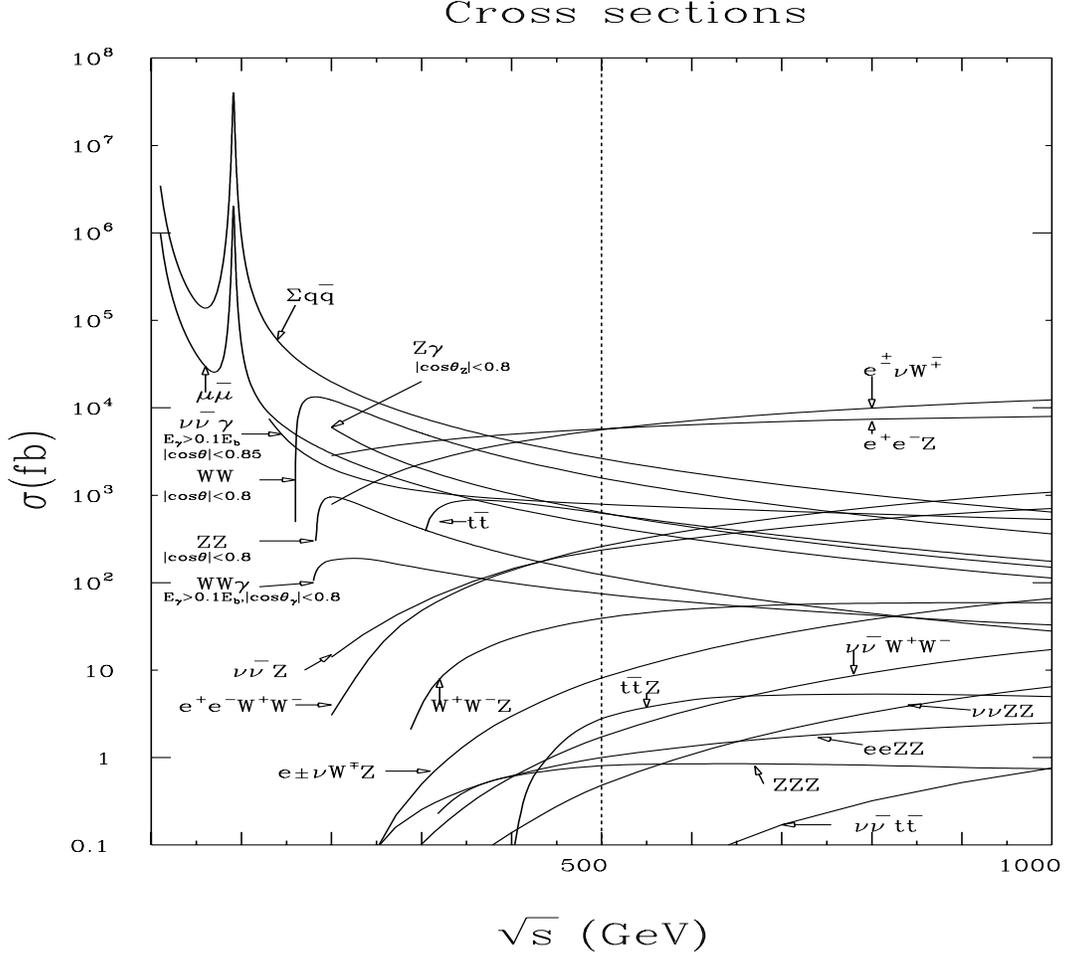,height=5.5in,width=5in,angle=90} }
\centerline{\parbox{5in}{\caption[*]{Cross sections for Standard Model
physics processes in $e^+e^-$ annihilation, as a function of center of
mass energy, from \cite{Miyahawaii}.}
\label{miyamoto}}}
\end{figure}
\begin{figure}[htb]
\leavevmode
\centerline{\epsfxsize=4in \epsfbox{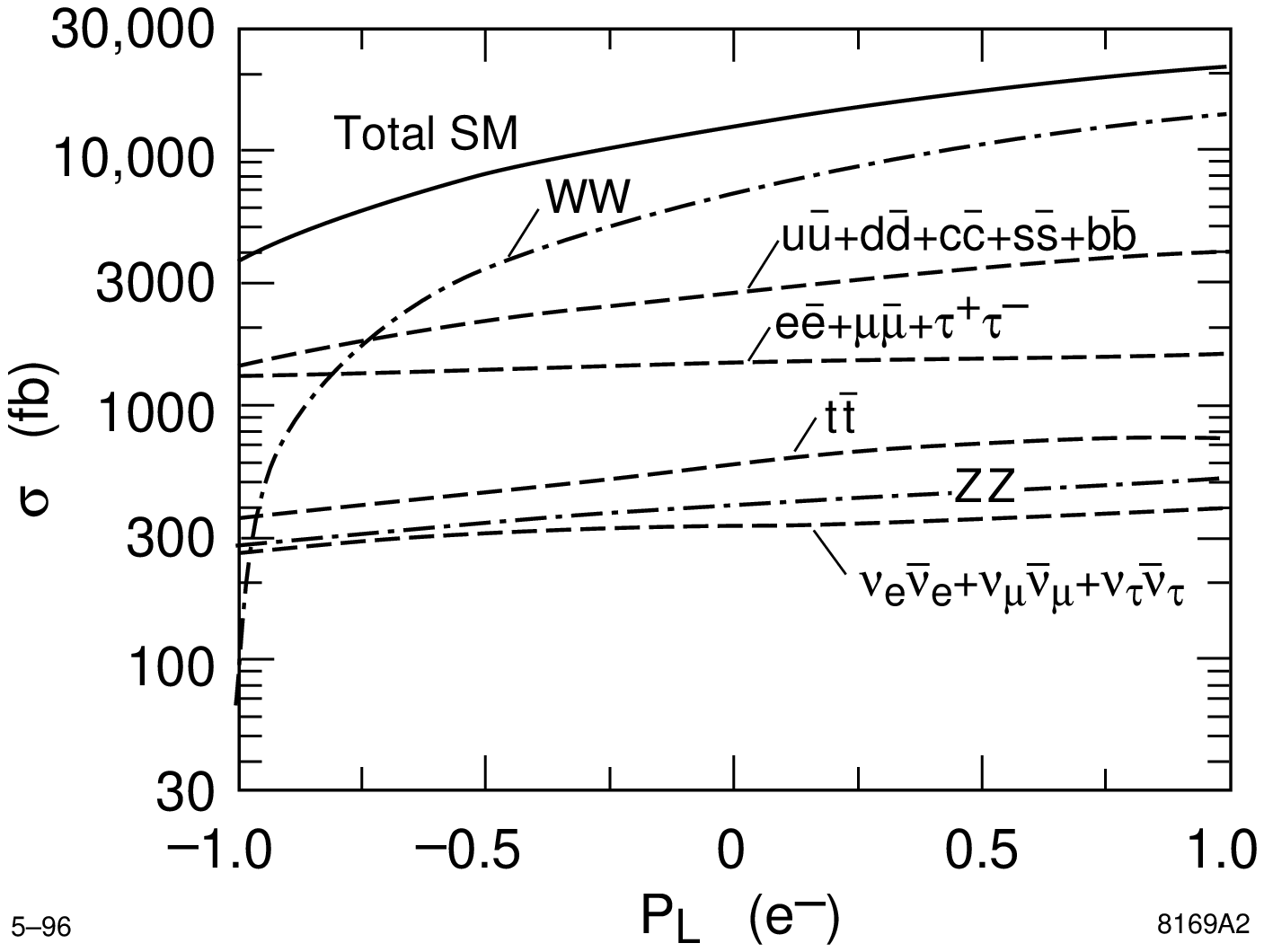}}
\centerline{\parbox{5in}{\caption[*]{Cross sections for Standard Model
physics processes in $e^+e^-$ annihilation at 500 GeV, as a function of
the electron longitudinal polarization.}
\label{nlc500}}}
\end{figure}

The cross sections of Standard Model processes at an $e^+e^-$ collider
are shown as a function of center of mass energy in Fig.~\ref{miyamoto}
\cite{Miyahawaii}.  From left to right across this plot, the familiar
$e^+e^-$ annihilation processes fall with energy according to the point
cross section for $e^+e^-\rightarrow \mu^+\mu^-$ in QED,
\begin{equation}
      1\ \hbox{\rm R} = {4\pi \alpha^2\over 3 s} = {87\ \mbox{fb}\over 
                s \ (\mbox{TeV}^2)} \ .
\label{Rdefin}
\end{equation}
At the same time, new processes involving pair production and multiple
production of weak interaction vector bosons become important.

Another view of the standard model backgrounds is given in
Fig.~\ref{nlc500}, where the cross sections for the dominant $e^+e^-$
annihilation processes are shown as a function of the degree of
longitudinal polarization. The curves were calculated using ISAJET 7.13
\cite{ISAJET}. The peripheral two photon, t-channel Bhabha scattering
and $e^{+}e^{-} \rightarrow Z^{\circ}\gamma$ processes are not shown;
the cross sections for these reactions are relatively independent of
polarization.  The reactions $e^+e^- \rightarrow e^+ \nu W^-, e^-
\bar\nu W^+,e^+e^- Z^0$ are also not shown.  The first of these is
present only for left-handed $e^-$; the other two depend only weakly on
$e^-$ beam polarization. The most troublesome source of background in
many of the physics analyses is the reaction $e^+e^-\rightarrow W^+
W^-$, whose special role we have already pointed out.  It is noteworthy
that the cross section for this process can be  reduced substantially
by adjusting the electron beam polarization.
 
\begin{figure}[htb]
\leavevmode
\centerline{\epsfxsize 2.5in\epsfbox{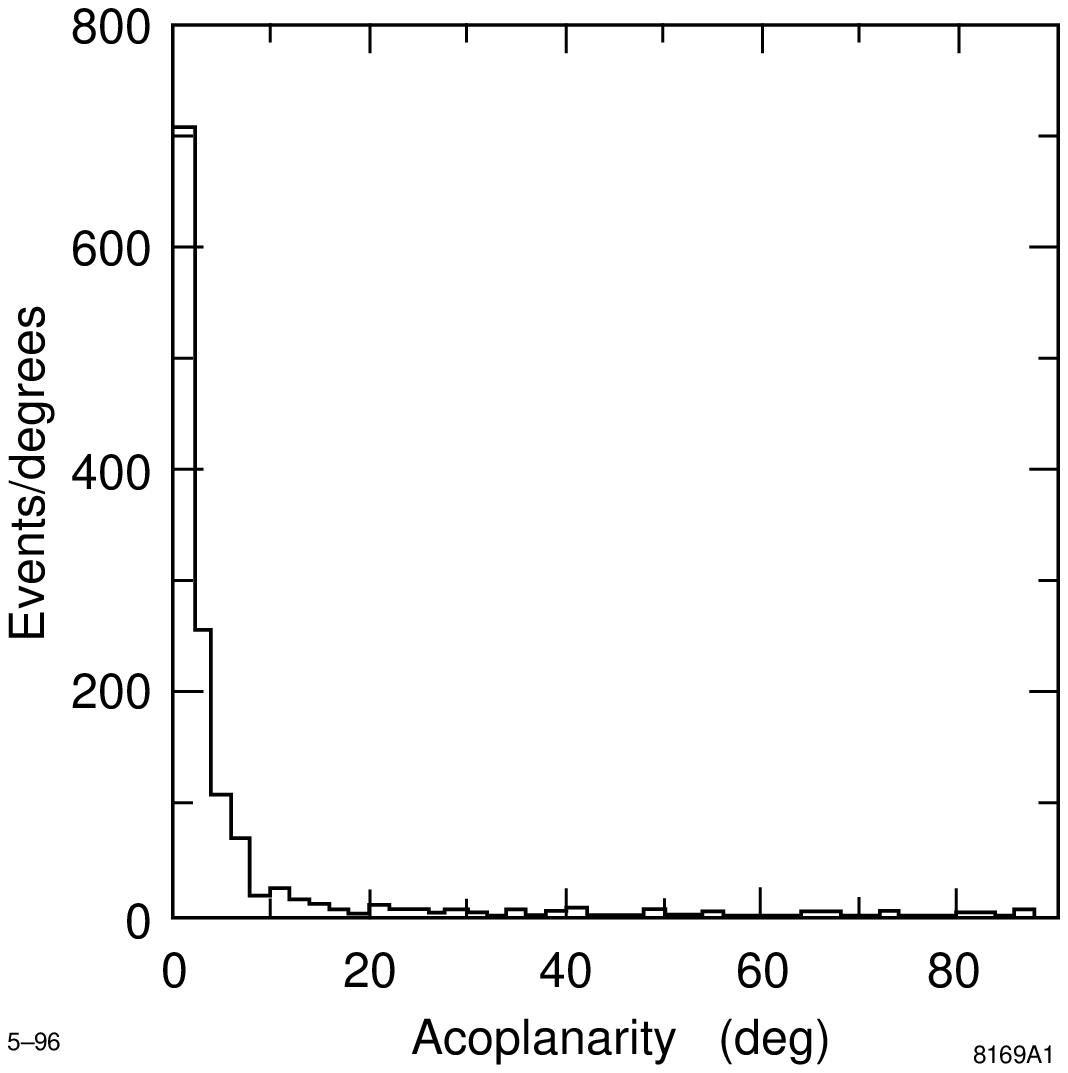} }
\centerline{\parbox{5in}{\caption[*]{Expectation for the acoplanarity
distribution in $e^+e^-\rightarrow W^+W^-$ events in which both $W$
bosons decay to hadrons.}
\label{acop}}}
\end{figure}

\begin{figure}[htb]
\leavevmode
\centerline{\epsfxsize=4in\epsfbox{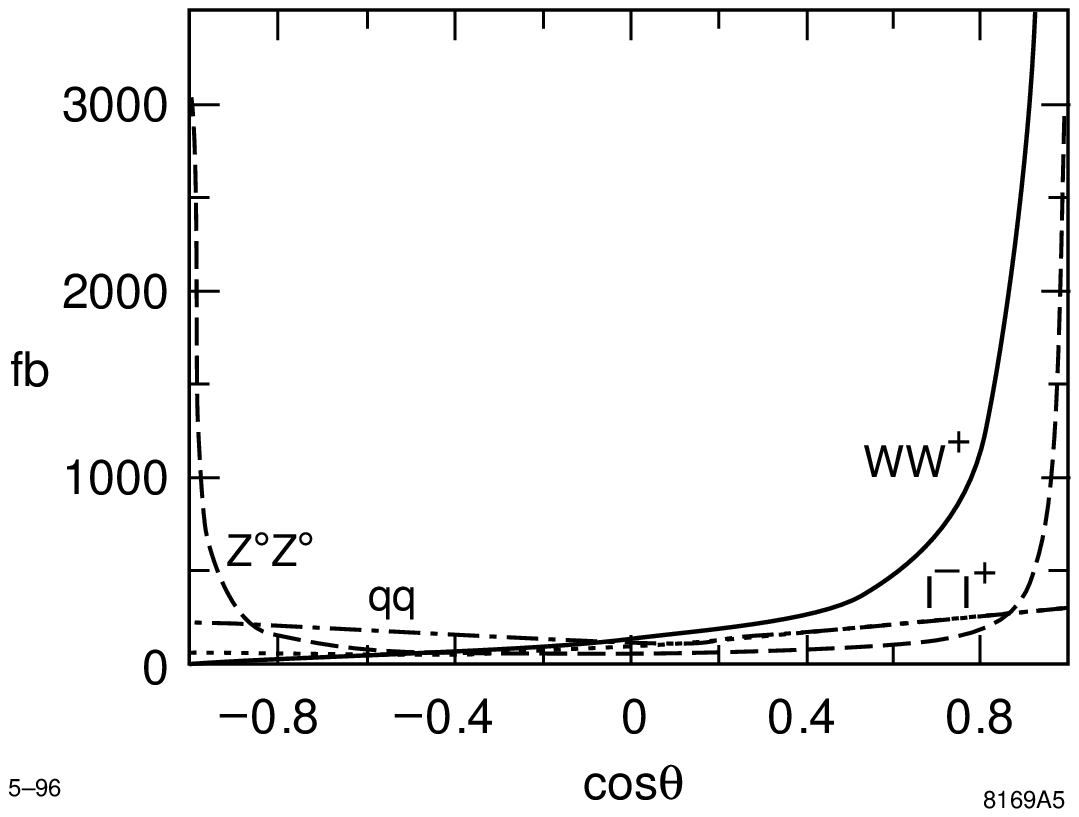} }
\centerline{\parbox{5in}{\caption[*]{Expectation for the $\cos\theta$
distribution due to standard model processes.}
\label{ctheta}}}
\end{figure}

Cuts on other quantities, such as the acoplanarity and production angle
will also be useful for removing standard model background.  The
distributions in these variables for standard model annihilation
processes are shown in Figs.~\ref{acop} and \ref{ctheta}.

In general, the two photon and $e^+e^- \rightarrow
Z^0\gamma$ processes are not important as backgrounds to
annihilation processes because they may be removed easily from the
data sample by low transverse momentum and multiplicity cuts
\cite{slac329}.  The cross section for Bhabha scattering is very
large in the forward direction but drops to a few units of R at
large angles.

\clearpage

\newcommand{\rar}{\rightarrow}
\newcommand{\lar}{\leftarrow}
\newcommand{\eett}{$e^+e^- \rightarrow t\bar{t}$}
\newcommand{\ttbwbw}{$t\bar{t} \rightarrow bW^+\bar{b}W^-$}
\newcommand{\wlnu}{$W^+ \rightarrow l^+ \nu$}

\section{Top Quark Physics}


The stage for the future of top physics has been set by the recent
discovery \cite{CDFDZtop} of top at Fermilab. The very large top mass,
$m_t\approx 174 \pm 8$ GeV/c$^2$, forces one to consider the
possibility that the top quark plays a special role in particle
physics. At the very least, the properties of the top quark could
reveal important information about the physics of electroweak symmetry
breaking. In this context, the determination of the  complete set of
top quark properties should be an important goal. A high-energy future
linear $e^+e^-$ collider provides a very impressive tool to carry out a
detailed top quark physics program.

The $t\bar{t}$ threshold region has a rich phenomenology which derives
from its mix of toponium and continuum structure.  Only in $e^-e^+$
collisions can this threshold structure be properly resolved, making
possible definitive measurements of the top mass and width and tests of
the QCD potential at very short range. Above threshold, the NLC makes
it possible to measure the complete set of top couplings to gauge
bosons, both for neutral current ($\gamma$ and $Z^0$) and charged
current interactions. The high electron-beam polarization available at
NLC plays an important role in such studies, simplifying the search for
anomalous top couplings and CP violating effects. A complete
understanding of electroweak symmetry breaking will require the
measurement of the Higgs boson couplings to fermions; of these, the
coupling to top is the most accessible. At the NLC, this quantity can
be measured by direct $t\bar t H$ production above threshold and also,
if the Higgs boson is light, by the effect of Higgs boson exchange on
the threshold properties. Finally, the NLC provides a relatively
clean final state and precise vertex detection which make it
straightforward to study the decays of the top.  All standard decay
modes can be reconstructed with reasonable efficiency, and exotic decay
modes, in those examples studied to date, can be readily identified.

The physics program for the top quark also imposes important
constraints on the NLC design.  The energy must be adjustable, to run
both at the $t\bar t$ threshold and at a point in the continuum about
100 GeV above threshold.  The study of the threshold region requires
that the center-of-mass energy spread be much smaller than the top
quark width, and that tails in the energy distribution be understood.
Experimenters must be able to determine both the absolute energy and
the differential luminosity spectrum.

\subsection{Top Production, Decay, and Measurement}
\label{sec-top-prod}

The large mass of the top quark causes it to have a very large decay 
width, and this exerts a decisive influence on its phenomenology. In the
Standard Model, the weak decay of top proceeds very rapidly via
$t\rightarrow bW$, resulting in a total decay width given by
\begin{equation}
\Gamma_t \approx (0.18) (m_t/m_W)^3 \ \mbox{\rm GeV}\  .
\label{EQwidth}
\end{equation}
For $m_t=180$ GeV/c$^2$ this lowest-order prediction is $\Gamma_t =
1.71$ GeV.  After first-order QCD and electroweak corrections
\cite{corwid}, this becomes $1.57$ GeV. Hence, top decay is much more
rapid than the characteristic time for hadron formation, for which the
scale is $\Lambda_{\rm QCD}$. This implies that the phenomenology of
top physics is fundamentally different than that of the lighter quarks.
For example, there will be no top-flavored mesons. While we lose the
familiar study of the spectroscopy of these states, we gain unique
clarity in the ability to reconstruct the properties of the elementary
quark itself. This may prove to be a crucial advantage toward
uncovering fundamental issues.

The top decay also provides a natural cutoff for gluon emission.
Indeed, in $t\bar t$ processes, the nonperturbative color strings
appear in fragmentation only after the tops decay and form along the
separating $b$  and $\bar b$ lines.  Hard gluons emitted from the top
and its product bottom quark can exhibit interference phenomena which
are sensitive to the value of $\Gamma_t$ \cite{Orrx,Orr}.

In the Standard Model, $| V_{tb} |\approx 1$, so that the decay mode
$t\rightarrow bW$ completely saturates the decay width. Then the
branching ratios are determined by the $W$ decay modes from the
$b\bar{b}W^+W^-$ intermediate state. This gives 6-jet, 4-jet + lepton,
and 2-lepton final states in the ratio 4:4:1, or, including QCD
corrections to the $W$ decay rates, BR$(t\bar{t}\rightarrow b\bar{b}q
q^\prime qq^\prime) = 0.455$;  BR$(t\bar{t}\rightarrow b\bar{b}
qq^\prime \ell\nu) = 0.439$;  BR$(t\bar{t}\rightarrow \ell\nu\ell\nu) =
0.106$, where $q=u,c$, $q^\prime = d,s$, and $\ell = e,\mu,\tau$.

The parton-like decay of top implies that, unlike other quarks, its
spin is transferred to a readily reconstructable final state.
Measurement of the $b\bar{b}W^+W^-$ final state therefore provides a
powerful means of probing new physics manifested by top with helicity
analyses. This is explored in Section~\ref{sec-top-couplings}. Another
implication of the large $m_t$ is the Standard Model prediction that
the decay $t\rightarrow bW$ produces mostly longitudinally polarized
$W$ bosons; the degree of longitudinal polarization is given by
$m_t^2/(m_t^2 + 2m_W^2)\approx 72\%$ for $m_t=180$ GeV/c$^2$. This
reflects the fact that the longitudinally polarized $W$ bosons contain
degrees of freedom from the electroweak symmetry breaking sector.

\begin{figure}[htb]
 \leavevmode
\centerline{\epsfxsize=4in\epsfbox{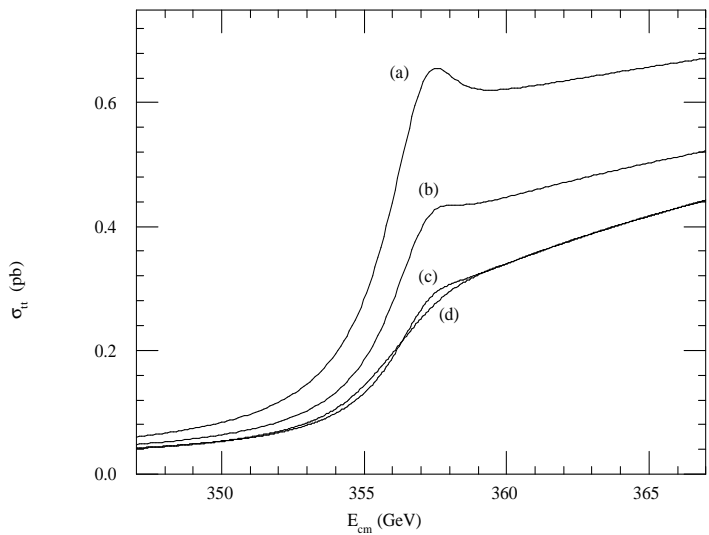}}
\centerline{\parbox{5in}{\caption[*]{Production cross section for
top-quark pairs near threshold for $m_t = 180$ GeV/c$^2$. The ideal
theoretical cross section is given by curve (a).  In curves (b), (c)
and (d), we add, successively, the effects of initial-state radiation,
beamstrahlung, and beam energy spread.}
\label{fig-top-thresh}}} 
\end{figure}

The $t\bar{t}$ cross section due to $s$-channel $e^+e^-$ annihilation
mediated by $\gamma,Z$ bosons increases abruptly just below threshold
(see Fig.~\ref{fig-top-thresh}), reaches a maximum at roughly 50 GeV
above threshold, then falls roughly proportional to the point cross
section, Eq. \ref{Rdefin}, as the energy increases. At $\sqrt{s}=500$
GeV the lowest-order total cross section for unpolarized beams is
$0.54$ pb; it is $0.74$ ($0.34$) for a fully left-hand (right-hand)
polarized electron beam. Hence, in a design year of integrated
luminosity (50 fb$^{-1}$) at $\sqrt{s}=500$ GeV we can produce $25,000$
$t\bar{t}$ events. The cross sections for $t$-channel processes,
resulting, for example, in final states such as $e^+e^-t\bar{t}$ or
$\nu\bar{\nu}t\bar{t}$, increase with energy, but are still relatively
small.  We will discuss these processes in Section 7.3.

The emphasis of most event selection strategies has been to take
advantage of the multi-jet topology of the roughly $90\%$ of $t\bar{t}$
events with 4 or 6 jets in the final state. Therefore, cuts on thrust
or number of jets drastically reduces the light fermion pair
background. In addition, one can use the multi-jet mass constraints
$M($jet-jet$)\approx m_W$ and $M($3-jet$)\approx m_t$. Simulation
studies \cite{Fujii} have shown that multi-jet resolutions of 5
GeV/c$^2$ and 15 GeV/c$^2$ for the 2-jet and 3-jet masses,
respectively, are adequate and readily achievable with LEP/SLC
detectors.  A detection efficiency of about 70\% with a signal to
background ratio of 10 was attained by selecting 6-jet final states
just above threshold. These numbers are typical also for studies which
select the 4-jet$+\ell\nu$ decay mode.

Of the backgrounds considered in this study, that from $W$-pair
production is the most difficult to eliminate. However, in the limit
that the electron beam is fully right-hand polarized, the $W^+W^-$
cross section is dramatically reduced.  Thus it is possible to use the
beam polarization to experimentally control and measure the background.
We note, though, that   the signal is also somewhat reduced by running
with a right-handed polarized beam. A possible strategy might be to run
with a right-handed polarized beam only long enough to make a significant
check of the background due to $W$ pairs. Another important technique
is that of precision vertex detection. The present experience with
SLC/SLD can be used as a rather good model of what is possible at NLC.
The small and stable interaction point, along with the small beam sizes
and bunch timing, make the NLC ideal for pushing the techniques of
vertex detection. This has important implications for top physics.
Rather loose b-tagging, applied in conjunction with the standard
topological and mass cuts mentioned above, should lead to substantially
improved top event selection efficiencies and purities.

\subsection{Threshold Physics}
\label{sec-top-thresh}

In Fig.~\ref{fig-top-thresh} we show the cross section for $t\bar{t}$
production as a function of nominal center-of-mass energy for $m_t =
180$ GeV/c$^2$.  In this discussion, $m_t$ is the pole mass in QCD
perturbation theory. The theoretical cross section, indicated as curve
(a), is based on the results of Strassler and Peskin \cite{Peskin},
using the  $q\bar q$ potential of QCD with $\alpha_s(M_Z^2)=0.12$ and 
Standard Model couplings to $\gamma$ and $Z$. To this curve, the
energy-smearing mechanisms of initial-state radiation, beamstrahlung,
and beam energy spread,  have been successively applied; curve (d)
includes all effects. The beam effects were calculated using NLC design
parameters.

The threshold enhancement given by the predicted cross section curve
of Fig. \ref{fig-top-thresh}a  reflects the Coulomb-like
attraction of the produced $t\bar{t}$ state due to the short-distance
QCD potential
\begin{equation}
 V(r)\sim -C_F\frac{\alpha_s(\mu)}{r} \ ,
\label{EQpot}
\end{equation}
where $C_F=4/3$ and $\mu$ is evaluated at the scale of the Bohr radius
of this toponium atom: $\mu\sim \alpha_s m_t$.  The level spacings of
the QCD potential, approximately given by the Rydberg energy, $
\sim\alpha_s^2 m_t$, turn out to be comparable to the widths of the
resonance states, given by $\Gamma_\theta\approx 2\Gamma_t$. Thus, the 
bound state exists, on average, only for approximately one classical
revolution before one of the top quarks undergoes a weak decay. The
level spacings of the QCD potential approximately given by the Rydberg
energy, $ \sim\alpha_s^2 m_t$, turn out to be comparable to the widths
of the resonance states, given by $\Gamma_\theta\approx 2\Gamma_t$.
Therefore the various toponium states become smeared together, as seen
in Fig. \ref{fig-top-thresh}, where only the bump at the position of
the 1S resonance is distinguishable. The infrared cutoff imposed by the
large top width also implies \cite{Fadin} that the physics is
independent of the long-distance behavior of the QCD potential. The
assumed intermediate-distance potential is also found \cite{Fujii} to
have a negligible impact. Hence, the threshold physics measurements
depend only on the short-distance potential (Eq.~\ref{EQpot}) of
perturbative QCD.

An increase of $\alpha_s$ deepens the QCD potential, thereby increasing
the wave function at the origin and producing an enhanced 1S resonance
bump. In addition, the binding energy of the state varies roughly as
the Rydberg energy $\sim \alpha_s^2 m_t$. So the larger $\alpha_s$ has
the combined effect of increasing the cross section as well as shifting
the apparent position of the threshold to lower energy. The latter
effect is also what is expected for a shift to lower $m_t$. Therefore,
there exists a significant correlation between the measurements of
$\alpha_s$ and $m_t$ from a threshold scan.

\begin{figure}[htbp]
 \begin{center}
  \begin{tabular}{c}
  \mbox{\epsfig{file=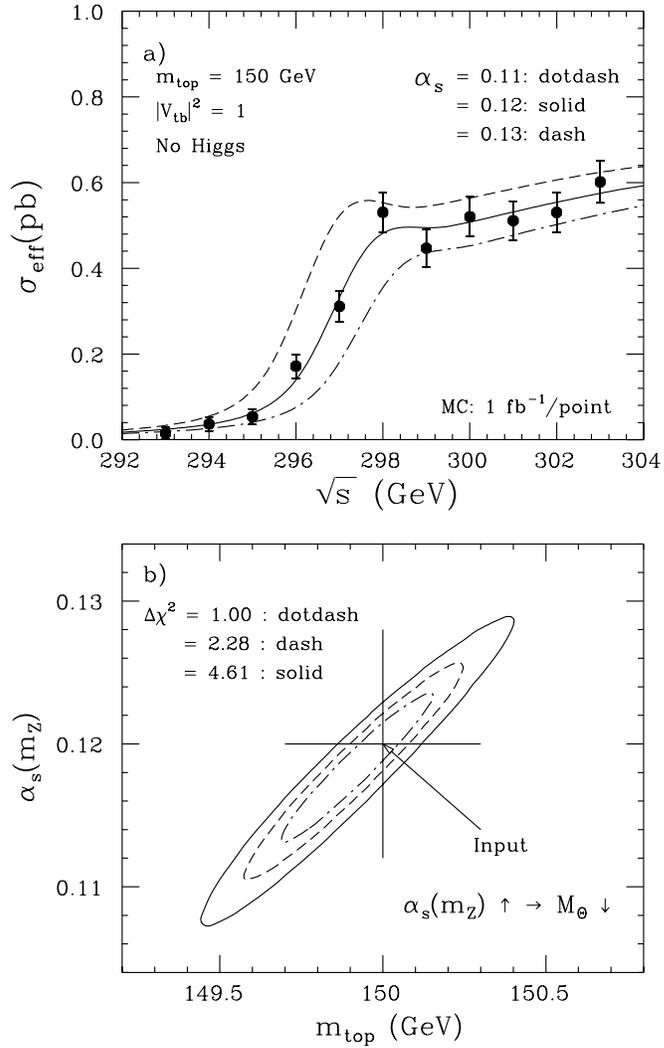,width=4.0in}}
  \end{tabular}
\centerline{\parbox{5in}{\caption[*]{(a) Top threshold scan; (b)
corresponding error ellipse for $m_t$ and $\alpha_s$. A value for $m_t$
of 150 GeV/c$^2$ was assumed.}
\label{fig-top-fujii}}} 
 \end{center}
\end{figure}

A number of studies have been carried out to simulate the measurement
of the $t\bar t$ threshold cross section. Figure~\ref{fig-top-fujii}a
depicts a threshold scan \cite{Fujii} for which an integrated
luminosity of 1 fb$^{-1}$ has been expended at each of 10 energy points
across the threshold, plus one point below threshold to measure
backgrounds. A value of $m_t=150$ GeV/c$^2$ was used.  No beam
polarization was assumed. A fit of the data points to the theoretical
cross section, including all radiative and beam effects discussed
above, results in a sensitivity for the measurement of $m_t$ and
$\alpha_s$ shown in Fig.~\ref{fig-top-fujii}b. The correlation between
these two parameters is apparent. Even for the modest luminosity
assumed here, the cross section measurement gives quite good
sensitivity to these quantities. If no prior knowledge is assumed, the
errors for $m_t$ and $\alpha_s$ are $200$ MeV/c$^2$ and $0.005$,
respectively. Conversely, the single-parameter sensitivity for $m_t$
approaches $100$ MeV/c$^2$  if $\alpha_s$ is known to much better than
2\% accuracy. We will describe a method for the precision measurement
of $\alpha_s$ in Section 10.1.  The theoretical systematic error due to
uncertainties in the $t\bar t$ threshold cross section is of order 200
MeV.

\begin{figure}[htb]
 \begin{center}
  \begin{tabular}{c}
  \mbox{\epsfig{file=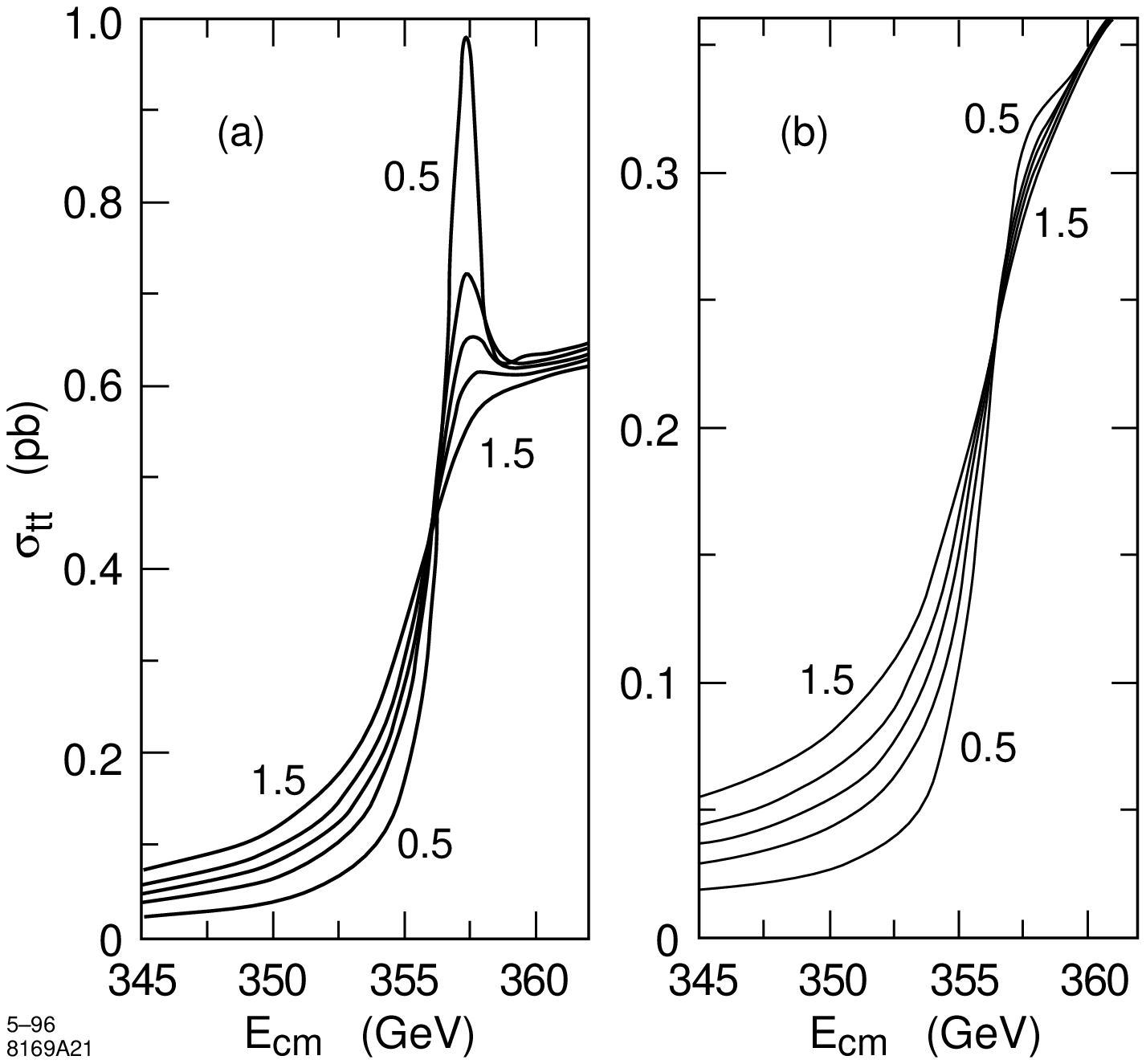,width=3.5in}}
  \end{tabular}
\centerline{\parbox{5in}{\caption[*]{Variation of the $t\bar t$
threshold cross section with the top width for $m_t=180$ GeV/c$^2$. The
curves correspond to values of $\Gamma_t/\Gamma_{SM}$ of 0.5, 0.8, 1.0,
1.2, and 1.5, presented in the order indicated, for  (a) the
theoretical cross section, and (b) the cross section after including
radiative and beam effects.}
\label{fig-top-wideall}}}
 \end{center}
\end{figure}

For a quarkonium state, we expect the cross section at the 1S peak to
vary with the total width roughly as $\sigma_{1S}\sim |V_{tb}|/\Gamma_t
$, and therefore is very sensitive to the width, as indicated in
Fig.~\ref{fig-top-wideall} for rather wide variations in $\Gamma_t$
relative to the Standard Model expectation. (It is noted that the
calculations of Figs.~\ref{fig-top-thresh} and \ref{fig-top-wideall}
use the uncorrected top width, so that the resonance structure will be
slightly more pronounced than what is shown.) After applying a
correction for initial-state radiation  and the beam-related energy
spread, the width is affected as shown in Fig.~\ref{fig-top-wideall}.
This implies that a scan strategy optimized for measuring $\Gamma_t$
would spend a relatively large fraction of running time below the 1S
peak. The threshold physics, combining the cross section information
with the momentum and asymmetry results, as discussed below, represents
what is most likely the best opportunity to measure $\Gamma_t$.

In addition to the QCD potential, the $t$-$\bar{t}$ pair is also
subject to the Yukawa potential associated with Higgs exchange:
\begin{equation}
 V_Y = - \frac{\lambda^2}{4\pi} \ \frac{e^{-m_H r}}{r} \ ,
\label{EQYukawa}
\end{equation}
where $m_H$ is the Higgs mass and $\lambda$ is the $t\bar t$-Higgs
Yukawa coupling, $\lambda =  m_t/v = \bigl[\sqrt{2}G_F\bigr]^{1/2}\,m_t
$. Because of the extremely short range of the Yukawa potential, its
effect is primarily to alter the wave function at the origin, and hence
to shift the level of the cross section. This exciting possibility is
discussed further in Section~\ref{sec-top-higgs}. The physics of the
threshold cross section is, in summary, expected to depend on the
following set of parameters:
\begin{equation} 
\sigma = \sigma (m_t,\alpha_s,\Gamma_t,m_H,\lambda) \ .
\label{EQparams}
\end{equation}

As we have discussed, the lifetime of the toponium resonance is
determined by the first top quark to undergo weak decay, rather than by
the annihilation process. This has the interesting implication that the
kinetic energy (or momentum) of the top quark as reconstructed from its
decay products reflects the potential energy of the top in the QCD
potential.  Hence, a measurement of the momentum distribution will be
sensitive to $\alpha_s$ and $\Gamma_t$. The theory \cite{topptheory}
and phenomenology \cite{Fujii,toppphen} of this physics has been
extensively studied. A convenient observable which has been used to
characterize the distribution is the position of the peak in the
reconstructed top quark momentum distribution.  The position of this
peak at a given center-of-mass energy is indeed found to be sensitive
to $\Gamma_t$ and the other parameters in Eq. \ref{EQparams}.

Yet another, quite different observable has been studied
\cite{Summur,Fujii} to help further pin down the physics parameters at
threshold. Top is produced symmetrically when produced in the 1S state.
The vector coupling of $t\bar t$ to the $\gamma$ and $Z$ can create S-
and D-wave resonance states. On the other hand, the axial-vector
coupling of the top quark to the $Z$  gives rise to P-wave resonance
states. Hence, there is naturally interference between S- and P-waves
which gives rise to a forward-backward asymmetry ($A_{FB}$)
proportional to $\beta\cos\theta$. Because of the large width of the
resonance states, due to the large $\Gamma_t$, these states do overlap
to a significant extent, and a sizable $A_{FB}$ develops. The value of
$A_{FB}$ varies from about 5\% to 12\% across the threshold, with the
minimum value near the 1S resonance. Since the top width controls the
amount of S-P overlap, we expect the forward-backward asymmetry to be a
sensitive method for measuring $\Gamma_t$.

In summary, a data set of 50 fb$^{-1}$ at threshold would provide
sensitivity to $m_t$ and $\alpha_s$ at the level of 120 MeV/c$^2$ and
0.0025, respectively. Similarly, the sensitivity to the total top decay
width is 5--10\%. Accelerator and detector designs have become
sufficiently stable to make possible calculations which incorporate the
systematics associated with luminosity spectra and backgrounds. This
would allow better determination of the limiting systematic errors at
threshold, which are presently estimated to be at or below the
sensitivities above. The measurement of the luminosity spectrum is
discussed in more detail in Section 13.

\subsection{Top Couplings}
\label{sec-top-couplings}

At the NLC, \eett\ above threshold will provide a unique opportunity to
measure simultaneously all of the top couplings. Due to its rapid weak
decay,  the top spin is transferred directly to the final state with
negligible hadronization uncertainties, therefore allowing the
helicity-dependent information contained in the Lagrangian to be
propagated to the final state. This final state, expected to be
dominated by $bW^+\bar{b}W^-$, can be fully reconstructed with good
efficiency and purity, so that a complete helicity analysis can be
performed.

The top neutral-current coupling can be generalized to the following
expression  for the $Z t \bar t$ or $\gamma t \bar t$ vertex factor:
\begin{equation}
{\cal M}^{\mu(\gamma ,Z)} = 
e\gamma^\mu\left[ Q_V^{\gamma ,Z}F_{1V}^{\gamma ,Z}
+ Q_A^{\gamma ,Z}F_{1A}^{\gamma ,Z}\gamma^5  \right]
+ {\frac{ie}{2m_t}}\sigma^{\mu\nu}k_\nu\left[
Q_V^{\gamma ,Z}F_{2V}^{\gamma ,Z}
+ Q_A^{\gamma ,Z}F_{2A}^{\gamma ,Z}\gamma^5 \right] \ .
\label{topcoupling}
\end{equation}
This expression reduces to the familiar Standard Model  tree level
expression when we set the form factors to  $ F_{1V}^\gamma = F_{1V}^Z
= F_{1A}^Z = 1$, with all others zero. The quantities $Q_{A,V}^{\gamma
,Z}$ are the usual SM coupling constants:
$Q_V^{\gamma}=\frac{2}{3}$, $Q_A^{\gamma}=0$, $Q_V^Z =
(1-\frac{8}{3}\sin^2\theta_W)/(4\sin\theta_W\cos\theta_W)$, and $Q_A^Z
= -1/(4\sin\theta_W\cos\theta_W)$. The non-standard couplings
$F_{2V}^{\gamma ,Z}$ and $F_{2A}^{\gamma ,Z}$ correspond to the
electroweak magnetic and electric dipole moments, respectively. While
these couplings are zero at tree level in the Standard Model, the
analog of the magnetic dipole coupling is expected to attain a value of
order $\alpha_s/\pi$ due to corrections beyond leading order. On the
other hand, the electric dipole term violates CP and is expected to be
zero in the Standard Model through two loops~\cite{Suzuki}. Such a
non-standard coupling necessarily involves a top spin flip, hence is
proportional to $m_t$.

\begin{figure}[htb]
 \begin{center}
  \begin{tabular}{c}
  \mbox{\epsfig{file=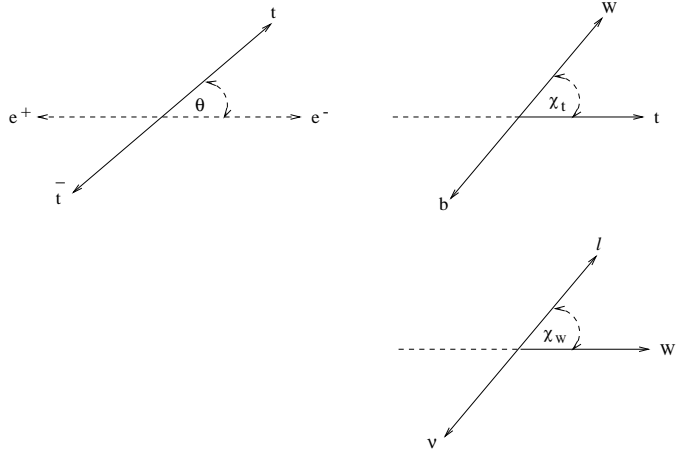,width=3.5in}}
  \end{tabular}
\centerline{\parbox{5in}{\caption[*]{Definitions of helicity angles.
(a) Production angle $\theta$ in $t\bar{t}$ rest frame; (b) $\chi_t$
measured in the top rest frame as shown; and (c) $\chi_W$ in the W rest
frame.}
\label{fig-top-helang}}} 
 \end{center}
\end{figure}

The form factors can be measured through their distinct dependencies on
the helicities of the $e^-$, $e^+$, $t$, and $\bar{t}$, which can be
accessed experimentally through the beam polarization and  the angular
distributions in the final state. The production and decay angles can
be defined as shown in Fig.~\ref{fig-top-helang}. The angle $\chi_W$ is
defined in the $W$ rest frame.  The analogous statement holds for the
definition of $\chi_t$. Experimentally, all such angles, including the
angles corresponding to $\chi_t$ and $\chi_W$ for the $\bar{t}$
hemisphere, are accessible. Given the large number of constraints
available in these events, full event reconstruction is entirely
feasible. To reconstruct $\theta$ one must also take into account
photon and gluon radiation. Photon radiation from the initial state is
an important effect, which, however, represents a purely longitudinal
boost which can be handled within the framework of final-state mass
constraints. Gluon radiation can be more subtle. Jets remaining after
reconstruction of $t$ and $\bar{t}$ can be due to gluon radiation from
$t$ or $b$, and the correct assignment must be decided based on the
kinematic constraints and the expectations of QCD.

\begin{figure}[htb]
 \leavevmode
\centerline{\epsfysize=3.5in\epsfbox{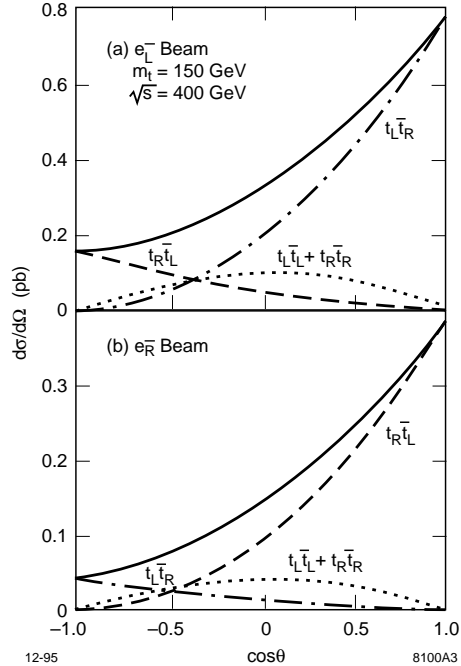}}
\centerline{\parbox{5in}{\caption[*]{Production angle for $t\bar{t}$
for the possible final-state helicity combinations, as indicated, for
100\% polarized beams with (a) left-hand polarized electrons, and (b)
right-hand polarized electrons. The complete cross sections are the
solid curves.}
\label{fig-top-prodang}}} 
\end{figure}

The distributions of the production angle $\theta$ for the SM in
terms of the various helicity states are given in
Fig.~\ref{fig-top-prodang} for left and right-hand polarized
electron beam. We see, for example, that for a left-hand polarized
electron beam, top quarks produced at forward angles are
predominantly left handed, while forward-produced top quarks are
predominantly right handed when the electron beam is right-hand
polarized. These helicity amplitudes combine to produce the
following general form for the angular distribution~ \cite{Yuan}: 
\begin{equation} 
\frac{d\sigma}{d\cos\theta} = \frac{\beta_t}{32\pi s}\left[
c_0\sin^2\theta + c_+(1+\cos\theta)^2 + c_-(1-\cos\theta)^2 \right]
\ , 
\label{angular} 
\end{equation}
where $c_0$ and $c_\pm$ are functions of the form factors of
Eq.~\ref{topcoupling}, including any non-standard couplings. The
helicity structure of the event is highly constrained by the
measurements of beam polarization and production angle.

For the measurement of the decay form factors, there are two
alternative methods that might provide higher statistics.  The first is
to measure the top quark decay distributions using polarized beams at
the $t\bar t$ threshold, making use of the fact that the spin of the
nonrelativistic top quarks follows the spin of the incident electron
and positron \cite{HarlanJK}. The second is to analyze the polarization
of top quarks above threshold using the beam axis boosted to the top
frame; this gives a very high polarization for the decay analysis
\cite{Parke}.

For the top charged-current coupling we can write the $Wtb$
vertex factor as
\begin{equation}
{\cal M}^{\mu,W}=\frac{g}{\sqrt{2}}\gamma^\mu\left[ P_L F_{1L}^W 
+  P_R F_{1R}^W \right]
+ {\frac{ig}{2\sqrt{2}\,m_t}}\sigma^{\mu\nu}k_\nu\left[
 P_L F_{2L}^W + P_R F_{2R}^W \right] ,
\label{chgcouple}
\end{equation}
where the quantities $P_{L,R}$ are the left-right projectors. In the
Standard Model, we have $F_{1L}^W = 1$ and all others zero. The form
factor $F_{1R}^W$ represents a right-handed, or $V+A$, charged current
component. As mentioned earlier, the case where the $W$ is
longitudinally polarized is particularly relevant for heavy top, and
the $\chi_t$ and $\chi_W$ distributions are sensitive to this behavior.

We now outline an analysis \cite{Frey} to measure or set limits
on the various form factors mentioned above. We consider a
modest integrated luminosity of 10 fb$^{-1}$, $m_t = 180$ GeV/c$^2$,
and $\sqrt{s} = 500$ GeV. Electron beam polarization is assumed to
be $\pm 80\%$. The decays are assumed to be $t\rightarrow bW$. In
general, one needs to distinguish $t$ from $\bar{t}$. The most
straightforward method for this is to demand that at least one of
the W decays be leptonic, and to use the charge of the lepton as the
tag. (One might imagine using other techniques, for example with
topological secondary vertex detection one could perhaps distinguish
$b$ from $\bar{b}$.)  So we assume the following decay chain:
\begin{equation}
t\bar{t}\rightarrow b\bar{b}WW \rightarrow 
b\bar{b}q\bar{q}^\prime\ell\nu ,
\end{equation}\label{EQdecay}
where $\ell=e,\mu$. The branching fraction for this decay chain is
29\%.

Since the top production and decay information is correlated, it is
possible to combine all relevant observables to ensure maximum
sensitivity to the couplings. In this study, a likelihood function is
used to combine the observables. We use the Monte Carlo generator
developed by Schmidt \cite{Schmidt}, which includes $t\bar{t}(g)$
production to ${\cal O}(\alpha_s)$. Most significantly, the Monte Carlo
correctly includes the helicity information at all stages. The top
decay products, including any jets due to hard gluon radiation, must be
correctly assigned with good probability. The correct assignments are
rather easily arbitrated using the W and top mass constraints. When the
effects of initial-state radiation and beamstrahlung are included, it
has been shown \cite{Yuan} that the correct event reconstruction can be
performed with an efficiency of about 70\%. The overall efficiency of
the analysis, including branching fractions, reconstruction efficiency,
and acceptance, is about 18\%.

After simple, phenomenological detection resolution and acceptance
functions are applied, the resulting helicity angles (see
Fig.~\ref{fig-top-helang}) are then used to form a likelihood which is
the square of the theoretical amplitude for these angles given an
assumed set of form factors. Table~\ref{table-top-couplings} summarizes
some of the results of this analysis. The upper and lower limits of the
top quark couplings in their departures from the Standard Model  values
are given at 68\% and 90\% CL. All couplings, with real and imaginary
parts, can be determined in this way. The right-handed charged-current
coupling is shown both for unpolarized and 80\% left-polarized electron
beam, whereas the other results assume 80\% left-polarized beam only.
We see that even with a modest integrated luminosity of 10 fb$^{-1}$ at
$\sqrt{s}=500$ GeV, the sensitivity to the form factors is quite good,
at the level of 5--10\% relative to Standard Model couplings. In terms
of absolute units, the 90\% CL limit of $F_{2A}^Z$ at $0.15$, for
example, corresponds to a $t$-$Z$ electric dipole moment of $8\times
10^{-18}$ e-cm.

\begin{table}[ht]
\centering
\caption[*]{Results from the global top quark form factor analysis
described in the text, for a data sample of 10 fb$^{-1}$ and 
$\sqrt{s} = 500$~GeV.}
\label{table-top-couplings}
\bigskip
\begin{tabular}{|c|c|r|r|} \hline \hline
Form Factor         & SM Value        & Limit       & Limit     \\
                    & (Lowest Order)  & 68\% CL     & 90\% CL   \\ 
\hline
$F_{1R}^W (P=0$)    & 0 &  $\pm 0.13$ &  $\pm 0.18$             \\
$F_{1R}^W (P=80\%$) & 0 &  $\pm 0.06$ &  $\pm 0.10$             \\
$F_{1A}^Z$          & 1 & $1\pm 0.08$ & $1\pm 0.13$             \\
$F_{1V}^Z$          & 1 & $1\pm 0.10$ & $1\pm 0.16$             \\
$F_{2A}^\gamma$     & 0 &  $\pm 0.05$ &  $\pm 0.08$             \\
$F_{2V}^\gamma$     & 0 &  $\pm 0.07$ &  $^{+0.13}_{-0.11}$     \\
$F_{2A}^Z$          & 0 &  $\pm 0.09$ &  $\pm 0.15$             \\
$F_{2V}^Z$          & 0 &  $\pm 0.07$ &  $\pm 0.10$             \\
$\Im (F_{2A}^Z)$    & 0 &  $\pm 0.06$ &  $\pm 0.09$             \\
\hline \hline
\end{tabular} 
\end{table}

\subsection{The Higgs-Top Yukawa Coupling}
\label{sec-top-higgs}

The coupling strength of the Higgs boson to a fermion is proportional
to the fermion's mass.  The Higgs-top coupling is consequently large
and may be unique among the Higgs-fermion couplings in that it is
accessible to direct measurement.  Such measurements have been
contemplated at LHC \cite{Atlas}, but they require efficient vertex
tagging in  high-luminosity running. The environment at NLC is much
cleaner, but the luminosity requirements are comparable.  With the
availability of large data sets ($> 50$ fb$^{-1}$), several approaches
are tractable at NLC: (1) for light to moderate mass Higgs bosons, the
$t\bar t$ production cross-section near threshold is sensitive to the
Higgs contribution to the $t\bar t$ potential; (2) for relatively light
Higgs, the yield of $t\bar t H$ events measures the Higgs-top coupling;
and (3) for Higgs masses exceeding the $t\bar t$ threshold, the Higgs
boson resonance can appear in $t\bar t Z$ events and exhibit the 
Higgs-top coupling.

\begin{figure}[htb]
 \leavevmode
\centerline{\epsfxsize=4in\epsfbox{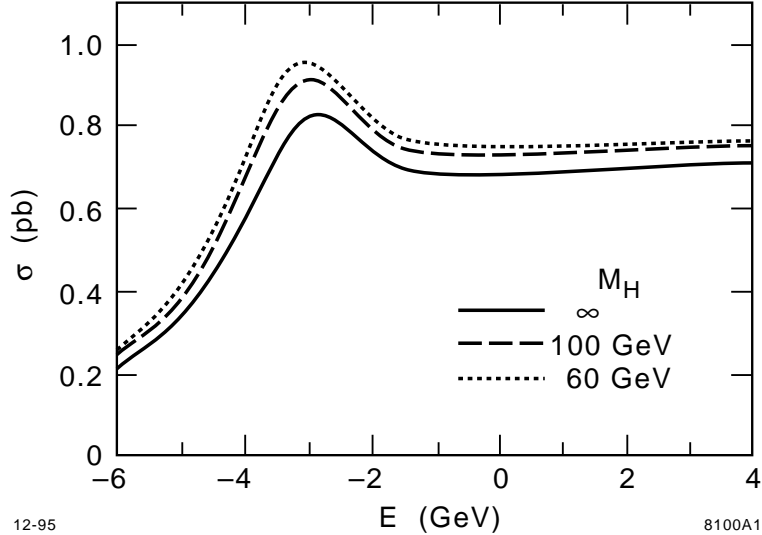}}
\centerline{\parbox{5in}{\caption[*]{Theoretical cross section as a
function of Higgs mass for $m_t=180$ GeV/c$^2$.}
\label{fig-top-higgs}}} 
\end{figure}

Threshold measurements have been discussed above for their intrinsic
interest and sensitivity to basic top parameters.  Here we note that
the presence of an additional attractive, short range force arising
from Higgs exchange increases the modulus of the toponium wavefunction
at the origin, and thereby enhances the cross-section.
Fig.~\ref{fig-top-higgs} shows the distinctive energy dependence of the
Higgs enhancement factor, which peaks at the 1S state \cite{Kuhn}.
Fujii \etal \cite{Fujii} have simulated a threshold scan of 10 points,
spaced at 1 GeV intervals, to determine the sensitivity to the
Higgs-top coupling strength.  Their results imply that a 10\%
measurement is possible with 100 fb$^{-1}$ evenly distributed over the
10 points for $M_H = 100$ GeV.  The enhancement is roughly inversely
proportional to the Higgs mass, so the scan would yield a 20\% (30\%)
measurement for a 200 (300) GeV Higgs mass.  An optimized scan will do
better.

The $t\bar t H$ events almost always result from ``Higgs-strahlung'',
radiation of the Higgs boson from one of the top quarks, so their yield
provides a measure of the square of the Higgs-top coupling.  The
cross-section for the process is small \cite{Djouadi}, in the $\cal
O$(1) fb range for a 500 GeV NLC and $M_H < 100$ GeV. Detection of the
events is challenging; they typically contain 8 jets, including 4 $b$
jets.  The process $e^+e^-\rightarrow t \bar t Z$ occurs at comparable
rate, and along with $e^+e^-\rightarrow t \bar t j j$ constitutes the
important background. Preliminary studies \cite{Fujii,EuroTop} show
that a 100 fb$^{-1}$ sample at 500 GeV will give a $\leq 15$\%
measurement of the Higgs-top coupling for $M_H\leq 100$ GeV.  At
$\sqrt{s} = 1000$ GeV, the sensitivity extends to over 200 GeV for a
measurement of similar accuracy.

The cross-section for $e^+e^-\rightarrow t \bar t Z$ is about 5 fb
between 500 and 1000 GeV center-of-mass energies.  When the Higgs mass
is above the $t \bar t$ threshold, this cross section is enhanced by
the process $\ee \to Z^0  H^0$, with Higgs decay to $t \bar t$. Fujii
\etal\ \cite{SSI-Fujii} have studied the process for $m_t = 130$ GeV,
and concluded that, with an integrated luminosity of 60 fb$^{-1}$ at
$\sqrt{s} = 600$ GeV, one could measure the top-Higgs coupling within
10\% for a 300 GeV Higgs.  For Higgs masses above $2 m_t$, the
cross-section is lower, and the increased width of the Higgs will make
isolating a signal in the $t\bar t$ invariant mass distribution more
difficult.  Even so, one could measure the Higgs-top coupling for a 400
GeV Higgs produced at $\sqrt{s} = 1000$ GeV within  about 35\% with a
data sample of 100 fb$^{-1}$.

The Higgs-strahlung process is also sensitive to deviations from the
Standard Model involving extended Higgs sectors. The $t\bar{t}H$ final
state can result from Higgs emission from the $t$ (or $\bar{t}$), or
from the intermediate $Z$. Interference between these sub-processes can
give rise to large CP violating effects in extended Higgs models. This
was studied in Ref.~\cite{CPttH} for a range of two-Higgs doublet
models, and it was found that for roughly 100 fb$^{-1}$ of data at
$\sqrt{s}=800$ GeV it would be possible to observe a significant CP
asymmetry in a number of final-state observables.

The Higgs-top coupling can be determined in the case that the Higgs is
very heavy ($M_H>400$ GeV/c$^2$) by measuring the rate of the process
$e^+e^-\rightarrow \nu\bar{\nu} t\bar{t}$ At $\sqrt{s}=1500$ GeV, the
cross section for this process is about 2 fb in the absence of a Higgs,
but will be enhanced by more than a factor of two for Higgs masses in
the range 400--1000 GeV/c$^2$. Preliminary studies by Fujii
\cite{SSI-Fujii} show that care is required to eliminate radiative
$t\bar{t}$, $e^+e^-t\bar{t}$, and $t\bar{t}Z$ backgrounds, but suggest
that the Higgs-top coupling can  be measurable up to $m_H=1$ TeV/c$^2$.
The case of a very heavy Higgs boson is discussed in more detail in
Section 7.3.

\subsection{Top Physics Reach of NLC and Hadron Colliders}
\label{sec-top-compare}

Table ~\ref{table-top-compare} summarizes the top physics reach of the
NLC and several hadron colliders. The Tevatron Upgrade (TeV$^*$) will
establish the baseline for top quark physics in the LHC/NLC era, and
will address many subjects of interest in top physics. Its reach has
been studied in a report by Amidei \etal\ \cite{Amidei}. The Atlas TDR
\cite{Atlas}  provides some information on the top physics reach at
LHC; this subject will certainly be developed further in the future.
The table at best represents what has been studied to date. If a
particular measurement at a particular machine has not yet been
analyzed, the corresponding entry has been left blank.  An ``X'' marks
measurements that cannot be made at a particular machine, by virtue of
excessive backgrounds, insufficient signal, or unavailable production
mechanisms.

\begin{table}[htb]
\caption{Top Physics at Future Facilities}
\label{table-top-compare}
\bigskip
\begin{tabular}{|c|c|c|c|c|c|} \hline \hline
Quantity & TeV$^*$ & TeV33 & LHC & NLC$(\sqrt{s}=360)$ & NLC$(\sqrt{s}=500)$   \\
         & (1 fb$^{-1}$)   & (10 fb$^{-1}$) & (100 fb$^{-1}$) & (50 fb$^{-1}$) 
         & (50 fb$^{-1}$)  \\  \hline
$\Delta m_t$          & 3.5 GeV/c$^2$ & 2.0 GeV/c$^2$  & 2 GeV/c$^2$ 
                      & 0.20 GeV/c$^2$ &               \\
$\Delta \Gamma_{t}$   &                &        &       & 6--8\%   &      \\
$\Delta a_t$          & X           & X    & X   &          & 4\%  \\
$\Delta v_t$          & X           & X    & X   &          & 5\%  \\
$\Delta V_{tb}$       & $14\%$        &  6\%   & X  &          &      \\
$\Delta V_{ts}$       & X           & X   & X?    &   ?      & ?  \\
$\Delta V_{td}$       & X           & X    & X   &  X     & X \\
$\Delta \lambda_t$    & X           & X    & ?     & 14\%     & 20\%   \\
$\Delta \alpha^t_s$   & $(\sigma_{t\bar{t}}=11\%)$ 
                      & $(\sigma_{t\bar{t}}=4\%)$      && 0.005    &  \\
$\Delta B(t \rar bW^0)$ & $4\%$       & 1.3\%          & & 1\% &      \\
$\Delta B(t \rar bW_R)$ & $2\%$       & $0.6\%$        & &  & 2\%      \\
$\Delta \delta$       & X          & X    & X     &   
                      & $<0.3\,e\hbar/2m_t$\\
$\Delta d$            & X          & X    & X     &     
                      & $<4 \times 10^{-18}$ e-cm   \\
$B(t \rar H^+b)$      & $<15\%$       & $<6\%$ & $<1.4\%$ &  & $<2\%$    \\
$B(t \rar \tilde{t}\tilde{\chi}^0)$   &        &       & &   & $< 1\%$ \\
$B(t \rar c\gamma)$   & $<0.3\%$      & $<0.04\%$      &  &   &       \\
$B(t \rar cZ)$        & $<1.5\%$      & $<0.4\%$       
       & $<5\times 10^{-5}$ &       & $< {\rm few}\cdot10^{-4}$  \\
$B(t \rar ch^0)$      &  &  &  & $<1\%$ &  \\  \hline \hline
\end{tabular}
\end{table}

The table demonstrates how crucial a role the NLC plays in obtaining a
complete  picture of top quark physics. NLC will provide the definitive
top mass measurement.  It will provide the only direct measure of the
top width; at hadron colliders, the total width can be inferred only
from a $V_{tb}$ measurement using the assumption that the top has no
unobserved exotic decays.  The NLC will measure the axial-vector and
vector electroweak couplings, some of the charged-current couplings
(expressed here as CKM elements), the top-Higgs coupling, and the
flavor specific strong coupling. Hadron colliders will also measure the
charged current couplings (although $V_{td}$ is probably impossible at
both hadron and $e^+e^-$ colliders), and the strong and electromagnetic
couplings, but not the couplings to the $Z$. The LHC may probe the
top-Higgs coupling by isolating $t\bar{t}H$ events, but only with
difficulty. The NLC can measure the top decay form factors, checking
for longitudinal $W$ production and searching for right-handed $W$'s,
as can the hadron colliders. Only the NLC can measure the electroweak
magnetic and electric dipole moments, because they depend on the
neutral current production mechanism.  We should note that LHC can be
sensitive to top-associated CP violation through more complicated
effective interactions \cite{PeskinCP, YuanCP}. Rare decays with
distinctive signatures can be sought in either environment, with the
advantage to hadron decays by virtue of the large statistical samples
anticipated. The more exotic decays, {\it e.g.} $t \rightarrow \tilde t
\tilde \chi^0$, are more sensitively sought in the clean environment of
the NLC.

\clearpage

\section{Higgs Boson Searches and Properties }

\subsection{Introduction}

Despite the extraordinary success of the Standard Model (SM) in
describing particle physics up to the highest energy available today,
the mechanism responsible for electroweak symmetry breaking (EWSB) has
yet to be determined. In particular, the Higgs
boson~\cite{higgs1,higgsrev,higgs2} predicted in the minimal Standard
Model and the theoretically attractive Supersymmetric (SUSY) Grand
Unified Theory (GUT) extensions thereof have yet to be observed. If
EWSB does indeed derive from nonzero vacuum expectation values for
elementary scalar Higgs fields, then one of the primary goals of
constructing future colliders must be to {\em completely} delineate the
Higgs boson sector. In particular, it will be crucial to discover all
of the physical Higgs bosons and determine their masses, widths and
couplings.  Conversely, if a fundamental Higgs boson does not exist, it
is essential to demonstrate this unambiguously.

The EWSB mechanism in the Standard Model is phenomenologically
characterized by a single  Higgs boson ($h_{SM}$) in the physical
particle spectrum.  The mass of the $h_{SM}$ is undetermined by the
theory, but its couplings to fermions and vector bosons are completely
determined. In SUSY theories, there are two Higgs doublets with vacuum
expectation values $v_1$, $v_2$.  These contribute mass terms for the
gauge bosons proportional to $(v_1^2 + v_2^2)$, masses for  down-type
fermions proportional to $v_1$, and masses for up-type fermions
proportional to $v_2$. In the Minimal Supersymmetric Standard Model
(MSSM)~\cite{Nillrev,HKrev} these two doublets give rise to five
physical Higgs bosons: $h^0$, the lighter of the two $CP$-even states;
$H^0$, the heavier $CP$-even state; the $CP$-odd $A^0$ boson, and a
pair of charged bosons $H^{\pm}$.  The mass of the minimal SM Higgs
boson is unspecified, but in the MSSM, there are tree-level relations
which determine the spectrum of masses in terms of one of the boson
masses (e.g., the mass of the $A^0$) and the ratio of the vacuum
expectation values, $v_2/v_1$. The CP-even and CP-odd neutral Higgs
bosons have nontrivial mixing angles $\alpha$ and $\beta$,
respectively, which affect their couplings and decays.  In particular,
$v_2/v_1 =\tan\beta$. Both masses and couplings receive further
radiative corrections which are functions of the SUSY Higgs mass
parameter, $\mu$, the scale of mass at which SUSY is broken,
$M_{SUSY}$, the mass of the top quark, and the $A_i$ parameters of the
soft supersymmetry-breaking interaction. More general models of the
Higgs sector, which also include electroweak singlets, are also
possible in SUSY theories.   Finally in non-supersymmetric
models with two Higgs doublets (2HDM), the Higgs bosons may have mixed
$CP$ character.

Supersymmetry has exciting implications for the discovery potential of
the Higgs bosons that it predicts.  In the MSSM, considering
renormalization group improved radiative corrections and assuming $m_t
= 180$~GeV with the stop mass less than 1~TeV, the lightest Higgs boson
must have mass $M_{h^0} \lsim 130$~GeV. An even more sweeping statement
can be made~\cite{sweep} that $M_{h^0}\lsim 150$~GeV for {\em any} SUSY
theory with a grand unification at high energy which includes the
elementary Higgs fields.

\subsection{Present and Future Limits}

The best direct limits on the SM Higgs boson come from searches at LEP,
with the present limit~\cite{LEPlim} being $M_{h_{SM}} > 65.2$~GeV at
95\% confidence level (C.L.). These limits can also be interpreted in
the framework of the MSSM to exclude the lightest SUSY Higgs with mass
less than approximately 45~GeV. Electroweak radiative corrections
including the top quark and the Higgs boson affect precision
electroweak measurements, and global fits~\cite{LEPfit} using data from
LEP, SLC, the Tevatron, and neutrino scattering give the relatively
weak limit implying that $M_{h_{SM}} < 300$~GeV (95\% C.L.).

\subsubsection{LEP2}

The limit on the Higgs boson mass will be improved in the near future
with the operation of LEP2. With an integrated luminosity of
150~pb$^{-1}$ in each of the four LEP detectors, expected from one year
of running at design luminosity, the 5$\sigma$ discovery reach can be
increased~\cite{LEP2} to about 95~GeV with running at center-of-mass
energies of 192 GeV scheduled for 1997. At the same energy and
luminosity, the process $e^+e^- \rightarrow hA$ can be discovered
(excluded) at a cross section of 65 (30) fb, when supersymmetric decay
channels are closed. The resultant exclusion region in the MSSM
parameter space can be found in Fig.~\ref{LHCfig}.  The possibility of
running at 205~GeV, which would result in an extension of limits close
to the MSSM bound, is currently being investigated.

\subsubsection{Upgraded Tevatron}

\begin{figure}[htb]
\leavevmode
\centerline{\epsfysize=2.5in \epsfbox{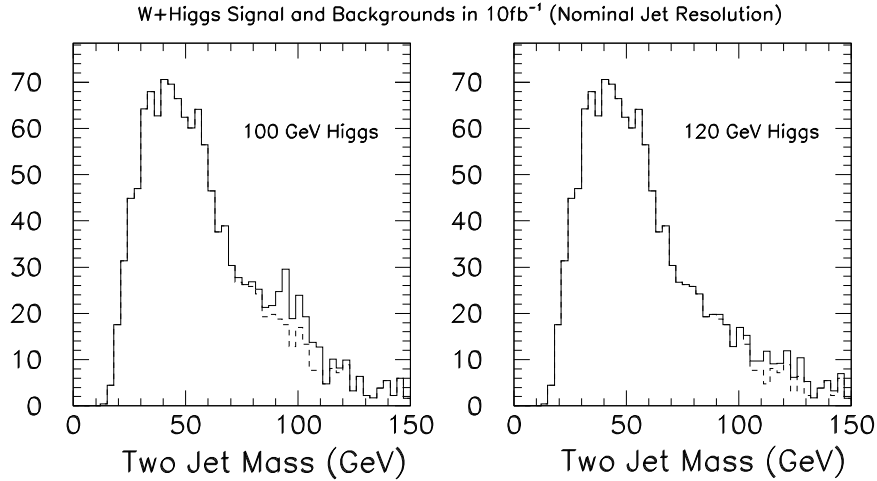}}
\centerline{\parbox{5in}{\caption[*]{The signal plus background mass
distributions for the $WH$ process with 10~fb$^{-1}$ of data at 2 TeV. 
The solid line is signal+background, the dashed line the sum of all
backgrounds.}
\label{figtev}}}
\end{figure}

The associated production of a Higgs boson and a $W$ or $Z$ boson, with
the Higgs decaying to $b \bar b$ and the $W$ or $Z$ decaying
leptonically, is a possible way to detect the Higgs in the mass range
60--130 GeV, at a high luminosity Tevatron collider~\cite{Amidei}. The
Higgs decay gives rise to 2 jets, thus one will use $b$ tagging to
reduce the large $W+2$~jet background.  It appears that the present
$b$ tagging capability at CDF is more than adequate to reduce this
background (at moderate Run II luminosities, e.g., $10^{32} \times$
cm$^{-2}$ sec$^{-1}$, 1 TeV $\times$ 1 TeV) if this capability is
extended to larger rapidities (as is planned in Run II for both CDF and
D0). After $b$ tagging, the largest background at Higgs masses below
100~GeV is QCD production of $W+b\bar{b}$ and top backgrounds for
masses above 100~GeV. Figure~\ref{figtev} shows the dijet mass
distribution for the sum of all these backgrounds, plus the $W+H$
signal for 10~fb$^{-1}$. An observation of the Higgs for masses below
100 GeV is possible after the Main Injector upgrade, and is  within
reach of the present Run II accelerator after several years of
data-taking. For higher mass Higgs bosons, these statistics are too
low; one would need about 25 fb$^{-1}$ to observe the 120 GeV
Higgs. This study assumed an approximate 20\% improvement in dijet mass
resolution obtained from applying  a clustering algorithm that reduces
the effect of gluon radiation at large angles to the jet.  This dijet
mass resolution and jet clustering is crucial in seeing the Higgs.  It
has been argued that the $h\to \tau^+ \tau^-$ and $Z\to \nu\bar\nu$
channels can be used to improve these results \cite{KandMr}.

\subsubsection{Large Hadron Collider}

At the Large Hadron Collider (LHC), detection of the SM $h_{SM}$ is
possible through the process $gg \rightarrow  h_{SM}  \rightarrow
\gamma \gamma$ for $M_{h_{SM}}  < 150$~GeV and through  $gg \rightarrow
h_{SM} \rightarrow ZZ^{(*)} \rightarrow 4\ell$ for $M_{h_{SM}} > 
130$~GeV. A heavy $h_{SM}$ is also detectable in the reaction $WW
\rightarrow h_{SM}$, with the Higgs decaying to $ZZ$ and also,
possibly, to $WW$.  The $\gamma \gamma $ channel that is crucial for a
light $h_{SM}$ demands an excellent electromagnetic calorimeter, and
much attention has been devoted to this in the LHC detector designs. 
For $M_{h_{SM}}  < 120$~GeV, it will also be possible to detect
$t\bar{t}h_{SM}$  and (possibly) $Wh_{SM}$ with $h_{SM} \rightarrow
b\bar{b}$, provided that the high $b$-tagging efficiency and purity
projections are realized. Detection of the $h_{SM}$ in the intermediate
mass region when $M_{h_{SM}} < 2M_W$ generally requires accumulating
data for at least a year when the LHC is run at full luminosity. This
should be contrasted with $e^+e^-$ collisions, where the $e^+e^-
\rightarrow Zh_{SM}$ mode will allow detection in the same mass region
in a matter of a few hours, assuming full instantaneous luminosity.

In the case of the MSSM for large $M_{A^0}$, the $h^0$ is similar to
the Standard Model Higgs $h_{SM}$. As for the $h_{SM}$, the $h^0$ is
straightforward to detect at an $\ee$ collider.  On the other hand, the
$H^0$ and $A^0$ do not resemble the Standard Model Higgs boson, and so
one must separately consider their production process.  We will show
below that the observability of $H^0$ and $A^0$ at an $\ee$ collider
depends only on the beam energy: for $\sqrt{s} > 2M_{A^0} - 20$~GeV,
these particles are found in the reaction $e^+e^- \rightarrow Z^*
\rightarrow H^0 A^0$.

\begin{figure}[htb]
\leavevmode
\centerline{\epsfxsize=4.5truein \epsfbox{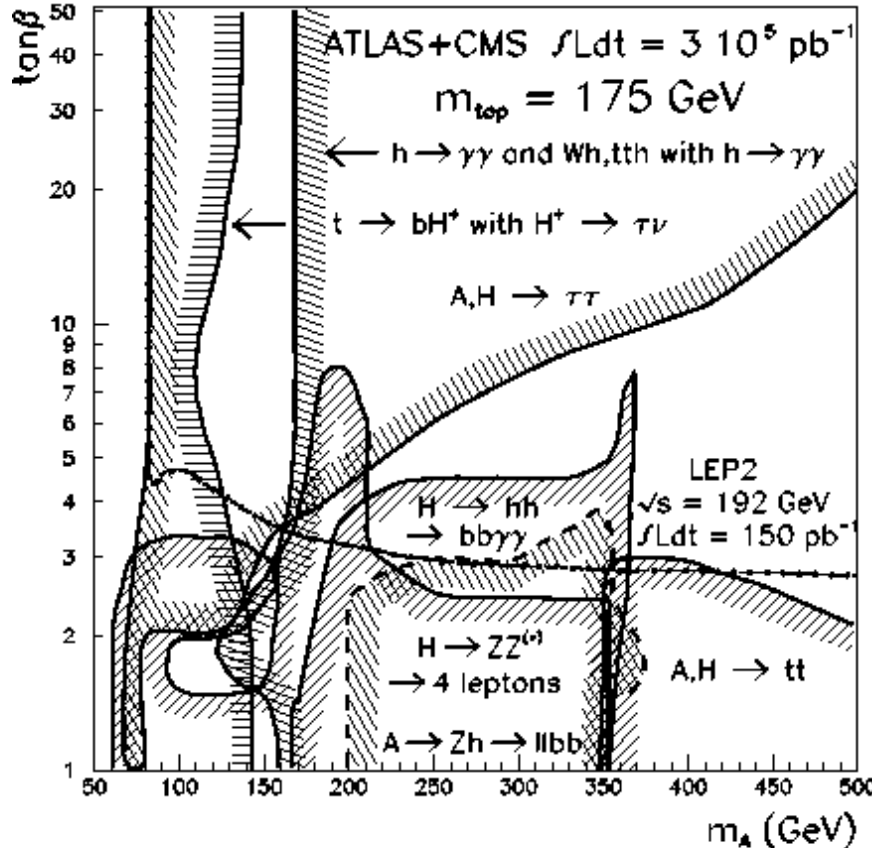}}
\centerline{\parbox{5in}{\caption[*]{Higgs discovery contours
(5$\sigma$) in the generic parameter space of the MSSM for ATLAS+CMS at
the LHC, for a multi-year run at design luminosity, $300$~fb$^{-1}$ per
detector, from  \protect\cite{higgs2}. Renormalization group improved
radiative corrections are included for $M_{h^0}$ and $M_{H^0}$,
assuming $m_{\tilde{t}} = 1$~TeV and no squark mixing.}
\label{LHCfig}}}
\end{figure}

The story at the LHC is much more complex. The reactions which can be
used to detect the Higgs particles of the MSSM, and their limits of
applicability in parameter space, are displayed in Fig.~\ref{LHCfig}
\cite{higgs2}. This figure represents the limit of the LHC capability,
summing the results of two detectors in a multi-year run at design
luminosity. For values of $M_A > 200$ GeV, the LHC can detect the $H^0$
and $A^0$ only in certain specific decay channels, shown in the figure,
whose availability depends on the value of $\tan\beta$.  Since this
figure summarizes a great deal of analysis, we must point out at least
a few of the assumptions which are used. The channel $A^0,H^0\to
\tau^+\tau^-$ can be used only when the branching ratio to $\tau$ is
enhanced by a large value of $\tan\beta$. For small $\tan\beta$, modes
with $b\bar b$ or $t\bar t$ in the final states require $b$ tagging
capabilities that will be challenging in the detection environment of
the LHC.  In addition, it should be noted that the process $A^0,H^0\to
t\bar t$ has so far been studied only at the level of the comparison of
cross sections for signal and background, and that, since the signal is
2--10\% of the background, an excellent knowledge of the $g g \to t\bar
t$ cross section is required.  Finally, though the $h^0$ should be
detected for the generic situation illustrated in Fig.~\ref{LHCfig},
there are regions of the full parameter space of the MSSM where the
$h^0$ would not be observed \cite{wellsco}.  Thus, it is unlikely that
the whole MSSM Higgs spectrum would be observed at the LHC, and it is
not possible to rule out the MSSM if none of its Higgs bosons are seen
at the LHC.

\subsection{Standard Model Higgs}

\begin{figure}[htb]
\leavevmode
\centerline{\epsfxsize=5in \epsfbox{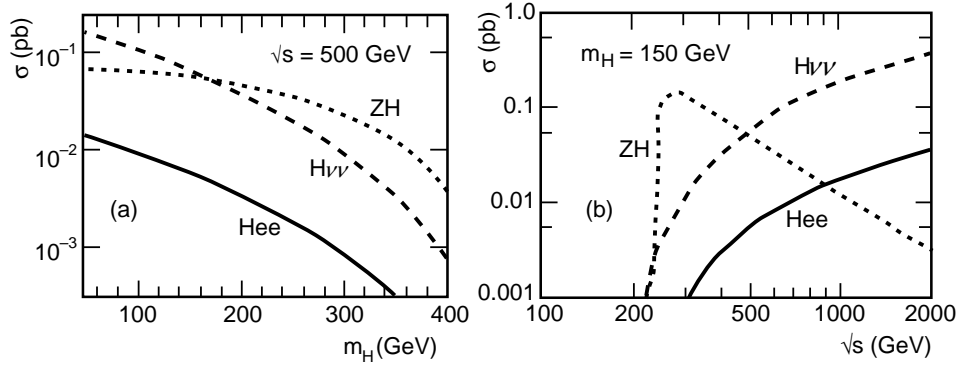}}
{\caption{Cross section for Standard Model Higgs boson production.}
\label{xsec}}
\end{figure}

The main production processes for the SM Higgs in $e^+e^-$ annihilation
are $\ee\to ZH$ and the gauge boson fusion processes $\ee \to
\nu\bar\nu H$ ($WW$ fusion) and $\ee \to \ee H$ ($ZZ$ fusion). The
cross sections for these processes are shown in Fig.~\ref{xsec}.  With
a typical integrated luminosity of 10 fb$^{-1}$ at $\sqrt{s} = 500$~GeV
with $M_H = 150$~GeV, about 1000 signal events would be expected before
cuts and branching ratios. Handy ``rules of thumb" are that the peak
for $ZH$ production occurs at $\sqrt{s} \approx M_Z + \sqrt{2}M_H$ and
that the cross-over for equal cross sections from the fusion and
bremsstrahlung mechanisms occurs at $\sqrt{s} \approx 0.6M_H + 400$~GeV.

\begin{figure}[htb]
\leavevmode
\centerline{\epsfysize=2.8truein \epsfbox{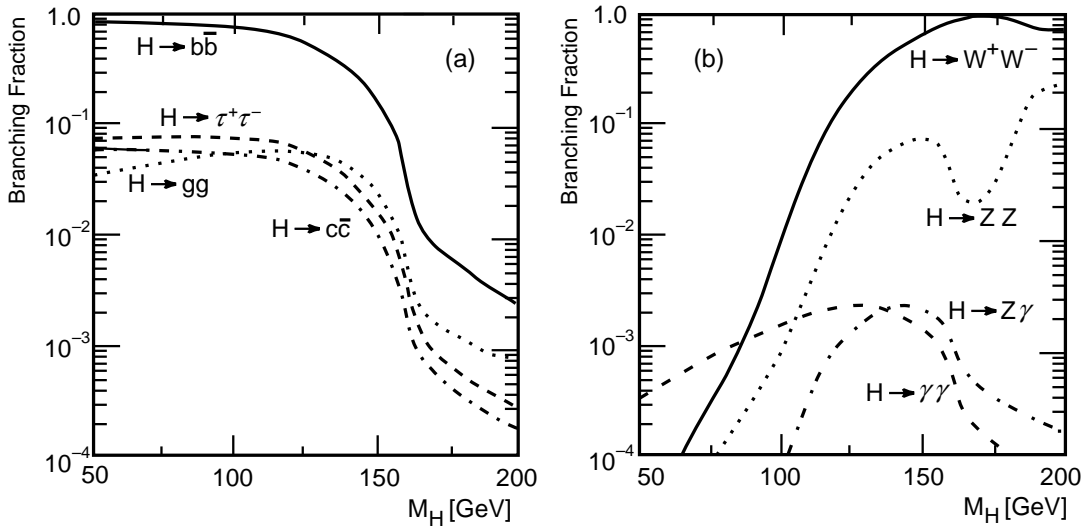}}
{\caption{Branching fractions of a SM Higgs boson.}
\label{decay}}
\end{figure}

The decay modes  of the Standard Model Higgs depend strongly upon its
mass.  The branching ratios of the Standard Model Higgs are shown in
Fig.~\ref{decay}. A very interesting region is the intermediate mass
Higgs with $M_Z < M_H < 2M_W$, which at a hadron collider is relatively
more difficult to detect than a heavy Higgs boson. Almost all of the
decays, both those to fermions and those into pairs of gauge bosons are
identifiable in $\ee$ experiments, and it should be possible to measure
individual branching ratios. For an intermediate mass Higgs, the
dominant decay channel is clearly $H^0 \rightarrow b \bar{b}$, with the
branching ratio for $H^0 \rightarrow W^+ W^{-(*)}$ growing with
increasing mass (even for $E_{cm} < 2M_W$ where one of the $W$'s must
be off shell).  This latter channel remains dominant for heavy Higgs
bosons, and is joined by $ZZ$ and $t\bar{t}$ modes when kinematically
accessible.

\subsubsection{Signal Topologies and Backgrounds}

\begin{figure}[htb]
\leavevmode
\centerline{\epsfysize=1.7truein \epsfbox{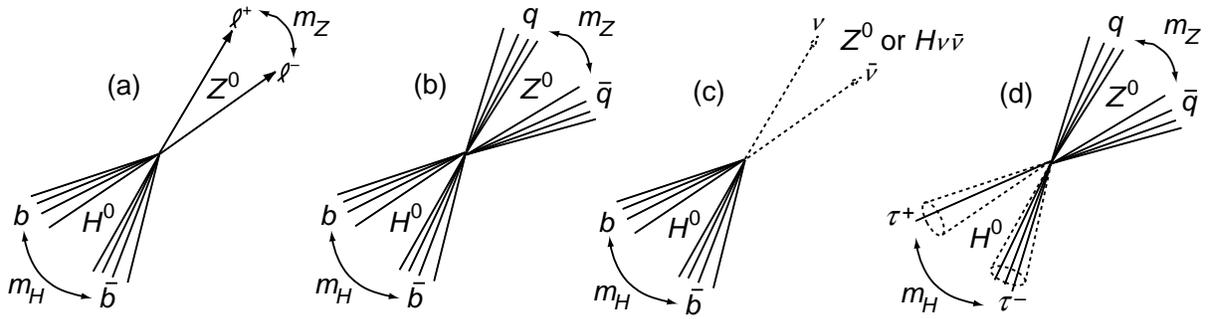}}
\centerline{\parbox{5in}{\caption[*]{Important signal topologies for an
intermediate mass SM Higgs boson.}
\label{topo}}}
\end{figure}

Typical signal topologies in the intermediate mass range are shown
in Fig.~\ref{topo}. The associated production $e^+e^- \rightarrow
Z^0 H^0$, is followed by standard decays of the $Z^0$ (10\%
$\ell^+\ell^-$, 20\% $\nu \bar{\nu}$, and 70\% $q\bar{q}$) and
decays of the $H^0$ mostly into $b\bar{b}$, occasionally into
$\tau^+ \tau^-$, and more rarely into $c \bar{c}$ and $gg$.
After straightforward cuts, the most serious backgrounds are due to 
irreducible Standard Model processes, $\ee\to ZZ$, $Z\nu\bar\nu$, and
$We\nu$.

\subsubsection{Experimental Studies}

The detection of Higgs bosons in $\ee$ collisions at high energy has
been studied extensively in simulations \cite{janot,gunionHH,other}.
The detector simulations that have been employed usually emulate
LEP/SLC-type detectors using  smeared four-vectors, while some consider
simulations of a more ambitious JLC-type detector~\cite{JLC}. Most
studies include the effects of beamstrahlung, the radiation of
photons in the intense electromagnetic fields of the beam-beam
collision.  Like more standard initial-state radiation, this effect is
usually taken into account in kinematic fits by allowing for an unknown
missing momentum along the beam axis.

In the topology of Fig.~\ref{topo}a, one can first identify two
leptons with invariant mass close to the mass of the $Z^0$, and then
investigate the remaining hadronic mass or use kinematic constraints
to study the missing or recoil mass in the event:
\[
M_{miss} = \sqrt{(\sqrt{s} - E_{\ell^+} - E_{\ell^-})^2
- (\vec{p}_{\ell^+} + \vec{p}_{\ell^-})^2}.
\]
This quantity has a large peak at the $Z^0$ mass from the irreducible
background process $e^+e^- \rightarrow Z^0 Z^0; \thinspace Z^0
\rightarrow \ell^+ \ell^-; \thinspace Z^0 \rightarrow q\bar{q}$. If
$M_H \approx M_Z$, the signal and the $ZZ$ background are kinematically
equivalent, and one would need $b$-quark tagging to distinguish the
signal.  Since  $Br(Z^0 \rightarrow b \bar{b}) \simeq 20\%$, while
$Br(H^0 \rightarrow b \bar{b}) \simeq 85\%$ at this mass, an analysis
in this worst case would require  50~fb$^{-1}$ of data.

\begin{figure}[htb]
\leavevmode
\centerline{\epsfysize=2.5truein \epsfbox{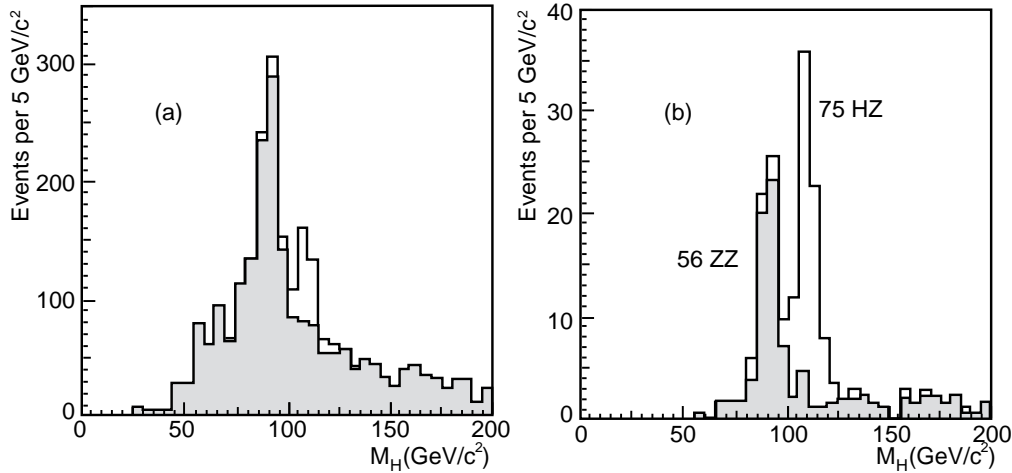}}
\centerline{\parbox{5in}{\caption[*]{Distribution of the invariant mass
of the Higgs jet pair in the four-jet topology (a) before and (b) after
$b$-quark tagging, for all known backgrounds (shaded histograms) and
for the signal ($M_H = 110$~GeV) (adapted from Ref.~10).}
\label{fourjet}}}
\end{figure}

The four-jet topology of Fig.~\ref{topo}b has been considered in a
number of studies, in particular,  in a comprehensive study by
Janot~\cite{janot} at $\sqrt{s} = 500$~GeV which assumed an integrated
luminosity  of 10~fb$^{-1}$.  After selection cuts, for $M_H =
110$~GeV, a small signal is observed above the background, which comes
mainly from $\ee\to W^+W^-$, $Z^0 Z^0$, and $qq(\gamma)$. This signal
is greatly enhanced, as shown in Fig.~\ref{fourjet}b, by requiring that
at least one of the jets forming the Higgs signal peak come from a
tagged $b$ quark. The  vertex-tagger is assumed to have the
conservative performance  $\epsilon_{b \bar{b}}=50\%$, $\epsilon_{c
\bar{c}}=2.5\%$, $\epsilon_{c \bar{s}}=0.3\%$, and $\epsilon_{q
\bar{q}}=0.1\%$, where the numbers give the efficiency for tagging a
particular quark combination. The importance of $b$ tagging is even
greater as one moves up in mass.

The missing energy topology  shown in  Fig.~\ref{topo}c can arise
either from the $ZH$ process, with $(Z^0 \rightarrow \nu \bar{\nu})(H^0
\rightarrow b \bar{b})$,   or from $WW$ fusion.  The resultant  events
will have  large missing energy, transverse momentum, and mass, plus
the presence of acoplanar jets. This distinctive signature offsets the
loss of the $Z$-mass constraint. The last topology of Fig.~\ref{topo}d
can be isolated by tagging two $\tau$ leptons either from their one- or
three-prong decays recoiling against a reconstructed $Z^0$ decaying
into $q \bar{q}$. Using a kinematically-constrained fit the missing
neutrinos from the $\tau$ decay can be taken into account.

The examples given so far rely primarily on the large branching ratio
for  $H^0 \rightarrow b \bar{b}$.  As the $H^0$ gets heavier, other
decay modes begin to become important.  For example, for $M_H =
140$~GeV, $Br(H^0 \rightarrow W^* W) \simeq 45\%$. The mode $(Z^0
\rightarrow q \bar{q})(H^0 \rightarrow W^* W)$ has been
investigated~\cite{hildreth} by demanding a six-jet event with  a
reconstructed $Z^0$ hadronic decay, one jet pair reconstructing to
$M_W$, and the last pair peaking at $m < M_W$, depending on $M_H$.
Similarly, for $M_H = 160$~GeV, the decay $H^0 \rightarrow W^+W^-$
dominates, and for production via fusion, $e^+ e^- \rightarrow H \nu
\bar{\nu}$, the result is an acoplanar pair of reconstructed $W$ bosons
and a total visible mass peaking at the $H$ mass which is expected to
be well above background without the need for $b$-tagging.

\begin{figure}[htb]
\leavevmode
\centerline{\epsfysize=3truein \epsfbox{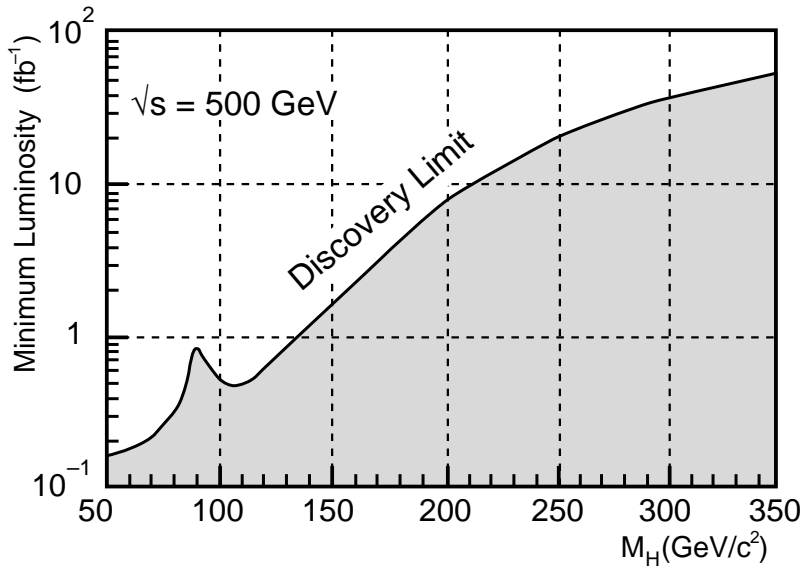}}
\centerline{\parbox{5in}{\caption[*]{Minimum luminosity needed to
discover a SM Higgs boson at a center-of-mass energy of 500 GeV.}
\label{minlum}}}
\end{figure}

Thus, studies have shown that with a detector similar to the
LEP/SLC detectors, with $b$-quark vertex tagging provided by silicon
microvertex detectors using present technology, an intermediate-mass SM
Higgs boson cannot escape detection at an $e^+e^-$ linear collider at
$\sqrt{s} = 500$~GeV. Figure~\ref{minlum} shows an
estimate~\cite{janot} of the minimum luminosity required to discover at
the $5 \sigma$ level a SM Higgs boson of a particular mass.  An
integrated luminosity of only 5~fb$^{-1}$ would be adequate to cover
the entire intermediate mass range, while 20~fb$^{-1}$ would allow a
reach in mass up to about $\sqrt{s}/2$.

Higher $\sqrt{s}$ would of course allow one to  probe for the existence
of much heavier Higgs bosons, mostly through $WW$ fusion and decay into
pairs of vector bosons.  At center of mass energies of 1--2~TeV,
different backgrounds such as $e^+e^-W^+W^-$ and $e^+ \nu_e W^- Z^0$
need to be addressed. Even though there exist older
studies~\cite{slac329,old} of searches for heavy Higgs bosons  in
1--2~TeV $e^+e^-$ collisions, these investigations need to be updated
with more detailed simulations and to keep abreast of theoretical
developments~\cite{djouadi} regarding backgrounds and the decay of
heavy Higgs bosons.  For very large Higgs boson masses, this study
becomes a part of the general problem of studying $WW$ scattering at
high energies; we discuss this problem in some detail in Section 7.2.

\subsection{Minimal Supersymmetric Standard Model Higgs}

In the framework of the MSSM, production of the lightest $CP$-even
state $h^0$ is similar to that of the SM higgs boson.  It is produced
by the $Zh$ and $WW$ fusion processes just described, with $h^0$
replacing $h_{SM}$.  In addition, new modes of production also open up,
involving the heavy Higgs bosons $H^0$ and $A^0$.  The various
production processes for  $h^0$ and $H^0$ in $\ee$ annihilation depend
on the mixing angles $\alpha$ and $\beta$ as indicated in
Table~\ref{tablea}.  Notice the sum rule: One process in each line
always has a substantial rate. As $m_A\to \infty$ in the MSSM,
$\cos(\beta-\alpha)\to 0$, and only the processes in the left-hand
column of the table occur.  In this limit, the rates of the $h^0$
production processes are identical to those for the Standard Model
Higgs.

\bigskip 
\begin{table}[ht]
\centering
\caption{Dependence of the cross section on Higgs boson mixing angles
    for various Higgs boson production processes in the MSSM.}
\label{tablea}
\bigskip 
\begin{tabular}{|c|c|} \hline \hline
$\sin^2(\beta-\alpha)$ & $\cos^2(\beta - \alpha)$\\ \hline
 $h^0Z^0$ & $H^0Z^0$ \\
 $h^0 \nu \bar{\nu}$ & $H^0 \nu \bar{\nu}$ \\
 $H^0A^0$\hfil       & $h^0A^0$ \\ \hline \hline
\end{tabular}
\end{table}
\bigskip 

\begin{figure}[htb]
  \leavevmode
\centerline{\epsfysize=3.1truein \epsfxsize=4in\epsfbox{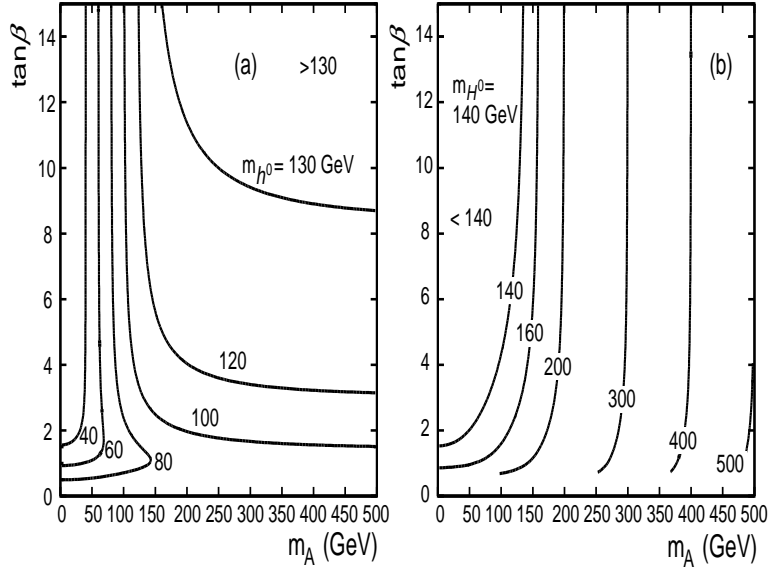}}
\centerline{\parbox{5in}{\caption[*]{Contours of the values of (a)
$M_{h}$ and (b) $M_H$ in the MSSM for $m_t = 180$~GeV, assuming
$m_{SUSY} = 1$ TeV and maximal top squark mixing.}
 \label{hmass}}}
 \end{figure}

The phenomenology of the SUSY Higgs bosons varies in a smooth way as
$M_A$ is varied.   The contours of Higgs mass over the MSSM parameter
space are shown in Fig.~\ref{hmass} \cite{HHHMMM}. If $M_A < 125$~GeV,
then $A^0$ and $h^0$ are close in mass; if $M_A > 125$~GeV, then $M_A
\simeq M_H$ and we begin to approach the large $M_A$ limit.   However,
if $M_A < 230$~GeV, then all of the MSSM Higgs bosons should still be
observable at the NLC with $\sqrt{s} = 500$~GeV.   If $M_A > 230$~GeV,
it is possible that, at the $\sqrt{s}=500$~GeV stage of the NLC, only
the lightest SUSY Higgs $h^0$ may be observable {\it and} it would have
production rates virtually indistinguishable from those of a minimal
Standard Model Higgs boson. The remaining Higgs states could be
discovered at higher $\sqrt{s}$, and there are also precision tests
available, to be described later, which could distinguish a Standard
Model Higgs from a supersymmetric Higgs. However, since the $h^0$ will
result in decay topologies similar to that of the SM Higgs, if this
lightest $h^0$ is not observed, then the MSSM is categorically ruled
out.  If the $h^0$ is not seen below 150 GeV, the more general
supersymmetry models incorporating grand unification are also excluded.

In general, for an intermediate mass boson, the branching ratio for a
minimal SM Higgs boson to $b\bar{b}$ is very close to that of the light
$CP$-even state $h^0$.  For large values of $\tan\beta$, the other
neutral MSSM Higgs bosons decay predominantly into $b\bar{b}$, with a
3\% branching ratio into $\tau^+\tau^-$.  This simple pattern becomes
more complex for smaller values of $\tan\beta$~\cite{djouadi} with
modes such as $H^0,A^0 \rightarrow t \bar{t}$ (for $M_{A^0} \approx
M_{H^0} > 2m_t$) and $H^0 \rightarrow h^0 h^0$ and $A^0 \rightarrow
Zh^0$ (for $M_{A^0} \approx M_{H^0} < 2m_t$) becoming more important.
Despite more complicated cascade decays into lighter Higgs states, the
bottom line remains clear: There should be plenty of jets from
$b$-quarks to tag and elucidate signals. A case in point is the
spectacular decay of $H^0 A^0$ into six $b$ jets $H^0 \rightarrow
h^0h^0, \thinspace h^0 \rightarrow b\bar{b}$. Most final states decay
into at least four $b$-quark jets, underlining the overwhelming
importance of $b$-tagging in experimental studies. An interesting case
deserving further study in simulations is the heavy Higgs decay into
$t\bar{t}^{(*)}$.

\subsubsection{Experimental Studies}

\begin{figure}[htb]
 \leavevmode
\centerline{\epsfysize=3.0truein \epsfxsize=4in\epsfbox{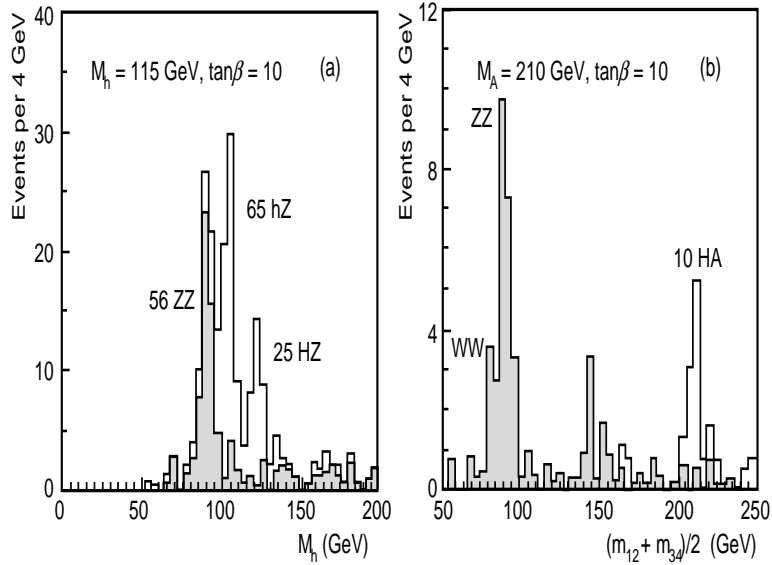}}
\centerline{\parbox{5in}{\caption[*]{(a) Jet masses recoiling from a
reconstructed $Z^0$ after $b$ tagging for $\sin^2(\beta - \alpha)
\simeq 0.5$, and integrated luminosity of 10$^{-1}$~fb; (b) average of
the two jet-pairs closest in invariant mass in an identified
$b\bar{b}b\bar{b}$ final state with $M_A = 210$~GeV (adapted from
\protect\cite{janot}).}
\label{mssmexp}}}
\end{figure}

All of the topologies of Fig.~\ref{topo} can be explored in the MSSM
with the $h^0$ taking the role of the Standard Model Higgs boson. A
repetition of the analyses described  earlier  either would observe a
single $h^0$ similar to that of the SM Higgs, or, if $\sin^2(\beta -
\alpha) \simeq 0.5$, would observe both the $h^0$ and $H^0$ states as
shown in Fig.~\ref{mssmexp}a~\cite{janot}.

\begin{figure}[htb]
\leavevmode
\centerline{\epsfxsize=3in \epsfysize=3.0in \epsfbox{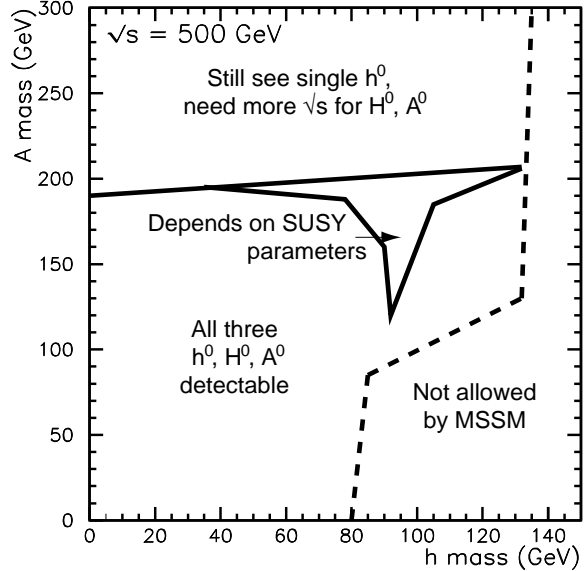}}
\centerline{\parbox{5in}{\caption[*]{Regions of simultaneous
detectability of $h^0$, $H^0$, and $A^0$ at center-of-mass energy of
500~GeV \protect\cite{Rosiek}.}
\label{rosiek}}}
\end{figure}

An identical preselection for a four-jet topology can be used to search
for $HA$. We require in this case that all four jets in $(H^0
\rightarrow b\bar{b})(A^0 \rightarrow b\bar{b})$ be tagged as $b$ jets.
For large enough $M_A$ and $\tan\beta$, we have $M_A \simeq M_H$, and
we can demand that the two jet-pair masses of the possible combinations
to be close to equal. Then a signal as shown in Fig.~\ref{mssmexp}b is
possible. Since all the neutral SUSY Higgs decay into $\tau^+ \tau^-$
at some level, it is possible to observe all three MSSM states in a
single analysis by looking at the invariant mass of both the
$\tau^+\tau^-$ and $q\bar{q}$ in the $\tau^+\tau^- q\bar{q}$ final
state.  In Fig.~\ref{rosiek}, we show the regions of plane of $M_A$
versus $M_h$ in which is it possible to observe all three neutral Higgs
states of the MSSM at a linear collider with $\sqrt{s} = 500$~GeV, or,
conversely, the region where only the $h^0$ can be 
observed~\cite{Rosiek}.

In theories with multiple Higgs doublets such as the MSSM, searches for
the charged Higgs bosons $H^{\pm}$ are also important.  An $e^+e^-$
collider should be able to better resolve the hadronic decays of the
$H^{\pm}$ compared to a hadronic collider, and all of the expected
final states $H^+H^- \rightarrow c\bar{s}\,\bar{c}s$,
$t\bar{b}\bar{t}b$, plus the easier topologies of $c\bar{s}\tau^- \nu$
and $\tau^+ \nu \tau^- \bar{\nu}$ should be observable. A detailed
simulation analysis~\cite{sopczak} at $\sqrt{s} = 500$~GeV with an
integrated luminosity of 10~fb$^{-1}$, which makes heavy uses of
$b$-tagging, has shown that one can establish a signal over the $e^+e^-
\rightarrow t\bar{t}$ and $W^+W^-$ background in all of these channels.
The  results show a detection sensitivity for charged Higgs bosons up
to about 210~GeV at $\sqrt{s} = 500$~GeV, independent of decay mode.
These conclusions can be strengthened further by adding the  decay $t
\rightarrow b H^+$. The bottom line is that if the $A^0$ is lighter
than about 200~GeV, then the $H^{\pm}$ should be observable also.

If one is considering SUSY Higgs bosons, one should allow for the
possibility of their decay into other SUSY particles.  As an example,
it is possible for $h^0$ or $H^0$ to decay into a pair of the lightest
neutralinos (mixtures of fermionic partners of the $Z$ and $\gamma$)
that would be stable and neutral.  The result would be an ``invisible"
decay of the Higgs.  This topology can be identified in the same way as
described previously by studying the missing mass in a $hZ$ or $HZ$
event where the $Z$ decays into a pair of electrons or muons.  Both
missing mass resolution and backgrounds (smaller direct $Z \nu
\bar{\nu}$ cross section) improve with lower center-of-mass energy, and
such an analysis would benefit from running at $\sqrt{s} \simeq
300$~GeV. The possibility of SUSY Higgs decays into other SUSY
particles~\cite{djouadi} such as SUSY partners of quarks and leptons
warrant further experimental simulations.

\subsection{Determination of Properties of Higgs Bosons}

\subsubsection{Mass Measurement}

\begin{figure}[htb]
\leavevmode
\centerline{\epsfysize=3truein \epsfbox{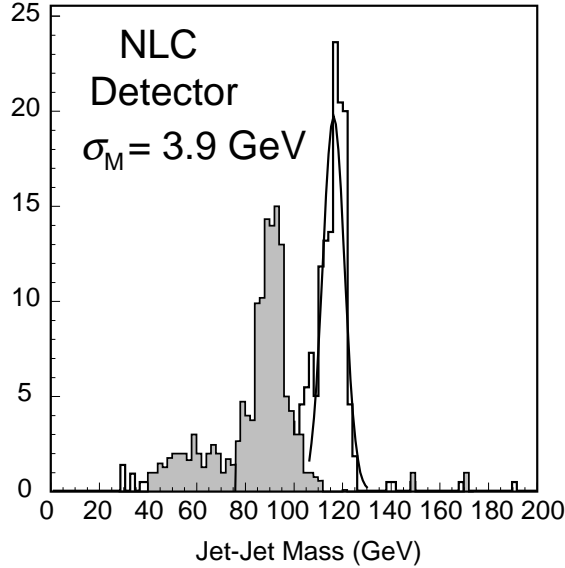}}
\centerline{\parbox{5in}{\caption[*]{Mass resolution of $\sigma_M
\simeq 3.9$~GeV in the jet-jet invariant mass for jets from Higgs decay
assuming the performance of a NLC detector (see text) for a simulated
signal (open histogram) of a 120~GeV SM Higgs boson and all known
backgrounds (shaded histogram) at 300~GeV with 10~fb$^{-1}$.}
\label{massres}}}
\end{figure}

To measure the mass of one or more of the possible Higgs bosons, one
would probably optimize the running conditions to have smaller
center-of-mass energy , to improve momentum resolution and to go to the
peak of the cross section.  For an intermediate mass Higgs,
$\sqrt{s} = 200$--300~GeV is appropriate.  Under these conditions, one
can precisely measure the recoil mass in $e^+e^- \rightarrow Z^0 h^0$
events opposite to the reconstructed leptonic decay $Z^0 \rightarrow
e^+e^-$ or $\mu^+ \mu^-$. Other modes, such as the four-jet topology,
can also be employed. In all cases,  kinematic fitting would be used to
constrain the leptons or jets from a $Z^0$ to reconstruct to $M_Z$ and
to allow for missing $E_{\gamma}$ along  the beam axis.  A typical
jet-jet mass resolution of $\sigma_M \simeq 2.0$~GeV can be achieved
assuming the excellent momentum resolution of $\sigma_{p_t}/p_t = 1
\times 10^{-4} \oplus 0.1\%$ envisaged for the JLC detector~\cite{JLC}.
For our NLC detector design, we could achieve $\sigma_M \simeq
3.9$~GeV, as shown in Fig.~\ref{massres}. The differences between
detector designs are much smaller when kinematic constrained fitting is
included in the analysis. The JLC-type detector has been estimated to
provide a estimated precision on the Higgs mass of approximately 0.1\%
for $\sqrt{s} = 300$~GeV, $\int L \cdot dt =30$~fb$^{-1}$, and a 2.0\%
full width beam energy spread~\cite{kawagoe}. On the other hand, the 
NLC-type detector gives $\Delta M_h \leq 160$~MeV up to $M_h \simeq
160$~GeV with 50~fb$^{-1}$ at $\sqrt{s} = 500$~GeV.

Once $M_H$ is known precisely, it can be used as an input to check the
experimental measurements of branching ratios and the production cross
section with Standard Model predictions.  Its value can also be
compared to the theoretical value obtained from precision electroweak
measurements, combined with the measurements of $M_W$ and $m_{top}$
expected from a linear collider.

\subsubsection{Cross Section Measurement}

Measuring the production cross section of the Higgs provides one way
of disentangling a SM Higgs boson from a SUSY Higgs, if one can 
observe the cross section suppression due to mixing
\[
\sigma(e^+e^- \rightarrow
h^0 Z^0) = \sin^2(\beta -\alpha) \cdot \sigma(e^+e^- \rightarrow
h^0_{SM} Z^0).
\] 
A distinct advantage of $e^+e^-$ linear colliders over hadronic
colliders is the ability to almost unambiguously tag the $Z^0$ in $hZ$
events and being able to study all of the decays $h^0 \rightarrow X$
with small backgrounds.  Both total absolute cross sections  and
individual Higgs branching ratios  can then be measured. By using
leptonic decays of the $Z^0$ and kinematical fitting, the absolute
production cross section can be measured~\cite{kawagoe} with a
precision of 7\% with an integrated luminosity of 30~fb$^{-1}$ and to
5\% with 50~fb$^{-1}$.  However it should be kept in mind that for a
large area of SUSY parameter space, the SUSY Higgs cross section is
less than 10\% different from the SM Higgs cross section.

\subsubsection{Spin-Parity and CP Determination}

In principle, the spin and parity of the Higgs boson can be found by
studying both the production angular distribution of the Higgs and also
the resulting angular distribution of the decay products of the $Z^0$
in its rest frame in $HZ$ events.  In the high energy limit, Table
\ref{tableb} shows the expected angular distribution of scalar (\eg,
$h^0$, $H^0$) and pseudoscalar (\eg, $A^0$) Higgs bosons. In the table,
$\theta$ is the production angle of the Higgs boson and $\theta_*$ is
the polar angle of the fermions from $Z$ decay measured in the $Z$ rest
frame. In practice, however, a purely CP-odd Higgs boson couples to
$ZZ$ only at the one-loop level, and then the $ZA$ cross section would
be very small.  For a Higgs boson that is a mixture of CP-even and
CP-odd components, the production would mainly be sensitive to the
CP-even part, and the angular distributions would not reveal the CP-odd
component~\cite{CPgunion}.

\bigskip 
\begin{table}[ht]
\centering
\caption{Expected angular distributions for Higgs bosons
with different spin-parity.}
\label{tableb}
\bigskip 
\begin{tabular}{|c|c|c|} \hline \hline
       &  Scalar, 0$^{++}$  & Pseudoscalar, 0$^{-+}$ \\ \hline
 $d\sigma(e^+e^- \rightarrow H Z^0)/d\cos\theta$
       & $\propto \sin^2\theta$ 
       & $\propto (1 - \sin^2\theta)$ \\
 $d\sigma(Z^0 \rightarrow f\bar{f})/d\cos\theta_*$
       & $\propto \sin^2\theta_* $ 
       & $\propto (1 \pm \cos\theta_*)^2$  \\ \hline \hline
\end{tabular}
\end{table}
\bigskip 

A much better way to determine the Higgs' $CP$ character is with
polarized $\gamma \gamma$ collisions~\cite{CPgunion2,gamgunion}. In
this technique, which we will discuss in detail in Section 9.2, the
Higgs boson is produced as an $s$-channel resonance.  Then it is
possible to study the angular correlations of the decay products of the
resonance in decays such as $H \rightarrow \tau^+ \tau^-$ and $t
\bar{t}$.  By spin analyzing the subsequent decays $t \rightarrow b
\ell \nu$ for top quarks from heavy Higgs boson decay, and $\tau
\rightarrow \pi \nu$ or $\tau \rightarrow \rho \nu$ for $\tau$'s from
intermediate mass Higgs decay, a $CP$-even state and $CP$-odd state can
be distinguished~\cite{CPgunion,stong}. This $CP$ state separation is
much better in the angular correlations between top quark decay
products.

\subsubsection{Branching Ratio Measurements}

\begin{figure}[htb]
\leavevmode
\centerline{\epsfysize=3truein \epsfxsize=4.5in  \epsfbox{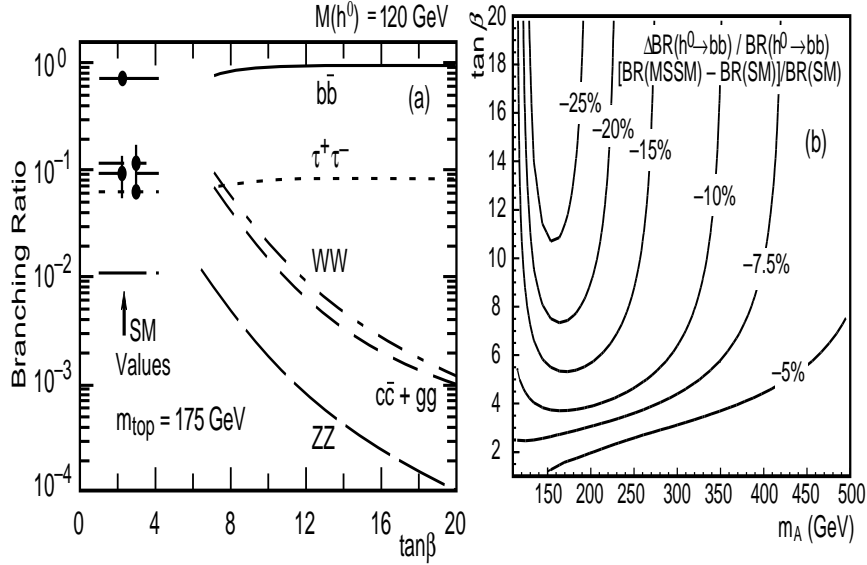}}
\centerline{\parbox{5in}{\caption[*]{(a) Expected errors on the
Standard Model branching fractions compared to those predicted for a
120~GeV $h^0$ MSSM Higgs boson; (b) contour lines of fractional
deviation of $Br(h^0 \rightarrow b\bar{b})$ ($m_t = 175$~GeV, $m_{SUSY}
= 1$~TeV).}
\label{brh}}}
\end{figure}

The measurement of the branching ratios of any observed Higgs boson is
an essential ingredient to understand the nature of the symmetry
breaking and to make predictions about other aspects of the Higgs
sector.  This is  especially when only a single neutral Higgs is
observed, which might be either the Standard Model Higgs or the
lightest neutral Higgs from SUSY. The clean environment in $e^+e^-$
annihilation permits one to tag a $Z^0$ in one hemisphere, and then
observe the decay $h^0 \rightarrow X$ in any  decay mode in the
opposite hemisphere. An example of such an analysis~\cite{hildreth} at
$\sqrt{s} = 400$~GeV, simulating an SLD-like detector, first identifies a
$Z^0$ in a $HZ$ event and then considers those decays where the
recoiling Higgs decays into jets.   The Higgs decays to two jets can be
separated by flavor by counting the number of tracks with a
significantly large impact parameter: $b_{norm} = b/\sigma_b > 3$,
where $b$ is the impact parameter and $\sigma_b$ is the error on $b$.
The decay $h \rightarrow W W^{(*)}$ is identified by demanding that the
event be consistent with containing six jets, and that a jet pair with
invariant mass close to the $W$ mass is found. With 50~fb$^{-1}$ of
data, $Br(h \rightarrow b\bar{b})$ can be measured to a statistical
precision of 7\%, and branching ratios into $WW^*$ and $(c\bar{c} +
gg)$ to 24\% and 39\% respectively. These relative errors are shown
superimposed upon the Standard Model values in Fig.~\ref{brh}a.  The
figures also shows the variation in branching ratios that one would
expect from the variation of $\tan\beta$.  In the MSSM, it is very
difficult to arrange such a large variation in $\tan\beta$ without a
compensatory variation in $\alpha$, but the figure shows the utility of
this measurement in Higgs studies in a more general context. For the
comparison of the Standard Model Higgs boson to the MSSM, one should
consult Fig.~\ref{brh}b, where the branching ratio of a light Higgs
boson into $b\bar b$ is compared for these two possibilities over the
MSSM parameter space.

\begin{figure}[htb]
\leavevmode
\centerline{\epsfysize=2.5truein \epsfbox{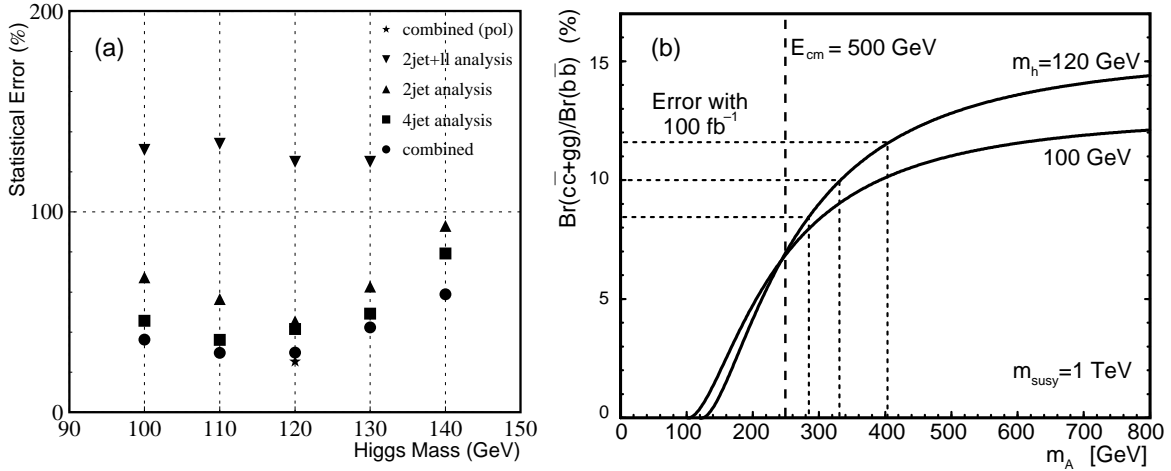}}
\centerline{\parbox{5in}{\caption[*]{(a) Statistical error with
50~fb$^{-1}$ of data on $Br(h \rightarrow c\bar{c}+gg)/Br(h \rightarrow
b\bar{b})$ as a function of Higgs mass; (b) implications for estimation
of $A^0$ mass.}
\label{brh2}}}
\end{figure}

An interesting quantity~\cite{kamoshita} is the ratio of branching
ratios to $c\bar c$ versus $b \bar b$.  At tree level, 
\[
\frac{Br(h \rightarrow c\bar{c})}{Br(h \rightarrow
b\bar{b})} \approx \frac{m_c^2}{m_b^2} \cdot \left(\frac{M_h^2 -
M_A^2}{M_A^2 - M_Z^2}\right)^2,
\]
where $m_c$ and $m_b$ are the $c$ and $b$ quark masses respectively.
(We should note that this formula can receive substantial radiative
corrections in some regions of the MSSM parameter space.) If the
branching ratios indicated are measured along with $M_h$, it is
possible to estimate $M_A$.  In a simulation study of this
measurement~\cite{nakamura} at $\sqrt{s} = 300$~GeV, $HZ$ events are
selected for each decay mode of the $Z^0$, and the decay mode of the
Higgs is determined using three-dimensional impact parameters.  Flavor
tagging is performed by selecting charged tracks that satisfy
$b/\sigma_b \geq 2.5$ and counting the number in each jet from the
Higgs decay. For $M_H = 120$~GeV, 50~fb$^{-1}$ of data, and assuming
90\% polarization of the electron beam, the statistical error on the
ratio of branching ratios $Br(h \rightarrow c\bar{c}+gg)/Br(h
\rightarrow b\bar{b})$ would be 20.4\%, varying with the Higgs mass as
shown in Fig.~\ref{brh2}a. This does not include a substantial
systematic uncertainty from  $m_c/m_b$, which we believe will be
reduced in the next few years through lattice gauge theory
calculations. Then, as shown in Fig.~\ref{brh2}b, this measurement
could be sensitive to $A^0$ masses up to 400~GeV, well above the maximum
kinematic reach of a 500 GeV collider.  Observation of the $A^0$ in
this way would help to plan the next step in energy.

For the decay $h\to \gamma\gamma$, it is difficult to measure the
branching ratio at an $\ee$ collider because this mode is relatively
rare.  However, it should be possible to measure the absolute partial
width $\Gamma(h\to \gamma\gamma)$ by exploiting the ability of an
electron collider to be run as a $\gamma\gamma$ collider.  This
measurement is discussed in Section 9.1.

\subsubsection{Determination of Higgs Total Width}

In the preceding sections, we have indicated many ways in which
measurements at the NLC can distinguish between the Standard Model
Higgs and the light Higgs $h^0$ of the MSSM.  There are also a number
of quantities at the LHC which are sensitive to this difference, as
outlined in \cite{higgs2}.  However, to obtain the complete set of
partial width of the Higgs boson in a model-independent way,
measurements from LHC must be combined it with data both from $e^+e^-$ and
$\gamma \gamma$ collisions at the NLC.  A possible procedure is the
following. First determine $Br(b\bar{b})$ from $Zh$ events and combine
with $\sigma(WW \rightarrow h) \cdot Br(b\bar{b})$ both measured at the
NLC to obtain the $WWh$ coupling.  Alternatively, a measurement of
$\sigma(e^+e^- \rightarrow Zh)$ at the NLC gives the $ZZH$ coupling and
the ratio of the $WWh$ and $ZZh$ couplings given by $M_W^2/M_Z^2 = \cos^2
\theta_W$ also gives the $WWh$ coupling. This coupling and a
measurement of $\sigma(Wh) \cdot Br(\gamma \gamma)$ at the LHC can be
used to determine $Br(\gamma \gamma)$.  Therefore one can combine
$Br(b\bar{b})$ with the $\gamma \gamma$ collider measurement of
$\sigma(\gamma \gamma \rightarrow h) \cdot Br(b\bar{b})$ to obtain
$\Gamma (h \rightarrow \gamma \gamma)$. We can then finally compute the
total width $\Gamma^{\rm tot}_h = \Gamma(h \rightarrow \gamma
\gamma)/Br(\gamma \gamma)$ and $\Gamma(h \rightarrow b \bar{b}) = Br(b
\bar{b}) \Gamma^{\rm tot}_h$. A simpler route exists using only
$e^+e^-$ data when the Higgs boson is heavy enough that the branching
ratio to $WW^*$ is relatively large, so that it can be measured
accurately.  In this case, we can simply measure $Br(h \rightarrow
WW^*)$ and infer $\Gamma^{\rm tot}_h = \Gamma(h \rightarrow
WW^{(*)})/Br(WW^{(*)})$. Although the accumulation of errors may be
significant, the basic point is that data from all three colliders or
from the NLC alone can be combined to complete a {\em
model-independent} determination of the properties of a light Higgs
boson.

\subsection{Summary}

From the studies described, the discovery of a Standard Model
intermediate-mass Higgs boson at an $e^+e^-$ linear collider at
$\sqrt{s} = 500$~GeV can be easily achieved with an integrated
luminosity of only 10~fb$^{-1}$. Such a machine allows the detection of
{\it at least} the lightest MSSM Higgs $h^0$, if not all three SUSY
neutral states.  If the lightest Higgs is not observed, then not only
is the Minimal Supersymmetric Standard Model ruled out, but also the
general idea that the Higgs boson is a fundamental particle up to the
unification scale is called into question. If the $A^0$ is not
kinematically accessible at $\sqrt{s} = 500$~GeV, then the measurement
of $h^0$ branching ratios can give hints of the values of $M_A$ and
$\tan\beta$ and tell us where to go next in energy. For definitive
evidence, $\sqrt{s} > 2M_A$ would still be needed.  For Higgs bosons
above the intermediate mass range, their decay into pairs of vector
bosons makes them straightforward to detect at the NLC as at the LHC;
however, more up-to-date experimental simulations are needed. Just as
important as its ability to discover the Higgs boson is the ability of
the linear collider to make precision measurements of the properties
and couplings of a Higgs boson.  Even if the Higgs boson is discovered
earlier at LEP2 or at the LHC, we will need the NLC to learn its
complete story.

\clearpage
\section{Supersymmetry}

To build a complete unified theory with a fundamental Higgs boson, one
is led to introduce supersymmetry, the symmetry between fermions and
bosons in space-time.  Supersymmetry is the only known principle with
sufficient structure to allow the construction of grand unified
theories in which fundamental scalar particles can  naturally be very
light compared to the unification scale.  Supersymmetric unification
models explain the values of the Standard Model coupling constants as
measured at $Z^0$ energies, and also incorporate a mechanism of
electroweak symmetry breaking associated with the heavy top quark.
General reviews of supersymmetric models can be found in \cite{Nillrev,
HKrev, HaberTASI, Murarev}. Supersymmetry also offers the more
speculative but tantalizing possibility of a connection between
phenomena observable at collider energies and string theory and other 
profound mathematical theories of the fundamental forces
\cite{PeskYuk,Arno}.

In this section, we will examine the manner in which supersymmetry
(SUSY) might manifest itself at a 0.5-1.0 TeV $\ee$ Linear Collider
(NLC). Our discussion here is part of a broader, and continuing,
investigation. At present, our study is being carried out within the
supersymmetry scenario based on the minimal supergravity model with
gauge coupling unification and radiative electroweak symmetry breaking
(SUGRA). We have calculated most of the relevant cross sections and
angular distributions for the production and decay of supersymmetric
particles, and we include a report of them as an appendix
\cite{CSECTANG}. Because of the difficulty of knowing where and how
supersymmetry will manifest itself,  we must study the phenomenology of
supersymmetry over a wide range of its parameters.  In this study, we
have chosen five points in the parameter space of the SUGRA model which
illustrate qualitatively different possibilities for the spectrum of
new particles \cite{CSECTANG}.  Because of the power of the
experimental tools offered by the NLC, our goals are much more
ambitious than simply to discover the existence of supersymmetry. We
would like to measure the masses of supersymmetric particles with
precision, and determine the underlying values of the basic parameters
of the theory.  In the most optimistic scenario, the extrapolation of
these parameters to the unification scale would give evidence into the
details of the fundamental unified model \cite{JLC1}.

\begin{figure}[htbp]
 \leavevmode
\centerline{\epsfxsize=5in\epsfbox{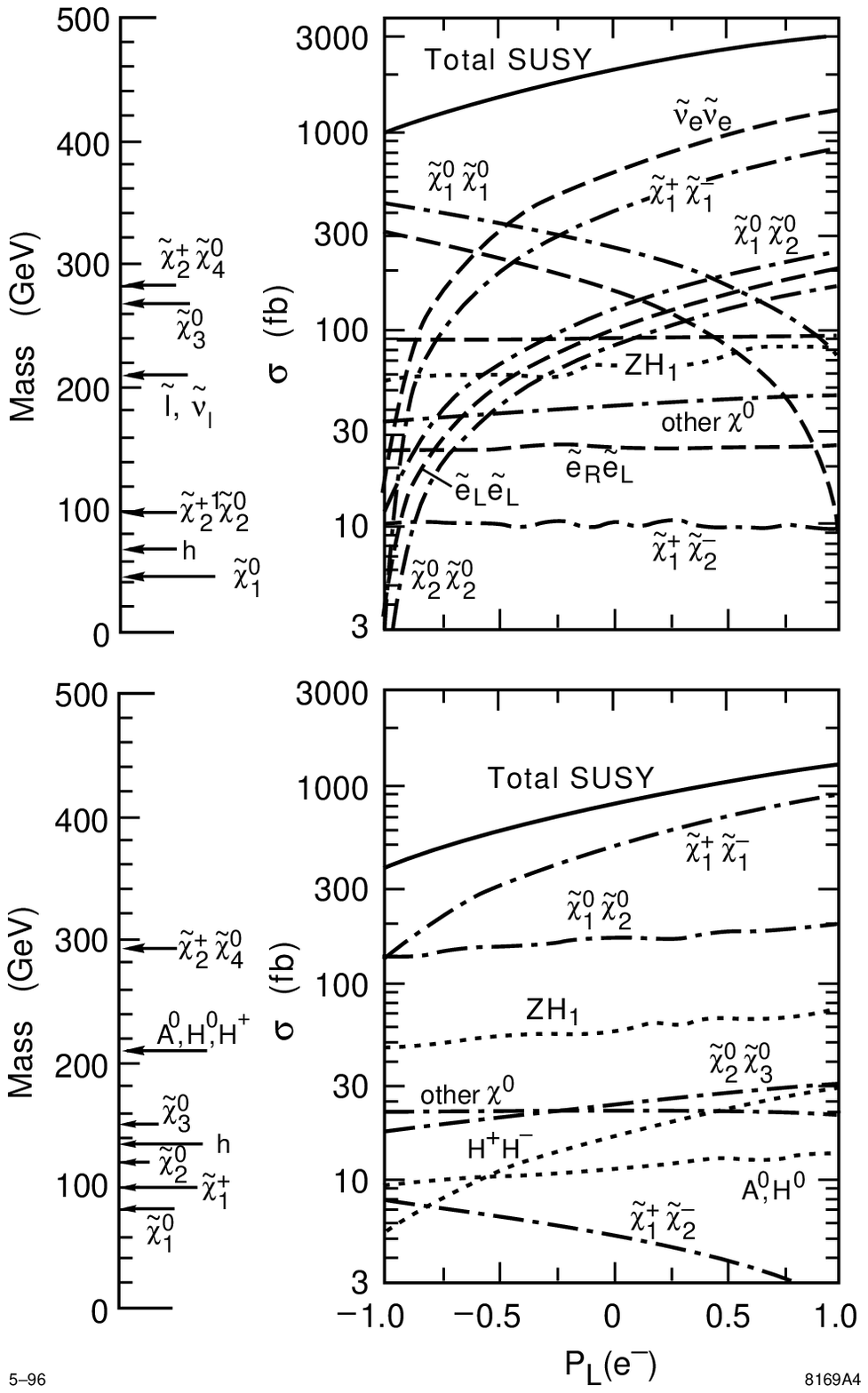}}
\centerline{\parbox{5in}{\caption[*]{The cross section of
Supersymmetric particle production at E$_{\rm cm}$ = 0.5 TeV as a
function of the electron polarization for points 3, 4 in our parameter
space.}
  \label{csect2a}}}
\end{figure}

\begin{figure}[htbp]
 \leavevmode
\centerline{\epsfxsize=4in\epsfbox{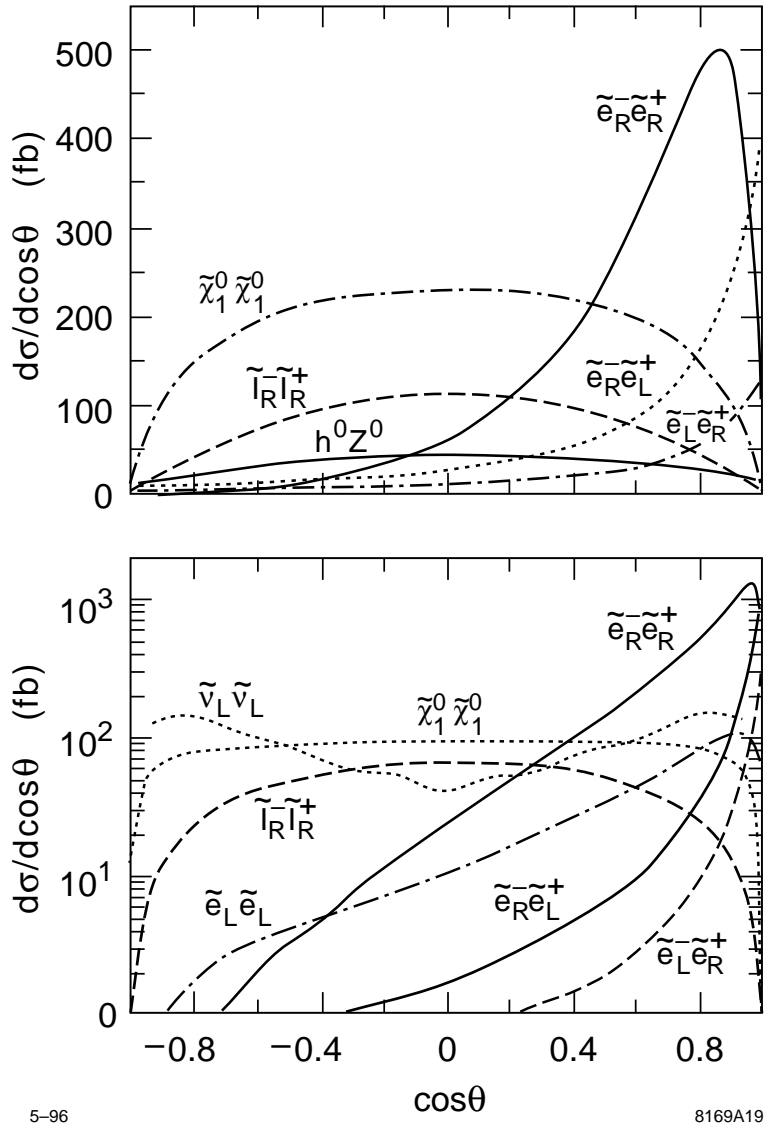}}
\centerline{\parbox{5in}{\caption[*]{The differential cross section of
supersymmetric particle production for an 80\% polarized right handed
electron at E$_{\rm cm}$ = 0.5 and 1.0 TeV for point 2 in our
parameters list.}
  \label{ssmang2}}}
\end{figure}

The number of supersymmetric particles is quite large. Hence, it is
typical that many of these particles will be produced in the same data
sample at a particular energy center-of-mass energy. One of the
properties of $e^+e^-$ linear colliders is that the electron can be
longitudinally polarized and its orientation can be changed at will.
Already, the SLC  provides an electron beam with 77\% polarization. We
expect that, in the future, this magnitude can be increased
substantially. The ability to  have  electron beams with high
longitudinal polarization is very useful to discriminate between the
various supersymmetric signals and to understand and remove the
Standard Model background processes \cite{HELAS}. This can be seen by
examining the standard model cross sections as a function of
polarization, shown already in Fig.~\ref{nlc500}, and comparing these
to the polarization-dependence of the cross sections for the
supersymmetric production processes, shown for two representative
points in Fig.~\ref{csect2a}.  Having a 90\% longitudinally polarized
electron in the right handed mode will reduce the production rate of
Standard Model background processes such as $W^+W^-$ pair production by
an order of magnitude while enhancing some of the supersymmetric
particle signals. The left-handed supersymmetric particles, like the
Standard Model background, are also suppressed, while the right-handed
ones are enhanced. This is seen in Fig. \ref{csect2a}. Hence, by varying
the polarization we can determine which of the supersymmetric particles
are giving us a particular signal. Polarization of the positrons could
give an additional advantage. For example, if we could collide  totally
right-handed electrons with totally left-handed positrons, all of the
standard model backgrounds from $\ee$ annihilation processes would
disappear, while the supersymmetric signals from $\tilde e^-_R \tilde
e^+_R$ production would remain. (Some background would also remain due
to two-photon reactions \cite{TWOGAMMA}.) An additional handle on our
ability to discriminate among the various supersymmetric signals is
their different angular distributions \cite{CSECTANG}. These
distributions for a typical case are shown in  Fig.~\ref{ssmang2}. A
third powerful discriminating tool is the adjustment of the
center-of-mass energy.  Once one has an estimate of the masses of the
lightest supersymmetric particles, it is advantageous to decrease the
energy of the collider so that only these lightest states are produced,
measure their properties at this lower energy, and then increase the
energy of the collider systematically.

\subsection{Supersymmetry Signals at the NLC}

In our study, we have generated signal and background processes using the
simulation program ISAJET \cite{ISAJET}. This program allows for both
electron and positron longitudinal polarization. The influence of the
detector is accounted for by smearing the generated momenta and
directions of the particles produced in the simulation with resolution
functions as described in Section 2.

The spectrum of supersymmetric particles for the five parameter sets
that we have chosen for detailed study are exhibited in
Table~\ref{supertable}. These spectra are computed consistently from a
supergravity model with $m_t = 180$~GeV.  The values of the underlying
parameters for these scenarios is given in \cite{CSECTANG}.

\begin{table}[htbp]
\centering
\caption[*]{Supersymmetric particle masses at five representative points
in the parameter space of phenomenological supergravity models.}
\label{supertable}
\bigskip
\begin{tabular}{|c|r|r|r|r|r|}\hline\hline
{\bf Parameter Set}         &  1\qquad &  2\qquad &  3\qquad &  4\qquad &
5\qquad \\
\hline
{\bf $\mbox{$\chi^0_1$}$} &  85.87 & 128.51 &  44.41 &  77.83 &  57.00 \\
{\bf $\mbox{$\chi^0_2$}$} & 175.24 & 257.05 &  96.73 & 115.29 & 111.21 \\
{\bf $\mbox{$\chi^0_3$}$} & 514.96 & 549.08 & 267.81 & 146.84 & 440.03 \\
{\bf $\mbox{$\chi^0_4$}$} & 523.78 & 556.07 & 284.18 & 292.40 & 460.12 \\
{\bf $\mbox{$\chi^{\pm}_1$}$} & 175.12 & 257.02 &  96.10 &  96.06 & 109.82 \\
{\bf $\mbox{$\chi^{\pm}_2$}$} & 522.82 & 555.72 & 282.52 & 292.45 & 457.09 \\
{\bf $h^0$      } &  84.86 &  92.24 &  68.82 & 130.58 & 102.15 \\
{\bf $H^0$      } & 766.47 & 698.24 & 389.39 & 201.72 & 619.21 \\
{\bf $A^0$      } & 762.32 & 693.30 & 381.75 & 200.00 & 616.45 \\
{\bf $H^{\pm}$  } & 765.70 & 697.05 & 388.81 & 214.75 & 620.52 \\
{\bf $\tilde q_L$} & 605.23 & 670.84 & 317.23 &1000.00 & 464.04 \\
{\bf $\tilde b_L$} & 516,58 & 621.43 & 272.31 &1000.00 & 384.70 \\
{\bf $\tilde t_1$} & 417.65 & 537.12 & 265.55 & 923.13 & 179.85 \\
{\bf $\tilde q_R$} & 605.23 & 670.84 & 317.23 &1000.00 & 464.04 \\
{\bf $\tilde b_R$} & 597.15 & 655.70 & 313.40 &1000.00 & 457.79 \\
{\bf $\tilde t_2$} & 547.20 & 655.21 & 328.15 &1099.35 & 495.72 \\
{\bf $\tilde l_L^-$} & 425.96 & 238.35 & 215.72 &1000.00 & 320.69 \\
{\bf $\tilde \nu_L$} & 421.43 & 230.16 & 206.63 &1000.00 & 314.65 \\
{\bf $\tilde l_R^-$} & 408.80 & 156.97 & 206.54 &1000.00 & 307.45 \\
{\bf $\tilde g$  } & 552.19 & 760.16 & 298.15 & 900.00 & 428.03 \\
\hline \hline
\end{tabular}
\end{table}

In each of the five cases, the lowest mass supersymmetric particle is 
the $\mbox{$\chi^0_1$}$.  In the class of models we discuss,
there is a conserved R-parity which implies that this particle is stable.
It then passes through the detector
without leaving a signal. This particle is in all cases  sufficiently 
massive  that it carries away significant missing energy. On the other
hand, Standard Model background processes can mimic this signal
because, as a result of their peaked differential cross-section in the
forward and backward directions \cite {CSECTANG}, many of the particles
in the final state go along the beam direction. Also it is possible
that neutrinos can carry away a sizable portion of the energy, or that
the event is mismeasured due to the detector resolution.

\begin{figure}[htbp]
\leavevmode
\centerline{\epsfxsize=4in\epsfbox{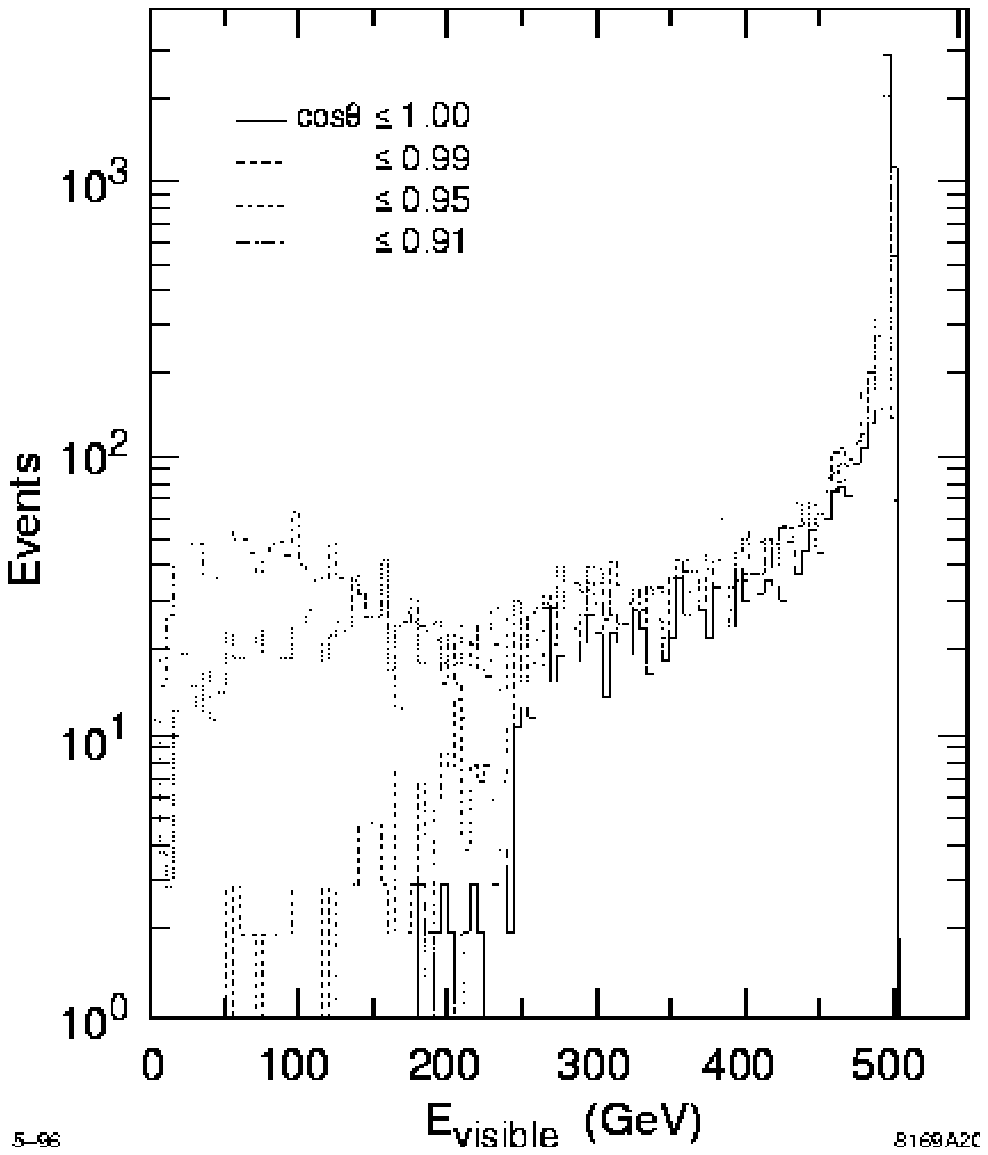}}
\centerline{\parbox{5in}{\caption[*]{The Visible Energy for Standard
Model final states $W^+W^-, Z^0Z^0, q \bar q$ for various calorimetry
coverage. We require that there be at least 3 particles in each
hemisphere.}
  \label{energy}}}
\end{figure}

In Fig. \ref{energy} we show the expected observed energy for the
Standard Model processes $e^+e^-\rightarrow W^+W^-, Z^0Z^0, q \bar q$
after a requirement that at least three particles be present in each
hemisphere. This allows us to avoid including the $e\nu W$ final state
and most of the events where the W's and Z's decay into leptons and
neutrinos; only a few events with $\tau$ in the final state remain. The
figure shows the effect of various assumptions about the calorimetric
coverage in cos($\theta$), to determine how much energy is lost in the
beam direction. We note that the tail of events with low visible energy
(E $\approx$ 100 GeV or less) begins to increase noticeably for
cos($\theta$) $<$ 0.97 or so. Hence good calorimetric coverage is
imperative if we want to use the low visible energy as a signal for
supersymmetric particles. We do not smear the particles in this plot to
determine the true loss of particles down the beam direction. The
detector design presented in this report has calorimetric coverage down
to cos($\theta$) = 0.99, which, as we will see, is sufficient to allow
us to see the signal due to supersymmetric particles.

\begin{figure}[htbp]
 \leavevmode
\centerline{\epsfysize=3in\epsfxsize=3.5in\epsfbox{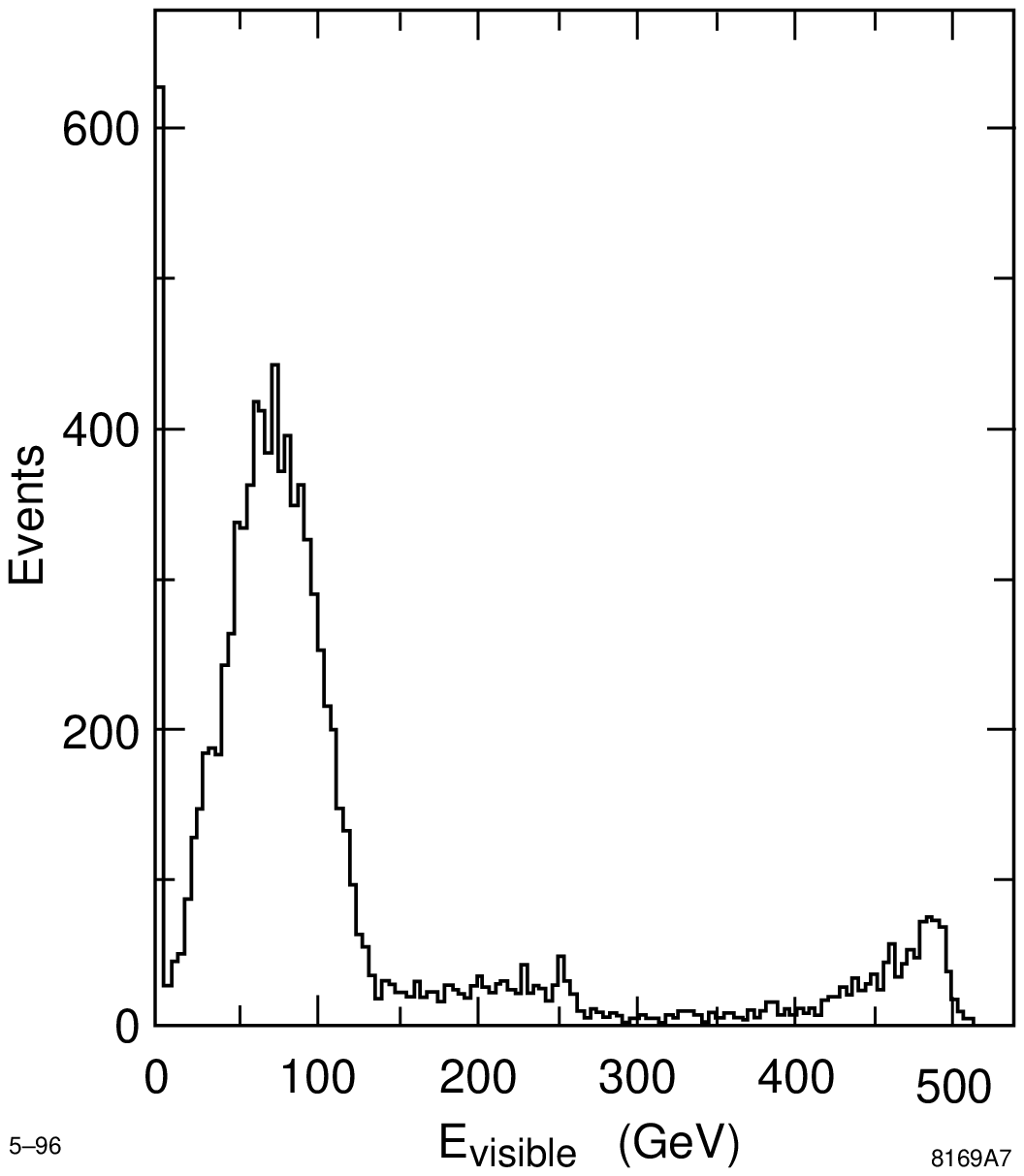}}
\centerline{\parbox{5in}{\caption[*]{The visible energy for the
supersymmetric processes defined by the SUGRA parameter set 4 after
detector resolution smearing. The peak at 0 is due to $\tilde\chi^0_1
\tilde\chi^0_1 $ and $\tilde\chi^0_2 \tilde\chi^0_1 \rightarrow \nu
\bar \nu \tilde\chi^0_1  \tilde\chi^0_1 $ processes. The broad small
peak at 500 GeV is due to $Z^0 h^0$, $H^+H^-$, and $H^0 A^0$ final
states.}
  \label{sig4}}}
\bigskip
 \leavevmode
\centerline{\epsfysize=3in\epsfxsize=3.5in\epsfbox{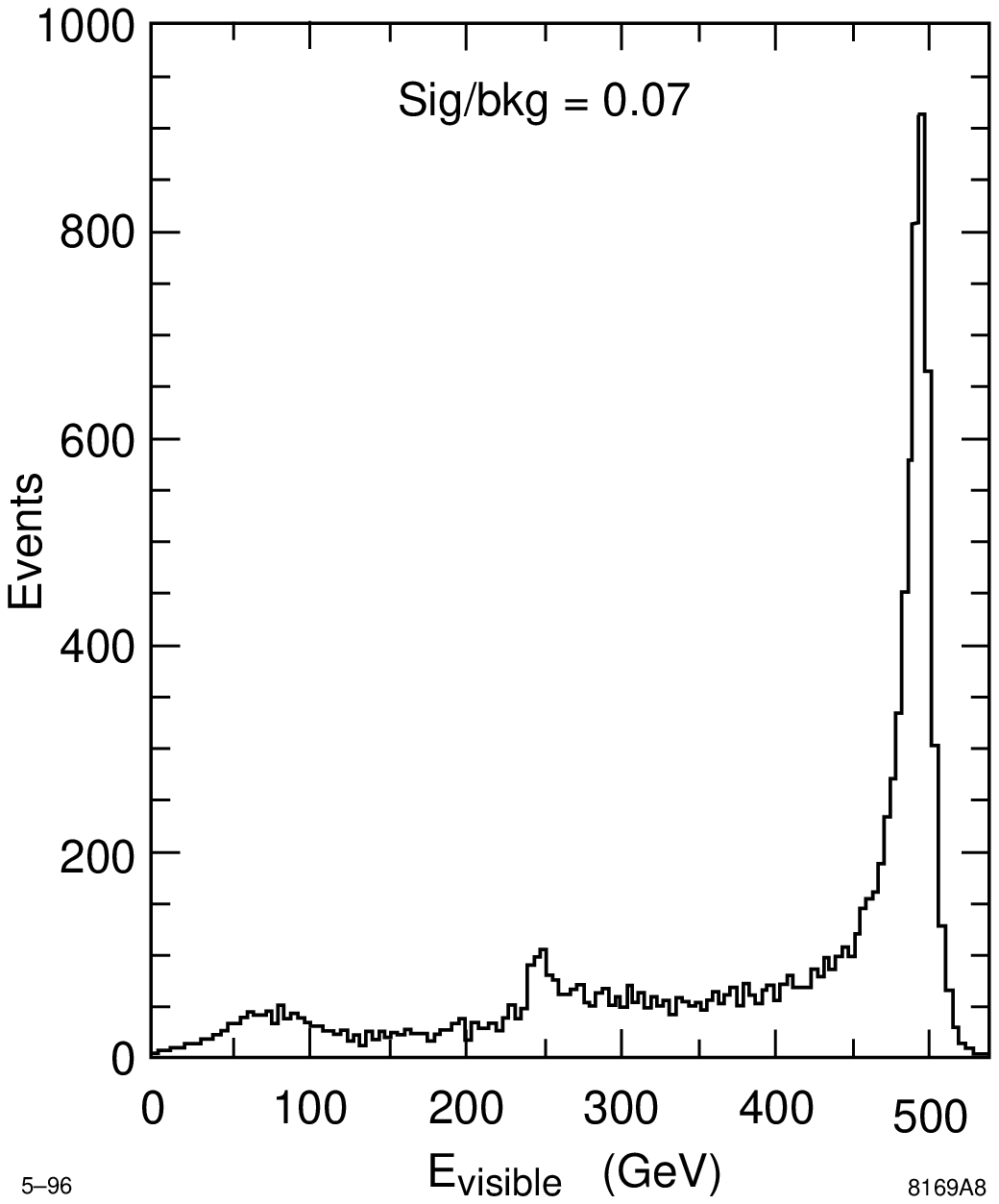}}
\centerline{\parbox{5in}{\caption[*]{The visible energy for the
normalized (relative cross-section) SUSY processes associated with
SUGRA parameters 4 and the Standard Model processes. The smearing due
to detector resolution is included.}
  \label{sigbkg4}}}
\end{figure}  

\begin{figure}[htb]
 \leavevmode
\centerline{\epsfxsize=3.5in\epsfbox{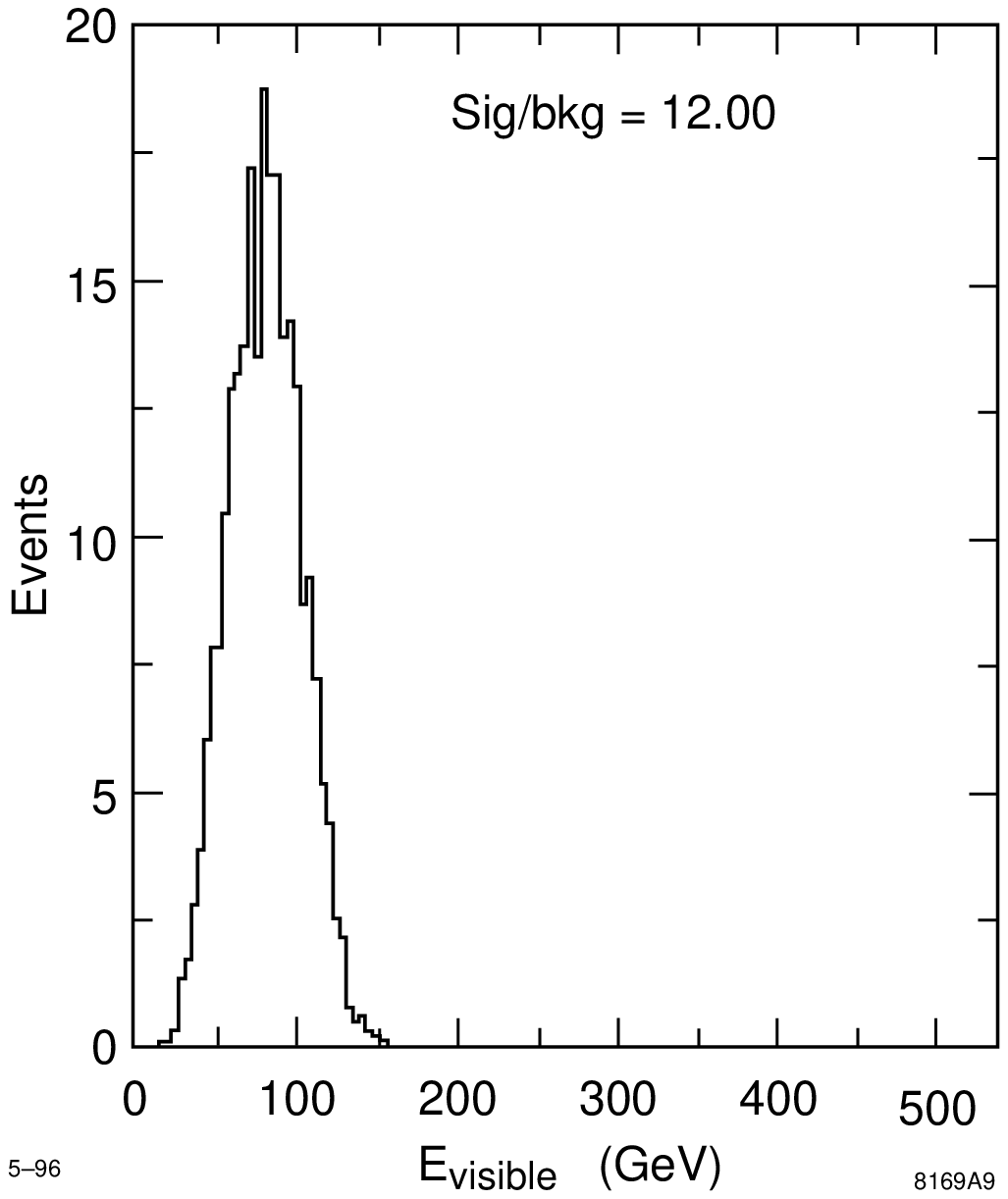}}
\centerline{\parbox{5in}{\caption[*]{The visible energy for the
normalized (relative cross-section) SUSY process $e^+e^-\rightarrow
\mbox{$\chi^{+}_1$} \mbox{$\chi^{-}_1$}$ associated with parameter set
4 and Standard Model  processes after cuts to enhance the signal over
background. The cuts are that there be only 1 broad jet with $ > $ 5
particles and with $E_{vis} < 80$ GeV in each hemisphere.}
  \label{w1w1}}}
\end{figure}

\begin{figure}[htbp]
 \leavevmode
\centerline{\epsfxsize=3.5in\epsfysize=4in\epsfbox{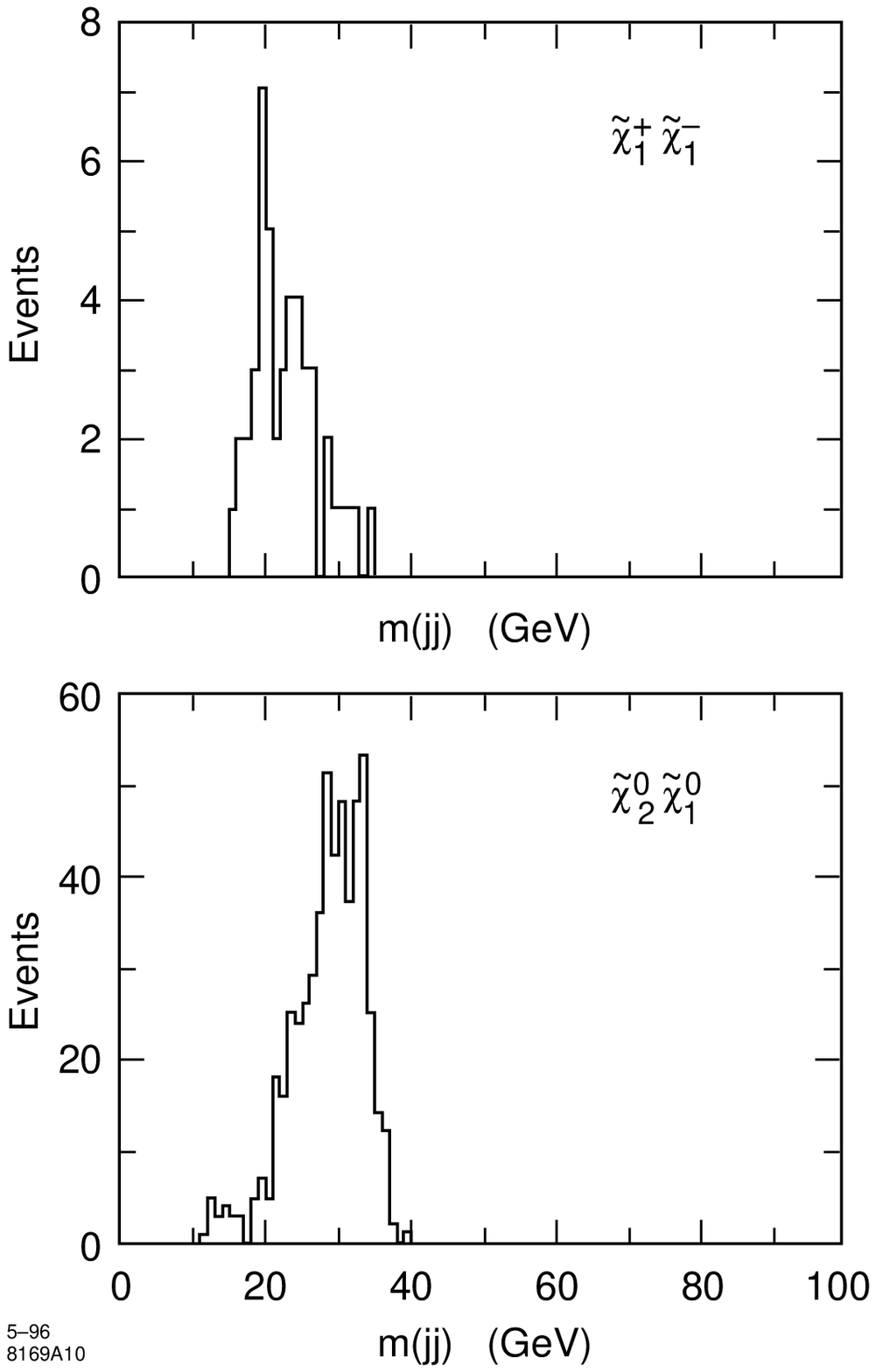}}
\centerline{\parbox{5in}{\caption[*]{The 2-jet mass distribution for
the processes $e^+e^- \to \mbox{$\chi^{+}_1$} \mbox{$\chi^{-}_1$}
(\mbox{$\chi^0_1$} \mbox{$\chi^0_2$}) \to q \bar q \mbox{$\chi^0_1$} q
\bar q \mbox{$\chi^0_1$} (\mbox{$\chi^0_1$} q \bar q
\mbox{$\chi^0_1$})$. The cuts require 2 jets with more than 1 particle
in each and both jets in one hemisphere only. $E_{visible} < 125, 70$
GeV for the two jets, and cos($\theta$)$<$ 0.85 for the thrust axis of
the event.}
  \label{mqqz2w1}}}
\bigskip
 \leavevmode
\centerline{\epsfxsize=3.5in\epsfysize=2.5in\epsfbox{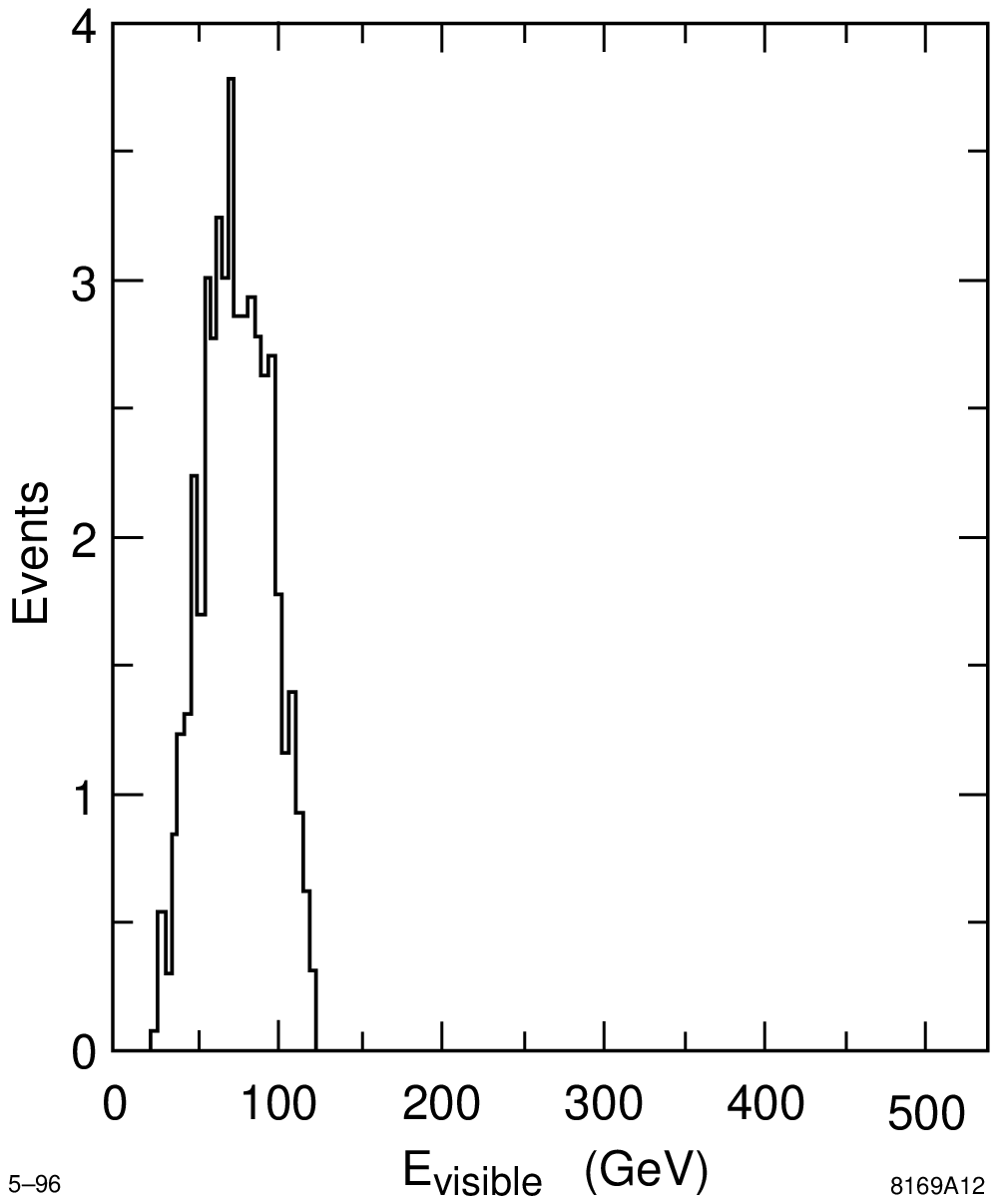}}
\centerline{\parbox{5in}{\caption[*]{The visible energy distribution
for the SUSY process $e^+e^-\rightarrow \mbox{$\chi^0_2$}
\mbox{$\chi^0_1$}$ associated with parameter set 4 and the Standard
Model background processes, after the cuts defined in the previous
figure  to enhance the signal over background.}
  \label{z1z2}}}
\end{figure}

We have made a study of the signals seen in the various sets of SUGRA
parameters \cite{CSECTANG}. We show here  some of the results for the
signal to background ratios for the SUGRA parameter set 4.  In this
scenario, the main signals  are due to the pair-production of the
charginos $\mbox{$\chi^{\pm}_1$}$ and the neutralinos
$\mbox{$\chi^0_1$}$, $\mbox{$\chi^0_2$}$. The visible (observed) energy
for these signals and the others \cite{CSECTANG} is shown in Fig.
\ref{sig4}. The signal is in the region of 100 GeV visible energy;
hence our requirement that there needs to be calorimetric coverage down
to small angles. In Fig. \ref{sigbkg4} we show the visible energy
distribution, properly normalized with the relative cross section for
background Standard Model events and signal from supersymmetric events.
A small bump can be observed in the region of small visible energy. To
isolate the signal due to chargino ($\mbox{$\chi^{+}_1$}
\mbox{$\chi^{-}_1$}$) production, we require that there be at least
five hadrons in each hemisphere. After additional cuts in the data to
enhance the signal, we are able to obtain a signal to background ratio
of 12 to 1 as shown in Fig. \ref{w1w1}. This signal is then used to
determine the masses of the $\mbox{$\chi^{\pm}_1$}$ and the
$\mbox{$\chi^0_1$}$ as described below. Similarly, with appropriate
cuts we obtain a signal for  $\mbox{$\chi^0_2$}\mbox{$\chi^0_1$}$
production with no background from the Standard Model processes,  but,
as shown in Fig. \ref{mqqz2w1}, a background of  $\approx$ 10\%
from $\mbox{$\chi^{+}_1$} \mbox{$\chi^{-}_1$}$. (We expect
to reduce this background with further analysis.)  The complete
observed signal is shown in Fig. \ref{z1z2}. This can then be used to
determine the masses of the $\mbox{$\chi^0_2$}$ and
$\mbox{$\chi^0_1$}$. It is interesting to check that the two
determinations of the $\mbox{$\chi^0_1$}$ mass agree. This self
consistency would give us confidence that we are seeing the
consequences of a consistent model and would encourage us to use the
resulting model parameters to predict the masses of the other
particles. The levels of signal to background seen in this analysis
are typical for scenarios of supersymmetric particle production
processes in e$^+$e$^-$ colliders.
 
\subsection{Superparticle Mass Measurements}

We will now discuss in more detail the measurement of the masses,
spins, and cross-sections of the various possible supersymmetric
signals. Some very beautiful studies on these issues have already been
reported in  \cite{JLC1,JLC2,TESTING}.  These papers
indicate that, indeed, linear collider experimentation provides very
powerful methods by which to measure the production and decay
parameters of the various supersymmetric particles. This should allow
us to uncover which of the various supersymmetric models is the correct
one. One recent study \cite{JLC1} has shown that, for slepton or
chargino pair-production, we can use the upper and lower limits of the
energy spectrum  of the secondaries from supersymmetric particle decays
to determine these particle  masses.  In addition, we can use the
angular distribution of the signal to say something about the spin of
the sparticles producing these distributions.  A threshold scan will
also differentiate between scalars and fermions by determining whether
the energy dependence follows a a $\beta$ or a $\beta^3$ law.

\begin{figure}[htb]
 \leavevmode
\centerline{\epsfxsize=3.5in\epsfbox{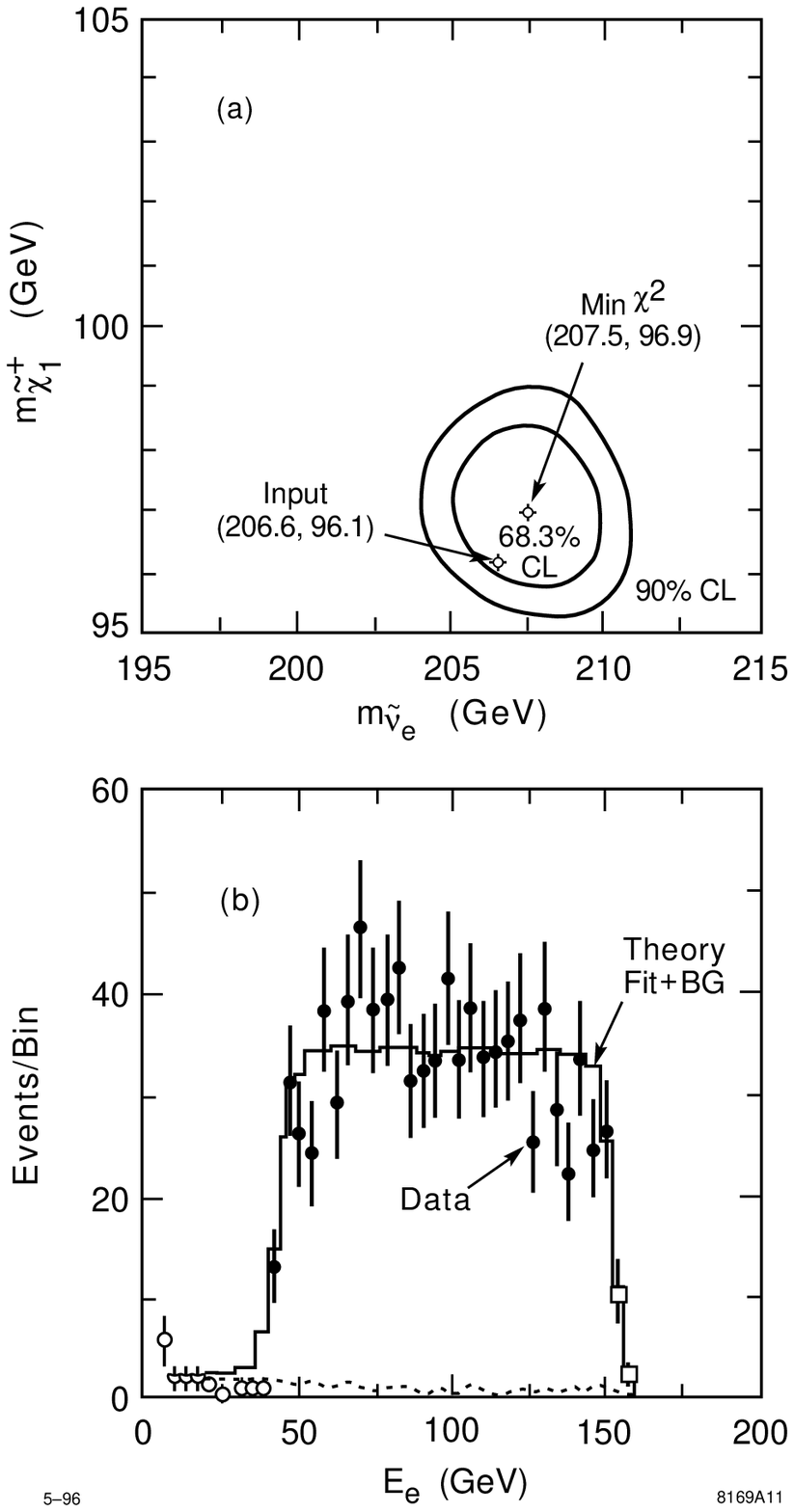}}
\centerline{\parbox{5in}{\caption[*]{The electron energy distribution
in the process $e^+e^- \rightarrow \tilde \nu_e \tilde \nu_e \to e^-
\mbox{$\chi^{+}_1$} e^+ \mbox{$\chi^{-}_1$} \to e^- \mu^{+-}
\mbox{$\chi^0_1$} e^+ \mbox{$\chi^0_1$}$ + 2 jets and the fit that
determines the mass of the $\tilde \nu_e$ = 207.5 $\pm$ 2.5 GeV and the
mass of the $\mbox{$\chi^{+}_1$}$ = 97.0 $\pm$ 1.2 GeV at the 68\%
C.L.}
  \label{sneutm}}}
\end{figure}

In this study,  we expand on this work by increasing the number of
cases that have been studied and attempt to determine how much these
measurements constrain the possible region of parameter space. Here we
will describe how well we can determine the masses of the
supersymmetric particles using the simulated resolution parameters in
our detector design. For brevity, we will only consider here the
parameters sets 3 and 5.  In case 3, the most important supersymmetry
cross section for an incident 95\% left-hand polarized electron beam is
that of of sneutrino pair production, $\tilde \nu_e \tilde \nu_e$, as
shown in Fig. \ref{csect2a}. The branching ratio for  $\tilde \nu_e \to
e^- \mbox{$\chi^{+}_1$}$ is 61\%. The $\mbox{$\chi^{+}_1$}$ decays
mostly to $W^+ \mbox{$\chi^0_1$}$. Hence, 5-10\% of the time we can
have a final state signal $e^{\mp} e^{\pm} \mu^{\pm}$ +2 jets. This
signal will have hardly any Standard Model background. The energy
distribution of the $e^{\mp}$ can be used to determine the $\tilde
\nu_e$ mass. The $e^{\mp}$ energy spectrum, based on a 20 fb$^{-1}$
data sample, is shown in Fig. \ref{sneutm}. The background is shown by
the dotted line. A fit to this energy spectrum leads to the following
values for the masses: $$M_{\tilde \nu_e} = 207.5 \pm
2.5~~(4)~~\hbox{\rm GeV} $$ $$M_{\mbox{$\chi^{+}_1$}} = 97.0 \pm
1.2~~(2)~~\hbox{\rm GeV} $$ at the 68\% (90\%) confidence level. These
results should be compared with the input values, given in the table,
of 206.6 and 96.1 GeV respectively.

\begin{figure}[htb]
 \leavevmode
\centerline{\epsfxsize=3.5in\epsfbox{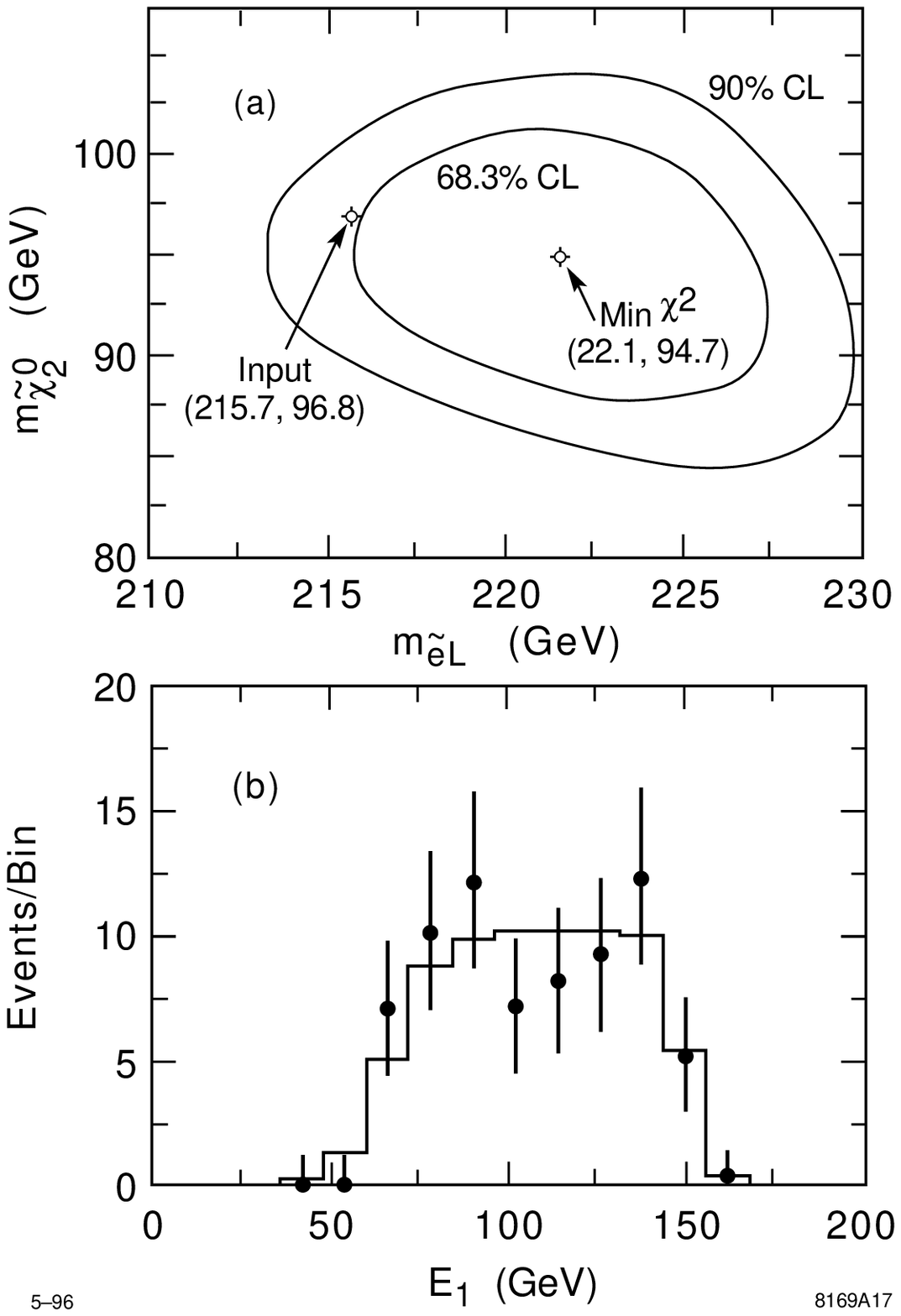}}
\centerline{\parbox{5in}{\caption[*]{The energy distribution of the two
highest energy leptons in the process $e^+e^- \to \tilde \ell_L^-
\tilde \ell_L^+ \to \ell^-\tilde\chi^0_2  \ell^+ \tilde\chi^0_2  \to
\ell^-Z^0\mbox{$\chi^0_1$} \ell^+Z^0 \mbox{$\chi^0_1$} \to$ six leptons
plus missing energy, and the fit that determines the mass of the
$\tilde \ell_L$ and the $\mbox{$\chi^0_2$}$.  Only those events were
considered in which the highest energy leptons were $\ee$ or
$\mu^+\mu^-$. The fit gives the mass values $M_{\tilde \ell_L} = 221.6
\pm 5.6$ GeV and $M_{\tilde\chi^0_2 } = 94.7 \pm 5.3$ GeV}
  \label{seleft}}}
\end{figure}  

Another interesting signal is the $\tilde e_L^+ \tilde e_L^-$ process
in order to determine the mass of the $\tilde e_L$ (left-handed
slepton). The useful signal is due to the decay chain $\tilde e_L^- \to
e^- \mbox{$\chi^0_2$} \to e^-Z^0 \mbox{$\chi^0_1$} \to
e^-\mu^-\mu^+\mbox{$\chi^0_1$}$. This leads to a final state with one
electron, one positron, four muons, and a small visible energy. We can
also consider the analogous decay chain for
$\tilde\mu_L^+\tilde\mu_L^-$.  We considered these two possibilities
together by isolating final states with 6 leptons and missing energy,
in which the highest energy leptons are either $\ee$ or $\mu^+\mu^-$.
This analysis assumes that $\tilde e_L^-$ and $\tilde\mu_L^-$ have the
same mass; with higher statistics, a mass splitting would be apparent.
The results for an effective 1 year run  (50 fb$^{-1}$) is shown in
Fig. \ref{seleft}. The fit to the lepton energy spectrum gives the
following values for the masses: $$M_{\tilde e_L} = 221.6 \pm
5.6~~(8)~~\hbox{\rm GeV} $$ $$M_{\mbox{$\chi^0_2$}} = 94.7 \pm 
5.3~~(10)~~\hbox{\rm GeV} $$ 
at the 68\% (90\%) confidence level, which are quite close to the input
values of 215.7 and 96.8 GeV in spite of the low statistics in this
sample.
Since $M_{\tilde \nu} - M_{\tilde e_L}$ is
determined by just $SU(2)$ symmetry, these measurements lead to a
model-independent constraint on the parameter tan($\beta$) according to
the equation: $$M^2_{\tilde \nu} - M^2_{\tilde e_L} = M^2_{W}\cos(2
\beta)\ . $$

\begin{figure}[htb]
 \leavevmode
\centerline{\epsfxsize=3.5in\epsfbox{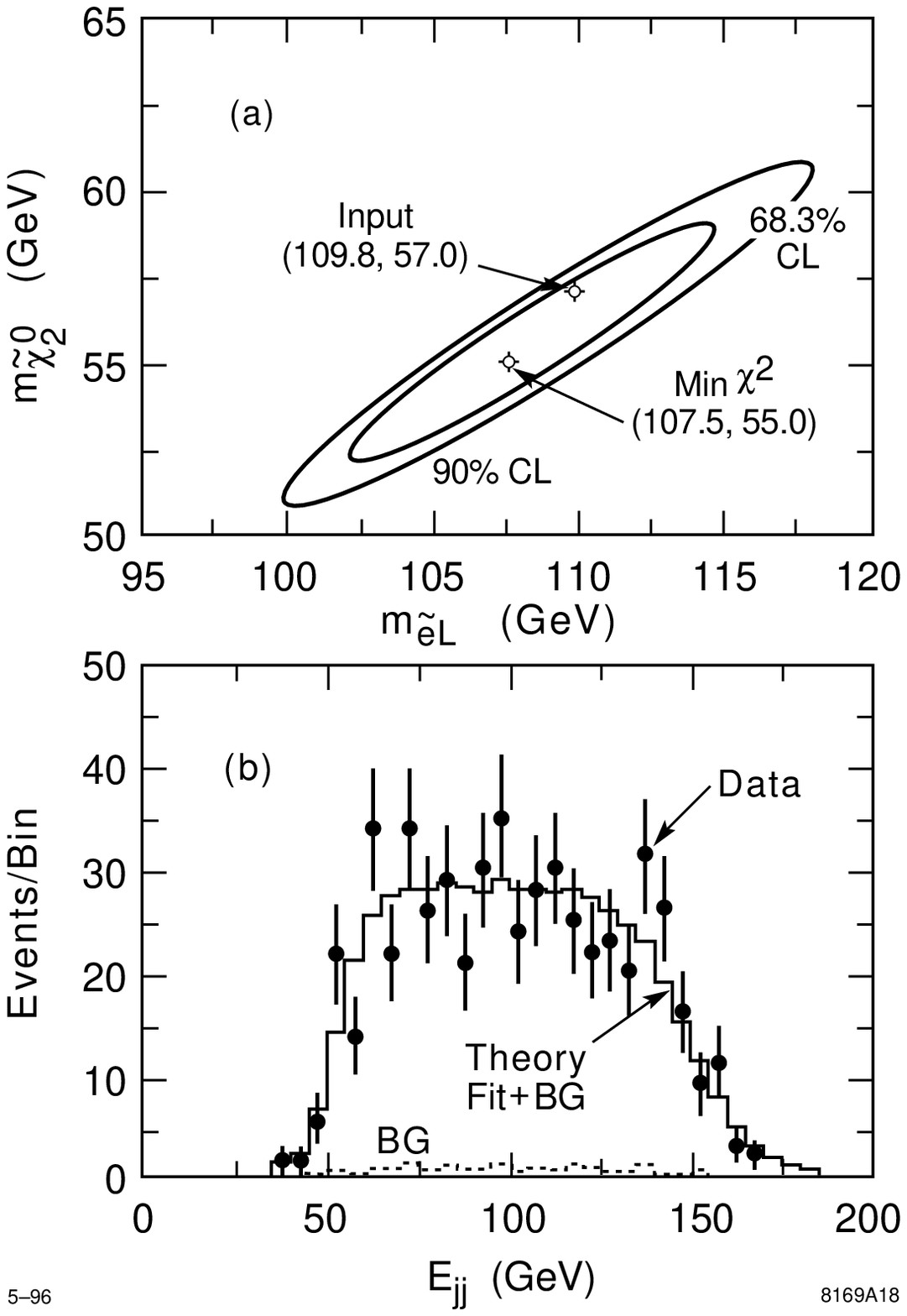}}
\centerline{\parbox{5in}{\caption[*]{The quark pair energy distribution
in the process $e^+e^- \to \tilde\chi^+_1 \tilde\chi^-_1 \to q \bar q
\tilde\chi^0_1  q \bar q \tilde\chi^0_1 $ and the fit to the
$\tilde\chi^+_1$, $\tilde\chi^0_1 $ masses. The fit gives the mass
values $M_{\tilde\chi^+_1} = 107.5 \pm 6.5$ GeV and $M_{\tilde\chi^0_1
} =  55.0 \pm 3.5$ GeV.}
  \label{Ejj}}}
\end{figure}
  
In the case of SUGRA parameter 5 we have a series of signals whose
masses can be determined. This point includes a low mass stop squark
$\tilde t_1$, a chargino $\mbox{$\chi^{+}_1$}$,  neutralinos
$\mbox{$\chi^0_1$}$ and $\mbox{$\chi^0_2$}$, and the light Higgs boson
$h^0$.  We measure the $\mbox{$\chi^{+}_1$}$ and $\mbox{$\chi^0_1$}$
masses by studying the production process e$^+$e$^- \to
\mbox{$\chi^{+}_1$}+ \mbox{$\chi^{-}_1$}$. For a 95\% left handed
polarized electron the cross-section is $\approx$ 0.75 pb so that for a
1 year run at our standard luminosity we get over $10^4$ events. The
$\mbox{$\chi^{+}_1$}$ decays into the 3 body final states $q \bar q
\mbox{$\chi^0_1$}$ and $e \nu \mbox{$\chi^0_1$}$,  with the branching
ratios predicted for the Standard Model $W$ boson decay into the
similar channels. Hence, about 68\% of the time it will decay into two
hadronic jets + $\mbox{$\chi^0_1$}$. To isolate this signal we use
similar cuts to those discussed above associated with Fig. \ref{w1w1}.
The resulting E$_{jj}$ values, the energies of each of the 2 jet
systems from the $\tilde\chi^+_1$ decays, has no sharp end point
behavior due to the 3 body nature of its decay. Hence we cannot easily
use the E$_{jj}$ spectrum to determine the masses. Since the combined
mass of the 2 jets, M$_{jj}$ does not vary much in this case, and since
we have a large sample of events, we can force two body kinematics on
this process by selecting a slice of M$_{jj}$ around a given value,
which in our case is chosen to be 30 GeV. Hence the E$_{jj}$
distribution follows approximately the two body kinematics of the
process $\tilde\chi^+_1\to \tilde\chi^0_1  + (jj)$ (30 GeV). The
E$_{jj}$ distribution and the best mass fit to the data is shown in
Fig.~\ref{Ejj}. The result is: $$M_{\tilde \chi^+_1} = 107.5 \pm 6.5\
\hbox{\rm GeV} $$ $$M_{\tilde\chi^0_1 } =  55.0 \pm 3.5\ \hbox{\rm GeV}
$$ at the 68\% confidence level. This is to be compared with the input
values of 109.8 and 57.0 GeV respectively.

\begin{figure}[htb]
 \leavevmode
\centerline{\epsfxsize=3.5in\epsfbox{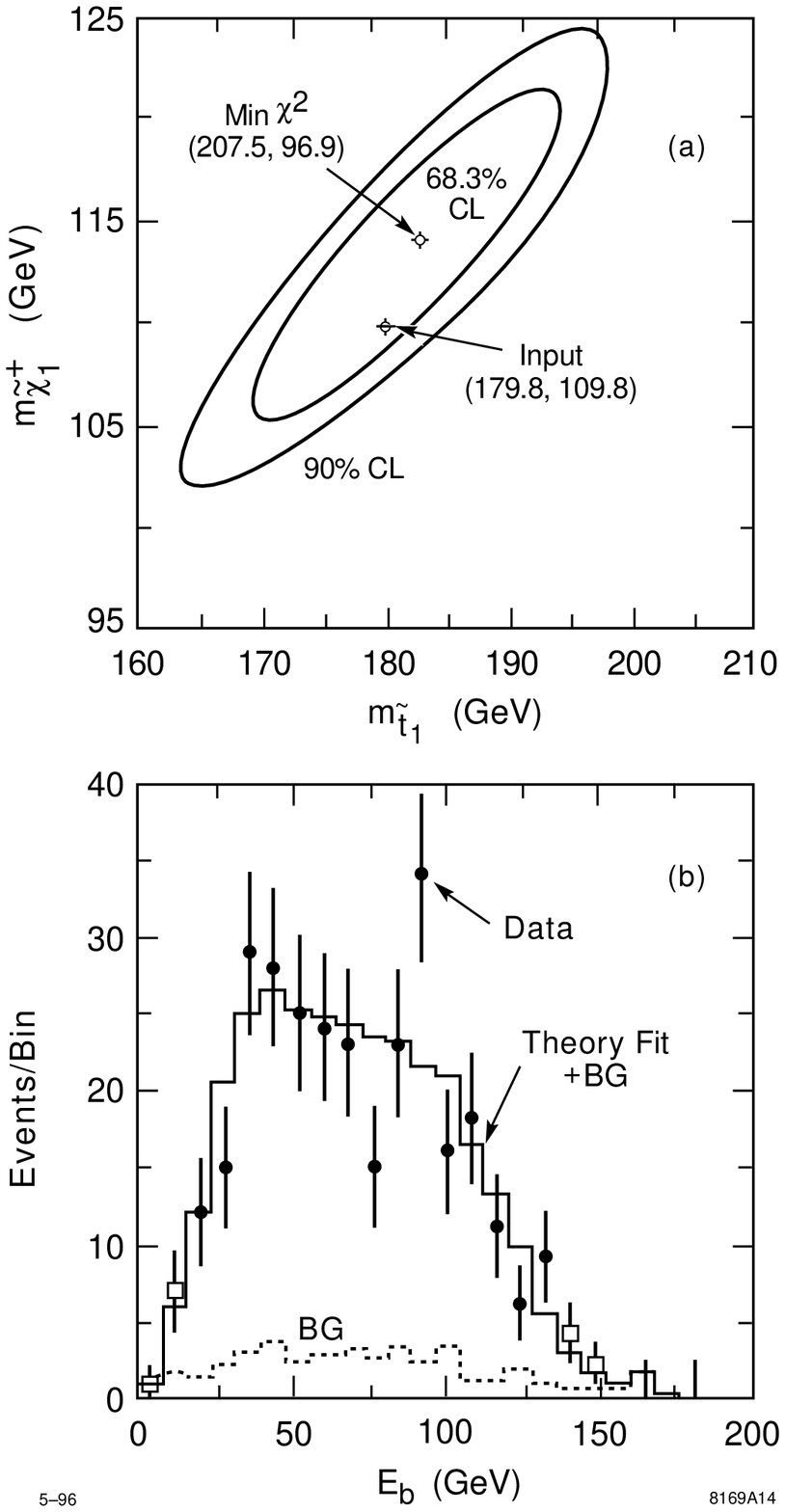}}
\centerline{\parbox{5in}{\caption[*]{The b-jet energy distribution in
the process $e^+e^- \to \tilde t_1 \bar {\tilde t_1} \to b
\tilde\chi^+_1 \bar b \tilde\chi^-_1$ and the fit to the $\tilde t_1
\tilde\chi^+_1$ masses. The fit gives the mass values $M_{\tilde t_1} =
182 \pm 11$GeV  and $M_{\tilde\chi^+_1} =  114 \pm 8$ GeV.}
  \label{stopm}}}
\end{figure}  

Finally, for parameter point 5, we have also studied determining the
mass of the $\tilde t_1$ (stop) quark. Here we note that the process
$e^+e^- \to \tilde t_1^+ \tilde t_1^- \to b \tilde\chi^+_1 \bar b
\tilde\chi^-_1$ occurs with a 100 \% branching ratio. Since this
cross section hardly depends on the electron polarization we study this
case with a 95\% right handed polarized electron ($P_L(e^-) = -0.9$) to
minimize the background from $WW$ pair production \cite{CSECTANG}. We
isolate the events with $\geq$ 5 jets, and we select from these events
with two  tagged $b$'s, and no isolated leptons or $\tau$ jets. Finally
we require a missing mass $>$ 140 GeV. For our standard 1 year run we
obtain a SUSY signal of 286 events with a $WW$ background of 36 events.
The energy distribution of the $b$-jets is shown in Fig. \ref{stopm}.
This distribution depends on the mass of the $\tilde t_1$ and the mass
of the $\tilde\chi^+_1$. The masses we obtain are: $$M_{\tilde
t_1} = 182 \pm 11 GeV $$ $$M_{\tilde\chi^+_1} = 114 \pm 8  GeV $$
to be compared with the input values of 180 and 110 GeV respectively. Other
interesting work on squark mass determination \cite{SQUARK} has also been
carried out.

We hope that in this short presentation we have indicated the
effectiveness of an 0.5-1.0 TeV $e^+e^-$ Linear Collider in determining
the masses of the Supersymmetric particles. We have not discussed how
to determine the spin of these. This we propose to accomplish where
possible by looking at their production angular distribution and by
looking at their production behavior as a function of the electron
longitudinal polarization. This work will continue in order to
determine further what additional parameters need to be determined to
be able to guarantee that the signals we observe are due to
supersymmetric particles.

\begin{figure}[htb]
 \leavevmode
\centerline{\epsfxsize=3.5in\epsfbox{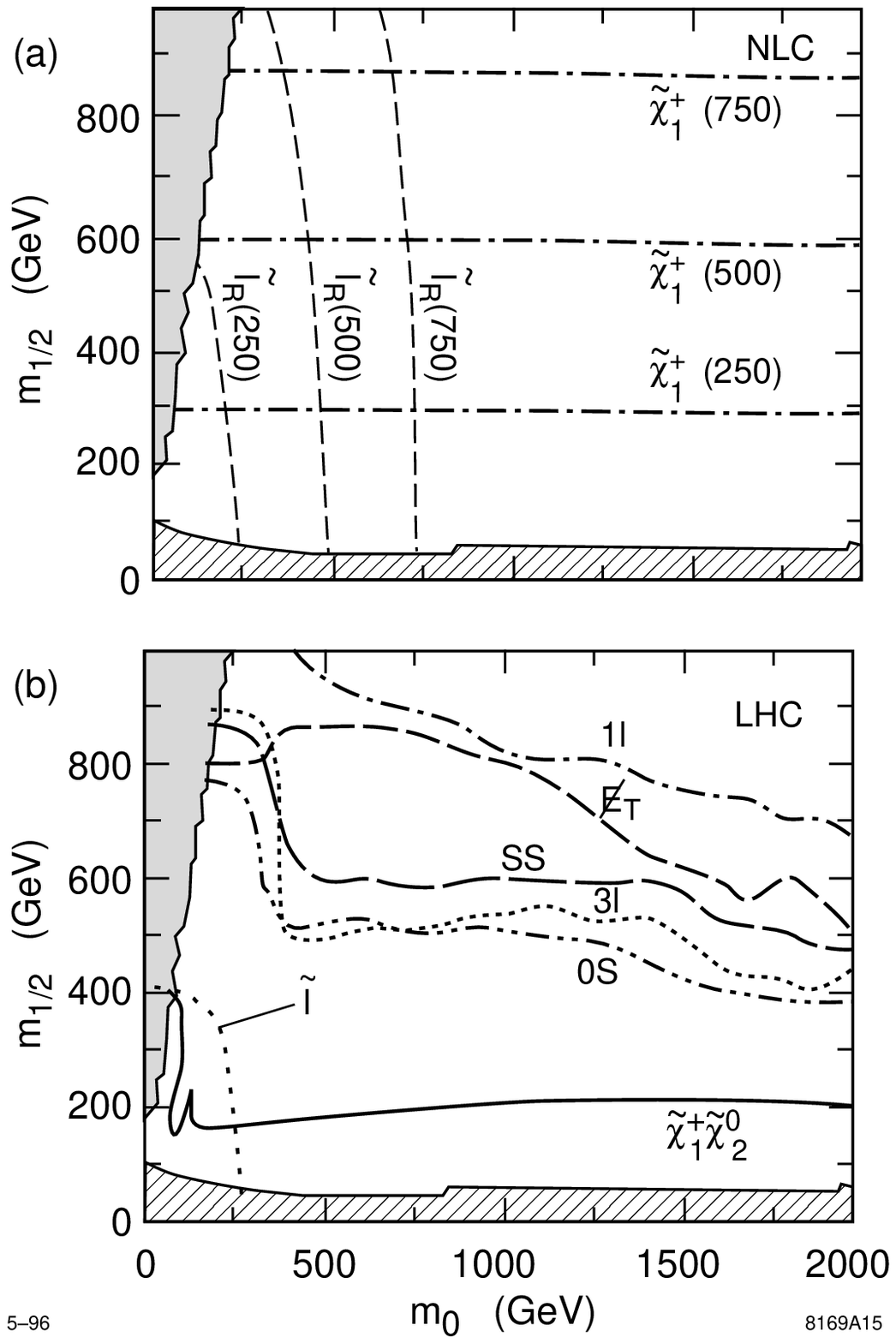}}
\centerline{\parbox{5in}{\caption[*]{Comparison of the reach of the
NLC(0.5 TeV), NLC(1.0), NLC(1.5) presented in the top figure, and the
LHC assuming an integrated luminosity of 10 fb$^{-1}$ in the bottom
figure. The contours are labeled $\tilde\ell$ for the slepton reach,
$\tilde\chi^+_1\tilde\chi^0_1 $ for the $\tilde\chi^+_1\tilde\chi^0_1
\rightarrow 3\ell$ reach, $1\ell$ for the reach via lepton + jets +
$E\llap/ _T$ events, $E\llap/ _T$ for events with multi-jets + $E\llap/
_T$, SS for same-sign dileptons + jets + $E\llap/ _T$ and $3\ell$ for
trilepton + jets +$E\llap/ _T$.}
  \label{compare}}}
\end{figure}  

\subsection{Supersymmetry Reach of the NLC and LHC}

To conclude, we make a brief comparison of the relative reach
capabilities of a 0.5-1.5 TeV $e^+e^-$ Linear Collider and the CERN LHC
$pp$ collider as to their ability to determine whether the observed
signals are due to supersymmetric particles. First of all, the NLC,
operating at E$_{\rm cm} \ge$ 250-300 GeV should be able to search for
the light Higgs boson, $h^0$, over the entire parameter space range of
the minimal supersymmetry model. If the NLC does not observe the $h^0$,
then this model must be ruled out. In addition, since the $h^0$ is
expected to behave very nearly like a Standard Model Higgs boson, even
if it is discovered, it may be difficult to tell if it is a SUSY or
Standard Model Higgs. Hence, discovery of the $h^0$ alone may not be
sufficient evidence for supersymmetry. On the other hand, the NLC has a
substantial ability to discover many of the superpartners.

In Fig. \ref{compare}, we show our estimates of the reach of NLC and
LHC into the SUGRA parameter space, defined by underlying mass
parameters $m_0$ and $m_{1/2}$.  In the top figure, we have plotted the
contours corresponding to $\tilde \ell_R$ and $\tilde\chi^\pm_1$
masses of 250, 500 and 750 GeV, approximately representing the reach of
NLC(0.5 TeV), NLC(1.0) and NLC(1.5) in observing these supersymmetric
particles. We also show the reach for supersymmetry recently calculated
\cite{BCPT} for the CERN LHC assuming 10 fb$^{-1}$ of integrated
luminosity. Comparing the two figures we note that the reach of the LHC
is larger than that of the NLC at 500 GeV, but its reach is comparable
to that of the NLC at 1 TeV.  It is important to note that the
reactions at NLC and LHC typically access different particles in the
supersymmetry spectrum, so the experiments at these colliders should be
considered cooperative rather than competitive.  In addition, precision
measurements of particle properties such as mass, spin, and mixing
angles will be much easier at the NLC \cite{JLC1} than at the LHC. The
LHC might be able to provide complementary information via squark and
gluino production channels which may not be accessible at the NLC.

\clearpage

\def\doubarrow#1{#1\hskip-0.5em^{^\leftrightarrow}}

\section{Anomalous Gauge Boson Couplings}

%

Although the Standard Electroweak Model has been verified to astounding
precision in recent years at LEP and SLC, one important component has
not been tested directly with significant precision: the non-Abelian
self couplings of the weak vector gauge bosons. Deviations of
non-abelian couplings from expectation would signal new physics,
perhaps arising from unexpected loop corrections involving propagators
of new particles. In addition, as will be discussed in Chapter 8,
precise measurements of $WWV$ couplings, where $V=\gamma$ or $Z$, can
provide important information on the nature of electroweak symmetry
breaking. Recent results from CDF and D0 indicate the presence of
triple gauge boson couplings, but have not yet reached a precision
better than order unity \cite{AGBI}. Upcoming measurements at LEP
II, at an upgraded Tevatron and at the LHC will improve upon this
precision considerably, but cannot the match the expected precision of
a 500 GeV NLC, much less that of a 1.0 or 1.5 TeV NLC.   There exist
indirect constraints on the anomalous couplings from the precision
electroweak measurements at the $Z^0$ resonances, in particular, from
the fact that loop diagrams involving weak vector bosons are seen to
take the values expected in the Standard Model. However, the
ambiguities in the calculation of these diagrams call for more direct
measurements \cite{Einhornsrev}.

In this brief report we restrict attention mainly to measurement of
possibly anomalous $WWV$ couplings via the process $e^+e^-\rightarrow
W^+W^-$, but much work has been done on other processes that involve
non-abelian couplings in $e^+e^-$\ annihilation, including $ZZ\gamma$,
$Z\gamma\gamma$, $WWZZ$, and $WWWW$ \cite{AGBI}. In addition, many of
these couplings can also be measured independently using the $e^-e^-$,
$e^-\gamma$\, and $\gamma\gamma$\ options for the NLC. We will describe
one common parametrization of anomalous $WWV$\ couplings, summarize
present and expected pre-NLC measurements of $WWV$\ couplings, and
discuss in more detail what can be done at the NLC.

\subsection{Parametrization}

In parametrizing anomalous couplings, we follow the notation of
Ref.~\cite{hpzh} in which the generic effective Lagrangian for
the $WWV$\ vertex is written:
\begin{eqnarray}
L_{WWV} / g_{WWV} 
& = &     i\,g_1^V (W^\dagger_{\mu\nu}W^\mu V^\nu 
                  - W^\dagger_\mu V_\nu W^{\mu\nu}) 
      + i\,\kappa_V W^\dagger_\mu W_\nu V^{\mu\nu} 
      + \frac{i\,\lambda_V}{M_W^2}W^\dagger_{\lambda\mu}
         W^\mu_{\>\nu}V^{\nu\lambda} \nonumber\\
& &   - g_4^V W^\dagger_\mu W_\nu(\partial^\mu V^\nu
                                    +\partial^\nu V^\mu) 
      + g_5^V\epsilon^{\mu\nu\rho\sigma}
        (W^\dagger_\mu\doubarrow{\partial}{}_\rho W_\nu)
        V_\sigma  \nonumber
      + \tilde\kappa_V W^\dagger_\mu W_\nu\tilde{V}^{\mu\nu} \\
& &   + \frac{i\,\tilde\lambda_V}{M_W^2}W^\dagger_{\lambda\mu}
        W^\mu_{\>\nu}\tilde{V}^{\nu\lambda} \ , 
\label{eqnaa}
\end{eqnarray}
where $W_{\mu\nu}\equiv\partial_\mu W_\nu -\partial_\nu W_\mu$,
$V_{\mu\nu}\equiv\partial_\mu V_\nu -\partial_\nu V_\mu$,
$(A\doubarrow{\partial}{}_\mu B)\equiv A(\partial_\mu B)-(\partial_\mu
A)B$, and $\tilde{V}_{\mu\nu} \equiv \frac{1}{2}
\epsilon_{\mu\nu\rho\sigma} V^{\rho\sigma}$. The normalization factors
are defined for convenience to be $g_{WW\gamma} \equiv -e$ and $g_{WWZ}
\equiv -e\cot\theta_W $. The 7 coupling parameters defined in Eq.
\ref{eqnaa} for each of $\gamma$ and $Z$ include the C and P violating
couplings $g_5^V$ as well as the CP-violating couplings
$g_4^V,\tilde\kappa_V,\tilde\lambda_V$. In most studies and in this
one, such terms are neglected. At tree level in the Standard Model $g_1^V
=\kappa_V=1$ and $\lambda_V=g_4^V=g_5^V=
\tilde\kappa_V=\tilde\lambda_V=0$. The couplings in Eq. \ref{eqnaa}
should properly be written as form factors with momentum-dependent
values. This complication is of little importance at an $e^+e^-$\
collider where the $WW$ center of mass energy is well defined, but it
must be borne in mind for other processes, particularly those
measurable at hadron colliders. 

We follow the common convention in defining $\Delta g_1^Z \equiv g_1^Z
- 1$ and $\Delta\kappa_V \equiv \kappa_V - 1$. The $W$ electric charge
fixes $g_1^\gamma (q^2\rightarrow0) \equiv 1$. In perhaps more familiar
notation, one can express the $W$ magnetic dipole moment as $\mu_W
\equiv \frac{e}{2\,M_W}(1+\kappa_\gamma+\lambda_\gamma)$ and the $W$
electric quadrupole moment as $ Q_W \equiv-\frac{e}{M_W^2}
(\kappa_\gamma-\lambda_\gamma)$.

In any model with new physics at high energy that couples to the $W$
boson, the anomalous couplings will be induced at some level.  A useful
way to represent this effect is to write an $SU(2)\times
U(1)$-invariant effective Lagrangian to represent the effects of the
new physics, and then to couple this to the weak vector bosons by
gauging the symmetry.  In the literature, this has been done using both
linear and nonlinear effective Lagrangians \cite{Einhornsrev,baggerdaw,aWudka}. 
Typically, the anomalous couplings predicted in such models are
suppressed by factors of $M_W^2/\Lambda^2$, where $\Lambda$ is a
multi-TeV scale, or by factors $\alpha_w/4\pi$. These
lead to typical values of the anomalous couplings below $10^{-2}$ and
make it difficult to observe these couplings before directly observing
the new physics itself.  In the Standard Model one expects loop
contributions of $O(10^{-3}$ \cite{TGVSMexpect}.  In the supersymmetry
model, one might expect loop contributions, depending on the values of
the supersymmetry parameters \cite{TGVSUSYexpect}.

For the present discussion, we use a linear realization with the
additional constraint of equal couplings for the U(1) and SU(2) terms
in the effective Lagrangian that contribute to anomalous triple gauge
boson couplings.  This gives an effective lagrangian with two free
parameters, which we will take to be $K_\gamma$ and $\lambda_\gamma$.
This restriction is called the `HISZ scenerio' \cite{HISZ}. This
restriction of the parameter space has recently been applied to
comparative studies of the anomalous $W$ couplings at colliders
\cite{AGBI}. It is important to note, however, that studies of $\ee\to
W^+ W^-$ at the NLC can also test this hypothesis by independently
determining the $\gamma$ and $Z$ couplings to the $W$ \cite{TLBFINLAND}.

\subsection{Present and Expected Pre-NLC Measurements}

The only present direct measurements of $WWV$\ couplings come from the
CDF and D0 Experiments \cite{AGBI} at the Tevatron, which have searched
for $WW$, $WZ$, and $W\gamma$ production. The $WW$ and $WZ$ searches
have yielded $O(1)$ candidates and the $W\gamma$ searches have yielded
$O(10)$ candidates, consistent with expectation. These observations
have led to limits on the corrections to the coupling parameters of
order unity. For example, D0 sets a 95\% CL range
$-1.8<\Delta\kappa_\gamma<1.9$ assuming $\lambda_\gamma=0$ or a range
$-0.6<\lambda_\gamma<0.6$ assuming $\kappa_\gamma=0$ (both limits
assume $\Lambda=1$ TeV in the appropriate form factors).

One expects significant improvement at LEP II \cite{LEPII}, once the
accelerator exceeds the $W$-pair threshold energy. Techniques similar
to those described below will be applied to $e^+e^-\rightarrow W^+W^-$\
events at c.m. energies ranging from threshold at $\approx$160 GeV to
$\approx$195 GeV. Assuming no anomalous couplings are observed after an
integrated luminosity of 500 pb$^{-1}$, 95\% CL limits on individual
couplings (all others set to zero) of $O(0.1)$ are expected.

After the Main Injector upgrade has been completed, it is expected that
the Tevatron will collect $O(1$-$10)$ fb$^{-1}$\ of data. (Further
upgrades in luminosity are also under discussion.) If 10 fb$^{-1}$\ is
achieved, it is expected \cite{AGBI} that limits on
$\Delta\kappa_\gamma$ and $\lambda_\gamma$ will be obtained that are
competitive with those from LEP II with 500 pb$^{-1}$.

Finally, one expects the LHC accelerator to turn on sometime before the
NLC and to look for the same signatures considered at the Tevatron. The
planned luminosity and c.m. energy, however, give the LHC a large
advantage over even the Main Injector Tevatron in probing anomalous
couplings. The ATLAS Collaboration has estimated \cite{Atlas} that with
100 fb$^{-1}$, one can obtain (in the HISZ scenario) 95\% CL limits on
$\Delta\kappa_\gamma$ and $\lambda_\gamma$ in the range 5-10 $\times$
$10^{-3}$.  We should note that these studies do not yet include
helicity analysis on the $W$ bosons.

\subsection{Measurements in $W$ Pair Production at the NLC}

The fact that the expected LHC limits improve dramatically upon those
of a high-luminosity Tevatron indicates the importance of
center-of-mass energy.  This reflects the fact that, in the effective
Lagrangian description, the anomalous couplings arise from
higher-dimension effective interactions.

The NLC has an added advantage over hadron colliders in reconstructing
$W$ pair events due to absence of spectator partons. To a good
approximation, full energy and momentum conservation constraints can be
applied to the visible final states. Thus an $e^+e^-\rightarrow
W^+W^-$\ event can ideally be characterized by five angles: the
production angle $\Theta_W$ of the $W^-$ with respect to the electron
beam, the polar and azimuthal decay angles $\theta^*$ and $\phi^*$ of
one daughter of the $W^-$ in the $W^-$ reference frame, and the
corresponding decay angles $\bar\theta^*$ and $\bar\phi^*$ of one of
the  $W^+$ daughters. In practice, initial-state photon radiation and
final-state photon and gluon radiation (in hadronic W decays)
complicate the picture. So does the finite width of the $W$.
Nevertheless, for the studies below, we will characterize
$e^+e^-\rightarrow W^+W^-$\ events by these five angles and fit
distributions in the angles to obtain values of anomalous couplings.

\begin{figure}[htb]
\leavevmode
\centerline{\epsfxsize=3in \epsfbox{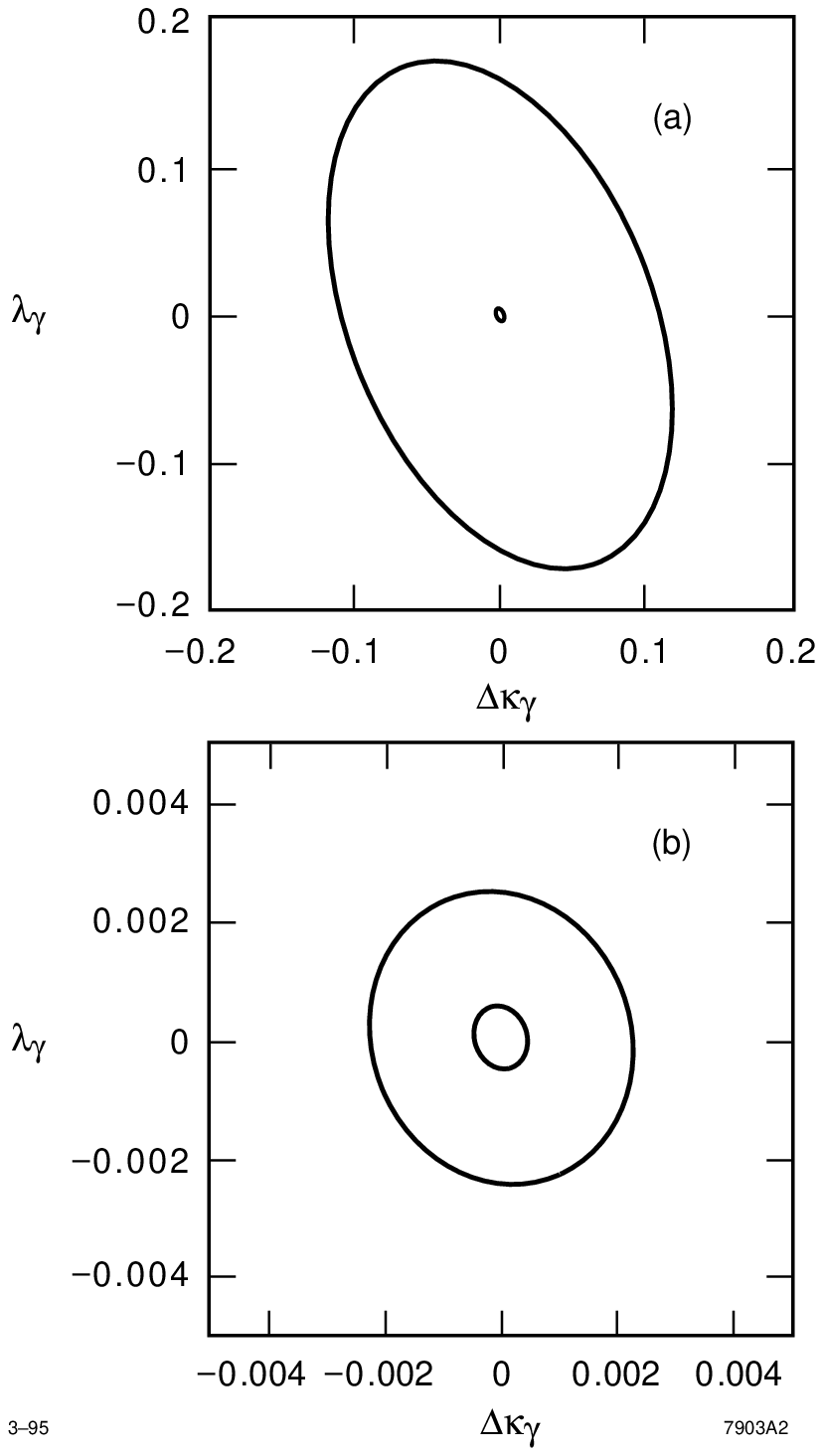}}
\centerline{\parbox{5in}{\caption[*]{95\% CL contours for
$\Delta\kappa_\gamma$ and $\lambda_\gamma$ in the HISZ scenario. The
outer contour in (a) is for ${\rm E}_{\rm CM}$\ = 190 GeV and 0.5
fb$^{-1}$. The inner contour in (a) and the outer contour in (b) is for
${\rm E}_{\rm CM}$\ = 500 GeV with 80 fb$^{-1}$. The inner contour in
(b) is for ${\rm E}_{\rm CM}$\ = 1.5 TeV with 190 fb$^{-1}$.}
\label{WWVFIGTWO}}} 
\end{figure}

At high energies, the $e^+e^-\rightarrow W^+W^-$\ process is dominated
by $t$-channel\ $\nu_e$ exchange, leading primarily to very
forward-angle $W$'s where the $W^-$ has an average helicity near minus
one. This makes the bulk of the cross section difficult to observe with
precision. However, the amplitudes affected by the anomalous couplings
are not forward peaked; the central and backward-scattered $W$'s are
measurably altered in number and helicity by the couplings. $W$
helicity analysis through the decay angular distributions provides a
powerful probe of anomalous contributions. Most studies, including
those discussed here, restrict attention to events for which
$|\cos\Theta_W| <0.8$. Not surprisingly, the most powerful channel to
use is one in which one $W$ decays leptonically (to $e\nu$ or $\mu\nu$)
and the other hadronically ($\approx30$\% of all $e^+e^-\rightarrow
W^+W^-$\ events) Although one gains statistics when both $W$'s decay
hadronically, one loses considerable discriminating power from being
unable to purely tag the charge of the $W$ daughter quarks. The channel
in which both $W$'s decay leptonically suffers from both poor
statistics and kinematic ambiguities due to two undetected neutrinos.
For the leptonic {\it vs}\ hadronic channel, the lepton energy carries
important information for kinematic reconstruction. For simplicity,
most studies have not yet attempted to incorporate
$W\rightarrow\tau\nu$ decays.

Figure~\ref{WWVFIGTWO} (taken from Ref.~\cite{TLBUCLA}) shows 95\% CL
exclusion contours in the plane $\lambda_\gamma$ {\it vs}\
$\Delta\kappa_\gamma$ in the HISZ scenario for different c.m. energies
and integrated luminosities (0.5 fb$^{-1}$\ at 190 GeV, 80 fb$^{-1}$\
at 500 GeV, and 190 fb$^{-1}$\ at 1500 GeV). These contours are based
on ideal reconstruction of $W$ daughter pairs produced on mass-shell
with no initial-state radiation. The contours represent the best one
could possibly do. A previous study \cite{TLBFINLAND} assuming a very
high-performance detector but including initial-state radiation and a
finite $W$ width found some degradation in these contours, primarily
due to efficiency loss when imposing kinematic requirements to suppress
events far off mass-shell or at low effective c.m. energies.
Nevertheless, one attains a precision of $O(10^{-3})$ at NLC(500) and
$O($few $\times10^{-4})$ at NLC(1500). Another nice feature of
coupling studies at NLC is the ability to disentangle couplings in
models more general than HISZ via tuning of the electron beam
polarization, as shown in Ref.  \cite{TLBFINLAND}.

A study  \cite{KR} undertaken for this workshop has examined the
effects of detector resolution on achievable precisions. One might
expect {\it a priori} that the charged track momentum resolution would
be most critical since the energy spectrum for the $W$-daughter muons
peaks at a value just below the beam energy, falling off nearly
linearly with decreasing energy. One might also expect the hadron
calorimeter energy resolution to be important in that it affects the
energy resolution of jets to be identified with underlying $W$-daughter
quarks.  Indeed, one finds that these resolutions do matter, but that
the nominal resolution parameters assumed for the purposes of this
workshop are quite adequate.  A significant source of degradation in
sensitivity to anomalous couplings comes from initial state photon
radiation, aggravated by beamsstrahlung, the presence of which is
difficult to establish in a single lepton-jets event, given the
undetected neutrino.  In this respect, the four-jets channel in which
both $W$'s decay hadronically, is more promising.  Further
investigation of this channel is in progress.

In summary, the studies reviewed in this section
confirm the power of the NLC to extract anomalous couplings. We expect
some degradation in coupling parameters precision from the ideal case
due to the underlying physical phenomena of initial state photon
radiation and the finite $W$ width and a smaller degradation from the
imperfection of matching detected particles to primary $W$ daughters,
but these effects are not serious and should be straightforward to
incorporate in a real measurement.

\subsection{Measurements in Other Reactions at the NLC}

In addition, the NLC allows measurements of non-Abelian gauge boson
couplings in other channels \cite{AGBI}. The process $e^+e^-\rightarrow
Z\gamma$ probes $ZZ\gamma$ and $Z\gamma\gamma$ couplings, and processes
such as $e^+e^- \rightarrow WWZ$ probe quartic couplings. The
$WW\gamma$ and $WWZ$ couplings can be probed independently via the
processes $e^+e^- \rightarrow \nu\bar\nu \gamma$ and $e^+e^-
\rightarrow \nu\bar\nu Z$, respectively.

Similar measurements can be carried out at $e^-e^-$, $e^-\gamma$\ and
$\gamma\gamma$\ colliders, where the expected reduction in luminosity
is at least partly compensated by other advantages \cite{TLBFINLAND,
CandCWWZ}. For example, the process $\gamma e^-\rightarrow W^-\nu_e$
probes the $WW\gamma$ coupling, independent of $WWZ$ effects. The
polarization asymmetry in this reaction reverses as the energy of the
collisions is varied, and the location of this zero-crossing provides a
sensitive probe of $\lambda_\gamma$ \cite{Brodsky}. The reaction
$\gamma\gamma\to W^+W^-$ also separates effects of the $\gamma$
couplings from those of the $Z$, and also probes the  4-boson
$WW\gamma\gamma$ vertex. The power of this facility is enhanced by its
ability to polarize both incoming beams \cite{TLBFINLAND}.

\subsection{Conclusions}

\begin{figure}[htb]
\leavevmode
\centerline{\epsfxsize=2.5in \epsfbox{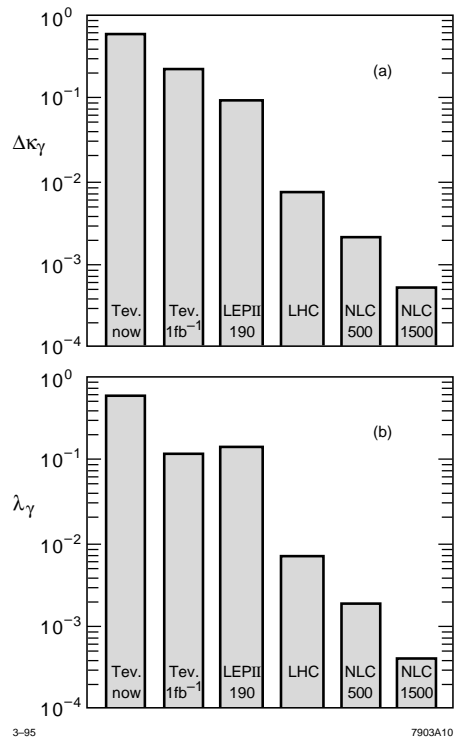}}
\centerline{\parbox{5in}{\caption[*]{Comparison of representative 95\%
CL upper limits on $\Delta\kappa_\gamma$ and $\lambda_\gamma$ for
present and future accelerators.}
\label{WWVFIGTHREE}}}
\end{figure}

Although there will have been a number of measurements of anomalous
coupling parameters from LEP II, the Tevatron, and the LHC before the
turn-on of the NLC, the precisions on the values of the couplings
attainable with the NLC will quickly overwhelm the previous
measurements. Moreover, the higher the accessible energy at NLC, the
more dramatic the improvement will be. Figure~\ref{WWVFIGTHREE} (taken
from Ref.~\cite{BDHS}) shows a useful comparison among these
accelerators. The enormous potential of the NLC is apparent.

\clearpage

\clearpage
\section{Strong $WW$ Scattering}


In this section we examine how well an $e^+e^-$ linear collider with a
center-of-mass energy of 500--1500~GeV can study the strongly interacting
Higgs sector. We also compare the estimated sensitivity of such a
collider with that of the LHC.

\subsection{The Reaction ${e^+e^-\rightarrow W^+W^-}$}
\label{sec:eeww}

Strong electroweak symmetry breaking affects the reaction
$e^+e^-\rightarrow W^+ W^-$ through anomalous couplings at the
$W^+W^-\gamma$ and $W^+W^-Z$ vertices and through enhancements in
$W^+_{\rm L}W^-_{\rm L}$ production due to $I=J=1$ resonances.  Here we
have used the symbol $W_L$ to denote a longitudinally polarized  W
boson. Anomalous couplings at the three-gauge boson vertices are
related to the chiral Lagrangian parameters $L_{9L}$ and
$L_{9R}$~\cite{baggerdaw}. A technipion form factor $F_T$ is used to
parameterize~\cite{peskinomnes} the strong $W^+_{\rm L}W^-_{\rm L}$
interaction in the $I=J=1$ state;  it is analogous to the
rho--dominated pion form factor in $e^+e^-\rightarrow \pi^+\pi^-$.
  
Whether one is measuring trilinear vector boson couplings or searching
for an enhancement in $W^+_{\rm L}W^-_{\rm L}$ production, the
experimental goal is the same:  disentangle the $W^+W^-$ polarization
states, and in particular isolate the polarization state $W^+_{\rm
L}W^-_{\rm L}$. We shall describe the results of a study that utilizes
a final-state helicity analysis of all observable final-state variables
in order to isolate $W^+_{\rm L}W^-_{\rm L}$
production~\cite{barkmorioka}.

\begin{figure}[htb]
\leavevmode
\centerline{\epsfxsize=3in\epsfbox{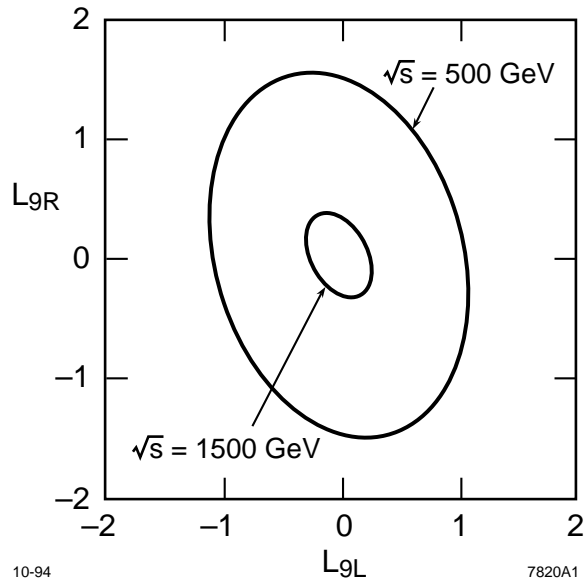}}
\centerline{\parbox{5in}{\caption[*]{The 95\% confidence level contours
for $\protect L_{9L}$ and $\protect L_{9R}$ at $\protect \sqrt{s}=500$
GeV with 80  fb$^{-1}$, and at $\protect \sqrt{s}=1500$ GeV
with 190 fb$^{-1}$.  The outer contour is for $\protect
\sqrt{s}=500$ GeV. In each case the initial state electron polarization
is 90\%.}
\label{lnllnr}}}
\end{figure}

The maximum likelihood method is used to fit for  chiral Lagrangian
parameters or for the real and imaginary parts of the technipion form
factor. Figure~\ref{lnllnr} shows the 95\% confidence level contours
for the chiral Lagrangian parameters $L_{9L}$ and $L_{9R}$ at
$\sqrt{s}=500$ GeV and at $\sqrt{s}=1500$ GeV.  The parameters $L_{9L}$
and $L_{9R}$ are normalized such that values of ${\cal O}(1)$ are
expected if the Higgs sector is strongly interacting.

\begin{figure}[htb]
\leavevmode
\centerline{\epsfxsize=3in\epsfbox{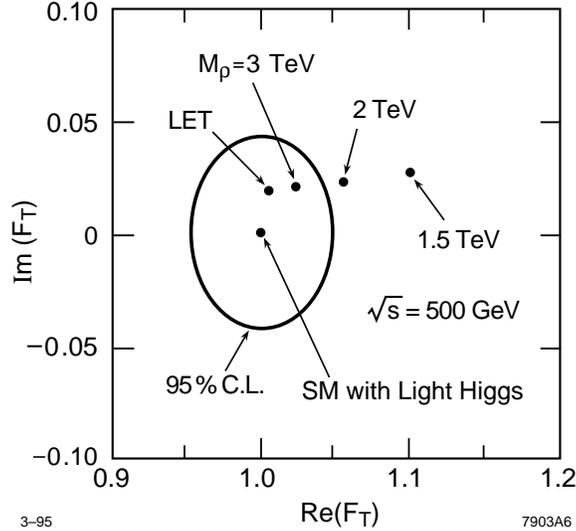}}
\centerline{\parbox{5in}{\caption[*]{95\% confidence level contours for
the real and imaginary parts of $F_T$ at $\protect\sqrt{s}=500$ GeV
with 80~fb$^{-1}$. The values of $F_T$ for various technirho masses are
indicated.}
\label{wlwllow}}}
\end{figure}

\begin{figure}[htb]
\leavevmode
\centerline{\epsfxsize=3in\epsfbox{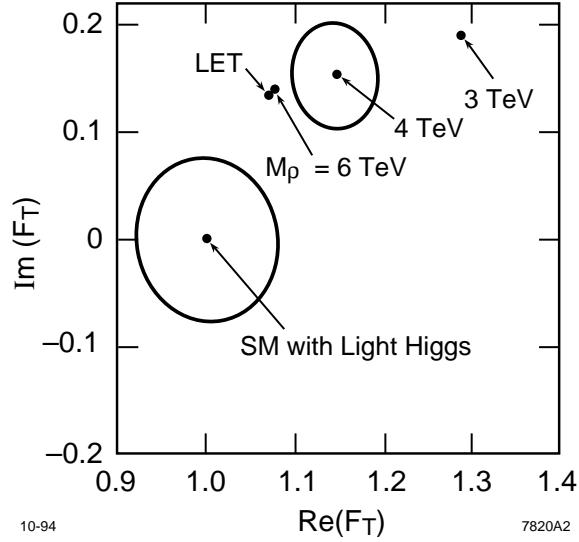}}
\centerline{\parbox{5in}{\caption[*]{Confidence level contours for the
real and imaginary parts of $F_T$ at $\protect\sqrt{s}=1500$ GeV with
190 fb$^{-1}$. The contour about the light Higgs value of $F_T=(1,0)$
is 95\% confidence level and the contour about the $M_\rho=4$~TeV point
is 68\% confidence level.}
\label{wlwlhigh}}}
\end{figure}

Figure \ref{wlwllow} shows the 95\% confidence level contour for the
real and imaginary parts of $F_T$ at $\sqrt{s}=500$~GeV.
The use of beam polarization is essential to obtain a strong constraint
on Im($F_T$).  Also
indicated are values of $F_T$ for various technirho masses.  The
infinite technirho mass point is labelled Low Energy Theorem (LET)
since in this limit there is residual $W^+_{\rm L}W^-_{\rm L}$
scattering described by the same low energy theorems that govern low
energy $\pi^+\pi^-$ scattering. We see that the NLC at
$\sqrt{s}=500$~GeV can exclude technirho masses up to about 2.5~TeV and
can discover technirho resonances with masses of more than 1.5~TeV. The
significance of the 1.5~TeV technirho signal would be $6.7\sigma$. A
1.0~TeV technirho would produce a $17.7\sigma$ signal.  For comparison,
the minimal technicolor model predicts a technirho mass of about 2.0
TeV.

Figure~\ref{wlwlhigh} contains confidence level contours for the real
and imaginary parts of $F_T$ at $\sqrt{s}=1500$ GeV. The non-resonant
LET point is well outside the light Higgs 95\% confidence level region
and corresponds to a 4.5$\sigma$ signal. The labeling of points here
deserves some comment.  We use the model of \cite{peskinomnes} to
describe the form factor $F_T$.  In this model, as the vector resonance
mass is taken to infinity, its effect on the form factor decreases, and
what is left behind is the residual scattering predicted by the LET. 
The values for high-mass technirho indicate this decoupling. With this
understanding, the  6~TeV, 4~TeV and 3~TeV technirho points
correspond to 4.8$\sigma$, 6.5$\sigma$, and $11\sigma$ signals,
respectively.   A 2~TeV technirho would produce a $37\sigma$ signal.

It might appear that the value of $F_T$, and hence the significance of
technirho signals, would be very sensitive to the technirho width when
$\sqrt{s}$ is much less than the technirho mass.  In the model we have
considered, however, this is not true. The results presented above were
obtained assuming that $\Gamma_\rho/M_\rho=0.22$\ .   If, for example,
the technirho width is reduced to $\Gamma_\rho/M_\rho=0.03$ then the
1~TeV signal at $\sqrt{s}=500$~GeV is reduced from $17.7\sigma$ to
$16.3\sigma$, the 1.5~TeV signal at $\sqrt{s}=500$~GeV is reduced from
$6.7\sigma$ to $6.4\sigma$, and the 4~TeV signal at $\sqrt{s}=1500$~GeV
is reduced from $6.5\sigma$ to $6.3\sigma$.

\subsection{The Reaction ${e^+e^-\rightarrow \nu\bar \nu W^+W^-} 
\rm{and}\ {\nu\bar \nu ZZ}$}
\label{sec:nnww}

The important gauge boson scattering processes $W^+_{\rm L}W^-_{\rm
L}\rightarrow W^+_{\rm L}W^-_{\rm L}$ and $W^+_{\rm L}W^-_{\rm
L}\rightarrow  Z_L Z_L$ are studied at the NLC with the reactions
$e^+e^-\rightarrow \nu\bar \nu W^+W^-$ and $e^+e^-\rightarrow \nu\bar
\nu ZZ$. We describe the results that Barger {\em et
al.}~\cite{bargerhan} have obtained by analyzing these processes.

\begin{figure}[htb]
\leavevmode
\centerline{\epsfxsize=3.0truein\epsfbox{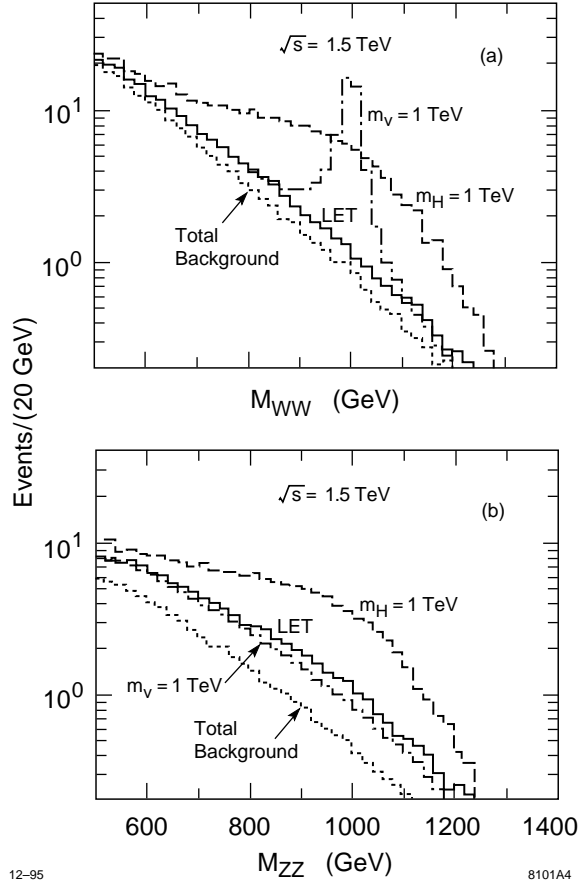}}
\centerline{\parbox{5in}{\caption[*]{Expected numbers of $\protect
W^+W^-,\ ZZ\rightarrow (jj)(jj)$ signal and background events after all
cuts for $\protect 200\ {\rm fb}^{-1}$ luminosity at $\protect
\sqrt{s}=1.5$~TeV: (a) $\protect e^+e^-\rightarrow \nu\bar \nu W^+W^-$,
(b) $\protect e^+e^-\rightarrow \nu\bar \nu ZZ$. The dotted histogram
shows total SM background. The solid, dashed and dot--dashed histograms
show signal plus background for the LET, SM, and CCV models,
respectively; CCS model results are close to the SM case.}
\label{sigbcknnww}}} 
\end{figure}

Barger {\em et al.} use several models to test the effectiveness of
their analysis of $e^+e^-\rightarrow \nu\bar \nu 
W^+W^-$ and  $\nu\bar
\nu ZZ$.  In addition to the Standard Model Higgs boson with mass
$m_H=1$~TeV, they use a Chirally--Coupled Scalar (CCS) model, a
Chirally--Coupled Vector (CCV) model, and the LET model. They utilize a
series of cuts to produce an event sample that is rich in the final
states $\nu\bar \nu W^+_{\rm L}W^-_{\rm L}$ and $\nu\bar \nu Z_L Z_L$.
Figure~\ref{sigbcknnww} shows the $M_{WW}$ and $M_{ZZ}$ distributions
after all cuts. The 1~TeV Higgs scalar resonance stands out in both the
$\nu\bar \nu WW$ and $\nu\bar \nu ZZ$ final states. The 1~TeV vector
resonance is prominent in the $M_{WW}$ distribution (of course such a
resonance would have been seen earlier as a $16\sigma$ signal at
$\sqrt{s}=0.5$~TeV, in $e^+e^-\rightarrow W^+W^-$). The LET signal is
larger for the final state $\nu\bar \nu ZZ$ than it is for $\nu\bar \nu
WW$.

\bigskip
\begin{table}[htb]
\centering
\caption[*]{Signal and background for $\protect e^+e^-\rightarrow
\nu\bar \nu W^+W^-$ and $\protect e^+e^-\nu\bar \nu ZZ$ with 100\%
initial state electron polarization, from Ref. \cite{bargerhan}.}
\label{tab:epemnnwwpol}
\bigskip
\begin{tabular}{|l|c|c|c|c|} \hline \hline
Signal $(S)$ or  & SM          & Scalar      & Vector      & LET \\
Background $(B)$ & $M_H=1$ TeV & $M_S=1$ TeV & $M_V=1$ TeV &     \\ 
\hline
$S(e^+e^-\rightarrow \nu\bar \nu W^+W^-)$ & 330 & 320 & 92  &  62 \\
$B$(backgrounds)          & 280 & 280 & 7.1 & 280 \\
$S/\sqrt{B}$              &  20 &  20 & 35  & 3.7 \\[2ex]
$S(e^+e^-\rightarrow \nu\bar \nu ZZ)$ & 240 & 260 & 72  &  90 \\
$B$(backgrounds)          & 110 & 110 & 110 & 110 \\
$S/\sqrt{B}$              &  23 &  25 & 6.8 & 8.5 \\ \hline \hline
\end{tabular}
\end{table}
\bigskip

The statistical significance of the signals for the different models is
given in Table~\ref{tab:epemnnwwpol} assuming 100\% initial-state $e^-$
polarization at $\sqrt{s}=1.5$~TeV and 200fb$^{-1}$ luminosity. Note
that the statistical significance of the LET signal is $8.5\sigma$ in
the $\nu\bar \nu ZZ$ channel.

\bigskip
\begin{table}[htb]
\centering
\caption[*]{Signal and background for $\protect e^-e^-\rightarrow \nu\bar \nu
W^-W^-$ with 100\% initial state electron polarization, from Ref.
\cite{bargerhanemem}.}
\label{tab:ememnnwwpol}
\bigskip
\begin{tabular}{|l|c|c|c|c|} \hline \hline
Signal $(S)$ or & SM & Scalar & Vector & LET \\
Background $(B)$ & $M_H=1$ TeV & $M_S=1$ TeV & $M_V=1$ TeV & \\ 
\hline
$S(e^-e^-\rightarrow \nu\bar \nu W^-W^-)$ & 110 & 140 & 140 & 170 \\
$B$(backgrounds)          & 710 & 710 & 710 & 710 \\
$S/\sqrt{B}$              & 4.0 & 5.2 & 5.4 & 6.3 \\ \hline \hline
\end{tabular}
\end{table}
\bigskip

The $I=2$ Goldstone boson scattering channel  can be probed at the NLC
through the reaction $e^-e^-\rightarrow \nu\nu W^-W^-$. Barger {\em et
al.}~\cite{bargerhanemem} have studied this reaction and they obtain
the statistical significances shown in Table~\ref{tab:ememnnwwpol}. An
initial state electron polarization of 100\% has been assumed for both
beams.

\subsection{The Reaction ${e^+e^-\rightarrow \nu\bar \nu t\bar t}$}
\label{sec:nntt}

If there is no light Higgs boson, the process $t\bar t\rightarrow
W^+W^-$ violates unitarity in the multi-TeV energy region. It is
natural then to ask if strong symmetry breaking can be detected through
the process $W^+W^-\rightarrow  t\bar t$. This process would be studied
at the NLC by observing the reaction $e^+e^-\rightarrow \nu\bar \nu
t\bar t$.

\begin{figure}[htb]
\leavevmode
\centerline{\epsfxsize=3.5truein\epsfbox{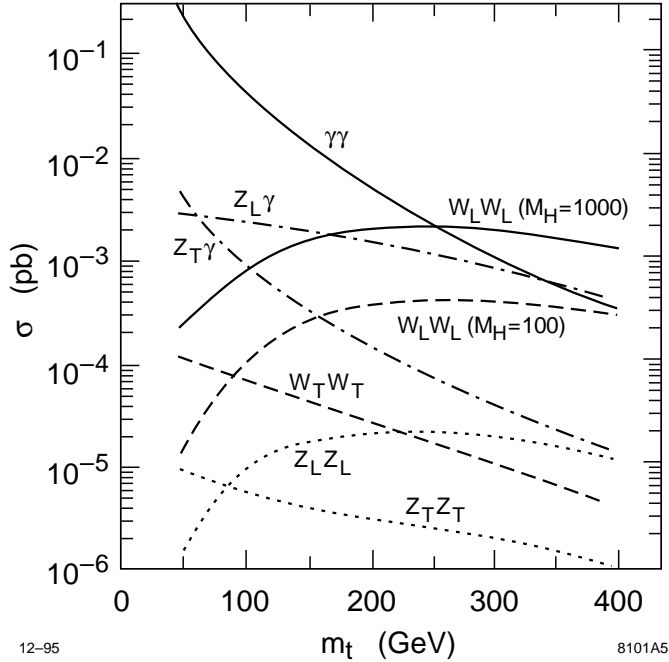}}
\centerline{\parbox{5in}{\caption[*]{Contributions from various
subprocesses to the total cross sections for $\protect
e^+e^-\rightarrow  e^+e^- t\bar t$ and $\protect e^+e^-\rightarrow
\nu\bar \nu t\bar t$.  The contributions are plotted as a function of
the top quark mass $\protect m_t$. The $e^+e^-$ center-of-mass energy
is 2~TeV.}
\label{signntt}}} 
\end{figure}

The total cross sections for $e^+e^-\rightarrow e^+e^- t\bar t$ and
$e^+e^-\rightarrow \nu\bar \nu t\bar t$ have been calculated by
Kauffman~\cite{kauffman} for an $e^+e^-$ center-of-mass energy of 2~TeV
(Fig.~\ref{signntt}). The cross sections for $\sqrt{s}=1.5$~TeV should
be similar. For $\sqrt{s}=2$~TeV and 200 fb$^{-1}$ luminosity we would
have 1200 events from $e^+e^-\rightarrow e^+e^- t\bar t$ and 60 (400)
events from $e^+e^-\rightarrow \nu\bar \nu t\bar t$ for $m_H=0.1\
(1.0)$~TeV.  A 1~TeV Higgs boson therefore produces a 30\% increase in
the sum of the cross sections for $e^+e^-\rightarrow e^+e^- t\bar t$
and $e^+e^-\rightarrow  \nu\bar \nu t\bar t$ before any cuts are
applied. The same 1~TeV Higgs boson produces only a 0.3\% increase in
the sum of the cross sections for $e^+e^-\rightarrow e^+e^- W^+W^-$ and
$e^+e^-\rightarrow \nu\bar \nu W^+W^-$\ .

The strong symmetry breaking signal can be further enhanced by
performing a helicity analysis on the $t\bar t$ final state to isolate
the polarization states $t_L \bar{t}_L$ and $t_R \bar{t}_R$
corresponding to helicity-flip pair production. 
This can be done by observing the $W$ energy distribution in $t$
decays.  Projecting out these
helicity combinations would be the analog of projecting out the
$W^+_{\rm L}W^-_{\rm L}$ and $Z_LZ_L$ final states in gauge boson
scattering.

\subsection{Statistical significances at LHC versus NLC}
\label{sec:statnlclhc}

The statistical significances of strong symmetry breaking signals at
the NLC and LHC are summarized in Table~\ref{tab:statsig}. The LHC
results are taken from the ATLAS design report~\cite{Atlas}. If an
entry is blank it may mean that the process is insensitive to the
corresponding model or that the analysis has not been done;  no
distinction is made between the two possibilities.  The entries for the
direct gauge boson scattering processes at the NLC ($W^+W^-\rightarrow 
ZZ$ and $W^-W^-\rightarrow  W^-W^-$) assume that the electron beam
always has 90\% left--handed polarization. For both the NLC and LHC
results it was assumed that the 1.5~TeV technirho had a width of
0.33~TeV.

\bigskip
\begin{table}[htb]
\centering
\caption[*]{Statistical significances of strong symmetry breaking signals 
at the NLC and LHC.}
\label{tab:statsig}
\bigskip
\begin{tabular}{|l|l|c|c||c|c|c|} \hline \hline
Collider & Process & $\sqrt{s}$ & ${\cal L}$&
$M_\rho=$& $M_H=$& LET \\
 &  & (TeV) & $(\rm{fb}^{-1})$ &
1.5 TeV & 1 TeV & \\
\hline
NLC & $e^+e^-\rightarrow W^+W^-$ & .5 & 80 & $7\sigma$ & -- & -- \\
NLC & $e^+e^-\rightarrow W^+W^-$ & 1.0 & 200 & $35\sigma$ & -- & -- \\
NLC & $e^+e^-\rightarrow W^+W^-$ & 1.5 & 190 & $366\sigma$ & -- & $5\sigma$ \\
NLC & $W^+W^-\rightarrow  ZZ$ & 1.5 & 190 & -- & $22\sigma$ & $8\sigma$ \\
NLC & $W^-W^-\rightarrow  W^-W^-$ & 1.5 & 190 & -- & $4\sigma$ & $6\sigma$
\\[2ex]
LHC & $W^+W^-\rightarrow W^+W^-$ & 14 & 100 & -- & $14\sigma$ & -- \\
LHC & $W^+W^+\rightarrow  W^+W^+$ & 14 & 100 & -- & $3\sigma$ & $6\sigma$ \\
LHC & $W^+Z\rightarrow  W^+Z$ & 14 & 100 & $7\sigma$ & -- & -- \\
\hline \hline
\end{tabular}
\end{table}
\bigskip

Some points about Table~\ref{tab:statsig} are worth noting:
\nopagebreak
\begin{enumerate}
\item 
Although the LHC can access larger $W^+W^-$ center-of-mass energies,
the statistical significance of the NLC signal is always larger. In
most cases, this is a small effect, but it makes a significant  point:
Electron beam polarization, smaller backgrounds, and the utilization of
the full $e^+e^-$ center-of-mass energy for vector resonances has
allowed the NLC to more than make up for its lower range of $W^+W^-$
center-of-mass energies.
\item 
The NLC has a special ability to detect vector resonances. At
$\sqrt{s}=500$~GeV the technirho mass reach of the NLC is equal to that
of the LHC.  At $\sqrt{s}=1500$~GeV the NLC is sensitive to strong
interaction effects in $I=J=1$ $W^+W^-$ scattering even when they are
nonresonant.
\item 
Signal significances for $W^+W^-\rightarrow  t\bar t$ were not included
in Table~\ref{tab:statsig} because detector simulations have not yet
been performed.  From the discussion in Section~\ref{sec:nntt},
however, it appears that this is a promising reaction for the study of
strong symmetry breaking at the NLC.  It is probably very difficult to
study $W^+W^-\rightarrow  t\bar t$ at LHC due to the large background
from $gg\rightarrow  t\bar t$.
\end{enumerate}

Finally it is important to remember that the significances shown in
Table~\ref{tab:statsig} include statistical errors only. Systematic
errors have largely been ignored in analyses so far, both for the LHC
and for the NLC.

\subsection {Conclusion}

An $e^+e^-$ linear collider with $\sqrt{s}=500-1500$~GeV would be an
effective partner to the LHC in the study of strong symmetry breaking. 
It provides important complementary capabilities for the discovery of
vector resonances and the extraction of chiral Lagrangian parameters.
The NLC and LHC are expected to have similar statistical errors for
scalar resonance and non--resonant signals in gauge boson scattering;
however,  the physics environments at the two machines are very
different, and the systematic errors  for the NLC analyses are probably
smaller. The NLC is probably the only machine that can study
$W^+W^-\rightarrow  t\bar t$.

\clearpage

\section{New Gauge Bosons and Exotic Particles}

%
%

In many extensions to the Standard Model, in particular those involving
an enlargement of the $SU(3)_c \times SU(2)_L \times U(1)_Y$ gauge
group or compositeness, new particles are expected to exist beyond
those associated with either supersymmetry or technicolor. The
properties and signatures of exotic particles have recently been
reviewed in \cite{rev} In almost all cases, these particles can be
directly produced and have their properties analyzed in $e^+e^-$
collisions given a sufficiently large center-of-mass energy. If such
energies are not available, the indirect influence of the existence of
many kinds of exotic particles can also be examined through precision
measurements at the NLC.

Perhaps the most well-studied of all exotic particles is the new
neutral gauge boson, $Z'$, which is a basic ingredient of all theories
with extended gauge sectors. (When the additional group factor is
non-abelian, new charged gauge bosons, $W'$, may also be present.) If
such a particle exists and is accessible at the TeV scale, it is
possible that it may first be directly produced via the Drell-Yan
process at the LHC, provided the $Z'$ couples to {\it both} quarks and
leptons as it does in most models. If the $Z'$ does not couple to
leptons it cannot be produced at the NLC; in this case, searches at the
LHC would also be difficult, since the effects of the $Z'$ would have
to be observed as perturbations of di-jet mass distributions. If the
$Z'$ couples only to leptons, then the NLC provides the unique means to
study it.

\begin{figure}[htbp]
\leavevmode
\centerline{\epsfxsize=5in \epsfysize=6.5in \epsfbox{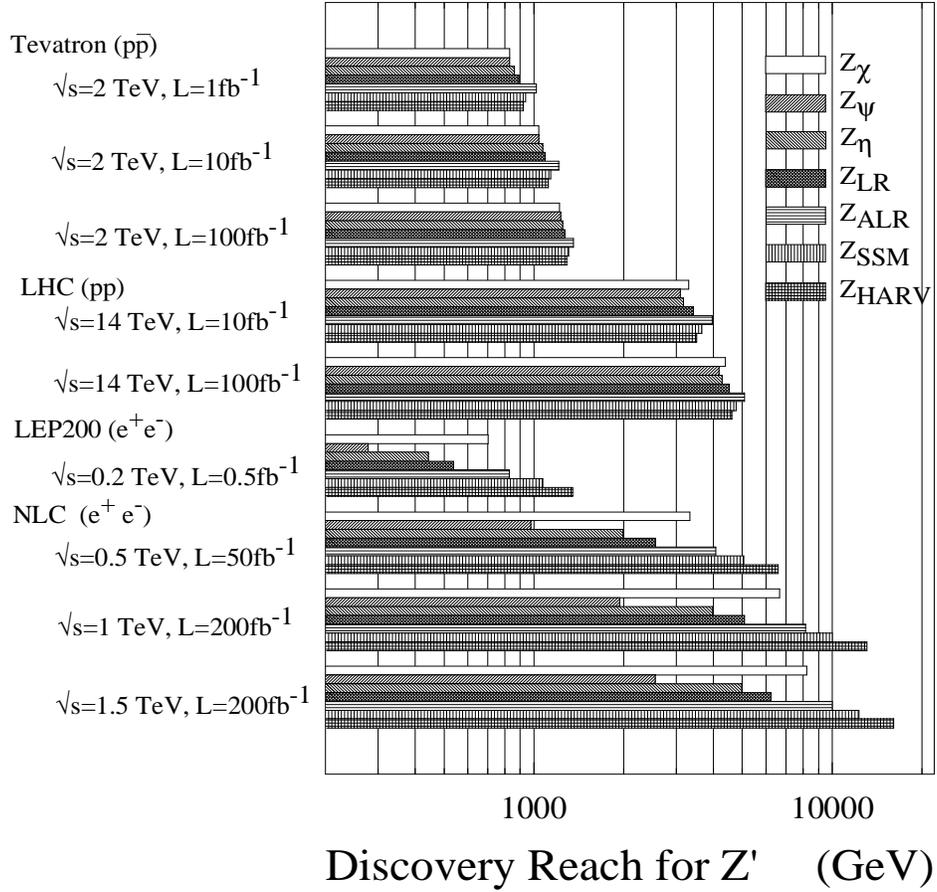}}
\centerline{\parbox{5in}{\caption[*]{Tevatron and LHC bounds are based
on 10 events in the $e^+e^-+\mu^+\mu^-$ channels; decays to SM final
states only is assumed. LEP and NLC bounds are $99\%$ CL using the
observables $\sigma_l,~R^{\rm had},~A_{LR}^l$ and $A_{LR}^{\rm had}$.}
\label{steve}}}
\end{figure}

It is possible that a $Z'$, though it exists, may not be seen at the
LHC due to its mass and/or the nature of its couplings to fermions. In
this case the NLC can extend the search reach for some models; in
addition the NLC may be able to determine both the mass and couplings
of the $Z'$. A `lucky' scenario would be one where the mass of the $Z'$
is less than the NLC center-of-mass energy; in this case, by sitting
on-resonance and repeating the LEP/SLC experimental program, we can
determine all of the $Z'$'s couplings to fermions. A much more likely
scenario is that $M_{Z'} > \sqrt {s}$ so that deviations in the cross
section and associated asymmetries for various flavors could be used to
look for {\it indirect} $Z'$ effects in a manner similar to the
observation of the $Z$ at PEP, PETRA, and TRISTAN below the resonance. 
In some cases, this indirect `reach' can be as large as $M_{Z'} \simeq
10 \sqrt {s}$. Figure {\ref {steve}} shows a comparison of the indirect
$Z'$ search capability at LEP and the NLC as well as the direct
production reach at the Tevatron and LHC for several extended models.

\begin{figure}[htbp]
\leavevmode
\centerline{\epsfxsize=3in \epsfbox{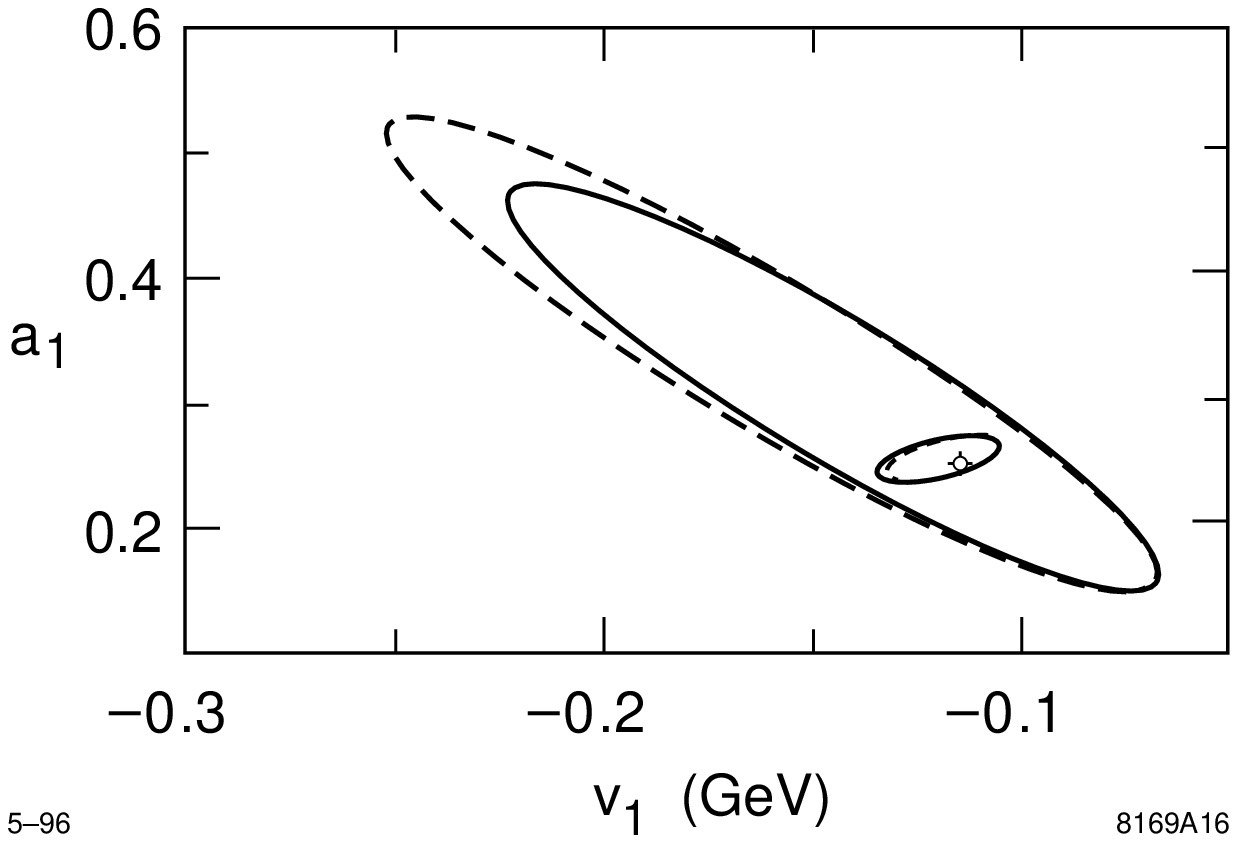}}
\centerline{\parbox{5in}{\caption[*]{Extracted leptonic couplings of a
1.53 TeV $Z'$ at the NLC, at $95\%$ CL, using data at 0.5(70 fb$^{-1}$),
0.75(100 fb$^{-1}$), and 1 TeV(150 fb$^{-1}$). All leptonic observables
were used and generation independence was assumed. The large ovals are
applicable when $M_{Z'}$ is unknown a priori and must be obtained from
the fit; the solid (dashed) curve corresponds to electron beam
polarization $P=90(80)\%$. The small ovals correspond to the scenario
in which  $M_{Z'}$ is known (for example, from the LHC); the two curves
represent the same two cases. The dot represents the input values
of the couplings.}
\label{newzcouplings}}}
\end{figure}

If a $Z'$ exists, we would like to know how it couples to the various
flavors so that we can identify the correct extended gauge structure.
If the $Z'$ is seen at the LHC and its mass is not much larger than
about 1 TeV, several analyses have shown {\cite {rev}} that the LHC and
a $\sqrt s=500$ GeV NLC will nicely complement each other in extracting
fermion coupling information. If the $Z'$ is more massive, then LHC
cannot perform coupling analyses using any of the presently available
techniques either due to backgrounds or statistical limitations.
However, if the mass of the $Z'$ is less than about 2 TeV, then the NLC
running in the $\sqrt s =0.5-1$ TeV range can obtain significant
coupling information even if the $Z'$ is missed by the LHC and its mass
is {\it a priori} unknown. (This value goes up to $M_{Z'}\simeq 3$ TeV
if $\sqrt s=1.5$ TeV energies are available.) A representative example
of such an analysis is shown in Fig. {\ref {newzcouplings}} {\cite
{tgr}}.  Similar results can be obtained for different values of the
input parameters as well as for the other fermion flavors (although
fewer observables are available), thus allowing the structure of the
extended gauge model to be determined by comparing the extracted values
of these couplings with specific model predictions.

In generic Standard Model  extensions, the new gauge bosons are usually
accompanied by additional fermionic and/or bosonic degrees of freedom
the most common of which are either vector-like or singlets with
respect to the Standard Model  electroweak group. In addition to new
quarks and leptons, some of these particles, such as leptoquarks (LQs)
and diquarks, may have atypical $B$ and $L$ number assignments. The
production and decay of both spin-0 and spin-1 LQs at $ep, e^+e^-$, and
$pp/p\bar p$ colliders has been studied by many groups {\cite
{rev,tgr}}. Allowing for both helicities in their interactions with
ordinary quarks and leptons there are 14 possible species of LQs with
reasonably strict constraints applicable to their intergenerational
couplings. (Low energy constraints also force LQ couplings to be
chiral.) At $ep$ colliders, LQ production rates are proportional to the
squares of unknown Yukawa couplings so they may easily be missed. By
contrast, at hadron colliders, the cross section is the same for all LQ
types of a given spin but the signal depends upon how the LQ decays.
Thus it will not be possible at hadron colliders to uniquely determine
which LQ type is being produced. On the other hand, at $e^+e^-$
machines, the production cross sections and asymmetries are determined
solely by the LQ electroweak quantum numbers and spin. Several studies
have shown {\cite {rev}} that the LQ species can found and identified
in both $e^+e^-$ and $\gamma e$ collisions. In the first case the `ID
reach' is $\simeq 0.45 \sqrt s$; this improves to $\simeq 0.8 \sqrt s$
in the $\gamma e$ mode if the Yukawa couplings are of order
electromagnetic strength. In a manner similar to the $Z'$ searches
discussed above, $t$ and $u$ channel LQ exchange can be probed for
indirectly via the process $e^+e^-\rightarrow q\bar q$ provided the
Yukawa couplings are sufficiently large. For Yukawa couplings of 
electromagnetic strength, the reach can exceed 6 TeV at the NLC.

\bigskip
\begin{table}[ht]
\centering
\caption[*]{Limits on various exotic particles, in GeV.
Run II is 2 fb$^{-1}$ at
2 TeV; NLC is 200 fb$^{-1}$; LHC is 14 TeV and 100 fb$^{-1}$. $Z'$ limits
at hadron colliders assume only SM final states are accessible. $Z'$
limits at $e^+e^-$ colliders are indirect. $^\star$ is the Tevatron
limit assuming the standard $W'$ search assumptions. $^\dagger$ limit
applies when none of the conditions in text are fulfilled; $^{\dagger
\dagger}$ result using $Z'$ bound and LR model relationships. $^+$ is
from single production. The limit is for first generation spin-0,1 LQs
at HERA and the Tevatron including Run 1b estimate; similar numbers
will hold for the 2nd generation case at hadron colliders. In the
dilepton, diquark and LQ cases, EM strength Yukawas are assumed.
$^{++}$ are indirect limits from $e^+e^-\rightarrow q\bar q$ with these
EM strength Yukawas. $^{\ast \ast}$ combines several search modes.
$^\ddagger$ is from $e^+e^-\rightarrow 2\gamma$.  $^\ast$ implies the
limit is highly decay mode dependent. The value of `60' as the present
limit for the last four entries reflect approximate LEP1.4 null
searches. New lepton limits are for both charged and neutral cases.
Many entries are estimates.} 
\label{limits}
\bigskip
\begin{tabular}{|l|c|c|c|c|c|} \hline\hline
Particle & Present & LEPII & TeV(Run II) & LHC & NLC(1 TeV)  \\ \hline
$Z'_{SM}$    &650 &1070 &1020 &4760 &10000 \\
$Z'_{\chi}$    &425 &703 &910 &4380 &6670 \\
$Z'_{\psi}$  &415 &278 &910 &4180 &1940 \\
$Z'_{\eta}$  &440 &443 &940 &4280 &3980 \\
$Z'_{LR}(\kappa=1)$ &445 &538 &970 &4520 &5090 \\
$Z'_{ALRM}$    &420 &820 &1100 &5080 &8150  \\
$W_R$      &720$^\star$,300$^\dagger$ &90& 990,400$^\dagger$ 
          &5310,1500$^{\dagger \dagger}$ & 800$^+$  \\
LQ(spin-0)& 180&90,160$^+$,700$^{++}$&300,400$^{++}$&1400,1700$^{++}$
             &500,900$^+$,6000$^{++}$ \\
LQ(spin-1)& 300 &90,160$^+$  &440,500$^{++}$&2200,2500$^{++}$  
             &500,900$^+$,?$^{++}$   \\
Axigluon                 & 200-930 &150? &1160  &5000?  & 800  \\
$E_6$ diquark            & 280-350 &90  & 200-570  &5000  &500   \\

$q^*$       & 90-750$^{\ast \ast}$ &90,180$^+$& 820&5000-6000?&500,900$^+$   \\
$e^*$          &60,127$^\ddagger$ &90,180?$^+$&500? &4000  &500,900$^+$   \\
new leptons    &60   &90  &??  &1000  &500   \\
new quarks     &60  &90  &$^\ast$  &$^\ast$  &500   \\
dileptons      &60  &90  &--  &--  &900$^+$   \\ \hline\hline
\end{tabular}
\end{table}
\bigskip

Table \ref{limits} shows a comparison of the search capabilities of
LEPII, the Tevatron, the LHC and NLC for a number of different exotic
particles. As is clear from the above, great care should be exercised
when comparing the capabilities of different machines to find and
explore the properties of exotics. The results in the table for
$e^+e^-$ colliders are indirect bounds for $Z'$'s but represent direct
production limits at hadron colliders. Limits on all other particles in
the case of $e^+e^-$ colliders are for direct production except where
noted. It is important to remember that in many other cases $e^+e^-$
colliders provide important indirect bounds as in both the $Z'$ and LQ
examples. For a $W'$ in the Left-Right Symmetric Model ($W_R$), a
number of additional assumptions are required before search reaches can
be quoted for hadron colliders; these are: ($i$) equal left- and
right-handed gauge couplings and CKM matrices, ($ii$) right-handed
neutrinos appear only as missing $p_t$, ($iii$) only Standard Model 
final states are accessible in decays. The possibility of surrendering
any or all of these assumptions has been analyzed and was found to lead
to downward, potentially drastic, alterations in search expectations
via the Drell-Yan process. Searches for new quarks at hadron colliders
are also subject to decay mode assumptions and are species dependent.
In some cases, for example, flavor changing neutral current decays are
at least as important as the more conventional charged current ones.

\clearpage

 
\section{$e^-e^-$, $e^-\gamma$, and $\gamma\gamma$ Interactions}
 
\begin{figure}[htb]
\leavevmode
\centerline{\epsfxsize=4in \epsfbox{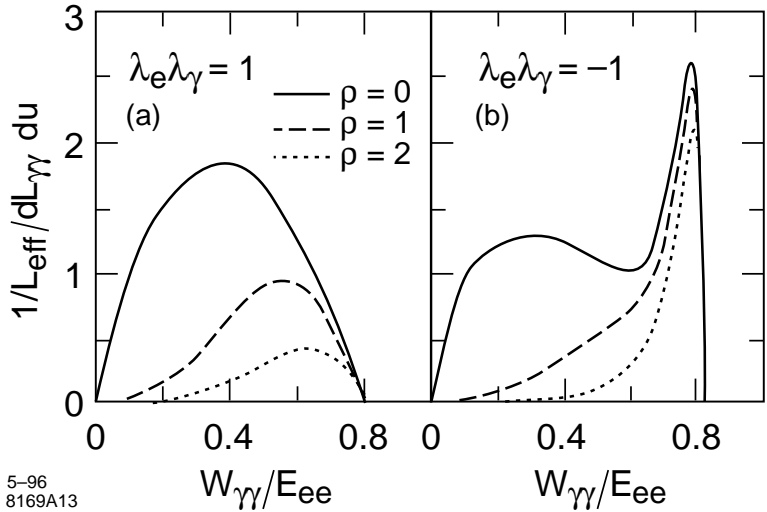}}
\centerline{\parbox{5in}{\caption[*]{$\gamma\gamma$ luminosity for two
different combinations of photon and electron polarization, and several
values of $\rho$, the ratio of the intrinsic spreads of the photon and
electron beams.}
\label{gammapol}}}
\end{figure}

Up to this point, we have concentrated on the reactions available at
the NLC in $\ee$ collisions.  However, it is possible with a linear
electron collider to study more general processes, and, in fact, to
collide $e^-$, $e^+$, and $\gamma$ beams in any combination.  A photon
beam with substantial brightness at high energy can be created by
Compton backscattering of a laser beam from the electron beam.  At
least for the $e^-$ and $\gamma$ combinations, the polarizations of the
two beams can be controlled independently.  The energy spectrum of the
backscattered Compton photon beam depends on the relative electron and
laser beam polarization in the manner shown in Fig.~\ref{gammapol}. Thus,
polarization is also useful in tuning the $\gamma$ energy spectrum, for
example, to produce a peak at a fixed energy.

The availability of these additional initial states at the NLC adds
significantly to the overall physics program.  We have already noted
their role in the study of Higgs bosons, anomalous electroweak
couplings, and strong $WW$ scattering. The exotic quantum numbers of
the $e^-e^-$ initial state and the quasi-hadronic nature of the photon
also permit searches not available to $e^+e^-$ annihilation. Table
\ref{tablec} illustrates the variety of physics issues that have been
discussed for each of these modes. The physics expectations for 
$e^-e^-$ and $\gamma\gamma$ colliders have been recently been reviewed
in specialized workshops \cite{ee95,ggLBL}.

In this report, we give special consideration to the possibility of
measuring the Higgs boson coupling to $\gamma\gamma$ by creating the
Higgs boson as a resonance in $\gamma\gamma$ scattering.  We will then
review a number of other physics topics which are special to $e^-e^-$
or $\gamma\gamma$ reactions.  We conclude this section with a
discussion of the expectations for the luminosity and the interaction
region design for these reactions.
 
\begin{table}[htb]
\centering
\caption[*]{A listing of topics that will be investigated at a
TeV-level linear collider. Check marks show which different initial
states contribute prominently to their study, highlighting the
unique contributions that $e^-e^-$, $e^-\gamma$, and $\gamma\gamma$
options will be able to make.}
\label{tablec}
\bigskip
\begin{tabular}{|l|ccccccccc|} \hline \hline
& QCD & $t\bar t$ & H & SUSY & Anomalous & $e^*$ & $\nu_M$
      & $Z'$ & $X^{--}$  \\[-8pt]
&     &      & && Couplings &&&&\\ \hline
$e^-e^-$    &      &      & $\surd$ & $\surd$ & $\surd$
                   &  $\surd$ & $\surd$ &$\surd$ &$\surd$\\
$e^+e^-$    &$\surd$ &$\surd$ & $\surd$ & $\surd$ & $\surd$ &&&$\surd$ &\\
$e^-\gamma$ &$\surd$ &      & $\surd$ & $\surd$ & $\surd$ &  $\surd$ &&&$\surd$
 \\
$\gamma\gamma$ & $\surd$    &$\surd$  & $\surd$ & $\surd$ &  $\surd$ &&&&\\
\hline \hline
\end{tabular}
\end{table}
\bigskip
 
\subsection{Higgs Boson Studies}
 
A photon linear collider provides a clean method to search for an
intermediate mass Higgs boson, via the reactions $\gamma\gamma$ $\to H
\to $$b\bar b$\ or ZZ.  It is complementary to searches using hadron
and $e^+e^-$\ machines, being sensitive to different models and
couplings. More importantly, a $\gamma\gamma$\ linear collider permits
a direct measurement of the two-photon width of the Higgs. The coupling
of the Higgs boson to two photons involves loops in which any charged
fermion or boson with couplings to the Higgs must contribute. The
dominant contributions come from species heavier than that Higgs boson.
Thus, a  measurement of the two-photon width is quite sensitive to new
physics even at higher mass scales.  Supersymmetric models,  and other
extensions of the Standard Model with more complicated Higgs sectors
typically predict Higgs spectra and two-photon couplings which differ
substantially from those of the Standard Model \cite{GHphoton}. In the
minimal supersymmetric model (MSSM), the  heavy neutral bosons $A^0$
and $H^0$ can be studied up to higher masses than in the $\ee$ mode
\cite{oddHiggs}. In technicolor models, in which Higgs bosons are
composite, there are often light CP-odd bound states with cross
sections for $\gamma\gamma$ production similar to those of elementary
Higgs bosons \cite{Tandean}. Finally, the dependence of the Higgs
production on the photon polarization tests the CP properties of the
Higgs boson, and may reveal CP violation in the Higgs sector
\cite{CPgunion2,Anlauf}.
 
\begin{figure}[htb]
 \leavevmode
\centerline{\epsfxsize=3.5in\epsfbox{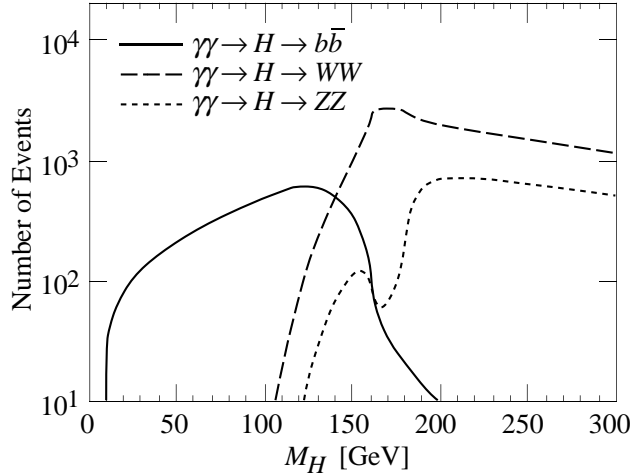}}
\centerline{\parbox{5in}{\caption[*]{Event rates for $\gamma\gamma$
$\to H \to $$b\bar b$\ and $\gamma\gamma$ $\to H \to WW,ZZ$, assuming a
luminosity function of $4 \times 10^{-2}$ fb$^{-1}$/GeV and
$\lambda_1\lambda_2=1$.}
\label{fig:NHiggs}}}
\end{figure}

For a Standard Model Higgs boson with mass below about 300~GeV, the
beam energy spread of a $\gamma\gamma$\ collider is much greater than the
total width of the Higgs boson, and so the number of $H \to X
(=$$b\bar b$$,WW,ZZ)$ events expected is
\begin{equation}
  N_{H \to
 X}=\left.\frac{dL_{\gamma\gamma,J_z=0}}{dW_{\gamma\gamma}}\right|_{M_H}
       \frac{8\,\pi^2\,\Gamma(H \to \gamma\gamma)\,B(H \to X)}{M_H^2}\, ,
\label{eqbbb}
\end{equation}
\noindent
where $W_{\gamma\gamma}$ is the two-photon invariant mass and the
initial-state photons have the same helicity.
Since the Higgs boson has spin zero, the initial photons must be in a
$J_Z=0$ state. Event rates for Higgs production and decay into the
three primary final states of interest are shown in Fig.
\ref{fig:NHiggs} \cite{Borden}.
                                                                             
In the intermediate mass region ($<150$ GeV) the Higgs decays
dominantly to $b\bar b$.  With vertex tagging to remove the backgrounds
from light quark production,  the most important backgrounds arise from
continuum production of heavy quarks.  These backgrounds are quite
large, but can be actively suppressed by exploiting the polarization
dependence of their cross-sections, which are dominated by the $J_z =
\pm 2$ configurations of initial photons, and by the use of angular
cuts. Additional event shape and jet width cuts must be used to
suppress radiative processes \cite{Borden2}.
 
Several additional potential backgrounds might fake the presence of a
Higgs boson with mass close to the $Z$ mass. The most important of
these is the reaction $e\gamma$ $\to eZ \to e$$b\bar b$, initiated by a
residual electron left over from the original Compton backscatter.  To
avoid this background, it will be necessary to sweep these spent
electrons away from the interaction point. The process $\gamma\gamma$
$\to f\bar{f}Z$, where the $f\bar{f}$ go down the beam pipe and the $Z$
decays to $b\bar b$\ provides a background of the same order of
magnitude as that of a $\approx 90~$GeV Higgs signal \cite{Ginzburg}.
These backgrounds are less important if the Higgs boson mass is already
known to be separated from that of the $Z$ by at least the experimental
mass resolution, for example, from the observation of the Higgs in
$\ee\to Z H$.
 
For Higgs boson masses above about 150 GeV, the dominant Higgs decay is
to $WW$, where one of the $W$'s may be virtual. However, the large
continuum cross section is not easily suppressed, so the $WW$ final
state will be a difficult one to use for doing Higgs physics.
Fortunately, the $ZZ$ or $ZZ^*$  decay channel can be utilized. Except
for Higgs masses between the $WW$ and $ZZ$ thresholds, the Higgs has a
branching fraction into this channel of approximately 1/3, while the
Standard Model cross section for $\gamma\gamma$ $\to ZZ$ is small.
Hadronic decays of the $Z$ bosons predominate, but the huge
$\gamma\gamma$ $\to WW$ cross section results in a large number of
background events in which both $W$'s are misidentified as $Z$'s. Thus,
to tag the Higgs unambiguously, it is necessary to require that at
least one $Z$ decays leptonically. For Higgs masses above about 350
GeV, the $\gamma\gamma \to ZZ$ continuum background becomes large and
makes detection of the Higgs very difficult \cite{Jikia}.
 
Detailed Monte Carlo studies \cite{Borden,MCHiggs} indicate that it
should be possible to measure the two-photon width of a Higgs to a
precision better than $10\%$ over most of the mass range up to 250 GeV.
Measurements of this quality should distinguish among many of the
competing models for Higgs production. Further discrimination will be
supplied if several Higgs bosons can be detected and their
$\Gamma_{\gamma\gamma}$ values obtained.
 
For $e^-e^-$ collisions, we note that the Standard Model Higgs signals
from the reactions $e^-e^- \rightarrow e^-e^-H, e^-e^-ZH,$ and
$e^-\nu_eW^-H$ have cross sections of order $10\%$ of the leading modes
for the $e^+e^-$ initial state. Moreover, in the intermediate mass
range where $H \rightarrow b\bar b$ decays dominate, there are
effective cuts which lead to very clean data samples \cite{HanEE}. For
$m_H >200$ GeV, the process $e^-e^- \rightarrow e^-e^-H$ via ZZ fusion
becomes a very favorable discovery channel for the Higgs: It overtakes
the process $e^+e^- \rightarrow ZH$  in cross-section with rising
$\sqrt{s}$, and the background rejection is more effective than in
$e^+e^- \rightarrow H \nu \nu$ via $WW$ fusion because there is no
missing momentum and both electrons can be detected \cite{Hikasa}. In
certain models with extended gauge symmetries, $e^-e^-$ collisions also
offer the possibility of double Higgs production ($H^- H^-$)
\cite{RizzoEE} or  doubly charged Higgs production ($H^{--}$)
\cite{GunionEE}.

\subsection{Other New Physics Signatures}
 
Experiments on $e^-e^-$, $e^-\gamma$, and $\gamma\gamma$ collisions can
also give additional insight into many of the other physics topics we
have discussed.  In the study of the top quark, for example, the
$e^-\gamma$ reactions offers the possibility of creating top quarks
singly in a well-controlled way through $e^-\gamma \rightarrow \nu\bar
t b$. This reaction has been suggested as a method for obtaining a very
precise value of $V_{tb}$, with an accuracy better than 3\%
\cite{Jikia2}.  The $\gamma\gamma$ and $e^-\gamma$ colliders are also
powerful tools for tests of QCD in the study of the photon structure
function and other manifestations of the photon hadronic structure
\cite{BrodskyHH}.

In supersymmetric theories, the reaction $e^-e^-\to \tilde e^-\tilde
e^-$ is a promising mode for the discovery and study of the selectron. 
In regions of parameter space where the neutralinos are heavy or
degenerate, this reaction has the advantage of exceptionally low
background \cite{Cuypers}.  In addition, since it proceeds by
neutralino  exchange, the cross section for this reaction can be used
to measure the neutralino masses and mixings. If the selectron is
heavy, the kinematic reach for its discovery can be extended by
searching for  single selectron production via $e^-\gamma \rightarrow
\tilde{e}\tilde{\gamma} \rightarrow e\tilde{\gamma}\tilde{\gamma}$
\cite{ChoudCuy}.

We have discussed the utility of $e^-\gamma$ and $\gamma\gamma$
collisions in the study of anomalous $W$ couplings at the Section 6.4.
In Section 7.2, we have pointed out the importance of the reaction
$e^-e^- \to \nu\bar\nu W^-W^-$ in the general program of studies of a
strongly interacting Higgs sector.  We should also point out the
possibility of studying $WW$ scattering at $\gamma\gamma$ colliders.
The probability for a photon to branch into a virtual $W^+W^-$ is large
at high energies.  Thus, it is interesting to observe the process
$\gamma\gamma\to W^+W^-W^+W^-$ and to use the spectator $W$'s at
relatively low transverse momentum to define the kinematics of a hard
$WW$ scattering process. It has been shown that it is possible to
extract a significant signal for the scattering of longitudinal $W$
bosons in $\gamma\gamma$ collisions at 1.5 TeV \cite{BrodskyHH,Cheung}.
 
In the study of exotic particles and  extended gauge groups, there are
many interesting reactions which are probed most easily in $e^-e^-$
scattering. If heavy Majorana neutrino exist and mix with the Standard
Model neutrinos, the reaction $e^-e^- \rightarrow W^-W^-$ might have a
significant cross section, estimated at 150 events/$100 fb^{-1}$ at a 1
TeV NLC  \cite{Heusch}. This process would directly probe lepton number
violating couplings. If the gauge group is just $SU(2)\times U(1)$, the
signal should appear only in the specific polarization state $e^-_L
e^-_L$. Unified models in which  the three generations are embedded
into a single representation of the gauge group predict the existence
of gauge bosons with dilepton quantum numbers.  These can be discovered
as s-channel peaks in $e^-e^-$ scattering \cite{Frampton}. An
$e^-\gamma$ collider can also be sensitive to dilepton gauge bosons
over almost the whole mass region less than $\sqrt{s}$, down to
coupling strengths as small as $10^{-4}\alpha$ \cite{Lepore}. Other
exotic states such as leptoquark bosons can be studied in $e^-\gamma$
and $\gamma\gamma$ reactions in a manner which complements their
observation in $\ee$ annihilation \cite{Nadeau}.  In particular, models
of  leptoquark bosons typically predict zeros in the angular
distribution for production in $e^-\gamma$, and the identification of
these features can be a powerful tool in discriminating different
models \cite{CLepto}.

Finally, M\o ller scattering at high energy is an ideal way to search
for lepton compositeness and other sources of lepton contact
interactions. For example, with 50 fb$^{-1}$ of data and 90\% polarized
beams,  the measurement of $e^-e^-\to e^-e^-$ can establish a 95\%
confidence limit on the electron compositeness parameter
$\Lambda_{LL}^+$ of 140 TeV \cite{BarklowEE}. Similarly, the
sensitivity to the presence of a $Z'$ boson is somewhat greater in M\o
ller scattering than in Bhabha scattering at the same energy and
luminosity \cite{Choudhury2,Cuypersee}, though $\ee$ annihilation
offers many additional channels to study. The $e^-\gamma$ option allows
for searches for excited electrons via $e^-\gamma \rightarrow e^*
\rightarrow e^-\gamma$ which are sensitive almost up to the kinematic
limit for couplings of size $10^{-5}\alpha$, and the $\gamma\gamma$
option offers similar sensitivity up to masses of   $0.7\sqrt{s}$ for
excited lepton searches in $\gamma\gamma\to ee^*, \mu^*, \tau\tau^*$
\cite{BoudjemaEE}.
 
\subsection{Accelerator, Lasers, and the Interaction Region}
 
As will be explained in Chapter 3, the NLC will include a bend before
the interaction region to remove muon and other lower-energy background
in the beam.  Thus, it is possible to have two interaction regions, one
on each side of the linear accelerator.  It is natural to design one of
these interaction region to be optimized for $e^-\gamma$ and
$\gamma\gamma$ collision.  Since these modes require $e^-e^-$ beams to
allow complete control of the polarization \cite{Telnov}, it is natural
to carry out $e^-e^-$ experiments also in this separate region and also
consider a final focus configuration optimized for $e^-e^-$ running. 
The detailed design of this interaction region has been begun in
\cite{ZDR}.  We now summarize some of the design considerations.
 
Of the three reactions to be studied in this interaction region,
$e^-e^-$ is the easiest to realize.  The major difference from $\ee$ is
that the like charges of the beams causes a repulsion which tends to
reduce the luminosity by about a factor of 3 relative to the $\ee$
design if no adjustment is made in  the final focus. In the
$\gamma\gamma$ and $e^-\gamma$ cases, there is an inefficiency in
converting an electron beams to a photon beam and preserving the tiny
spot size required to achieve high luminosity. On the other hand,
photon beams do not have a strong beam-beam interaction with
beamstrahlung radiation and $\ee$ pair creation, so it is possible to
recover part of what is lost by colliding more intense beams and by
making the beams round rather than flat, as required for $\ee$
operation.  For an $\ee$ collider capable of producing a luminosity of
$5\times 10^{33}$ cm$^{-1}$sec$^{-1}$, a reasonable goal would be to
obtain luminosities of $2.5\times 10^{33}$ cm$^{-1}$sec$^{-1}$ in
$e^-e^-$,  $3\times 10^{33}$ cm$^{-1}$sec$^{-1}$ in $e^-\gamma$,  and 
$10^{33}$ cm$^{-1}$sec$^{-1}$ in $\gamma\gamma$. In the $\gamma$
reactions, this goal refers to the luminosity within a 10\% band width
in center of mass energy.
 
To produce a $\gamma$ beam by Compton backscattering from an electron
beam,  the final focus must accommodate a system of mirrors designed to
focus light from intense laser beams onto Compton conversion points
located about 5 mm from the interaction point. The laser required for
the Compton conversion must have tens of kilowatts of average power,
compressed to a peak power of 1 Terawatt in a pulse matched to the
electron bunch. A wavelength of about 1 micron is required, with close
to 100\% polarization. The laser pulse timing should match that of the
electron beam. Such a laser could be built by either combining diode
pumping and chirped pulse amplification in solid state lasers or by a
free electron laser driven by an induction linac and using chirped
pulse amplification. We note that the SLAC experiment E-144 \cite{E144}
has succeeded in creating Terawatt laser pulses, at a repetition rate
of about 1 Hz, and has demonstrated their collisions with the beam spot
of the Final Focus Test Beam.
 
Round beams are required both to maximize the luminosity and to match
the laser beam profile at the conversion point.  Focussing with
$\beta_x^* = \beta_y^* \approx 0.5$ mm would produce a beam spot size
$\sigma_x^{CP} = 718$ nm, $\sigma_y^{CP} = 91$ nm at the conversion
point.  For these conditions, the Compton conversion efficiency is
about 65\%  per beam, so the luminosity in $\gamma\gamma$ collisions is
necessarily less then 40\% of the  geometric luminosity. Only about
20\% of this is in the spectral peak at high energy shown in
Fig.~\ref{gammapol}, so the resulting luminosity in high-energy
$\gamma\gamma$ collisions is about 10\% of the geometrical expectation.
After Compton scattering, the degraded electrons continue towards the
interaction point and must be deflected when  $e^-e^-$ or $e^-\gamma$
collisions are not desired. The simplest way of doing this is to bring
the beams together with a small vertical offset ($\approx 1 \sigma_y$),
and then rely on the mutual repulsion of the electrons to bend these
away. There are alternative proposals which involve using a small
sweeping magnet or plasma lens near the interaction point; these will
require more study to see if they can be implemented without degrading
luminosity and detector performance. The dispersion of the degraded
electrons also requires  a somewhat larger crossing angle (currently
estimated as 30 mrad) at the interaction point than is planned for 
$e^+e^-$ collisions.
 
While the inclusion of  the hardware of the interaction region puts
some special requirements on the second detector, the physics goals
demand performance comparable to that in $e^+e^-$. Certain backgrounds
may be more severe, especially those due to the spent electron beams.
Luminosity monitoring will require special, low-angle detectors for M\o
ller scattering in $e^-e^-$ mode and for $\gamma\gamma \rightarrow
\ell^+\ell^-$.

\clearpage

\section{Precision Tests of QCD}


Tests of QCD are both enriching and essential to the program of
measurements to be made at the NLC. Since QCD {\it is} our theory of
strong interactions it would be irresponsible not to test it at the
highest energy scales available in different hard scattering processes.
In addition, the precise determination of the strong coupling
$\alpha_s$\ is key to a better understanding of high energy physics. 
For example, the current precision of $\alpha_s(m^2_z)$\ measurements,
limited to 5--10\%, results in the dominant uncertainty on our
prediction of the energy scale at which grand unification of the
strong, weak and electromagnetic forces takes place \cite{qcd1}.
Measurements of hadronic event properties at high energies, combined
with existing lower energy data, would allow one to test the gauge
structure of QCD by searching for anomalous `running' of observables,
such as the 3-jet event rate, and to set limits on models which predict
such effects, for example those involving light gluinos. Gluon
radiation in $t\bar t$\ events is expected to be strongly regulated by
the large mass and width of the top quark; $t\bar tg$\ events will
hence provide an exciting new domain for QCD studies. Conversely,
measurements of gluon radiation patterns in $t\bar tg$\ events may
provide valuable additional constraints on the top quark decay width.
In addition, searches could be made for anomalous chromo-electric and
chromo-magnetic moments of top quarks, which modify the rate and
pattern of gluon radiation and for which the phase space increases as
the c.m. energy is raised. Finally, polarized electron beams could be
exploited to allow tests of symmetries using multi-jet final states.

\subsection{Precise Measurement of $\alpha_s$}

Tests of QCD can be quantified in terms of the consistency of values of
the yardstick $\alpha_s(m_z^2)$\ measured in different experiments.
Measurements of $\alpha_s(m_z^2)$\ have been performed over a range of
$Q^2$ from roughly 1 to $10^4$ GeV$^2$ \cite{qcd2}, and are consistent
within the errors; an average yields $\alpha_s(m^2_z)$\ = $0.117\pm0.006$,
implying that QCD has been tested to a precision of about 5\%, which is
modest compared with the achievement of sub-1\% level tests of the
electroweak theory. This is due primarily to the {\it theoretical
uncertainties} that dominate the measurements. These uncertainties are
due to both the restriction of perturbative QCD calculations to low
order, and non-perturbative (`hadronization') effects that are
presently incalculable in QCD. We consider whether a measurement of
$\alpha_s(m^2_z)$\ at the 1\%--level of precision is possible at the NLC
by extrapolation of a recent measurement from $\ee$ annihilation at the
$Z^0$ resonance by the SLD Collaboration, based on 15 hadronic event
shape observables measured with a sample of 50,000 hadronic events
\cite{qcd3}:
\begin{equation}
\alpha_s(m_Z^2)\quad=\quad 0.1200\pm 0.0025 \;{\rm (exp.)} \pm 0.0078\;
{\rm (theor.)} \ .
\label{eqccc}
\end{equation}
The experimental error is composed of statistical and systematic
components of about $\pm 0.001$ and $\pm 0.002$ respectively, and the
theoretical uncertainty has components of $\pm 0.003$ and $\pm 0.007$
arising from hadronization and missing higher order terms, respectively.

Based on this experience, we can estimate the errors to be expected for
a similar measurement of $\alpha_s$ at a 500 GeV NLC:
 
{\bf Statistical error:} At design luminosity, the NLC would deliver
roughly 200,000 $\qq$ (q=u,d,s,c,b) events per year implying that a
statistical error on $\alpha_s(m^2_z)$\ of about $\pm$ 0.0005 could be
obtained. Cuts for rejection of W$^+$W$^-$ and $t\bar t$\ backgrounds,
based on kinematic information as well as beam polarization and
b-tagging, will not substantially reduce the $\qq$ sample size.
 
{\bf Systematic error:} This results primarily from the uncertainty in
modeling the jet resolution of the detector. The situation may be
improved at the NLC both from an improved calorimeter and from the
naturally improved calorimeter energy resolution for higher energy
jets. It is not unreasonable to suppose that the current systematic
error of $\pm0.002$ could be reduced by a factor of two, but more
convincing demonstration of this point would require a simulation of
the detector, as well as the event selection and analysis cuts
\cite{qcd4}.
 
{\bf Hadronization uncertainty:} Since jets of hadrons, rather than
partons, are observed in detectors, it is necessary to correct hadronic
distributions for any smearing and bias effects that occur in the
hadronization process. Such corrections are usually estimated from
Monte Carlo simulations incorporating hadronization models.  In $Z^0$
decays, they are typically at the level of 10\% \cite{qcd3}. 
One can
argue  that non-perturbative corrections to jet final
states in $\ee$ annihilation can be parametrized in terms of inverse
powers of $\sqrt s$ \cite{qcd5}, and that for a generic observable $X$ the
ratio of non-perturbative to perturbative QCD contributions is
dominated by a term of the form:
\begin{equation}
 \frac{\delta X^{\rm non-pert}}{X^{\rm pert}}
\quad\sim\quad \frac{\log s}{\sqrt s}\ . 
\label{eqfff}
\end{equation}
Increasing $Q$ from 91 GeV to 500 GeV decreases this ratio by a factor
of 5, implying that hadronization corrections should be of order 2\% at
NLC. Assuming that these corrections can be estimated to better than
$\pm50$\%, the hadronization uncertainty on $\alpha_s(m^2_z)$\ should be
less than $\pm 0.001$.
 
{\bf Uncertainty due to missing higher orders:} Currently perturbative
QCD calculations of hadronic event measures are available complete up
to ${\cal O}(\alpha_s^2)$. Since the data contain knowledge of all
orders one must estimate the possible bias inherent in measuring
$\alpha_s(m^2_z)$\ using the truncated QCD series. It is customary to
estimate this from the dependence of the fitted $\alpha_s(m^2_z)$\ on the
QCD renormalization scale, yielding a large and dominant uncertainty of
about $\pm 0.007$ \cite{qcd3}.
At 500 GeV this uncertainty will be reduced only
slightly to $\pm0.006$.
 
From this simple analysis it seems reasonable to conclude that
achievement of the luminosity necessary for `discovery potential' at
the NLC will result in a $\qq$ event sample of sufficient size to
measure $\alpha_s(m^2_z)$\ with a statistical uncertainty of better than
1\%. Construction of detectors superior in performance to those in
operation today should enable reduction of systematic errors to the 1\%
level. Hadronization effects should be significantly smaller and imply
a sub--1\% uncertainty. The missing ingredient for an overall 1\%-level
$\alpha_s(m^2_z)$\ measurement at 500 GeV is the calculation of ${\cal
O}(\alpha_s^3)$\ perturbative QCD contributions, which should be
actively pursued.

\subsection{Energy Evolution Studies}

The non-Abelian gauge structure of QCD implies that the strong coupling
decreases roughly as $1/\ell n\,s$. Existing hadronic final states data
from $\ee$ annihilation at the PETRA, PEP, TRISTAN, SLC and LEP
colliders span the range $14\leq \sqrt s\leq 91$ GeV, although
hadronization uncertainties are large on the data below 25 GeV
\cite{qcd2}. A 1.5 TeV NLC would increase the lever-arm in
$1/\ell n\,s$ by almost a factor of two, allowing detailed study of the
energy evolution of QCD observables proportional to $\alpha_s$, such as
the rate of 3-jet production $R_3$. This would provide not only a test
of the fundamental structure of the SU(3)$_c$ group, but also a
search-ground for new physics that might produce `anomalous' running.
One such possibility is the existence of a light, electrically neutral
colored fermion that couples to gluons, for example, a light gluino.
The existence of such a particle would manifest itself via a
modification of gluon vacuum polarization contributions involving
fermion loops, effectively increasing the number of light fermion
flavors  $N_f$ entering into the QCD $\beta$-function. For the case of
a light gluino, which leads to a 10\% increase in the value of $R_3$ at
500 Gev, a 1\% measurement of $\alpha_s$ would allow this effect
to be detected with a significance of many
standard deviations.
 
However, data from experiments at different $e^+e^-$\ colliders
contributing  to this analysis, some of which were recorded more than
10 years ago, were treated differently by the various experimental
groups, and have relatively large systematic errors that are at least
partly uncorrelated from point to point. It is clear that the precision
of searches for anomalous running of QCD observables at NLC would be
improved significantly if new data were taken at lower c.m. energies
with the {\it same} detector and analysis procedures. Table
\ref{tab:QCDevents} shows the number of $\qq$ events delivered per day
at various c.m. energies by the NLC operating at design luminosity;
more luminosity would be delivered per day than was recorded in total
by the original lower energy colliders! The ability to maintain high
luminosity at low center of mass  energies presents a formidable
challenge to the design of the collider, but even running at lower
luminosity could deliver substantial data samples, especially at the
$Z^0$, although the high event rates would present extreme requirements
on the triggering and data processing capabilities of the detector.
 
\bigskip
\begin{table}[htb]
\centering
\caption{Number of $\qq$ events per day delivered by NLC at design luminosity}
\label{tab:QCDevents}
\bigskip
\begin{tabular}{|c|r|} \hline \hline
CM Energy $Q$ (GeV) &  $\qq$ events/day \\ \hline
500                 &  1750 \\ 
91                  &  20,000,000 \\
60                  &  75,000 \\ 
35                  &  150,000 \\  \hline\hline
\end{tabular}
\end{table}
\bigskip

\subsection{Symmetry Tests Using Beam Polarization}

For polarized $e^+e^-$\ annihilation to three hadronic jets one can
define the triple-product $\vec{S_B}\cdot(\vec{k_1}\times \vec{k_2})$
($B$ =$\gamma$,$Z^0$), which correlates the boson polarization vector
$\vec{S_B}$ with the normal to the three-jet plane defined by
$\vec{k_1}$ and $\vec{k_2}$, the momenta of the highest- and the
second-highest-energy jets respectively. The triple-product is even
under reversal of $CP$, and odd under $T_N$, where $T_N$ reverses
momenta and spin-vectors without exchanging initial and final states.
Standard Model  $T_N$-odd contributions of this form at the $Z^0$
resonance have been investigated \cite{qcd7} and are found to be of
order $\beta$ $\sim$ $10^{-5}$; the first experimental study of this
quantity has been made by SLD \cite{qcd8}, yielding limits: $-0.022 <
\beta < 0.039$. Above the $Z^0$ the dominant Standard Model
contributions remain smaller than 2 parts in $10^5$, presenting a
background-free observable for new contributions from beyond the
Standard Model, for example, due to rescattering of new gauge bosons
that couple only to baryon number \cite{qcd9}.
 
\subsection{Gluon Radiation in $t\bar t$\ Events; Anomalous Couplings}
 
The large mass and decay width of the top quark serve to make the study
of gluon radiation in $t\bar t$\ events a new arena for testing QCD.
The mass $m_t$ acts as a cutoff for collinear gluon radiation, and the
width $\Gamma_t$ acts as a cutoff for soft gluon radiation, allowing
reliable perturbative QCD calculations to be performed. We have noted
in Section 3.1 the influence of the large top quark width on the rate
and pattern of gluon radiation in $t\bar t$ events.

The existence of anomalous couplings of top quarks to gluons could
manifest itself via a modification of the rate and pattern of emitted
gluon radiation, beyond effects such as those just discussed. A
parametrization of anomalous couplings in the strong-interaction
Lagrangian may be written:
\begin{equation}
{\cal L}^{q\bar qg}\quad=\quad g_s \overline{q} T_a \left( \gamma_{\mu} +
\frac{i \sigma_{\mu\nu} k^{\nu}}{2 m_t} \left(\kappa - i \tilde{\kappa}
\gamma_5\right) \right) q G_a^{\mu}
\label{eqddd}
\end{equation}
where $\kappa$ and $\tilde{\kappa}$ represent anomalous
`chromomagnetic' and `chromoelectric' dipole moments, respectively. The
chromoelectric moment gives rise to CP-violating effects.
Their effects on gluon radiation have been calculated to the leading
order in $\alpha_s$ \cite{RizzoQCD}.  At $\sqrt{s}$ = 500 GeV,
limits of $|\kappa| \leq 0.03$ appear reachable;
at  $\sqrt{s}$ = 1.5 TeV, one may additionally constrain
 $|\tilde\kappa| \leq 0.2$.   These should be compare to a limit
of  $|\kappa| \leq 0.1$ which would be obtained from a measurement of 
the $t\bar t$ cross section at the Tevatron with 10 times the current 
integrated luminosity.  A similar limit on $\kappa$ from LHC would 
require a $t\bar t$ cross section measurement to better than 20\%.

\clearpage
\newpage
\section{Design of the NLC Detector}


Now that we have described the expected physics program of the NLC, we
must discuss in more detail what experimental facilities should be
necessary.  In this section, we will present a sample design of an NLC
detector and discuss the general issues which constrain this design.
This detector takes into account the particular features of the NLC
accelerator and its machine-related backgrounds.  A more complete
discussion of the machine/detector interaction is given at the end of
Chapter 3.

\begin{figure}[htb]
\leavevmode
\centerline{\epsfysize=3in\epsfbox{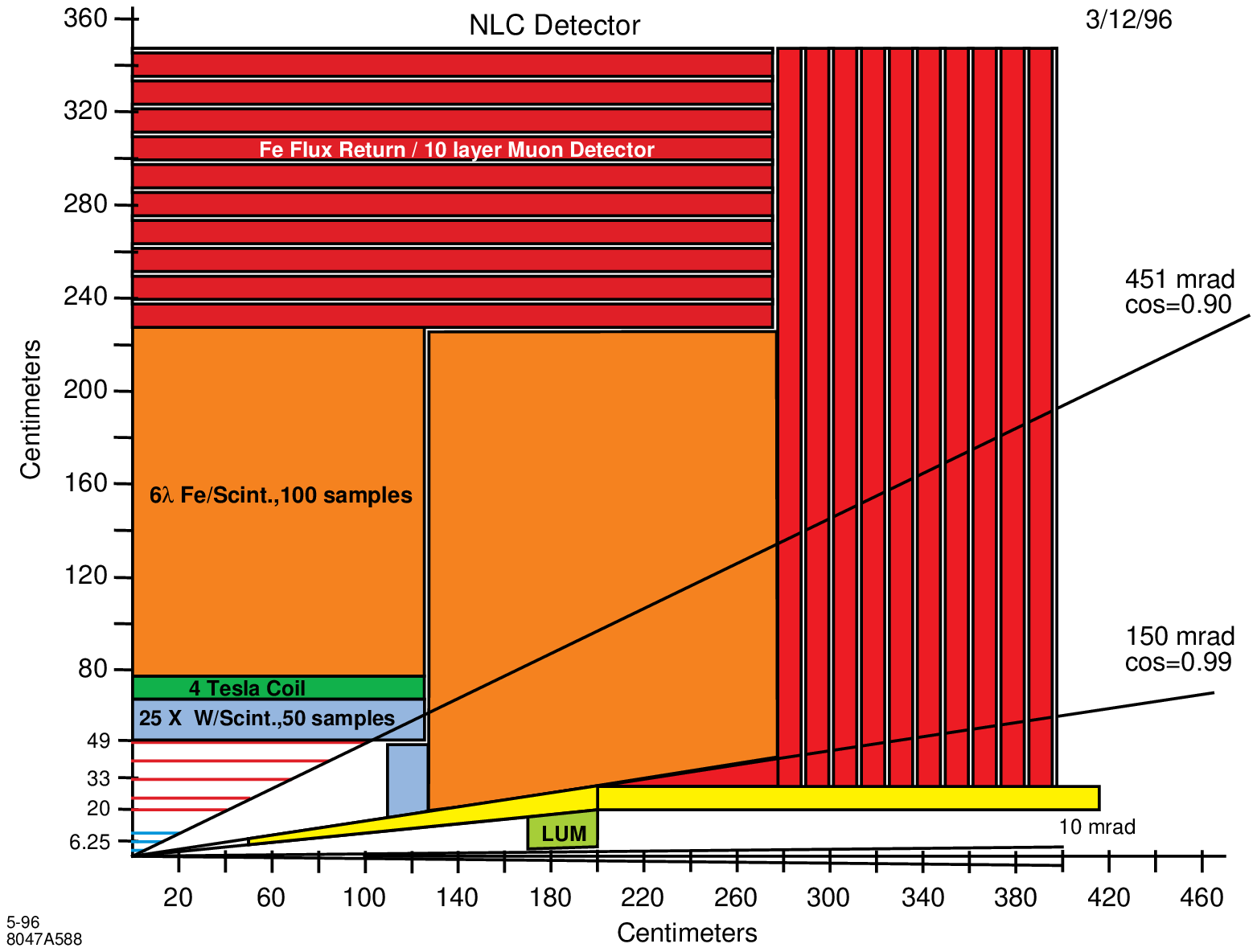}}
{\caption{NLC Detector.}
\label{fig:NLD}}
\end{figure}

A general plan of a detector for the NLC is shown, in quadrant view, in
Fig.~\ref{fig:NLD}. This detector  is at a very early conceptual design
stage. The architecture is a ``standard" solenoid, and is just one
possible approach. A 3 or 4 layer CCD vertex detector surrounds a
beryllium vacuum pipe. The  momentum measurement is carried out by
layers of silicon  microstrip detectors in a 4 Tesla field, which is 
produced by a superconducting solenoid outside of the electromagnetic
calorimeter. A hadronic calorimeter is located outside the coil, and
the structure is wrapped in laminated iron for a flux return and muon
identification system. The final focus quadrupoles are located inside
the detector at an $L^*$ of 2m.  A summary of the components and
expected performance of the detector is given in Table~\ref{nlcdetect}.

\begin{table}
\centering
\caption{Basic Components of the NLC Detector}
\bigskip
\begin{tabular}{|l|c|} \hline\hline
Vertex & $3-5$ CCD Layers \\
       & $2.6\oplus (13.7/p\sin^{3/2}\theta)$\\ \hline
Tracking & 5 Si Layers\\
         & $|\cos\theta|\leq 0.9$\\
         & $\delta p_T/p_T =2\times 10^{-4}$ at 100 GeV at 
$\cos\theta=0.9$\\ \hline
Coil     & 4T, 2--3 $X_0$\\ \hline
EM Calorimeter & $25X_0$, 50 samples \\
               & $\sigma_E/E = 
               \left(\frac{12\%}{\sqrt E}\right)\oplus
               1\%$ \\[1em] \hline
Hadronic Calorimeter & Fe Scintillator \\
                     & $6\lambda$, 100 samples\\
                     & $\sigma_E/E = 
                     \left(\frac{45\%}{\sqrt{E}}\right)\oplus
                     2\%$ \\[1em] \hline
Return Flux & $6\lambda$, 10 samples\\ \hline\hline
\end{tabular}
\label{nlcdetect}
\end{table}
\bigskip

\begin{figure}[tb]
\leavevmode
\centerline{\epsfysize=3in\epsfbox{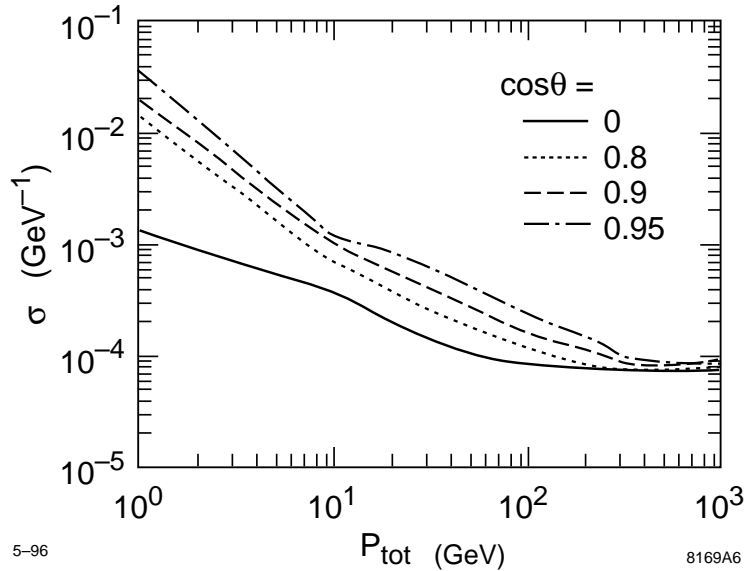}}
{\caption{NLC Detector tracking parameters.}
\label{fig:NLD_trackpars}}
\end{figure}

The small volume 4 Tesla field is well matched to the expected
backgrounds, cost, and technology. The small size of the tracking
system eases the mechanical problem of stabilizing  the final focus
elements relative to one another at the required nanometer level.
However, this design increases the complexity of the elements inside
the detector. For example, the final quadrupoles, envisioned as
permanent magnets, would have to be shielded by a small super
conducting solenoid. The tracking system is modeled as 5 layers of Si
microstrips with 5 micron spatial resolution. The expected momentum
resolution, including multiple scattering, is shown in
Figure~\ref{fig:NLD_trackpars}. Agreement that this performance is a
reasonable match to the physics goals is probably the most important
question for the overall  detector design. The cylindrical layers
extend to $\cos\theta = 0.9$. This leaves sufficient space for the ends
of the support structure, electronics, cabling, and cooling. End cap
trackers could be added for improved coverage. Experience with SLC/SLD
confirms the idea that backgrounds from a linear collider can be very
severe for conventional wire chambers. A silicon strip system would be
extremely robust.

The electromagnetic calorimeter would be inside the coil, 25 radiation
lengths thick, and with angular segmentation in the range of 30 to 40
mrad$^2$. Possible structures include tungsten silicon-diode sampling
towers and BGO or perhaps more exotic crystals. The crystals would be
interesting not so much for their good resolution but for their lack of
interconnections. The calorimetry is extended by endcaps to $\cos\theta
=0.99$.

The superconducting coil has an inner radius of 70 cm and a half length
of 1.2 m. Since there are only weak constraints on the thickness of the
coil,  given that it is outside the electromagnetic calorimeter, and
since it is a relatively small coil, it should be a reasonable coil to
design and build. This design will be made more challenging, however,
by its required integration with the quadrupole shields and supports.

The hadronic calorimeter is assumed to be an iron scintillator sampling
system, 6$\lambda$ thick and 100 samples. At this time, no magnetic
modeling of the structure has been done. It is unclear whether the
advantages of an iron based calorimeter in containing the magnetic flux
would be outweighed by the structural problems.

Finally, the detector is wrapped in a flux return 6$\lambda$ thick with
10 layers for muon detectors. Questions of overall geometry for
internal access and vibration control are completely open.

\begin{figure}[tb]
\leavevmode
\centerline{\epsfxsize=4in\epsfysize=3in\epsfbox{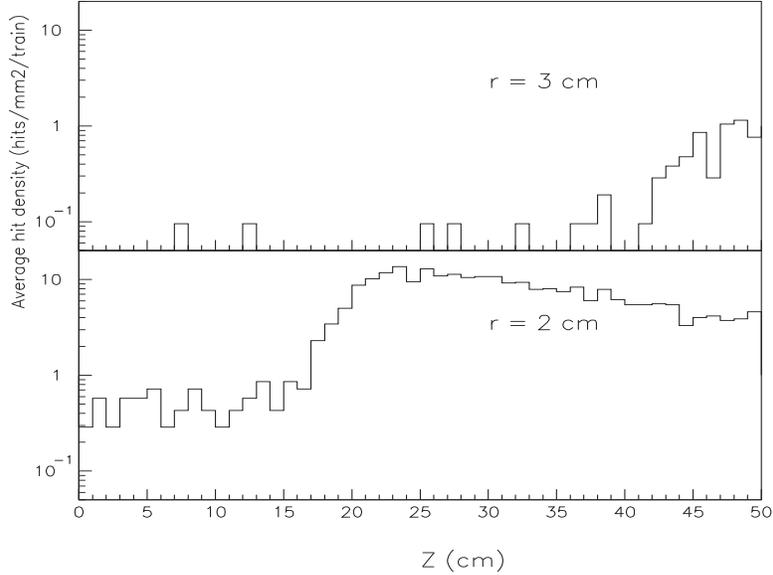}}
\centerline{\parbox{5in}{\caption[*]{Electron pair hit density per
mm$^2$ per train of 90 bunches, computed with the Monte Carlo program
ABEL. As the pairs leave the IP in a 4 Tesla field, hits are scored 
(a) at $r=3$ cm,  and (b) at $r=2$ cm.}
\label{fig:abel_hits}}} 
\end{figure}

The crossing of the two beams at the interaction point produces an
irreducible $\ee$ pair background at small $p_\perp$. These particles
can be kept away from a small radius vertex detector by the relatively
high 4 Tesla magnetic field. A reasonable vertex detector for this
environment would be a successor to the SLD vertex detector using large
area CCDs with 20 micron square pixels. The CCDs could be read out in a
time equivalent to a small number of beam crossings, and the beams
could be suppressed subsequent to a trigger if necessary. At SLD the
innermost layer of the vertex detector, at $r=2.5$ cm with $B=0.6$ T,
sees 0.4 hits/mm$^2$ summed over its 19 machine pulse readout time.
Vertex hits are successfully attached to tracks extrapolated from a
conventional drift chamber whose closest wire is at $r=22$ cm. Thus a
conservative estimate for the tolerable background is 1 hit/mm$^2$.
Were the NLC vertex detector to extrapolate hits from its own high
resolution outer layers into an inner layer, the much smaller search
area would allow a much larger (10-20 hits/mm$^2$) background to be
tolerated.

Fig.~\ref{fig:abel_hits} shows the hit density from the $\ee$ pair
background, per bunch train (a series of 90 bunch collisions), at a
radius of 3 cm  and at 2 cm. The results are shown as a function of $z$
for a 4 Tesla field. At $r=3$ cm the background is nonexistant out to
$z=45$ cm. At $r=2$ cm the background is similar to that tolerated by
an existing detector out to $z=17$ cm, a length adequate to provide
vertexing coverage well beyond the $\cos\theta=0.90$ currently used in
the physics simulations.

In addition to the irreducible background of pairs emanating from the
interaction point, the vertex detector will see a background of very
low energy backscattered particles resulting from the interaction of
the pairs that have been curled up by the field and gone forward to
strike the quad faces, septum, and luminosity monitor. In the current
interaction region design the calculated rate is 2.0 hits/mm$^2$/train
at $r=3$ cm, flat in $z$, and roughly constant down to $r=2$ cm, then
rising at lower radii. This background rate, already sufficiently low
at $r=2$ cm for efficient vertexing, is very sensitive to the details
of the interaction region design. Efforts to reduce the flux of these
soft particles, by, for instance, the use of low-$Z$ coatings, are
underway.

As experience in the collider operation is gained, one may want to
consider vertex detection at smaller radii. Even at the very aggressive
radius of 1 cm, in a 4 Tesla field the pair background hit rate remains
below 10/mm$^2$/train out to $z=3$ cm. Moreover, since the tracking
system radius largely drives the scale of the entire detector, a small
tracker permits consideration of relatively small $L^*$ as a possible
future upgrade of the final focus design. The small $L^*$ is plausible
because the axial extent of the tracker is small, about 1m for
$\cos\theta < 0.9$. This leads to significant improvements in required
tolerances in the final focus system.

\newpage

\section{Physics Processes  which Constrain Detector Performance}

In general, the requirements of the NLC physics program do not put
strong constraints on the detector design and performance.  For most
physics analyses, the performance of the detectors constructed for SLC
and LEP would be quite sufficient.  These detectors already include the
basic requirements for the study of $e^+e^-$ annihilation at high energy:
calorimetry with good energy resolution, segmentation, and charged
lepton identification; tracking with high momentum resolution, good
forward angular acceptance and overall granularity; and excellent
secondary vertex tagging.  However, it is always interesting to
consider advanced detectors with improved performance in one or another
of these areas.  In this section, we enumerate those physics topics
which put exceptional or unusual demands on the detector and which
might motivate a more advanced detector design.

A basic benchmark for detector performance is that given by standard
model Higgs production in the all-hadronic mode: $e^+e^-\rightarrow
Zh\rightarrow jj + b\bar b$.  This sets a standard of hadronic energy
resolution, angular coverage for high acceptance, and $b$-tagging
efficiency.  The recognition of the $Z$ in its hadronic decay is
required, more generally, for high-statistics branching ratio
determinations for the Higgs boson.

For the Higgs boson of the Minimal Supersymmetric Model, it is possible
for the Higgs boson to decay by mechanisms very different from those of
the Standard Model.  The interesting mode $e^+e^-\rightarrow
hA\rightarrow b\bar b b\bar b$ places a premium on angular coverage and
efficiency of $b$-tagging.  More generally, it is important in this
theory to be able to recognize the production of the Higgs boson in
$e^+e^-\rightarrow Zh$ making as few assumptions as possible on the
final state to which the Higgs decays. Such a model-independent search
for the Higgs boson is possible if the $Z$ can be reconstructed as an
$\ell^+\ell^-$ pair superimposed on any reasonable Higgs boson final
state.  This analysis constrains the momentum resolution and two-track
separation of the tracking device.

Finally, though the decay of the Higgs to $\gamma\gamma$ is rare, one
might hope to observe the branching ratio for $h \rightarrow
\gamma\gamma$ from the process $e^+e^-\rightarrow Zh$.  This would
demand a high-quality electromagnetic calorimeter, one with at least 3
times better resolution than is called for in our design \cite{KGCP}.

For studies of supersymmetry, we have shown that calorimeter coverage
down to $\cos\theta = 0.99$ is essential to reduce the standard model
backgrounds to events with low visible energy.  In addition, we have 
seen that the mass measurements for supersymmetric particles typically
involve the measurement of sharp kinematic endpoints in lepton or jet
energy distributions.  The calorimetric energy resolution must be 
sufficiently good that the measurement of these endpoints is limited only
by initial state radiation and beamstrahlung.

For studies of the $W$ boson vertices in $e^+e^-\rightarrow W^+W^-$, it
is necessary to completely reconstruct the final state kinematics by
measurement of a lepton + 2 jet system.  This would seems to demand
high-quality tracking. However, it has been shown that the tracking
requirements are less severe if kinematic fitting is used to determine
the jet energies \cite{KR}.

In the study of $WW$ scattering, the reactions $e^+e^-\rightarrow
\nu\bar\nu WW$ and $e^+e^- \rightarrow \nu\bar\nu ZZ$ give signatures
for distinct models of a strongly interacting Higgs sector, as
explained in Section 7.2. Thus, it is necessary to distinguish these
processes experimentally. Because the event rate is rather small, it is
necessary to rely on detection of $W$ and $Z$ in their hadronic decay
modes. Thus, it is important that the calorimetric mass resolution for
$W$ and $Z$ be sufficiently small that these particles can be
distinguished on the basis of their masses.  The analysis discussed in
Section 7.2, which assumed a calorimeter resolution
essentially the same as that of our model detector, found and
incorporated a probability of about 10\% for confusing each $W$ for a
$Z$ or vice versa.  In addition, this analysis requires good forward
coverage for high-energy electrons and positrons, to remove background
from the processes $e^+e^-\rightarrow e^+e^- W^+W^-$ and $e^+e^-
\rightarrow e\nu WZ$ which are induced by virtual photons.

Finally, we note that the study of the top quark threshold is affected
by the energy spread of the $e^+e^-$ collisions, which is determined by
the accelerator design. For this study to be carried out optimally, the
beam energy spread from beamstrahlung should be kept to the level of
initial state radation, about 3\%, and the intrinsic beam energy spread
of the accelerator should be kept to 0.1\%.  These requirements are
quite reasonable for the machine design discussed in Chapter 3.  In
addition, since the differential luminosity spectrum depends on the
beam parameters, it is necessary to measure this directly.  This
requires the measurement of acollinearity angles in small-angle Bhabha
scattering to better than 1 mrad for $\theta \sim 200$--500 mrad.  The
detector design in Section 11 includes an  appropriate electromagnetic
shower detector placed inside the synchrotron radiation mask.

\newpage
 
\section{Conclusions}
 
The Standard Model of strong and electroweak interactions represents a
tremendous scientific achievement. It provides an elegant and well
tested description of fundamental forces based on underlying local
symmetry concepts. However, despite those successes, some basic
questions remain unanswered. Addressing those issues should guide us to
a deeper understanding and better appreciation of Nature's laws.
 
What is the origin of electroweak symmetry breaking and mass
generation? That outstanding problem drives high energy physics and
establishes its experimental goals. Fortunately, resolution of that
puzzle appears to be within reach of the next generation of hadron and
lepton colliders, the LHC and the NLC\null.
 
In this report, we have described some of the physics goals and
capabilities of the NLC, emphasizing both its unique features and
complementarity to the LHC program planned at CERN\null. The envisioned
project would start with an initial variable center-of-mass energy up
to about 500 GeV and be upgradable to 1--1.5 TeV\null. It would feature
a highly polarized $e^-$ beam as well as $e^-e^-$, $\gamma e^-$ and
$\gamma\gamma$ collider options.
 
The first phase of the NLC guarantees extensive physics capabilities
and interesting discovery potential. Sitting at and above the $t\bar t$
threshold will allow unprecedented precision measurements of the top
quark's mass, partial decay widths, gauge and Higgs boson couplings
etc. Its very large mass may be a signal of the top quark's special
role in elementary particle physics and electroweak symmetry breaking.
We must, therefore,  explore its properties and search for additional
anomalous features. In that endeavor, the LHC will be a top quark
factory, capable of producing and studying samples of order $10^7$ tops
in a number of characteristic signatures.  The NLC will complement such
studies as a precision instrument which is sensitive to the broad range
of top couplings to strong and weak gauge bosons.  Similarly, studies
of gauge boson pair production at the NLC will provide high precision
measurements of triple-gauge boson couplings at the level of 1--0.1\%
(with increasing center-of-mass energy). At that level of sensitivity
anomalous properties could  become unveiled, particularly if there is
new strong dynamics at $\sim1$ TeV\null.
 
If a fundamental Higgs scalar exists and has mass below $350$ GeV, it
will be found and studied in the first phase of the NLC via $e^+e^-\to
ZH$. That mass reach is extremely broad and encompasses a most
important domain. If the Standard Model is perturbative up to
grand-unification energies, then $m_H\lsim200$ GeV\null. If
supersymmetry is manifested at the electroweak scale, then the lightest
scalar satisfies $m_h<150$ GeV\null. The first stage of the NLC will be
definitive in confirming or negating those expectations. A higher mass
Higgs ($>350$ GeV) would be easily accessed at the LHC  or higher
energy phases of NLC. Additional Higgs scalars of SUSY scenarios, or
pseudo-Goldstone bosons of dynamical symmetry breaking schemes, would
be unveiled in pair-production as their thresholds are passed
at the NLC. The clean environment of that facility would allow detailed
studies of their decay patterns. A compelling feature of $e^+e^-$
colliders is the ability to not only produce new particles, but to
thoroughly explore their properties.
 
If supersymmetry manifests itself near the electroweak scale, a
plethora of new particles, the supersymmetric partners of quark,
leptons, and gauge bosons, awaits  discovery.  Unfolding that entire
spectrum and its properties will require  the full capabilities of the
LHC and NLC. The first direct evidence for  production of gluinos and
squarks should be found at LHC, if not already at the Tevatron. On the
other hand, the study of color-singlet superpartners is best carried
out at $\ee$ colliders.  Since these particles are typically lighter,
they determine the decay patterns of the particles produced in strong
interactions.  Thus, the NLC will be essential for unraveling the
entire spectrum and precisely measuring the new particle properties. 
The study of superpartners will be greatly facilitated by use of the
high $e^-$ polarization both for suppressing backgrounds and enhancing
signals. High precision determinations of the SUSY parameters will provide
important insight and constraints on the mechanism of SUSY breaking,
as well as a potential new window to the physics of grand unification
and superstrings.
 
If new  strong dynamics is responsible for electroweak symmetry
breaking, it may be difficult to uncover and fully explore. In such a
scenario, the NLC will be an extremely valuable instrument. It will
probe the new dynamics via $W^+W^-\to W^+W^-$ and $W^+W^-\to t\bar t$
scattering, and through anomalous gauge boson couplings and other
probes in $\ee\to W^+ W^-$. Such studies will become ever more
revealing as the higher energy upgrades are attained. In addition, new
strong dynamics is likely to imply a wealth of spectroscopy at very
high energies. The NLC, particularly its highest energy phase, should
be a very important facility for discovering and studying such states.
 
The new energy domain opened up by the NLC will also allow systematic
searches for new particles such as $Z^\prime$ bosons, heavy fermions,
leptoquarks, etc. If such particles are found, the $e^-$ polarization
will be extremely useful for deciphering their properties.
 
The physics capabilities of the NLC can be greatly expanded by the
$e^-e^-$, $\gamma e^-$, and $\gamma\gamma$ collider options possible at
such an accelerator. Those collision modes will complement $e^+e^-$
studies and could prove particularly useful to provide new observables
of supersymmetric particles, to study the Higgs boson as an $s$-channel
resonance, and to explore $W^-W^-$ scattering.
 
During the twentieth century, elementary particle physics emerged as a
scientific discipline. Since its inception, progress and advancement
have followed our ability to probe short distances via high energy
collisions. By pushing to even higher energies, we should be able to
continue the advancement. In this report, we have reviewed the
questions that are now at the frontier of particle physics, and we have
shown that the NLC has a central role is answering every one.  We thus
view the NLC as a forefront  facility for particle physics, an
essential tool for  probing the hidden symmetries of Nature, a well
tuned instrument for the start of our next millennium.

\newpage
\newcommand{\journal}[4]{{\sl #1}\ {\bf #2}, #3\ (#4)}
\newcommand{\npb}[3]{{\sl Nuc.~Phys.}\ {\bf B#1}, #2\ (#3)}
\newcommand{\prd}[3]{{\sl Phys.~Rev.}\ {\bf D#1}, #2\ (#3)}
\newcommand{\plb}[3]{{\sl Phys.~Lett.}\ {\bf B#1}, #2\ (#3)}
\newcommand{\cpc}[3]{{\sl Comp. Phys. Comm.}\ {\bf B#1}, #2\ (#3)}

\chapter{A Zeroth-Order Design for the Next Linear Collider}
\label{chap:overview}

The Next Linear Collider consists of a set of subsystems---injectors,
linacs, beam delivery lines, and interaction regions.  These are responsible
for creating intense and highly condensed beams of positrons and
polarized electrons, accelerating them to high energy, focusing them to
small spots, and colliding them in an environment that allows sensitive
particle detectors to operate for physics. We introduce these various
parts of the collider in this section, and then provide a more detailed
discussion of each.  A comprehensive design study of the NLC is
given in Ref \cite{oneref}.

\section{Overview of Collider Systems}
\label{sec:over} 

\begin{figure}[htbp]
\leavevmode
\centerline{\epsfbox{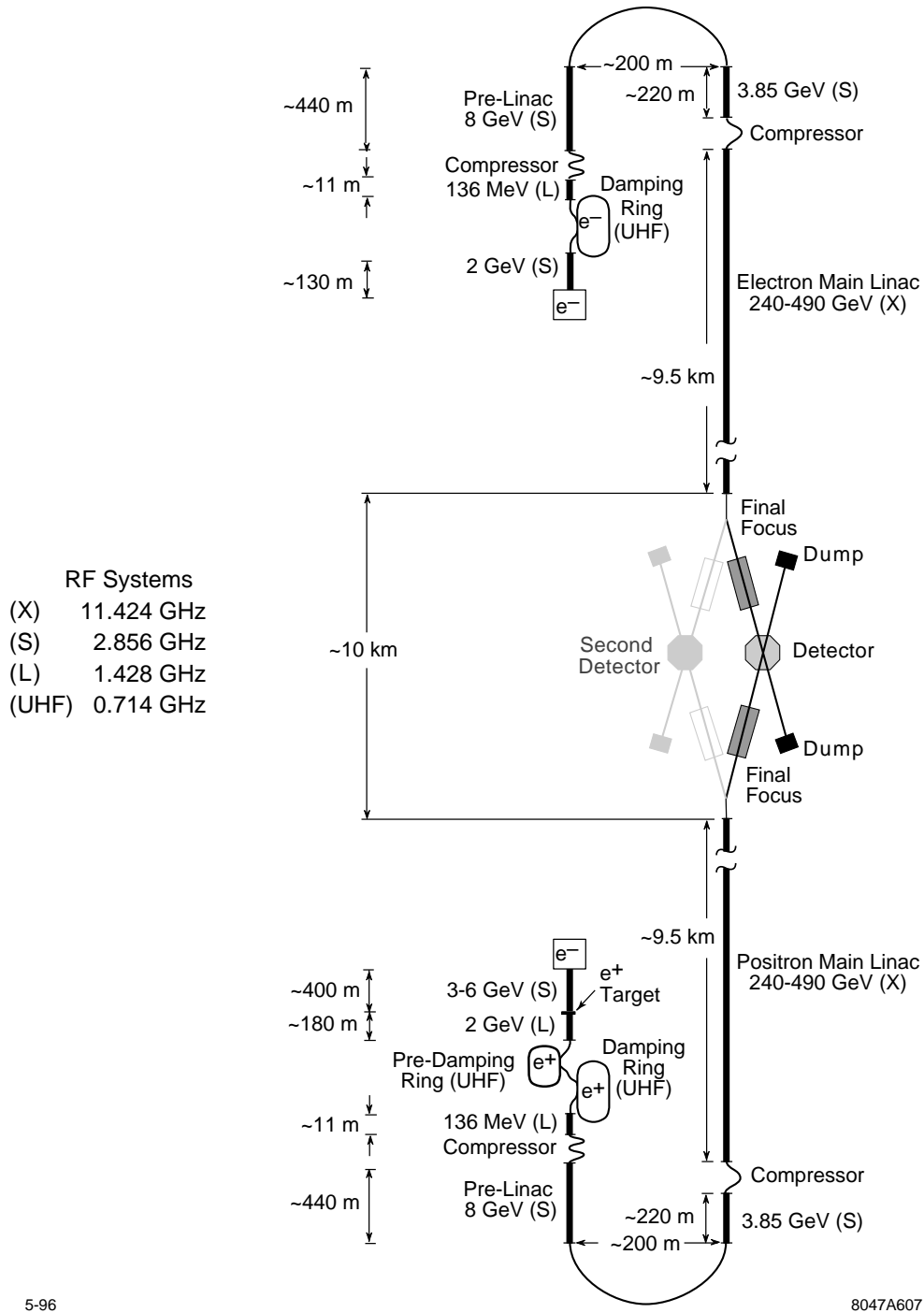}}
{\caption[*]{Schematic layout of NLC systems (not to scale).}
\label{fig:schem}}
\end{figure}

A schematic of the NLC is shown in Fig.~\ref{fig:schem}. The physical
footprint of the collider complex is approximately 30 km in length and
less than l km wide. This includes all beam transport lines in the
injectors and linacs necessary to obtain 1 TeV cms energy, and all
space needed in the beam delivery sections to accomodate 1.5 TeV cms
energy.  To reach 1.5 TeV, however, it may be necessary to extend the
``trombones'' in the layout of the collider to provide additional
length for the main linacs. Rather simple beam-transport lines can be
used to do this without moving the injectors or damping rings.

\subsection{Injectors and Damping Rings} 

The electron injector system for the
collider is copied from the system presently operating reliably at the
SLC.  This includes a polarized photocathode electron gun, and a
bunching system that operates at a subharmonic of the
main linac rf frequency. The new challenge for the NLC injector is that
it must produce trains of 90 bunches spaced by 1.4 ns at the machine
pulse repetition rate of 120--180 Hz.  The bunched beam is accelerated in an
S-band linac to 2 GeV, then injected into a damping ring designed to
condense the transverse phase space of the beam through radiation
damping.  The required vertical emittance, $\gamma \varepsilon_y = 3
\times 10^{-8}\mbox{m-rad}$, is two orders of magnitude smaller than
achieved in the SLC damping rings.  This places tight constraints on
the alignment of ring magnets, but many lessons have been learned from
operation of the SLC damping rings, and several modern synchrotron light
sources also provide valuable experience.  Wiggler magnets, similar to
that used in the SSRL ring at SLAC, are included in the NLC design to
reduce damping times, and the vacuum and rf systems required for the
NLC ring are similar in size and complexity to those of the Advanced
Light Source at Lawrence Berkeley National Laboratory.

\begin{figure}[htb]
\leavevmode
\centerline{\epsfbox{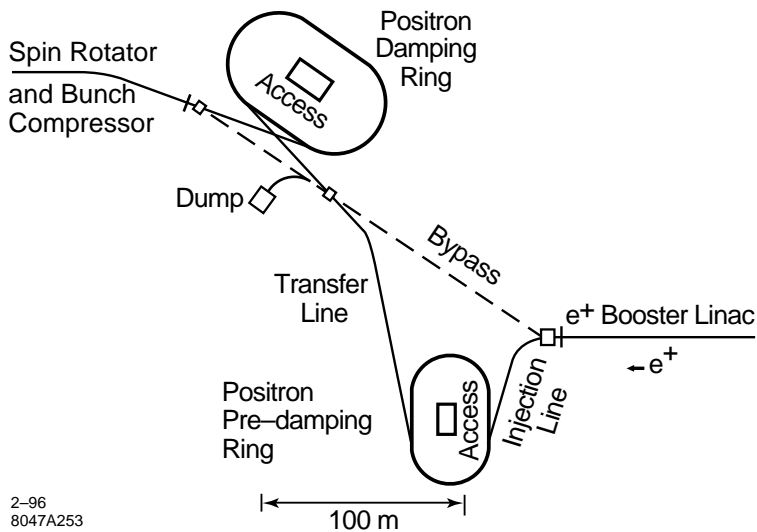}}
{\caption[*]{Layout of the positron target area and damping rings.}
\label{fig:posdr}}
\end{figure}

The techniques for positron production planned for the NLC are also
largely copied from those in use at the SLC. Positrons are created in
electromagnetic showers produced by targeting 3--6 GeV electrons onto a
rapidly rotating metallic disk.  The high-emittance positron beam is
focused in an extended solenoid, and accelerated to 2 GeV in a
large-aperture L-band linac. This leads to a good capture efficiency.
To further improve efficiency and ease operational tuning tolerances,
the beam is cycled through a large-aperture pre-damping ring prior to
injection into a main damping ring identical to that used for the electron
beam.  The overall layout also includes transport lines that will allow
the drive electron beam to bypass the positron-production target and
pre-damping ring (Fig.~\ref{fig:posdr}).  This will allow the study of
$e^-e^-$ collisions (with both beams polarized).

\subsection{Bunch Length Compressors} 

The lengths of the electron and positron bunches become too great in
the damping rings for the beams to be successfully accelerated in the
main X-band linacs, so they must be reduced.  This is done in stages
(Fig. \ref{fig:schem}). A
first compressor, located immediately after the damping rings, reduces
the bunch length from 4 mm to 0.5 mm to optimize injection into an
S-band linac that accelerates the beams to 10 GeV. The bunch lengths
are then further compressed to 100--150 $\mu$m in a two-stage
compressor that also bends through a 180$^\circ$ arc, reversing the
direction of travel; this arc allows for future upgrades of the main
linac length and permits feedforward of measurements of the bunch
charge and trajectory made in the damping rings and compressors. This
ability to preset the energy gain and injection orbit to the main linac
is an important feature of the design.

\subsection{Main Linacs}  
The main linacs of the NLC will use normal-conducting traveling-wave
copper structures powered with 11.424 GHz microwaves.  The required
power is generated by klystrons in 1.2-$\mu$s pulses which are then
compressed into shorter 0.24-$\mu$s pulses of higher peak power in a
passive rf transformer (SLED-II). A train of 90 electron (or positron)
bunches is accelerated in the main linac by each rf pulse. While the
number of particles in each of these bunches is less than that
in the SLC, the total charge accelerated by each rf pulse is more than
an order of magnitude greater. This multibunch design and the
correspondingly larger fraction of energy transferred from the power
sources to the beam is one of the important differences between the SLC
and the NLC. The linac focusing lattice is designed to allow the
center-of-mass energy to be increased from 300 GeV to 1 TeV; further
upgrade to 1.5 TeV will require modifications. Diagnostic stations,
located at five positions along each of the two main linacs, will
include laser wire scanners to measure the transverse phase space,
beam-based feedbacks to correct for centroid shifts of the bunch train,
multibunch BPMs and high-frequency kickers to measure and correct
bunch-to-bunch position errors, and magnetic chicanes to provide
noninvasive energy and energy spread measurements. These feedbacks must
maintain bunch trajectories to within a few microns of a tuned
reference orbit in order to prevent unacceptable emittance growth.
 
\subsection{The Beam Delivery System} 
\begin{figure}[htb]
\leavevmode
\centerline{\epsfxsize=4in\epsfbox{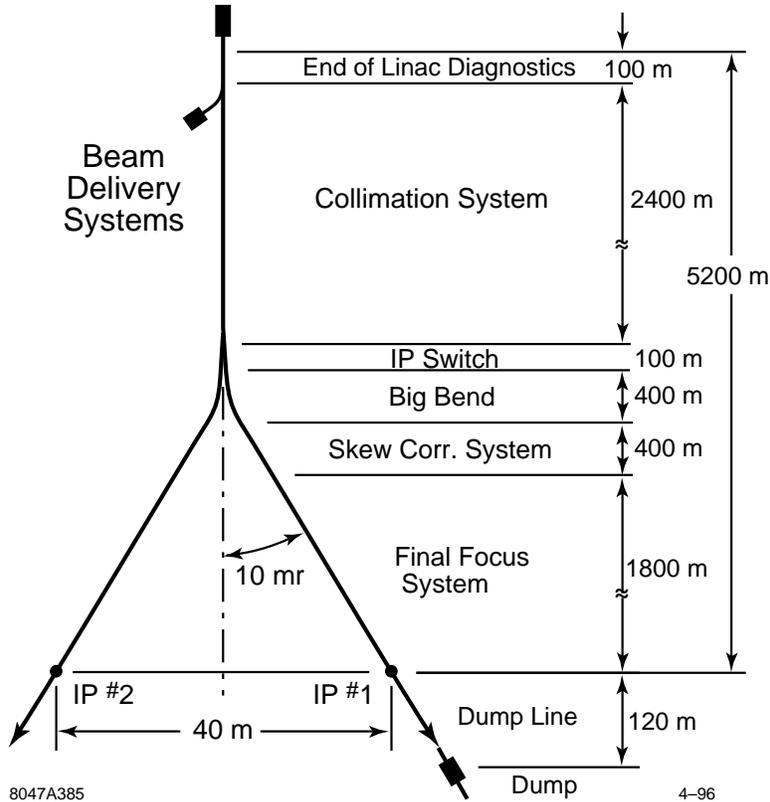}}
{\caption[*]{Schematic layout of the beam delivery system.}
\label{fig:bdsys}}
\end{figure}

The beam delivery system consists of a
collimation section, a switchyard, the final focus, the interaction
region, and the beam extraction and dump.  These are 
shown schematically in Figure~\ref{fig:bdsys}.

After acceleration in the main linac, the beams enter collimation sections where
particles at the extremes of the energy and transverse phase space
are eliminated by collimators. Collimation is performed for both the
horizontal and vertical planes at two betatron phases that differ by
90$^\circ$ to effectively cut the beam in both position and angle. 
This primary collimation is then followed by secondary collimation in
both planes and both phases to remove particles that are scattered by
the edges of the primary collimators. These collimation sections are
rather extensive, but our experiences at the SLC and FFTB have proven
the need to perform this function well.

A passive switching section is used to direct the beams to one of two
possible interaction points (IPs). The beam is deflected in the
switchyard by 10 mrad to produce a 20-mrad crossing angle between the
two beams at each IP. This deflection introduces little emittance
growth, but greatly reduces detector backgrounds and provides the IP
crossing angle required to collide the closely spaced bunches in each
train.

The NLC final focus follows the design of the Final Focus Test Beam 
(FFTB) at
SLAC. The optics consist of a matching section with appropriate
instrumentation to measure the incoming beam phase space, horizontal
and vertical chromatic correction sections, a final transformer, a
final doublet, and a diagnostic dump line for the exiting beam. The
final doublets require active stabilization to maintain beam
collisions. Measurement of the final spot size is a particular
challenge, but can be performed with an advanced laser fringe monitor
similar to that used in the FFTB. In-situ tuning of the spot sizes and
luminosity can also be done using measurements of the beam-beam
deflections and other techniques developed at the SLC.

\subsection{Interaction Region}  
Two interaction regions are included in the layout of the NLC.  These,
of course, must share the available luminosity, but it will be possible
to install two complementary experiments. As with all colliders, the
interaction region will be a very crowded location.  The final
quadrupole magnets of the machine optics must be positioned as closely
as possible to the collision point, and high-Z masking must be
installed to protect elements of the experimental detector.  The
detector itself will require clear access to as much of the volume
around the interaction point as possible.  The design presented in this
document includes quadrupoles 2 m from the interaction point, and
complete access for detector elements at polar angles greater than 150
mrad.  Calorimetric measurement of Bhabha scattering at smaller polar
angles should also be possible, and is expected to provide a precise
determination of the luminosity-weighted center-of-mass energy spectrum.

The NLC is designed to collide electrons and positrons at a 20-mrad crossing
angle.  This prevents the tightly spaced bunches of one beam from being
disturbed, as they approach the interaction point, by bunches in the
opposing beam that are leaving the interaction point.  To avoid loss of
luminosity due to this crossing angle, it is necessary to use a pair of
deflection rf cavities to ``crab'' the beams so that the bunches
collide head on. This is a new task not encountered at the SLC, but a
system with reasonable specifications has been designed.

\newpage

\section {Polarized Electron Injector}

The electron injector complex includes the polarized electron source,
bunching system, and linac that accelerates the beam to 2 GeV.

\begin{figure}[htb]
\leavevmode
\centerline{\epsfbox{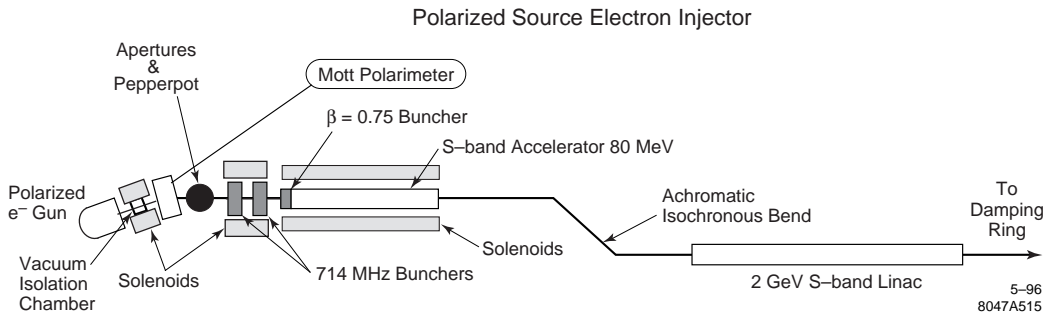}}
{\caption[*]{The polarized electron injector.}
\label{fig:esrc}}
\end{figure}

\noindent
Figure \ref{fig:esrc} shows a schematic of the injector for the
polarized electrons, and a summary of the corresponding design
parameters is given in Table~\ref{tab:parama}. The polarized  electron
source consists of a polarized high-power laser and a high-voltage DC
gun with strained-lattice gallium-arsenide photocathode.  Many of the
performance requirements for the NLC injector are similar to those in
the SLC, and the design of the NLC injector is based on the presently
operating SLC injector. The main difference is that, while the SLC
polarized electron injector produces one pair of electron bunches
separated by 60 ns at 180 Hz repetition rate, the NLC injector must
produce long trains of many bunches separated by only 1.4 ns. This
requires greater pulse currents to be extracted from the photocathode.

\begin{table}[htb]
\centering
\caption[*]{NLC polarized electron injector parameters up to the
damping ring.}
\label{tab:parama}
\bigskip
{\footnotesize 
\begin{tabular}{|l||c|c|c||c|c|c||c|c|} \hline \hline
  &  \multicolumn{3}{c||}{NLC-I}     &
     \multicolumn{3}{c||}{NLC-II}    &
     Overhead  &  SLC   \\ \cline{2-4}\cline{5-7}
\multicolumn{1}{|c||}{\raisebox{4pt}[0pt]{Parameters}}
&  a  &  b  &  c  &  a  &  b  &  c  &  20\%\  &  achieved \\
\hline
N/bunch at IP (10$^{10}$) & 0.65 & 0.75 & 0.85 & 0.95 & 1.1 & 1.25 & 1.5 & \\
N/bunch at Gun  & 1.2 & 1.4 & 1.6 & 1.75 & 2.0 & 2.3 & 2.8 & 8.8 \\
I$_{avg}$ at gun (A) & 1.4 & 1.7 & 1.9 & 2.0 & 2.4 & 2.7 & 3.2  &  \\
N/bunch at limiting aperture   &
   0.95  &  1.15  &  1.3  &  1.4  &  1.63  &  1.85  &  2.2  & \\
N/bunch at 80 MeV in 18 ps &
   0.83  &  0.97  &  1.1  &  1.2  &  1.4  &  1.6  &  1.96  & \\
I$_{avg}$ at 80 MeV (A)  &
   0.95  &  1.11   &  1.26  &  1.37  &  1.58  &  1.8  &  2.2  &  \\
N/bunch at damping ring   &
  0.75  &  0.88  &  1.0  &  1.1  &  1.28  &  1.45  &  1.7  & 5.5 \\
\hline
\hline
\multicolumn{1}{|c||}{\raisebox{-4pt}[]{Parameters}} &  
   \multicolumn{3}{c|}{\raisebox{-4pt}[]{NLC Required}}  &
    \multicolumn{4}{c|}{\raisebox{-4pt}[]{Simulation}}  &  
        SLC \\
\multicolumn{1}{|c||}{}  &  
   \multicolumn{3}{c|}{}  &  
   \multicolumn{4}{c|}{}  &  
       no train \\
\hline
\hline
Bunch train durations (ns)    &
     \multicolumn{3}{c|}{126}  &
     \multicolumn{4}{c|}{126}   &  \\
Bunch separation (ns)         &
     \multicolumn{3}{c|}{1.4}  &
     \multicolumn{4}{c|}{1.4}    &  \\
Pulse FWHM at gun (ps)       &
     \multicolumn{3}{c|}{700}  &
     \multicolumn{4}{c|}{700}   &  2000  \\
Pulse FW  after bunching (ps)  &
     \multicolumn{3}{c|}{18}  &
     \multicolumn{4}{c|}{18}   &  18  \\
   $\varepsilon_{\rm n,rms}$ at 40 MeV (10$^{-4}$ m-rad)  &
     \multicolumn{3}{c|}{1.0}  &  
     \multicolumn{4}{c|}{0.5}   &  1 to 1.3  \\
   $\Delta$E/E$_{\rm FW}$ at entr. to ring  (\%) &
     \multicolumn{3}{c|}{$\pm$0.6}        &
     \multicolumn{4}{c|}{$\pm$0.6}         &  $\pm$0.6 to 0.8  \\
   Train-to-train intensity jitter (\%)  &
     \multicolumn{3}{c|}{$<$0.5\%}  &
     \multicolumn{4}{c|}{negligible w/ intensity}  & 
     0.8 at gun  \\
 &   \multicolumn{3}{c|}{}  &
     \multicolumn{4}{c|}{jitter limiting aperture}  & 
     0.5 bunch  \\
\hline\hline
\end{tabular}}
\end{table}
\bigskip

The SLC polarized source delivers high-intensity pairs of electron
bunches with beam polarizations of 80\%\ or better.  With
maintenance of ultra high vacuum conditions, cathode lifetimes have
exceeded thousands of hours.  System availability now approaches 99\%.
The basic technologies of the SLC injector will be utilized for the
NLC.  Most importantly, the existing design of the polarized gun could
be duplicated for the NLC with few changes.

The NLC requires two electron injectors---one to produce polarized
electrons that will collide with positrons, and another to
produce electrons that will be used to make positrons. An
S-band linac accelerates the polarized electrons for collisions to 2
GeV prior to injection into the damping ring, while a similar but
longer linac accelerates the electrons for positron-production to
higher energy.  The number of positrons needed to satisfy the
luminosity requirements of the collider is greater at 1 TeV cms energy
than at 500 GeV, so the energy of the electrons for positron production
is increased from 3.11 GeV to 6.22 GeV as the cms energy is increased.
The positron injector beam line includes the space needed for this
upgrade. The phase space of the 2-GeV polarized electron beam is
matched to the damping ring lattice in a special transition region,
while the higher energy beam for positron production is transported to
one of two target systems. (See Section \ref{ps}.) The beam optics of
both electron injectors are identical up to the 2-GeV point so that it
is straightforward to replace the thermionic electron gun used for the
positron drive beam with a polarized gun in the event that a second
polarized electron source is needed for $e^-e^-$ or $\gamma\gamma$
collisions.

The polarized electron source could use a conventional 120-kV DC
gun designed to produce a 126-ns-long train of 90 bunches spaced 1.4 ns
apart with $2.8 \times 10^{10}$ electrons per bunch. Research efforts
to develop a polarized rf gun are underway. If successful, it may be
possible to benefit from the lower emittances produced by an rf gun to
improve the performance of the injector complex.

A vacuum isolation region and a $20^\circ$ bend protect the polarized
electron gun from the downstream environment (Fig.~\ref{fig:esrc}). The
bend also allows the beam to be steered to a Mott scattering 
polarimeter for polarization measurements. The intensity jitter of the beam
at the gun may be close to the allowed threshold of 0.5\%\ rms from
train to train due to the incoming laser intensity jitter on the
photocathode. An aperture 24 cm downstream of the $20^\circ$ bend will
be used to reduce the electron beam intensity jitter of the bunched
beam. Two 714-MHz standing-wave subharmonic bunchers and an S-band
buncher and accelerator section compress the beam from the gun such
that 83\%\ of the charge is captured in $18^\circ$ S-band (17.5 ps).
Simulations with the computer code PARMELA were made of the beam optics
from the gun through the second accelerator section, where the beam
energy reaches 80 MeV. The energy spread, calculated without including
the effect of beam loading in the long bunch train, is expected to be
$\pm 0.6\%$ for a single bunch.

Beam loading effects in the subharmonic bunchers are mitigated by using
cavities with low $R/Q$, injecting the beam during the fill time of the
cavity, and changing the phase of the input rf just before beam time.
Axial magnetic field focusing will be used on the first two accelerator
sections (up to 80 MeV) where the beam is expected to have larger
energy spread than in the rest of the injector linac.  At 80 MeV there
will be a doubly achromatic bend to connect the low energy beam line to
the main injector beam line.  This bend allows another low-energy
injector beam line in the future which could be used for a
polarized rf gun. Wire scanners on both sides of the achromatic bend
will measure emittance in the low-energy beam line and at the entrance
to the injector linac after the bend. Scrapers at the high-dispersion
point will clip away tails and a wire scanner following the scrapers
will measure energy spread.  From 80 MeV to 2 GeV, the linac will use
3-m-long accelerator sections with beam loading compensation. Each rf
module will consist of four 3-m accelerator sections powered by two
S-band klystrons similar to SLAC 5045 klystrons. Quadrupole doublets
between each section, and some wrapped around the first few sections,
will be used to control the transverse size of the beam.

It is planned to test the design and technologies planned for in the
NLC injector at the NLC Test Accelerator.  Of
particular interest will be tests of the beam loading compensation
techniques intended to be used in both the subharmonic bunchers and the
injector accelerator sections.

Our experience with polarized photocathodes leads us to expect that, with
the average  beam current required during the NLC bunch train, the beam
polarization will be less than the 80\% specified for the collider if no
improvements are made to the characteristics of the SLC source. Tests
of the cathodes in use at the SLC indicate that, if these cathodes were
driven to produce the charge required for the NLC, the beam
polarization would be no more than 60--65\%. The important factor is
the charge density (per unit area) at the surface of the cathode. There
are several straightforward ways to avoid this limit.  The area of the
cathode can be doubled or even tripled; however, the beam emittance
will increase proportionally, requiring larger apertures in the
bunchers.  It is also possible, albeit somewhat inefficient, to simply
build two guns and merge their beams at one of several points in the
transport optics. More elegant solutions with super-strained
photocathode materials are being investigated and have shown promising
results.  There remains work to be done, but beam polarizations of at
least 80\% will be available at the NLC.
\clearpage

\section{Positron Source}
\label{ps}

Positrons are created for the NLC by bombarding a thick high-Z metallic
target with 3--6 GeV electrons. The design of the NLC positron system
is based on the similar system used in the SLC, which has
demonstrated excellent reliability over many years of operation. One of
the key design goals for the NLC positron source is to produce a beam
with twice the intensity required at the interaction point.  The total
number of positrons required for the bunch train is then an order of
magnitude greater than the number of positrons in the single SLC bunch.
The limiting factor in the design of the positron system is the
survivability of the production target, and several steps are taken to
assure that the required charge can be produced and successfully used.

\begin{figure}[htb]
\leavevmode
\centerline{\epsfbox{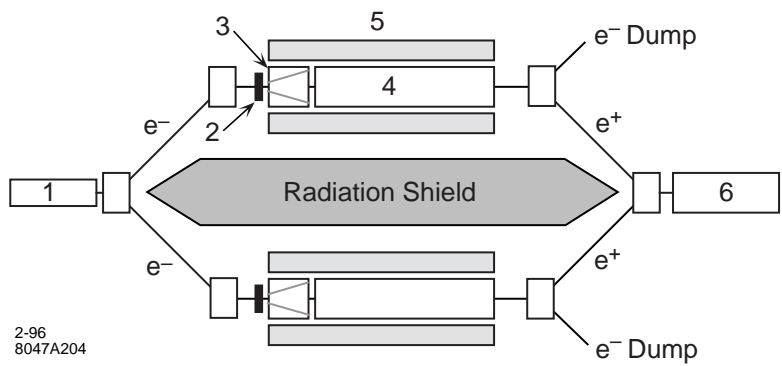}}
\centerline{\parbox{5in}{\caption[*]{Layout of the NLC positron
source with two side-by-side positron production and capture systems:
(1) drive electron accelerator, (2) positron target, (3) flux
concentrator, (4) L-band capture accelerator, (5) tapered-field and
uniform-field solenoids, and (6) positron booster linac.}
\label{fig:psys}}}
\end{figure}

The NLC positron source consists of three subsystems, a 3.11-GeV
accelerator for the drive electron beam (upgradable to 6.22 GeV), a
positron production and collection system, and a positron linac to
accelerate the positrons to 2 GeV.  A layout of the NLC positron
source is depicted in Fig.~\ref{fig:psys}.

The source design is constrained by failure of the target material due
to excessive stress generated by the impact of a single bunch train. To
overcome this limit, we have chosen to use a tapered solenoid and
large-aperture L-band linac to capture the produced positrons. This
leads to a 16-fold increase in the transverse phase space acceptance
relative to that of the SLC S-band design.   The use of L-band  also
permits a larger-area drive-beam spot on the target.  The net effect of
these changes is to maintain the energy density in the target during
the bunch train at values comparable to those in the SLC target. A
summary of key parameters in the source design is given in
Table~\ref{tab:psrc}.

\begin{figure}[htb] 
\leavevmode 
\centerline{\epsfbox{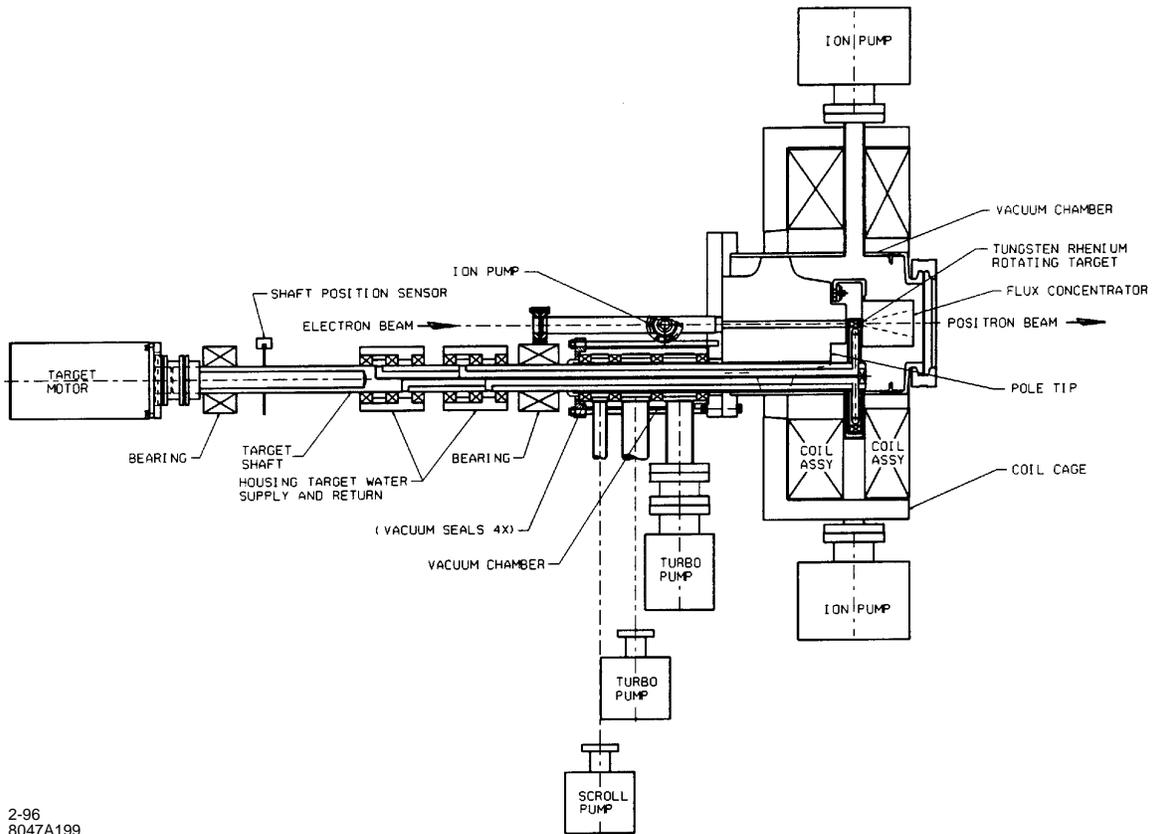}}
{\caption[*]{Schematic of the NLC positron target system.}
\label{fig:ptar}} \end{figure}

A conceptual design of the NLC positron target system is shown in
Fig.~\ref{fig:ptar}. The target material is chosen to be
$W_{75}$Re$_{25}$ because of its excellent thermo-mechanical
properties, its high-Z characteristic, and its excellent performance in
the SLC positron source. The target thickness is chosen to be four
radiation lengths based on considerations of both the positron
production yield and the single pulse energy deposition density in the
target.  The ring-shaped target, with a radial thickness of 1 cm and
and a medium radius of about 10 cm, is rotated at a rate of 120 rpm to
distribute the beam pulse impacts around the target ring.  A
three-stage differential pumping assembly using radiation-resistant
seals that rely on tight clearances (a few microns) between rotating
and stationary seal surfaces is used to seal the rotating target shaft
for the target vacuum chamber.  Cooling tubes are located in a copper
casting in close contact with the inner ring face
in order to adequately cool
the target, which absorbs about 23 kW of beam power.

\begin{table}[htbp]
\centering
\caption[*]{NLC Positron Source Parameters.}
\label{tab:psrc}
\bigskip
{\footnotesize 
%
%
{\begin{tabular}{|ll|c|r|r|r|r|}
\hline\hline
 & {Parameter} & {Unit} & 
    {SLC 94} & {SLC max} & 
    {NLC-I} & {NLC-II}  \\
 &  &  &  &  {design}  & {(500 GeV)} & 
     {(1 TeV)} \\
\hline
\hline
\multicolumn{7}{|l|}{General Parameters:}  \\ 
\hline
& Electrons per pulse at IP & [$10^{10}$] &  3.5 & 5 & 76.5 & 112.5 \\
& Bunches per pulse & & 1 & 1 & 90 & 90         \\
& Pulse duration & (ns) & 0.003 & 0.003 & 126 & 126   \\
& Bunch spacing & (ns) & - & - & 1.4 & 1.4         \\
 & Repetition frequency & (Hz) & 120 & 180 & 180 & 120 \\
\hline
\multicolumn{7}{|l|}{Drive Electron Beam:}    \\
\hline
& Energy & (GeV) & 30 & 30 & 3.11 & 6.22           \\
& Electrons per bunch & ($10^{10}$) & 3.5 & 5 & 1.5 & 1.5  \\
& Electrons per pulse & ($10^{10}$) & 3.5 & 5 & 135 & 135  \\
& Beam power & (kW) & 20.2 & 47 & 121 & 161        \\
& RMS beam radius & (mm) & 0.8 & 0.6 & 1.2 & 1.6 \\
 & 
    Beam energy density per pulse & (GeV/mm$^2$) & $5.2 \times 10^{11}$ 
      & $13.3 \times 10^{11}$ & $9.3 \times 10^{11}$ 
      & $10.4 \times 10^{11}$       \\
\hline
\multicolumn{7}{|l|}{Positron Production Target:}      \\
\hline
& Material & & W$_{75}$R$_{25}$ & W$_{75}$R$_{25}$
    & W$_{75}$R$_{25}$ & W$_{75}$R$_{25}$    \\
& Thickness & (R.L.) & 6 & 6 & 4 & 4   \\
& Energy deposition per pulse & (J) & 37 & 53 & 126 & 188 \\
& Power deposition & (kW) & 4.4 & 9 & 23 & 23 \\
& Steady-state temperature & ($^\circ$C) & 100 & 200 & 400 & 400 \\
\hline
\multicolumn{7}{|l|}{Positron Collection:}    \\
\hline
& Accel. frequency & (MHz) & 2856 & 2856 & 1428 & 1428 \\
& Accel. gradient & (MV/m) & 30 & 30 & 25 & 25        \\
& Minimum iris radius & (mm) & 9 & 9 & 20 & 20        \\
& Edge emittance (allowing for 
      &  (m-rad) 
      & 0.01 & 0.01 & 0.06 & 0.06            \\
& 2 mm clearance)  &  &  &  &  &  \\
& Positron yield per $e^-$ &  &  2.5$^a$   
   & 2.5$^a$
   & 1.4                
   & 2.05      \\
& Positrons per bunch & ($10^{10}$) & 8.7 & 12.5 & 2.1 & 3.1 \\
& Positrons per pulse & ($10^{10}$) & 8.7 & 12.5 & 189 & 279  \\
\hline
\multicolumn{7}{p{15cm}}{\footnotesize 
    $^a$~The actual yield immediately following the capture section 
    is 4, but decreases to 2.5 after two 180$^\circ$ bends 
    and a 2 km transport line.}   \\
\end{tabular}}}
\end{table}
\bigskip

For the drive beam accelerator, a thermionic gun coupled to two
subharmonic bunchers and one S-band buncher injects  a properly bunched
electron beam into an S-band linac, which accelerates the electrons to
3.11 GeV (or to 6.22 GeV for higher cms energies). The drive beam is
then directed into one of the two parallel beam lines for positron
production and collection.  The collection system includes a 240-MeV
high-gradient L-band accelerator embedded in a uniform-field solenoid
and an adiabatic phase space transformer consisting of a flux
concentrator and a tapered field solenoid.  The capture system has an
invariant transverse phase space admittance of 0.06 m-rad. Yield
simulations for the positron production and collection system are
performed by using the EGS-4 and ETRANS codes.  The positron beam is
injected into a 1.75-GeV L-band linac consisting of 5-m-long detuned
accelerating structures in a scaled FODO lattice, while the electrons
that are generated and captured along with the positrons are dumped.
Various types of beam diagnostic and control devices are installed at
strategic places throughout the positron source for tuning and beam
stabilization purposes.

Multibunch beam-loading in the positron source is compensated primarily
by injecting the beam during the filling time of the accelerator
structure. In the positron capture accelerator, where beam loading is
particularly severe, the compensation can be performed by shifting the
phase of the rf during the beam pulse, in addition to delaying the rf
pulses relative to the beam.

The dual-vault positron source design shown in Fig.~\ref{fig:psys} aims to
boost the operating efficiency of the positron source by providing an
on-line spare such that NLC operation may be continued in the event of
a component failure in the  positron production and collection system. 
Without the on-line  spare, failure of any component near the positron
target during  a physics  run  would cause extended machine downtime as
the very high radiation levels would prevent prompt maintenance.

\newpage

\section{Damping Rings}

The NLC damping rings are designed to damp the incoming electron and
positron beams to the small emittances needed for collisions.  The
rings have three purposes: (1) damping the incoming emittances in all
three planes, (2) damping incoming transients and providing a stable
platform for the downstream portion of the accelerator, and (3)
delaying the bunches so that feedforward systems can be used to
compensate for charge fluctuations.

To meet these goals, we have designed three damping rings: two
identical main damping rings, one for the electrons and one for the
positrons, and a pre-damping ring for the positrons. The pre-damping
ring is needed because the emittance of the incoming positrons is much
larger than that of the electrons. Each damping ring will store
multiple trains of bunches at once. At every machine cycle, a single
fully damped bunch train is extracted from the ring while a new bunch
train is injected.  In this manner, each bunch train can be damped for
many machine cycles.

The parameters of the two main damping rings are similar to the present
generation of synchrotron light sources and the B-Factory colliders
that are being constructed in that they must store high-current beams
($\sim$1\,A) while attaining small normalized emittances.
Table~\ref{4tab:TBcomppar} compares the NLC ring parameters with those
of the SLAC B-Factory Low-Energy Ring (PEP-II LER), the Advanced Light
Source (ALS) at Lawrence Berkeley National Laboratory, and the
Accelerator Test Facility (ATF) damping ring being constructed at KEK
in Japan to verify many of the damping ring design concepts. In
particular, the stored beam currents are less than half of what the
PEP-II LER has been designed to store while the emittance, energy, and
size of the rings are similar to those of the ALS and the ATF.

\begin{table}[htb]
\leavevmode
\centering
\caption{Comparison of NLC main damping rings with parameters of other
rings.}
\label{4tab:TBcomppar}
\bigskip
\begin{tabular}{|l|cccc|}
\hline \hline
& NLC MDR   & PEP-II LER & LBNL ALS & KEK ATF \\
\hline
Energy (GeV)   &   2 &    3.1 & 1.5  & 1.5   \\
Circumference\ (m)     &  220  & 2200   &  200 & 140   \\
Current (A)    &  1.2  &  3    & 0.6 & 0.6   \\
$\gamma\epsilon_x$ ($10^{-6}$m-rad) & 3   & 400  & 
   10 & 4 \\
$\gamma\epsilon_y$ ($10^{-6}$m-rad) & 0.03 & 16  & 
   0.2 & 0.04  \\
\hline\hline
\end{tabular}
\end{table}
\bigskip

Thus, these other rings will be able to test and verify many of the
accelerator physics issues that will arise in the NLC damping rings. In
particular, strong coupled-bunch instabilities will be studied in the
high current B-factories while issues associated with the very small
beam emittances, such as intrabeam scattering and ion trapping, will be
studied in the ALS and ATF damping rings.  In addition, the PEP-II LER
and the Advanced Photon Source (APS) at Argonne National Laboratory
will be able to study the photoelectron-positron instability that is
thought to arise in positron storage rings.

These similarities with other rings have also simplified the design
process and we have been able, and will continue, to benefit from
experience at these other accelerators. For example: the damping ring
rf system is based on those developed for the SLAC B-Factory and the
ATF damping ring, the multibunch feedback systems are based upon the
feedback systems which were designed for the SLAC B-Factory and
successfully verified on the ALS, and the vacuum system is similar to
that used by the ALS. Furthermore, the design uses ``C'' quadrupole and
sextupole magnets similar to those designed at the ALS and the APS at
Argonne National Lab, a high-field permanent magnet wiggler very
similar to a design developed and installed at Stanford Synchrotron
Radiation Laboratory (SSRL), and a double kicker system for extraction
similar to one to be tested in the ATF.

The NLC damping ring complex is designed to operate with the parameters
listed in Table~\ref{4tab:DesignReq}. These design parameters exceed
the requirements of all presently considered NLC upgrades. The rings
produce electron and positron beams with emittances 
$\gamma\epsilon_x\leq3 \times 10^{-6}$ m-rad and $\gamma\epsilon_y
\leq3 \times 10^{-8}$ m-rad, at repetition rates as high as 180 Hz. The
beams in the damping rings consist of multiple trains of 90 bunches
with a maximum single bunch charge of $1.3 \times 10^{10}$. (This is 5\%
higher than the required charge at the IP to allow for beam loss due to
collimation.) To provide operational flexibility, we have designed the
rings to operate with a peak current roughly 20\% higher than the
nominal peak current.  In addition, the rings have been designed to
operate over an energy range of 1.8 to 2.2 GeV with a nominal energy of
1.98 GeV.  This will allow the damping rate to be increased if
necessary.

\begin{table}[htb]
\leavevmode
\centering
\caption{Requirements for NLC damping ring complex.}
\label{4tab:DesignReq}
\bigskip
\begin{tabular}{|lc|}
\hline \hline
$\gamma\epsilon_{x\,eq}\ /\ \gamma\epsilon_{x\,ext}\ (10^{-6}$ m-rad) &
     3.0 / 3.0 \\
$\gamma\epsilon_{y\,eq}\ /\ \gamma\epsilon_{y\,ext}\ (10^{-8}$ m-rad) &
   2.0 / 3.0 \\
Charge/Bunch & 
   $1.57\times10^{10}$ \\
Bunches/Train &  90 bunches/train \\
    Bunch Spacing (ns) & 1.4 \\
   Repetition Rate (Hz) & 180 \\
\hline\hline
\end{tabular}
\end{table}
\bigskip

At present, most of the studies have concentrated on the main damping
rings. Although relatively little of the detailed engineering has been
performed, we have a good outline of the design and the problems we may
encounter.  In particular, we have identified and described solutions
for the most difficult issues: these are the dynamic aperture, the
vertical emittance, the impedance and instabilities, and the stability
and jitter in both longitudinal and transverse phase space.  As stated,
much of the design rests on work being performed for the B-factories
presently being constructed and the ATF damping ring.

In most aspects, the pre-damping ring has relatively loose
requirements.  The emittance and damping time requirements are not
severe.  The required beam stability is not very significant.  The two
difficult items are attaining the dynamic aperture and the
injection/extraction kicker systems. At this time, the design is still
in a preliminary stage but we have addressed the difficult items just
mentioned.

\subsection{Main Damping Rings.}
\label{4subsub:mainring}

\begin{figure}[htb]
\leavevmode
\centerline{\epsfbox{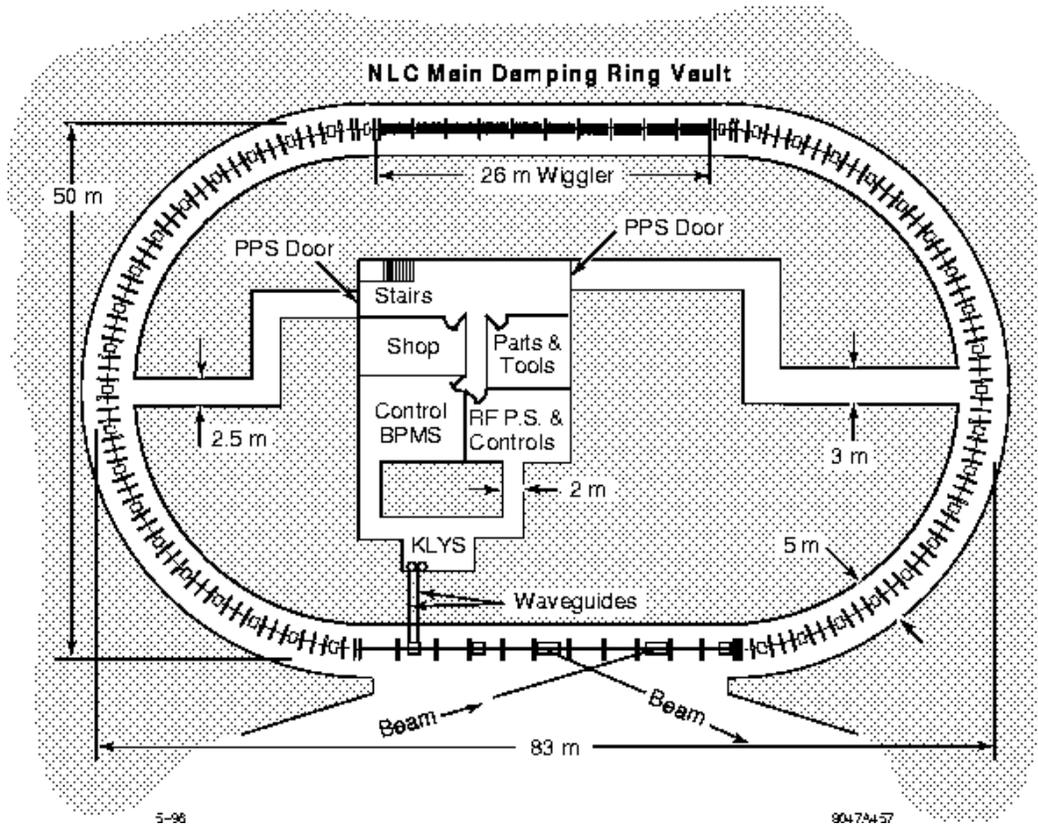}}
\centerline{\parbox{5in}{\caption[*]{Layout of main damping ring.}
\label{4fig:MDRLayout}}}
\end{figure}

The NLC main damping rings are roughly 223 m in circumference and
measure 50 m by 80 m. They have a nominal energy of 1.98 GeV and
the required damping is achieved using both high field bending magnets and
roughly 26 m of damping wiggler. The rings are designed in a racetrack
form with two arcs separated by straight sections; the layout is
illustrated in Figure~\ref{4fig:MDRLayout}.

The main damping rings are designed to damp beams with injected
emittances $\gamma\epsilon_{x,y}=1\times10^{-4}$ m-rad to emittances of
$\gamma\epsilon_{x}=3\times10^{-6}$ m-rad and
$\gamma\epsilon_{y}=3\times10^{-8}$ m-rad. The rings will operate at
180 Hz.  They provide sufficient damping to decrease the injected emittance
by over four orders of magnitude.  These parameters are summarized in
Table~\ref{4tab:MDRPar}.

\begin{table}[htb]
\leavevmode
\centering
\caption[*]{Parameters for main damping rings and the pre-damping ring.}
\label{4tab:MDRPar}
\bigskip 
\begin{tabular}{|lcc|}
\hline \hline
 & MDR & PDR \\
\hline
\hline
Circumference (m)         & 223  & 171  \\                                     
Energy (GeV)          & \multicolumn{2}{c|} {1.8--2.2, 1.98
nominal} \\       
Max.\ Current (A)         & \multicolumn{2}{c|} {1.2}   \\                                   
Max.\ Rep.\ Rate  (Hz)        & \multicolumn{2}{c|} {180} \\                                    
Bunch Trains $\times$ Bunches per train  &$4\times90$  &$3\times90$ \\                    
Train/Bunch Separation (ns)   &  60/1.4  & 64/1.4 \\                                     
$\nu_x$, $\nu_y$, $\nu_s$ & 23.81, 8.62, 0.004  & 10.18, 8.18, 0.018 \\                        
$\gamma\epsilon_{x\,eq}$ (m-rad)     & 2.56  & 7.7 \\                  
$\gamma\epsilon_{x}$, $\gamma\epsilon_y$ (with intra-beam scatt.) &
$3.1\times10^{-6}$,
$1.5\times10^{-8}$   & $4.5\times10^{-5}$ (coupled)   \\ 
$\sigma_\epsilon$, $\sigma_z$     & 0.09\%, $4.0$ mm  & 0.1\%, $7.5$ mm \\                                            
$\xi_{x\,,{\rm uncorr}}, \xi_{y\,,{\rm uncorr}}$ & $-$41.8, $-$22.4   &
$-$13.9, $-$10.0 \\    
$\tau_{x}$, $\tau_y$, $\tau_z$  (ms)
& 4.06, 4.62, 2.50    & 4.44, 6.15, 2.73  \\                       
$U_{SR}$ (kV/turn)                  & 644           & 371 \\        
$J_x$                      & 1.14              & 1.39 \\       
$\alpha_p$                 & 0.00047                   & 0.0051 \\                                  
RF Voltage, Frequency         & 1 MV, 714 MHz & 2 MV,  714 MHz \\                                    
Lattice                    & 40 TME Cells  & 30 FOBO Cells \\                             
\hline\hline
\end{tabular}
\end{table}
\bigskip

The main damping ring lattice is based on detuned Theoretical Minimum
Emittance (TME) cells which were chosen because of eased requirements
on the combined-function bending magnets. The chromaticity is corrected
with three families of sextupoles and the dynamic aperture is more than
sufficient. The damping is performed using both high-field bending
magnets and ten 2.5-m sections of damping wiggler.

The rings operate with four trains of 90 bunches. The bunch trains are
injected onto and extracted from the closed orbit using pulsed kickers
and DC septa. To avoid coupled-bunch instabilities the rf cavities are
based on those of the PEP-II B-Factory and a transverse bunch-by-bunch
feedback system is used. As stated, the rings are designed to operate
with maximum bunch charges of $1.57\times10^{10}$ particles; this is
roughly 20\% more than the maximum needed at the IP.  In addition, the
electron source has been designed to provide additional charge to allow
for at least 10\% losses during injection into the electron damping
ring.  Similarly, the positron source has been designed to produce at
least 20\% additional charge to provide for losses during injection
into the pre-damping ring.

Finally, because the rings must generate extremely small beam
emittances, there are tight jitter and alignment tolerances.  Extensive
effort has been made to include cancellations and tuning procedures in
the design that will ease the tolerances to reasonable levels.  In
particular, all of the quadrupoles and all of the sextupoles will have
independent power supplies and remote magnet movers.  This will
facilitate beam-based alignment as well as matching of the lattice
functions.

\subsection{Positron Pre-Damping Ring}

\begin{figure}[htb]
\leavevmode
\centerline{\epsfbox{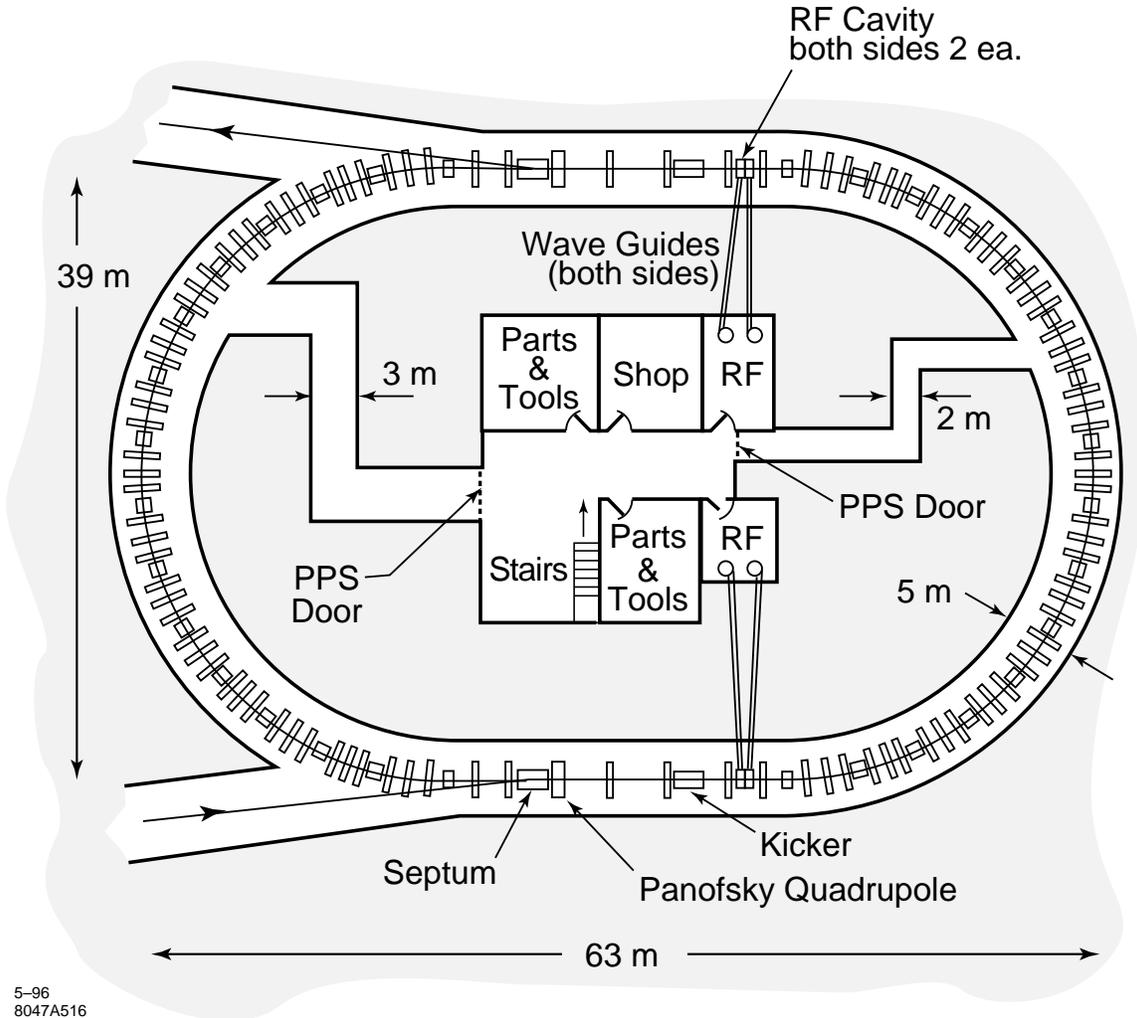}}
\centerline{\parbox{5in}{\caption[*]{Layout of the positron pre-damping ring.}
\label{4fig:PDRLayout}}}
\end{figure}

The design of the pre-damping ring for positrons is similar to the SLC
damping rings, except the lengths are scaled by a factor of five. The
ring has a racetrack form with dispersion-free straight sections for
injection and extraction.  The ring is roughly 60 meters by 40 meters
and is illustrated schematically in Figure~\ref{4fig:PDRLayout}.

In the design, the injection and extraction regions are in the straight
sections on opposite sides of the ring. To reduce the requirements on
these components, the systems are designed to occupy most of these
regions. To minimize rf transients during injection and extraction, a
new bunch train will be injected one half turn after a train is
extracted. Furthermore, the rf cavities are placed downstream of the
injection kicker and upstream of the extraction kicker so that the
injection/extraction process will not interrupt the beam current seen
by the rf cavities.

The positron pre-damping ring is designed to damp the large
emittance positron beam from the positron source to an emittance of
roughly $\gamma\epsilon_{x,y}=1\times10^{-4}$ m-rad; these parameters
are summarized in Table~\ref{4tab:MDRPar}. At this point, the positrons
are injected into the positron main damping ring where they are
damped to the desired final emittances. The pre-damping ring allows us
to decouple the large aperture requirements for the incoming positron
beams from the final emittance requirements of the linear collider.

The pre-damping ring does not need to produce flat beams. Thus, to
maximize the damping of the transverse phase space, the ring has
horizontal damping partition number J$_x\approx1.4$, and operates on
the coupling difference resonance. This increases the damping in both
the horizontal and vertical planes. Furthermore, like the main damping
ring, the ring damps multiple trains of bunches at once, the number of
which is determined by the ring circumference.

The magnets and vacuum system are being designed to provide sufficient
aperture to accept a 2-GeV beam with an edge emittance of
$\gamma\epsilon_{x,y}=0.09$ m-rad and momentum spread of $|\delta
p/p|\leq1.5\%$ plus 2-mm clearance for misalignments and mis-steering.
Given the nominal injected edge emittance of
$\gamma\epsilon_{x,y}=0.06$ m-rad, this provides a substantial margin
for injection and internal mismatches. In addition, the injector
specifications assume that roughly 20\% of the delivered charge is lost
at injection into the pre-damping ring.  The pre-damping ring is
designed to operate with a maximum bunch charge that is roughly 20\%
greater than the maximum required at the IP.

Like the main damping rings, all of the quadrupoles and all of the
sextupoles will have independent power supplies and remote magnet
movers.  This will facilitate beam-based alignment as well as matching
of the lattice functions which is especially important in the
pre-damping ring because of the limited aperture.
\clearpage

\section{Spin Rotators and Bunch Compressors}

The purpose of the spin rotators is rotate the polarization vector,
which is vertically oriented in the damping rings, into any arbitrary
direction so that the beam will be longitudinally polarized at the IP.
The rotators are located at the exit of the damping rings and are then
followed by the bunch compressors.  A spin rotator will be installed on
the electron side and space will be left for an installation on the
positron side to allow for a future upgrade to polarized positrons or
polarized electrons for $\gamma\gamma$ or
$e^-e^-$ collisions. The rotators are based on pairs of solenoids
separated by a horizontal arc to allow full control over the spin
orientation; this is similar in concept to the original SLC spin
rotator system.  Solenoids were chosen to control the spin because the
other alternate, namely a snake, requires a vertical bending chicane
which must be unreasonably long to prevent synchrotron radiation from
increasing the vertical emittance. Of course, the problem with
solenoids is that they couple the beam which, with flat beams, leads to
an increase in the projected vertical emittance; this is the reason the
original solenoid system is not presently used in the SLC.  Our present
design uses pairs of solenoids which are optically separated so that
the coupling is fully cancelled. There are very tight tolerances on the
quadrupole fields between the sextupoles but an extensive skew
correction system has been included to ease the tolerances as well as
correct any residual coupling from the damping rings.

\begin{figure}[htb]
\leavevmode
\centerline{\epsfbox{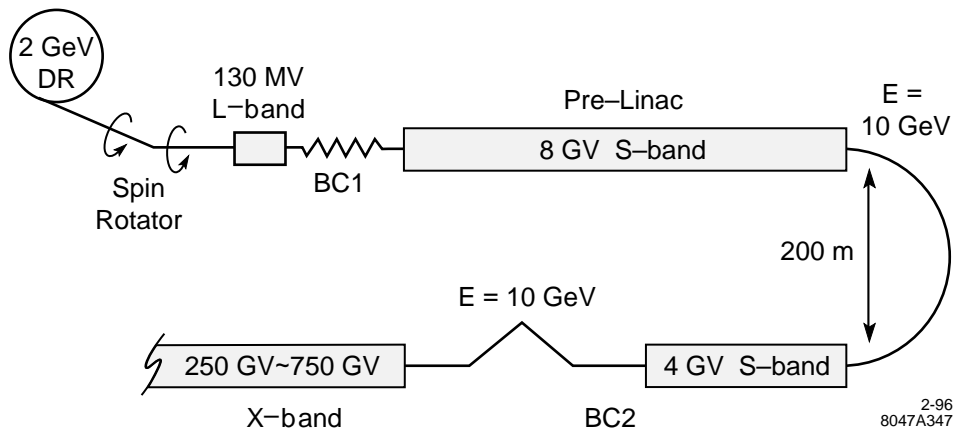}}
\centerline{\parbox{5in}
{\caption[*]{Schematic of the NLC bunch compressor system.}
\label{5fig:bclayout}}}
\end{figure}

The NLC bunch compressors must compress the bunches from the damping
rings with 4-mm lengths to the 100--150 $\mu$m bunch lengths
required in the main linacs and final foci.  To perform the
compressions, a two-stage compressor system has been designed: the
first stage follows the damping rings and the second stage operates at
a beam energy of 10 GeV at the exit of 8-GeV S-band prelinacs.
The compressor system is illustrated schematically in Figure
\ref{5fig:bclayout}.

The bunch compressor system was designed with three primary goals:

\begin{itemize}
\item 
The compressor system must compress a bunch with length
$\sigma_{z}\approx 4$\,mm from the damping rings to the 100--150$\mu$m
length needed in the main X-band linacs and the final foci. This
requires compressing the bunch length, and correspondingly increasing
the energy spread, by a factor of 30--40.
\item 
The bunch compressor system should be able to compensate for
bunch-to-bunch phase offsets induced by the beam loading in the damping
rings; if not corrected, these phase errors would become energy errors
at the end of the linacs.  This requires that the compressor system
rotate the longitudinal phase space by $\pi/2$, translating the phase
errors into injection energy errors.
\item 
The bunch compressor system should provide a trombone-like arm,
reversing the direction of the beam before injection into the main
X-band linacs. This allows extension of the linac lengths to upgrade
the beam energy at the IP without significant modification to the
injection systems. It also allows for abort systems to dump the beam
and feed-forward systems to correct the beam trajectory, phase, and
energy before injection into the main linacs.
\end{itemize}

The two-stage system has a number of advantages over a single stage
compressor. In particular, it keeps the rms energy spread less than
roughly 2\%, it is optically more straightforward, and the bunch
lengths are more logically matched to the acceleration rf frequency so
that the energy spread due to the longitudinal wakefields can be
cancelled locally. The disadvantage of the two-stage design is that it
is more complex and lengthy than a single-stage compressor. The first
stage rotates the longitudinal phase by $\pi/2$ while the second stage
performs a $2\pi$ rotation.  In this manner, phase errors due to the
beam loading in the damping rings and energy errors due to imperfect
multibunch energy compensation in the 8-GeV S-band prelinacs do not
affect the beam phase at injection into the main linac.

Assuming an rms energy spread of $\sigma_\delta=1\times10^{-3}$, the
low energy compressors compress the damping ring beam to a bunch length
of $500\mu$m.  The final bunch length is roughly independent of the
incoming bunch length and the system has been designed to accept an
initial bunch length of 5 mm; this is 25\% larger than expected from
the damping ring. The bunch compressors consist of a 139-MV L-band 
(1.4-GHz) rf section followed by a long period wiggler which generates the
$R_{56}$ needed for the bunch compression.

The high-energy bunch compressors follow the prelinacs.  Assuming an
rms bunch length of $\sigma_z=500\mu$m, they compress the beam to a
bunch length between $100$ and $150\mu$m depending on the NLC
parameters.  As stated, they are telescopes in longitudinal phase
space, rotating the longitudinal phase space by $2\pi$. The bunch
compressors consist of a 180$^\circ$ arc which is followed by a
4-GeV S-band (2.8-GHz) rf section and a chicane. Parameters of the
high-energy bunch compressors are listed in Table~\ref{5tab:HBCparm}
for four different NLC parameter sets.

\begin{table}
\leavevmode
\centering
\caption[*]{Parameters for 10-GeV compressor.}
\label{5tab:HBCparm}
\bigskip
\begin{tabular}{|l|cccc|}  
\hline\hline
 & NLC-I a & NLC-I c & NLC-II a & NLC-II c \\
\hline
 Energy & \multicolumn{4}{c|}{10 Gev} \\
Initial $\sigma_{z}$, $\sigma_{\epsilon}$  &
\multicolumn{4}{c|}{500$\mu$m, 0.25 \%} \\
Final $\sigma_{z}$ & 100$\mu$m & 150$\mu$m & 125$\mu$m & 150$\mu$m \\
Final $\sigma_{\epsilon} $ & 1.5 \% & 1.2 \% & 1.3 \% & 1.4 \% \\ 
180$^\circ$ arc $R_{56}$ & 0.211 & 0.147 & 0.184 & 0.162 \\
Arc Length & \multicolumn{4}{c|}{330 m} \\
$V_{rf}$ (GV) & 3.85 & 3.50 & 3.73 & 3.61 \\ 
$f_{rf}$ & \multicolumn{4}{c|}{2.8 GHz} \\ 
$L_{rf}$ & \multicolumn{4}{c|}{250 m} \\ 
Chicane $R_{56}$ & \multicolumn{4}{c|}{36 mm} \\ 
Chicane Length & \multicolumn{4}{c|}{100 m} \\ 
\hline\hline
\end{tabular}
\end{table}
\bigskip

One of the driving philosophies behind the NLC compressor design has
been to utilize naturally achromatic magnetic lattices wherever the
beam energy spread is large.  In particular, the optics is chosen so
that quadrupoles are not placed in regions of large dispersion and
strong sextupoles are not needed.  This choice arises from experience
with the SLC bunch compressors which are based on second-order
achromats where quadrupoles are located in dispersive regions and
strong sextupoles are used to cancel the chromatic aberrations. 
Unfortunately, the SLC design is extremely difficult to operate and
tune because of the large nonlinearities and the sensitivity to
multipole errors in the quadrupoles; over the years additional
nonlinear elements have been added (skew sextupoles and octupoles) to
help cancel the residual aberrations but the tuning is still difficult.
To further facilitate the tuning in the NLC design, we have explicitly
designed orthogonal tuning controls and diagnostics into the system. 
We feel confident that these considerations will make the system
relatively straightforward to operate and tune.

As discussed, the rf acceleration throughout the compressor system
(including  the 8-GeV prelinac) is
performed with relatively low frequency rf. In particular, the first
bunch compressors use L-band (1.4 GHz) and the prelinac and second
bunch compressor use S-band (2.8-GHz).  Although these
systems are, in general, longer and more expensive than higher
frequency accelerator systems, we feel that their superior acceptances
are needed at the lower
beam energies. In particular, we have designed the systems so that they
have small beam loading and relatively loose alignment tolerances. This
is important to provide the reliability and stability that is desired
in the bunch compressor system. Of course, such lightly-loaded systems
are inefficient and could not be used to accelerate the beams to very
high energy.

Finally, although the tolerances on components in the bunch compressor
systems are not nearly as tight as in the main linacs or the final
foci, we have chosen to adopt the same methods of beam-based alignment
and tuning. In particular, to ease the alignment procedures, all of the
quadrupoles will be mounted on remote magnet movers and will be powered
by individual power supplies.  In addition, each quadrupole will
contain a BPM with a resolution of $1\mu$m.  Similarly, all of the
accelerator structures will be instrumented with an rf BPM to measure
the induced dipole modes and each girder will be remotely movable;
depending on location, the girders will support either one or two
structures.

\clearpage

\section{Main Linacs: Design and Dynamics}                                      
\label{sec:7linac}                                                             
                                                                                                                                                                
The main linacs are designed to initially accelerate electrons and
positrons from an energy of 10 GeV to 250 GeV, and to be built so that
with reasonably expected near-term improvements in the X-band rf
system, the beam energy can be increased over time to 500 GeV.  The
infrastructure and layout of the collider is designed to allow the
energy of the beam to be further increased to as much as 750 GeV if
suitably efficient rf sources, such as cluster klystrons, grided
klystrons, or two-beam accelerators can be developed. The linac must
accelerate trains of 90 electron or positron bunches with intensities
that vary from $0.75 \times 10^{10}$ to $1.1 \times 10^{10}$ particles
per bunch as the energy is increased. The spacing between bunches in a
train is determined by the damping ring frequency to be 1.4 ns. The repetition
rate of the accelerator must be reduced from 180 Hz to 120 Hz as the
energy is increased to maintain the linac AC power consumption below
200 MW.
                                                                                
There are two major design issues for the main linacs: the efficient
acceleration of the beams using X-band technology as described in the
next section, and the preservation of the small beam emittances
that are required to achieve the desired collision luminosity. In this
section, we discuss the design and operation of the linacs emphasizing
the beam dynamics issues.
                                                                              
The NLC linac design builds on the experience that has been gained from
the operation of the SLC, in particular, the development of
beam-based alignment, tuning, and feedback techniques. The NLC design
also draws on the SLC operational experience in such areas as the
transport of flat beams, the use of BNS damping, and linac energy
management. In areas where we cannot directly verify NLC operational
issues with the SLC, we have built test facilities. These include ASSET
for X-band structure wakefield measurements, ASTA and NLCTA for X-band
accelerator technology testing, and the FFTB for testing of components
such as quadrupole movers that will be used in the main linacs as well as in
the final focus. Of these facilities, the NLCTA will provide the most
direct test of the basic linac design in its ability to accelerate
multibunch beams and to control the effects of beam loading.
 
Parameters of the main linacs are listed in
Table~\ref{tab:LIparameters} for NLC-I, which has a 500 GeV
center-of-mass energy, and NLC-II, which has a 1 TeV center-of-mass
energy. The length of each of the two main linacs in NLC-I is 8.8 km.
This will increase only slightly in the upgrade to NLC-II, with the
bulk of the energy gain coming from an increase in the unloaded
gradient from 50 to 85 MeV/m. This increase in gradient will be
performed adiabatically by upgrading and doubling the number of
klystrons which supply the rf power. It will be accompanied by an
increase in beam charge which will leave the fractional beam loading
roughly constant.
 Since the
total linac length required for NLC-II is not much longer than that for
NLC-I, we will probably construct the full NLC-II linac length
initially and use the additional length in NLC-I to attain beam
energies slightly higher than 250 GeV.

\begin{table}[htbp]                                                                                
\centering
\caption[*]{NLC Linac Parameters}                                                         
\label{tab:LIparameters}
\bigskip                                                             
\begin{tabular}{|l|ccc|ccc|}                                                   
\hline \hline                                                                         
    & \multicolumn{3}{c|} {\textbf{NLC-I}} &                                       
      \multicolumn{3}{c|} {\textbf{NLC-II}}  \\                                               
  & a & b & c & a & b & c \\                                                    
\hline                                                                          
N [$10^{10}$] & 0.65 & 0.75 & 0.85 & 0.95 & 1.10 & 1.25 \\                      
Bunches per Train & \multicolumn{3}{c|}{90} 
              & \multicolumn{3}{|c|}{90} \\                                                   
Repetition Rate [Hz]  
              & \multicolumn{3}{c|}{180} 
              & \multicolumn{3}{|c|}{120}  \\                                                 
$\sigma_z [\mu m]$ & 100 & 125 & 150 & 125 & 150 & 150 \\                        
Unloaded Gradient [MV/m] 
              & \multicolumn{3}{c|}{50} 
              & \multicolumn{3}{|c|}{85}  \\                                                  
Multibunch Loading [\%] & 25.5 & 29.4 & 33.34 & 22.0 & 25.5 & 28.9 \\           
Multibunch Loading [MV/m] & $-$12.8 & $-$14.7 & $-$16.7 &                       
         $-$18.7 & $-$21.5 & $-$24.6 \\                                               
Single-bunch Loading$^\ast$                                                     
   [MV/m] & $-$0.2 & $-$0.3 & $-$0.4 & $-$0.4 & $-$0.5 & $-$0.6 \\          
Average $\phi_{RF}^\dagger$                                                     
   [$^\circ$] & $-15$ & $-15.5$ & $-16$ & $-7.5$ & $-8.2$ & $-11.2$ \\                
BNS Overhead [\%] & 3 & 3 & 3 & 3.5 & 3.5 & 3.5 \\                                    
Feedback Overhead [\%] & 2 & 2 & 2 & 2 & 2 & 2 \\                               
Repair Margin [\%] & 3 & 3 & 3 & 3 & 3 & 3 \\                                   
$L_{\rm acc}$ [m] & \multicolumn{3}{c|}{8150} &                                              
      \multicolumn{3}{|c|}{8834}  \\                                                
Number of sections in Structure & \multicolumn{3}{c|}{4528} &                                            
      \multicolumn{3}{|c|}{4908}  \\                                                
$L_{\rm total}$ [m]  & \multicolumn{3}{c|}{8807} &                                           
      \multicolumn{3}{|c|}{9550} \\                                                 
 $E_{\rm max}^\ddagger$                                                             
   [GeV] & 266.7 & 250 & 232.2 & 529 & 500 & 468 \\                              
\hline                                                                          
\multicolumn{7}{l}{                                                             
   $^\ast$Single-bunch loading includes only HOM losses.}\\              
\multicolumn{7}{l}{                                                             
   $^\dagger$Average rf phase for 0.8\% FWHM                                      
   energy spread.}               \\                                                         
\multicolumn{7}{l}{                                                             
   $^\ddagger$Including 10 GeV initial energy.}      \\                              
\end{tabular}
\end{table}
\bigskip

For both NLC-I and NLC-II, three different parameter sets are listed.
These correspond to the same rf power and linac length, but assume
different bunch charges and bunch lengths. The three parameter sets
reflect a tradeoff in bunch charge versus beam size where the
luminosity is kept nearly constant. This is the main tradeoff that
remains in a linear collider design when all constraints are taken into
account. Larger beam sizes are desirable since they loosen many of the
tolerances. However, increasing the beam charge to compensate the loss
in luminosity produces problems related to beam power, and generally
leads to tighter wakefield-related tolerances. The three parameter sets
each represent compromises among these competing effects: together they
span the range of operating conditions that we consider reasonable.

The design normalized emittance of the bunches entering the main linacs
is $\gamma \epsilon_x$ = $3.6 \times 10^{-6}$m-rad, $\gamma \epsilon_y$
=  $4.0 \times 10^{-8}$m-rad, and the energy is 10 GeV.
During the acceleration of the bunches to $\sim$250 GeV (NLC-I)
or $\sim$500 GeV (NLC-II), it is necessary to preserve the low transverse
emittance and maintain the small final beam energy spread; these
requirements are listed in Table~\ref{tab:LIemitreq}. Note that we have
allowed for over 100\%\  vertical emittance growth. This lets us set
fairly conservative tolerances on the accelerator structure alignment.
However, we believe that we will do better than these tolerances as we
gain more experience with X-band accelerator construction and operation.
                                                                                
\bigskip
\begin{table}[htb]
\centering
\caption{NLC Linac Emittance Requirements}                                             
\label{tab:LIemitreq}
\bigskip                                                                
\begin{tabular}{|c|ccc|ccc|}                                                   
\hline \hline                                                                          
    & \multicolumn{3}{c|}{NLC-I} & \multicolumn{3}{c|}{NLC-II}  \\                         
  & a & b & c & a & b & c \\                                                    
\hline                                                                          
N [$10^{10}$] & 0.65 & 0.75 & 0.85 & 0.95 & 1.10 & 1.25 \\                      
$\sigma_z [\mu m]$ & 100 & 125 & 150 & 125 & 150 & 150 \\                        
$\gamma\epsilon_{x\,inj} [10^{-6}$m-rad]  &                                     
   \multicolumn{3}{c|}{3.6} & \multicolumn{3}{c|}{3.6}  \\                                              
$\gamma\epsilon_{y\,inj} [10^{-8}$m-rad]  &                                     
   \multicolumn{3}{c|}{4} &  \multicolumn{3}{c|}{4} \\                                              
$\sigma_{\epsilon\,inj} [\%]$ &                                                 
   \multicolumn{3}{c|}{$<$1.5} &  \multicolumn{3}{c|}{$<$1.5}  \\                                   
\hline                                                                          
$\gamma\epsilon_{x\,ext} [10^{-6}$m-rad] &                                      
   \multicolumn{3}{c|}{4} &  \multicolumn{3}{c|}{4}  \\                                         
$\gamma\epsilon_{y\,ext} [10^{-8}$m-rad] &                                      
   7 & 9 & 11 & 9 & 11 & 13  \\                                                   
$\Delta E/E_{FWHM\,ext}$ single bunch [\%]   &                                  
   \multicolumn{3}{c|}{0.8} & \multicolumn{3}{c|}{0.8} \\                                           
$\Delta E/E_{FWHM\,ext}$ train [\%]  &                                          
   \multicolumn{3}{c|}{1.0} &  \multicolumn{3}{c|}{1.0} \\                                          
\hline \hline                                                                         
\end{tabular}
\end{table}
\bigskip                                                                  
                                                                                
\subsection{Linac Layout and Site Requirements}                                                         
\label{sec:design}                                                              
                                                                                
Although the layout of the main linacs is fairly simple, an array of
X-band accelerator structures interleaved with a FODO quadrupole
lattice, one faces a number of challenges to accelerate low-emittance
bunch trains without significantly degrading the beam phase space. The
X-band structures that provide a high-acceleration gradient also
heavily load the beam, so careful control of the rf pulse shape will be
needed to achieve the small beam energy spread that is required at the
end of the linac. The X-band structures also have the drawback that
strong transverse wakefields, both long-range (bunch-to-bunch) and
short-range (intra-bunch), are generated when the bunch train passes
off-axis through the structures. Although damping and detuning of the
dipole modes of the structure will be used to suppress the long-range
wakefield, and autophasing will be used to offset the short-range
wakefield effect on betatron motion, the structures will still have to
be aligned precisely to limit emittance growth from both short-range
and long-range wakefield effects. The small bunch sizes also make the
beam sensitive to other emittance growth mechanisms such as dispersion
from quadrupole misalignments, focusing from the ions generated in the
residual gas of the beam line vacuum chambers, and dilution from beam
trajectory jitter, like that caused by vibrations of the quadrupole
magnets.
                                                                                
Most of the deleterious effects will be limited by design. For example,
the quadrupole magnet power supplies will be chosen to be stable enough
not to cause significant beam trajectory jitter. For some of the
effects, however, we will rely mainly on measurements of the beam
properties to control emittance growth. In particular, beam trajectory
measurements will be used to adjust the positions of the quadrupoles
and structures to minimize the emittance growth caused by their
misalignment. For this purpose, each quadrupole and structure in the
linacs will contain a beam position monitor (BPM), and will be
supported by movers that can be remotely controlled ({\it e.g.}, like
the magnet movers in the FFTB).
                                                                                
A beam-based approach will also be used in a trajectory feedback system
to suppress low-frequency jitter of the bunch train as a whole, and if
necessary, bunch-to-bunch position variations within the train (special
multibunch BPMs will be used in this case). These feedbacks will be
included in each of five instrumentation regions that will be located
along the linacs. These regions will also include high-dispersion
sections so that bunch energy and energy spread can be measured and
used for rf control feedback. As with the operation of the SLC, if the
feedback and alignment algorithms fail to suppress emittance growth to
the desired level, non-local types of tuning will be used, such as
orbit bumps. To aid in this tuning, and to monitor the beam phase
space, the instrumentation regions will also contain beam size monitors.
                                                                                
\begin{figure}[htbp]
\leavevmode
\centerline{\epsfxsize=6.5in\epsfbox{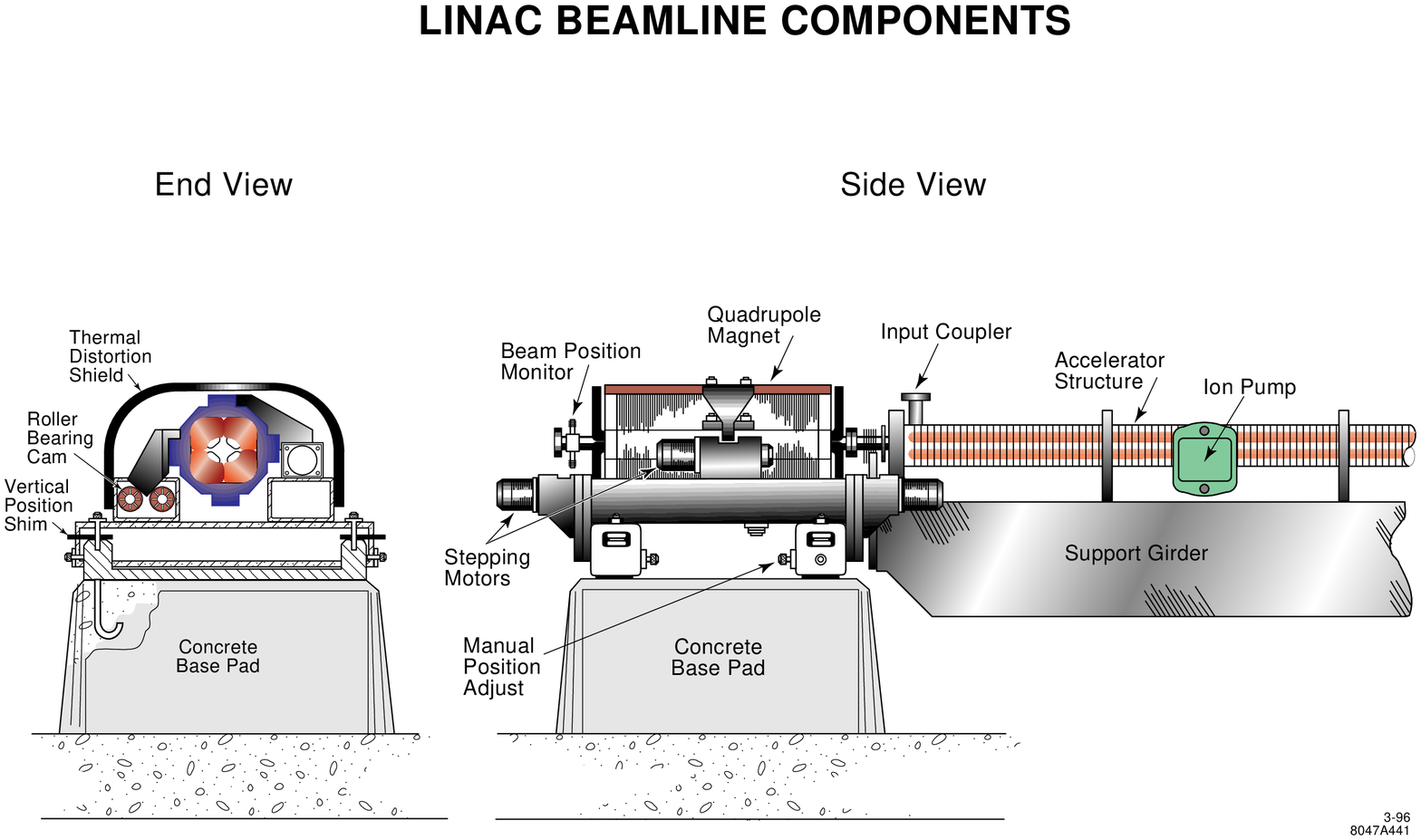}}
\centerline{\parbox{5in}{\caption[*]{End view and side view of the
beam line near the beginning of the linacs, where there is a
quadrupole between each pair of accelerator components.}
\label{fig:LIstrongbacka}}} 
\end{figure}
                                                                                
\begin{figure}[htbp]
\leavevmode
\centerline{\epsfbox{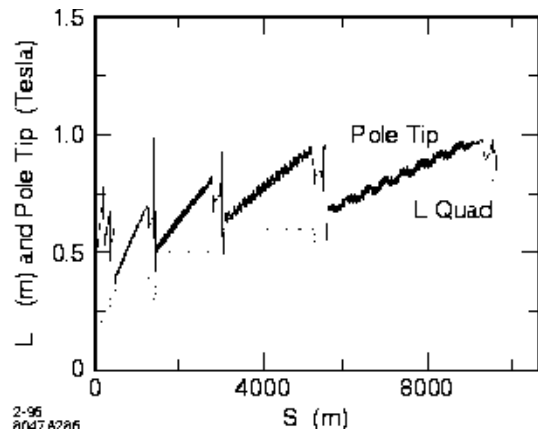}}
\centerline{\parbox{5in}{\caption[*]{Quadrupole lengths (dotted)                                                   
and pole-tip fields (solid) in NLC-IIb lattice.}                               
\label{fig:LIquadtwo}   }}                                                             
\end{figure}

\begin{figure}[htbp]
\leavevmode
\centerline{\epsfxsize=3.5in\epsfbox{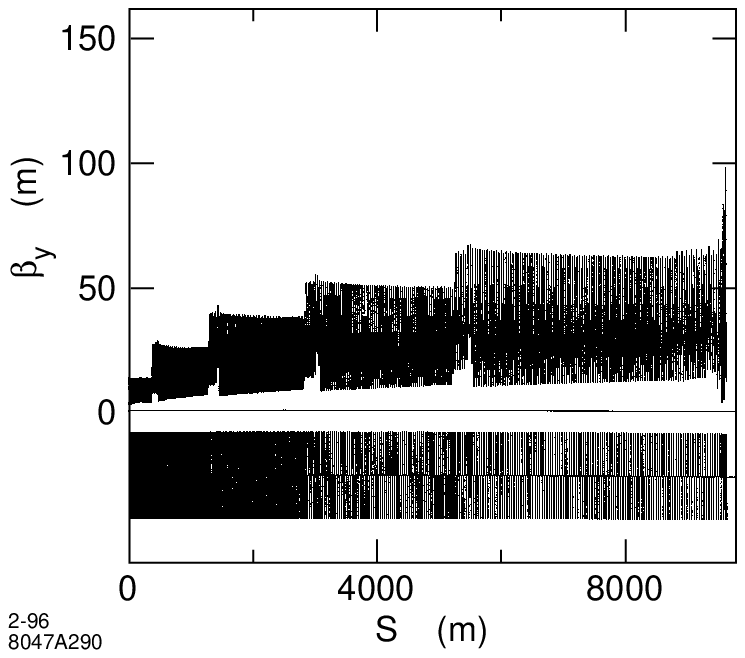}}                                                              
\centerline{\parbox{5in}{\caption[*]{Vertical beta function in the
NLC-II linacs.} 
\label{fig:LIbetaY} }}
\end{figure}

Near the beginning of each linac, there will be a quadrupole magnet
after each structure pair, these components are shown in
Fig.~\ref{fig:LIstrongbacka}. The separation between quadrupoles will
increase in two-structure increments, from one pair of structures to
five pairs at the end of the linacs. The quadrupole lengths and pole
tip fields are shown in Fig.~\ref{fig:LIquadtwo}, the vertical beta
function in Fig.~\ref{fig:LIbetaY}.  The  phase advance per cell ranges
between about 70$^\circ$ and 90$^\circ$, with somewhat stronger
focusing in the vertical plane. As part of the machine protection
system, there will be a titanium spoiler after each structure. For
normal beam operation, it will do little to prevent damage to the beam
pipe, but for tuning, it allows a single high-emittance bunch to
intercept the beam pipe without causing damage.

\begin{figure}[htbp]
\leavevmode
\centerline{\epsfxsize=3.5in\epsfbox{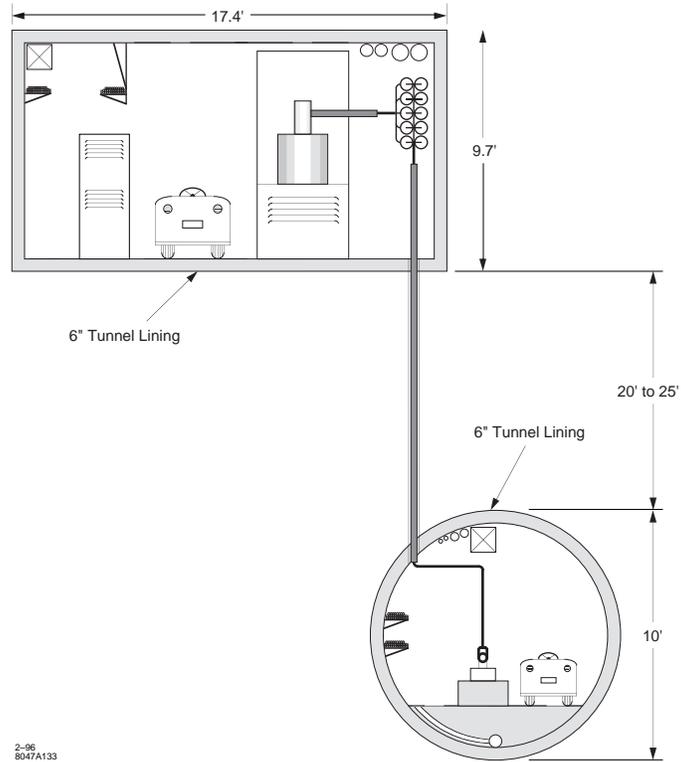}}
\centerline{\parbox{5in}{\caption[*]{Klystron gallery and beam line
tunnel layout (``SLAC type'').  This is just one of several possible
layouts for housing the accelerator and rf components.} 
\label{fig:LIslactype} }}
\end{figure}

Since the main linacs are the largest components of the NLC, the
impacts of the site characteristics on the cost and operation of the
linacs will be the dominant technical factors in choosing a location to
build the accelerator. As an example we have considered a layout
similar to that of the present SLAC accelerator
(Fig.~\ref{fig:LIslactype}). Placing the klystron gallery underground
will help stabilize the gallery air temperature to within the
1$^\circ$C required. In the beam line area, we want a 0.1$^\circ$C
stability, which should not be too difficult to achieve since the beam
line tunnels will be well isolated. The beam line is best located at
least 20~ft below ground to provide natural radiation shielding. Access
shafts to the surface will probably be required about every kilometer
to provide cooling tower and substation connections, and to allow
maintenance crews to reach the beam line. A flat terrain is desirable
to minimize the depth of such access shafts and would also facilitate
the establishment of a global coordinate system for alignment on the
surface and the subsequent transfer of the reference system to the beam
line. Hardness of the ground, seismic activity on all scales, and
cultural noise such as that produced by traffic and heavy machinery,
will all be factors in the site selection.

\subsection{Operation and Tuning of the Main Linacs}                                                    
\label{sec:dynam}

Initial set up of the machine for operation includes creating an
optimal single bunch energy spread profile for the BNS damping (or
``autophasing'') which is the introduction of a correlated energy
spread along each bunch to offset the short-range wakefield effect on
betatron motion.  We wish to do this with minimum energy loss due to
the phasing, and we also want the bunches at the end of the linac to
have small enough energy spread to fit into the final focus bandwidth.
The bunches are run behind the crest early in the linac, and then
sufficiently in front of the crest later in the linac to remove enough
energy spread to fit within the final focus bandwidth.

Another necessity for initial setup is multibunch energy compensation,
since otherwise there would be a drop in energy of between 20 and 30\%\
between the head and tail of the multibunch train.  The method of
compensation must keep the energies of all the bunches the same to
within a few tenths of a percent.  This will be done by pre-filling the
accelerator structure with an rf pulse shaped to simulate the steady
state beam-loaded rf field profile.  This profile is then automatically
maintained if the incoming rf field is kept constant during the passage
of the train, assuming systematic variations of the bunch charges are
kept small enough.  This places fairly stringent tolerances on the
injectors (and other upstream effects on the bunch charges).

To maintain high luminosity it will be required to keep both the
emittance and beam position jitter small at the interaction point. Due
to the tiny vertical beam size, pulse-to-pulse beam position jitter
generated by the movements of the quadrupole magnets could potentially
degrade the luminosity.  The ground motion frequencies of concern are
those above about 0.1 Hz where the feedback systems that will be used
to stabilize the beam may not produce the suppression that is required.
The spatial characteristics of the motion are important as well. The
effect of quadrupole vibration on the beams is highly suppressed if the
motion is correlated over distances longer than the betatron
oscillation lengths of the beams.

Since few data exist on the spatial properties of ground motion, we did
a series of seismic measurements at SLAC.  We found that the motion can
be well modeled by plane waves that are isotropic in direction.  These
surface waves are dispersive and the dependence of the phase velocity
on frequency was inferred from the data by assuming the plane-wave
model.

\begin{figure}[htb]
\leavevmode
\centerline{\epsfxsize=3.5in\epsfbox {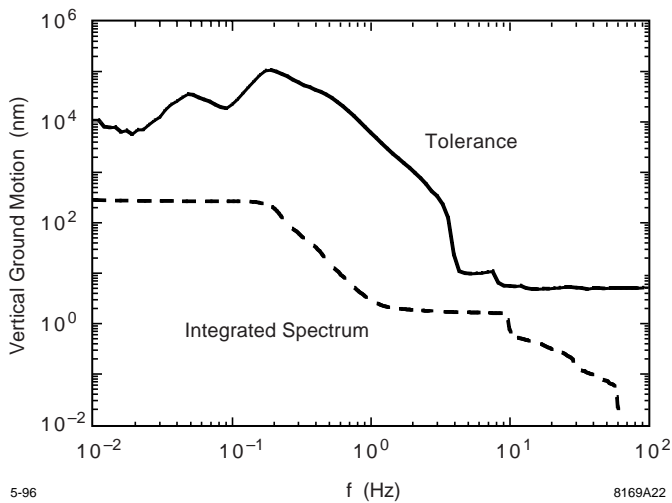}}
\centerline{\parbox{5in}{\caption[*]{Integrated vertical ground motion
spectrum (dashed) and tolerance on the vertical rms ground motion of
the linac (solid) for a 1.5\% luminosity reduction.}
\label{fig:LIfintegspectrum}}} 
\end{figure}

In addition to this wave-like motion, there can be an uncorrelated
component which, as a worst case, we take equal to the electronic noise
of the seismometers that we used for our measurements. The tolerance on
ground motion as a function of frequency was then calculated, and is
shown in Fig.~\ref{fig:LIfintegspectrum} for waves plus the
uncorrelated noise component. The tolerance includes the suppression
expected from a trajectory feedback system with performance similar to
that achieved at the SLC. The net effect on the luminosity is obtained
by integrating the power spectrum weighted by the inverse square of the
tolerance.  For frequencies greater than 0.01 Hz, this yields a
luminosity reduction of about 0.1\%.  Our conclusion is that ground
motion in the linacs should not be a problem as long as care is taken
to avoid introduction of substantial vibrations locally in the
accelerator housing, ({\it e.g.} water pumps or other equipment.
                                                                                                                                                                 
\begin{figure}[htb]
\leavevmode
\centerline{\epsfbox{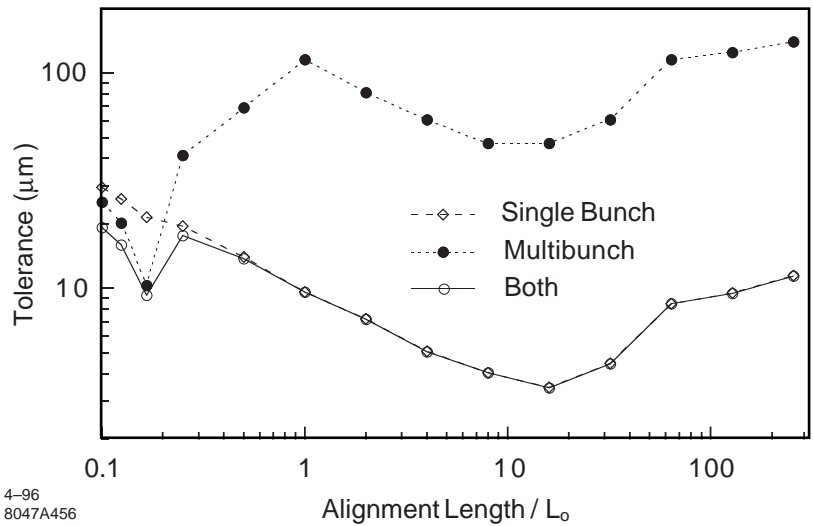}}
\centerline{\parbox{5in}{\caption[*]{Tolerance on the structure
alignment for a 50\% emittance growth in the NLC-IIb linacs as a
function of alignment length, when considering only single-bunch
effects, only multi-bunch effects, and both effects together. $L_0$ is
the 1.8 m structure length.}
\label{fig:LIfilosho}  }}                                                              
\end{figure}
\bigskip

Quadrupole and accelerator structure alignment will be a key aspect of
NLC operation, since quad offsets will lead to dispersive emittance
growth and structure offsets will lead to wakefield related emittance
growth. Initial alignment of the quadrupole magnets can be done with
sufficient accuracy by conventional methods, but obtaining and
maintaining the required beam quality will require additional
beam-based techniques. Alignment procedures that use measurements of
correlations between orbit trajectories and quadrupole strengths and
mover settings have been developed at the SLC and FFTB.  Adaptation of
these techniques to the NLC linacs is straightforward, and can be used
to determine the offsets between the beam position monitors and the
quadrupoles, as well as the misalignments of the quadrupoles with
respect to a smoothed line. The BPM trajectory readings  can then be
used to return the quadrupoles to smoothed locations at periodic
intervals. The accelerator structures will be aligned by measuring the
dipole wakefield signals generated by the passage of the beam.  It is
precisely these signals that must be nulled to eliminate wakefields
that damage the beam phase space.   Additional suppression of emittance
growth in the linac is possible with the use of non-local correction
methods; trajectory bumps, fast kickers, {\it etc.}, can be used to
cancel the emittance dilutions after they have accumulated.  These are
not preferred methods since they are sensitive to variations in beam
transport properties between the sources of the dilution and the
correction point, but they are in extensive use presently at the SLC.
                                                                               
Another major concern is the long-range wakefield in the accelerator
structures, which if not heavily suppressed, will resonantly amplify
the betatron motion of the bunch train by many orders of magnitude.
Much work to date has been devoted to designing an accelerator
structure that sufficiently suppresses the long range transverse wake
fields.  The outcome is the damped detuned structure (DDS), which
incorporates both detuning and damping of the wakefield. This brings
the amplification due to wakefields of incoming betatron oscillations
under control, but still leads to rather tight structure alignment
tolerances on some length scales. Figure~\ref{fig:LIfilosho} shows
tolerances on the structure alignment as a function of alignment length
scale, assuming the wakefield from the DDS. (For instance, 0.1 on the
horizontal axis of this figure corresponds to the case where each group
of $\sim 20$ cells in each of the $\sim 200$-cell accelerator
structures is separately misaligned.) Non-local correction methods can
be used to loosen these tolerances somewhat. As was noted earlier, we
allow for a vertical emittance growth from all sources of over 100\%,
even though we expect to do somewhat better than the tolerances that
would lead to this amount of growth.  The total emittance budget is
given in Table~\ref{tab:LIemitbudget} for the NLC-IIb linac design.

\bigskip
\begin{table}[htb]
\centering                                                                          
\caption[*]{NLC-IIb Linac Emittance Growth Budget}                                             
\label{tab:LIemitbudget}
\bigskip                                                                
{\small \begin{tabular}{|llc|c|l|}                                                   
\hline \hline                                                                         
Source       &      &       & $\Delta\epsilon_y / \epsilon_y$ 
&   Dynamics   \\                         
\hline                                                                            
\multicolumn{2}{|l}{Quad alignment}   &  &  40\%    &  Incoherent
dispersion \\                      
  &   BPM resolution    & $1 \mu m$ &   &  \\                      
  &   BPM to quad alignment & $2 \mu m$ &                     
  &              \\
\hline                      
\multicolumn{2}{|l}{Quad drift between alignment}
  &   &  10\%  & Coherent dispersion       \\                      
& Steering period  &  30 min  &  & \raisebox{1.5ex}{+ wakes}  \\                      
\hline                                                                                                                                                         
\multicolumn{2}{|l}{Structure alignment}   & &   &             \\                      
  &    Internal alignment & $15 \mu m$ & 25\% &  Long-range wakes   \\                      
  &    Beam measurement accuracy & $15 \mu m$ &  50\%  
  &  Short-range wakes  \\
\hline
\multicolumn{2}{|l}{Other ({\it e.g.}, quad roll, ion effects, and rf
deflections)}  &  &   50\%             &             \\
\hline
 &Total     &  &  175\%             &             \\                                                                                                                                                         
\hline \hline
\end{tabular}}
\end{table}
\bigskip

\clearpage

\section{The RF System for the Main Linacs}
\label{sec:rfsystem}

The basic design of the NLC main linacs rests on global experience
gained from the design, construction, and 30 years of operation of the
3-km-long SLAC linac, which is powered at a frequency of 2.856 GHz
\cite{see91, see93}. Since its initial operation in 1966, the SLAC
linac has been continuously upgraded for higher energy, higher
intensity, and lower emittance.

The radio frequency (rf) system for the NLC main linacs is similar in
character to the SLAC linac. The SLAC linac is currently energized by
240 high-power S-band klystrons. The klystron peak power and pulse
duration are, respectively, 65 MW and 3.5 $\mu$s. The power from each
klystron is compressed by a SLED pulse compressor, and then split to
feed four, 3-m-long, constant-gradient, S-band accelerator structures
operating in the 2$\pi$/3 mode.

\begin{figure}[htb]
\leavevmode
\centerline{\epsfbox{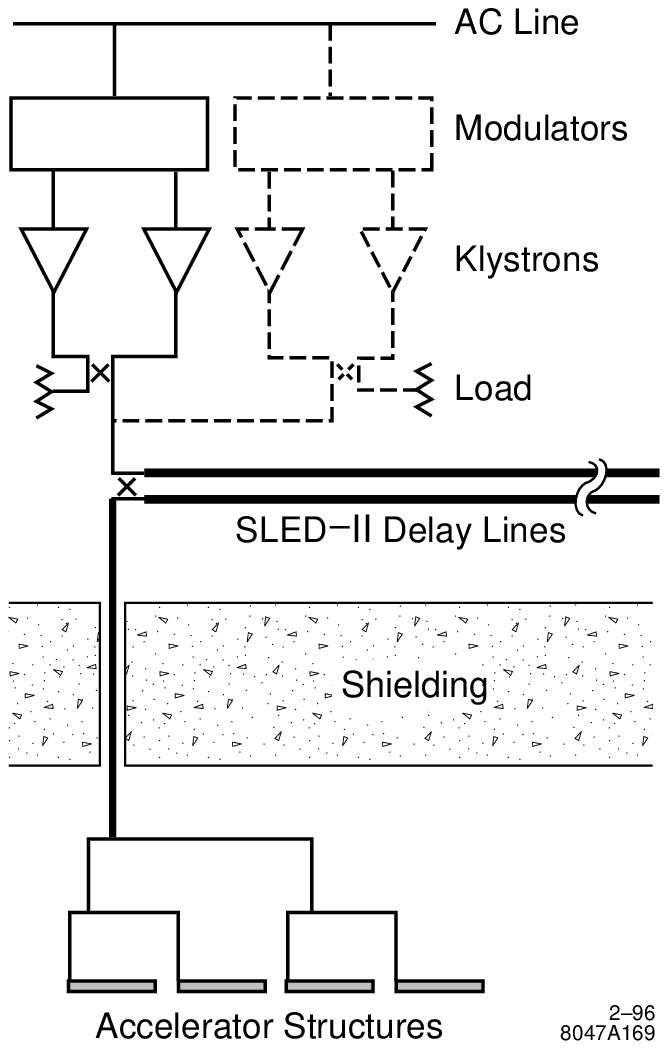}}
\centerline{\parbox{5in}{\caption[*]{Schematic of one module of the
high-power rf system for the main linacs. Dashes indicate additions for
the beam-energy upgrade from 250 GeV to 500 GeV. }
\label{fig:one}}}
\end{figure}

When the SLAC linac was built, the accelerating gradient was 7 MV/m.
The original design included a future upgrade path in which the number
of klystrons would be quadrupled. The upgrades that were eventually
implemented involved replacing each of the initial 24-MW klystrons with
a single higher-power klystron (first with 35-MW, XK-5 tubes and, later
on, with 65-MW, 5045 tubes), and adding a SLED pulse compressor
downstream of each klystron. For present-day SLC operations, fully
upgraded with 240 SLEDed, 65-MW klystrons, the accelerating gradient
has been tripled, to 21 MV/m, and the maximum beam energy is 60 GeV
(for unloaded, on-crest operation).

The rf power system for the NLC's two high-gradient linacs that
accelerate the electron and positron beams separately from 10 GeV to
250 GeV in the initial design, and to 500 GeV or more after the
upgrade, operates at 11.424 GHz. This system includes all the hardware
through which energy flows, from the AC line to the accelerator
structures. Figure~\ref{fig:one} shows one module of the rf system
schematically.

Electrical energy is transformed at each stage shown in the diagram: 
the modulator converts AC power into high-voltage pulsed DC, the
klystron transforms pulsed DC into high peak power rf, the SLED-II
pulse-compression system increases the peak power by about a factor of
four (at the expense of a reduced pulse width), and the accelerator
sections convert rf power into beam power. Because of the high average
rf power required to drive the accelerator structures, it is important
that the highest possible efficiency be maintained for the processing
and transmission of energy at every stage of the rf system.

The primary technical choice for the rf system is the use of the
11.424-GHz frequency. This frequency, high in the X-band (8.2--12.4
GHz), is exactly four times the operating frequency of the existing
SLAC 60-GeV linac. The choice of such a high frequency, relative to
existing high-energy linacs, allows higher accelerating gradient,
shorter linac length, and lower AC power consumption for a given beam
energy. Considering the size, weight, cost, and availability of
standard microwave components, we have chosen a frequency in the X-band
for a design that is upgradable from an initial 250-GeV beam energy to
500 GeV or more.

\renewcommand{\arraystretch}{1.15}
\begin{table}[htbp]
\begin{center}
\caption[*]{NLC main linac rf system parameters.}
\label{tab:linacs}
\medskip
{\scriptsize \begin{tabular}{|l|ccc|}
\hline \hline
  & NLCTA & 500-GeV         & 1-TeV  \\
 &  Achieved  & Design Goal     & Upgrade        \\
\hline
\underline {General Parameters}  &  &  &  \\
\qquad Frequency (GHz)   &  11.4  & 11.4 & 11.4   \\
\qquad Accel. Gradient (MV/m), Unloaded/Loaded  &  67/$-$  & 50/35.3  & 85/63.5  \\
\qquad Overhead Factor,$^{[1]}$ F$_{OH}$  &  &  1.20  &  1.15 \\
\qquad Active Linac Length$^{[2]}$ (m) &    &  16,300  &  17,700 \\
\qquad Total Linac Length = 1.08 $\times$ Active Length (m)   
     &  &  17,600  &  19,100  \\
\qquad Number of Modulators        &  &2264        &4908     \\
\qquad Number of Klystrons         &  &  4528        &9816      \\
\qquad Peak Power per meter of Structure (MW/m)  &  & 50   &145       \\
\qquad Rf Pulse Length at Structure (ns)      & 150   &  240  &  240   \\
\qquad Repetition Rate (Hz)     &60  &  180  &  120         \\
\qquad Particles per Bunch ($10^{10}$)   &  &  0.75  &  1.10   \\
\qquad Number of Bunches per Pulse    &   & 90   &  90    \\
\qquad Peak Beam Current (A)    &  &  0.86        &1.26          \\
\qquad Total Average RF Power at Structure $^{[3]}$ (MW)  & 
            &  34.1    &71.3  \\
\underline {Klystron}  &  &     &          \\
\qquad Output Power (MW)       &  50, 75  &  50     & 75     \\
\qquad Pulse Length ($\mu$s)   &  2.0, 1.1  &  1.2    &0.96    \\
\qquad Microperveance $(\mu A/V^{3/2})$   &  1.2  &  0.6   & 0.75  \\
\qquad Electronic Efficiency$^{[4]}$ (\%)   &  48  &  57  &  60   \\
\qquad Beam Voltage (kV)       &  440  &  465  &  487    \\
\qquad Focusing               &  Electromagnet  &  PPM  &  PPM  \\
\qquad Cathode Loading (A/cm$^2$)  &  7.4  &  7.4  &  7.6   \\
\underline {Modulator} (Blumlein PFN, transformer ratio 7:1) &  & & \\
\qquad PFN Voltage (kV)   &48  &  68  &  71          \\
\qquad Pulse Rise Time (ns) &400  &  275  &  175   \\
\qquad Net Modulator Efficiency (\%)   &58  &  72   &  75    \\
\underline {RF Pulse Compression}  &  &  &  \\
\qquad System Type &  SLED-II  & SLED-II & BPC/DLDS \\
\qquad Compression Ratio     &  5--7  &  5  &  4      \\
\qquad Intrinsic Efficiency  (\%)   &  80--69  &  80  &  100  \\
\qquad Pulse Compression Efficiency (\%)    & 73--64  &  76.5  &   93      \\
\qquad Power Transmission Efficiency (\%)   &   90  &  94    &  94   \\
\qquad Net Pulse-Compression Efficiency (\%),     &    &   &   \\
\qquad \quad Including Power Transmission Loss  &  66--58  &  72 & 87.5 \\
\qquad Net Power Gain      &3.3--4.1  &  3.6    & 3.5      \\
\underline {Net RF System Parameters} &  &   &      \\
\qquad Total AC Power (MW), Excl. Aux.   &  &  116     &  181        \\
\qquad Rf System Efficiency (\%), Excl. Aux.   & 19 &  29.6  & 39.4      \\
\qquad Total Auxiliary Power$^{[5]}$ (MW)    &  & 5.4    & 12    \\
\qquad Total AC Power (MW), Incl. Aux.   &   &  121   &  193 \\
\qquad Rf System Efficiency (\%), Incl. Aux.   &  &  28.2  &37.0 \\
\qquad Average Beam Power (MW), Excl. Injected   &  &  9.3    & 18.6    \\
\qquad AC-to-Beam Efficiency (\%)   &  &  7.7  & 9.4       \\[2ex]
\hline\hline 
\end{tabular}}
\end{center}
{\scriptsize \baselineskip9pt
$^{[1]}$ Includes overhead for BNS damping, feedback, 
  and stations off for repair.\\
$^{[2]}$ Active length = F$_{OH}$
  (E$_0 - 20$  GeV)/(Loaded Gradient).\\
$^{[3]}$ Assumes 3\% of klystrons and modulators are off 
  (repair margin) or running off beam time (on standby).\\
$^{[4]}$ Given by simulated efficiency less 5 percentage 
  points for 500-GeV design; equal to
  simulated efficiency for 1-TeV design.\\
$^{[5]}$ Includes klystron cathode heater power,
   thyratron heater and reservoir power, and
    power for modulators on standby (0.5\%).\par}
\end{table}

\renewcommand{\arraystretch}{1.3}

The general parameters of the high-power rf system and its major
subsystems (klystrons, modulators, rf pulse compressors, and the
accelerator structure itself) are specified in Table~\ref{tab:linacs}.
The set of parameters has been optimized to provide high acceleration
gradient (35--64 MV/m) for trains of bunches with a moderate charge per
bunch (1.2--1.8~nC). This optimization keeps single-bunch wakefields
under control and reduces the beamstrahlung at the collision point to
tolerable levels. The upgrade to 500-GeV beam energy (1-TeV
center-of-mass energy) is accomplished by doubling the number of
modulators (as shown by dashed lines in Fig.~\ref{fig:one}), and by
replacing each 50-MW klystron with a pair of 75-MW klystrons. The total
active length of linac must also be increased from 16,300 m to 17,700 m.
The upgrade plan also includes improvements in the modulator and pulse
compression systems to increase the rf system efficiency.

The design of the high-power X-band rf system for the NLC is based on
specific experience gained from building X-band prototypes and
operating them at high power, and on an rf systems-integration
test---the Next Linear Collider Test Accelerator (NLCTA)---which is
currently under construction at SLAC. The goals of the NLCTA
project~\cite{SLAC93,rut93} are to integrate the technologies of X-band
accelerator structures and high-power rf systems, to demonstrate
multibunch beam-loading energy compensation and suppression of
higher-order beam-deflecting modes, to measure any transverse
components of the accelerating field, and to measure the growth of the
dark current generated by rf field emission in the accelerator.

The NLCTA's 160-kV electron gun has been commissioned. The 70-MeV,
X-band injector module and two of the three 130-MeV, X-band linac
modules of the NLCTA are expected to become operational in 1996, with
each module powered by a single 50-MW klystron. The first accelerator
physics experiments are planned for 1996--1997. The last linac rf
module (including the fifth and sixth 1.8-m-long X-band sections) is
planned to be installed in 1997. It is expected that an upgrade from
50-MW klystrons to 75-MW klystrons will occur gradually as the new
higher-power tubes become available through the klystron development
program.

\subsection{Klystrons} 

\begin{figure}[htb]
\leavevmode
\centerline{\epsfbox{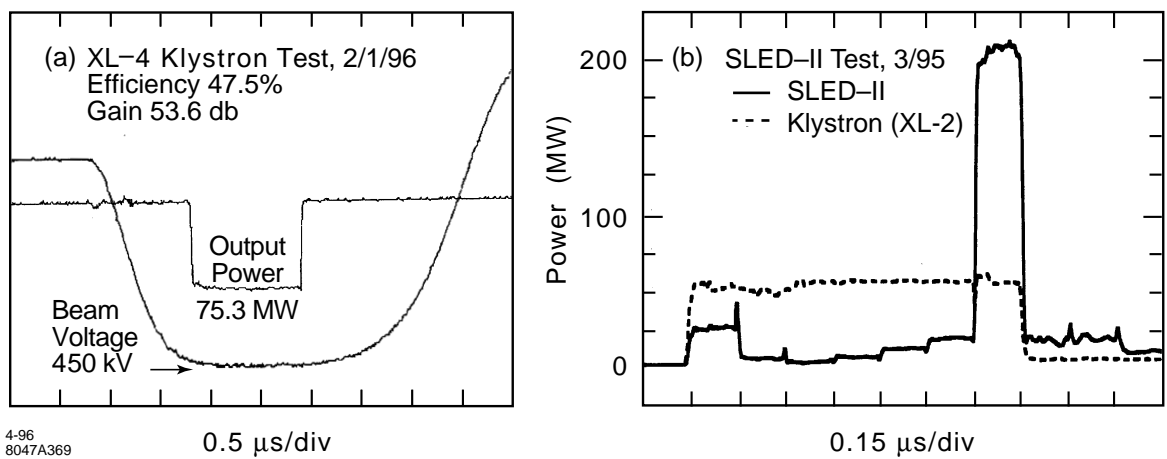}}
\centerline{\parbox{5in}{\caption[*]{High-power tests of (a) an X-band
klystron and (b) a SLED-II pulse compressor.}
\label{fig:two}}}
\end{figure}

Obtaining the X-band peak power for the NLC has required the
development of klystrons capable of delivering peak power significantly
greater than previously achieved by commercially available X-band
sources. Both the peak power and the pulse length required for the NLC
design and upgrade have already been achieved by solenoid-focused
X-band klystrons at SLAC. Four prototypes now produce 50-MW
pulses, over 1.5 microseconds long, with performance characteristics
that are correctly modeled by computer codes.  The most recent
prototype produces 75-MW pulses, one microsecond long, and exhibits
electronic efficiency of 48\% (Fig.~\ref{fig:two}). These klystrons
will be used to power the NLC Test Accelerator.

\begin{figure}[htbp]
\leavevmode
\centerline{\epsfbox{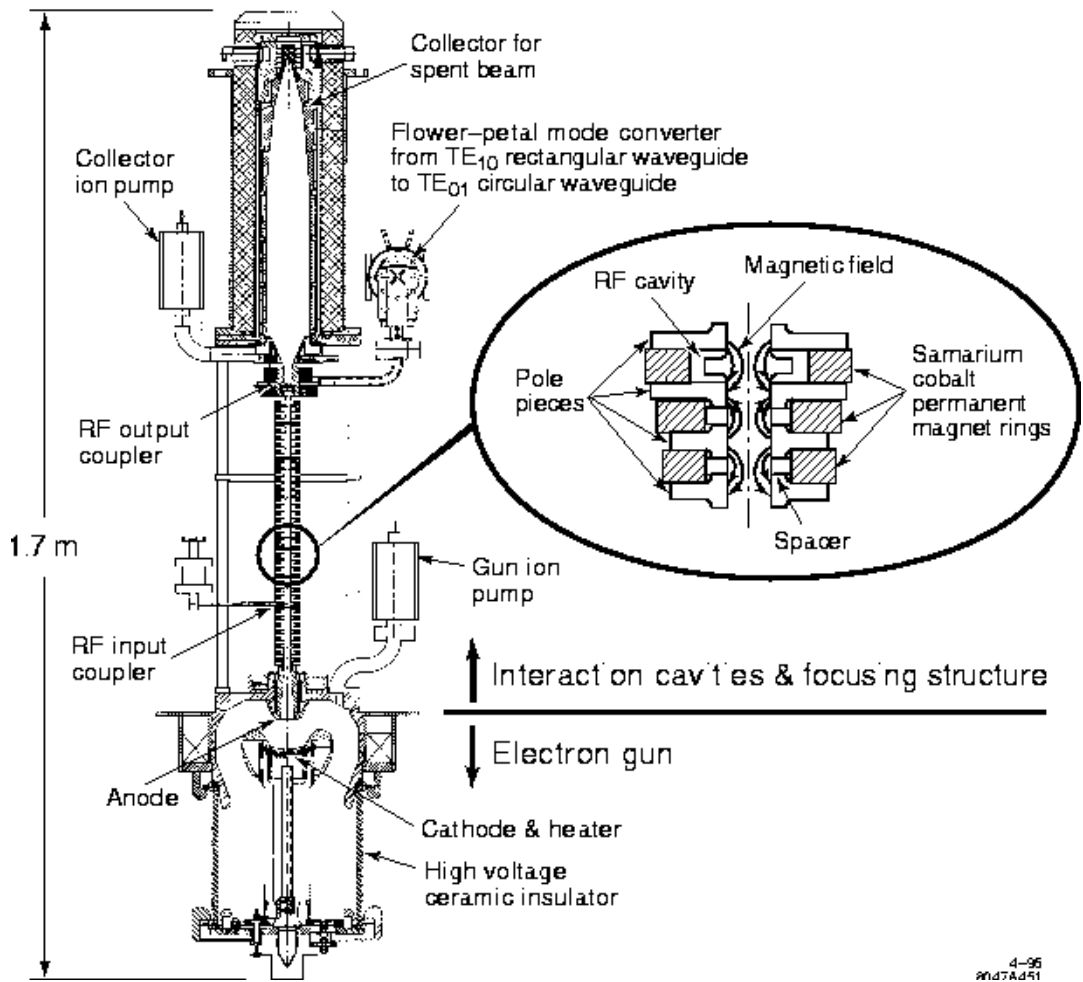}}
\centerline{\parbox{5in}{\caption[*]{Preliminary layout of the X5011
PPM klystron, with water jacket removed.  The insert shows one
representative buncher cavity, to illustrate the use of oversize
permanent magnets.}
\label{fig:three}}}
\end{figure}

The solenoid which focuses the electron beam in the prototype klystrons
has a weight of 750 kg and a power consumption of 20 kW.  Currently
nearing completion is the first prototype of a 50-MW klystron which is
focused instead by a periodic permanent magnet (PPM) array of
samarium-cobalt ring magnets weighing about 9 kg.  It is this klystron,
which operates at a higher voltage and lower beam current for
compatibility with PPM focusing, which is slated as the prototype for
the NLC.  Based on computer projections, the tube, designated X5011, is
expected to operate at about 57\%\ efficiency with 50 MW of peak output
power. This first, high-power PPM klystron, which is shown in
Fig.~\ref{fig:three}, was fabricated and tested as a beam tester (a
klystron without rf cavities) to validate the beam focusing. Beam
transmission in excess of 99.9\%\ was achieved without the magnet
trimming which is customarily required for traveling-wave tubes which
are PPM focused.  The klystron prototype with rf cavities has been
fabricated and is being readied for beam and rf tests.

\subsection{Modulators} 

Each high-power rf station consists of a pair of PPM-focused klystrons
(50 or 75 MW) and a pulsed-DC energy delivery system (modulator) that
is tightly integrated in design with the electron guns of the
klystrons. The pulsed-DC energy delivery system includes a
single-thyratron switch, a Blumlein pulse-forming network (PFN), a
high-efficiency power supply for charging the PFN's capacitance, and a
pulse transformer. Using a Blumlein PFN allows for a relatively low
transformer turns ratio (7:1), which yields a reasonably fast rise time
(0.3 $\mu$s), and hence, an improved efficiency.

The technology for standard line-type high power pulse modulators is
very mature.  The SLAC linac has been in operation for
about 30 years using about 240 such modulators to drive the high-power
klystrons. The efficiency of these modulators for converting wall-plug
power into klystron beam power in the flat-top portion of the high
voltage output pulse is about 60\%.  Similar modulators have been
constructed for the NLC Test Accelerator, and are currently under test.
Parameter values for these modulators are listed under the ``NLCTA
Achieved'' column in Table \ref{tab:linacs}.  The net efficiency for
the NLCTA modulator is 58\%.

A key consideration for a modulator in a linear collider application is
efficiency.  A 1\%\ relative decrease in modulator efficiency would
result in an increase by between one and two megawatts in the AC line
power required for the NLC rf system.  It is clear that efficiencies
for the SLC and NLCTA modulators are too low for consideration for the
NLC.

An important component in determining the modulator efficiency is the
pulse transformer.  The pulse forming network (PFN) delivers a
more-or-less rectangular pulse to the primary of the pulse transformer
when the thyratron switch is triggered.  The voltage pulse delivered to
the klystron, however, contains a considerable amount of wasted energy
in the rise and fall-time portions of transformer output pulse.  In
practical pulse transformer designs, it is observed that the rise time
(and fall time) is determined in large measure by the transformer turns
ratio:  a high turns ratio implies a longer rise time, and vice versa.
For the NLCTA modulator, a pulse transformer with a turns ratio of 21:1
is required to step up a 48-kV PFN voltage to the 500 kV required by
the NLC klystron.  This ratio can be reduced to about 14:1 by
increasing the PFN charging voltage from 48 to 72 kV.  This, of course,
requires a thyratron with a higher voltage hold-off capability.  A
further reduction in transformer ratio to 7:1 can be achieved by using
a voltage-doubling, or Blumlein, PFN configuration.  Simulations show
that a pulse-shape efficiency of about 80\%\ can be achieved for this
turns ratio.  Combining this with an estimated efficiency of 93\%\ for
the DC power supply which charges the PFN, and assuming an additional
3\%\ is lost in the voltage drop across the thyratron and other series
resistances, the net efficiency for such a modulator design is estimated
to be about 72\%.

The R\&D task for the NLC modulator group is to show that such an
efficiency can be obtained in practice in a prototype modulator.
Several possibilities are under consideration for the Blumlein PFN
using both lumped capacitors and smooth transmission lines.
Appropriate R\&D is underway to design and construct prototype PFN
configurations.  Techniques for improving pulse transformer efficiency
are also under investigation.

\subsection{RF Pulse Compression and Power Transmission}
\label{sec:pulse}

The 1.2-$\mu$s-long X-band klystron pulses will be compressed by a
factor of five---to 0.24$\mu$s---using the SLED-II rf pulse compression
technique~\cite{wil90}. SLED-II is a modification of the SLED~\cite{far74}
compressor is used to increase the rf power into the 60-GeV SLAC linac by a
factor of about 3.

A SLED-II pulse compressor works by storing microwave energy in a pair
of high-$Q$ resonant delay lines for most of the duration of the
klystron pulse. The round-trip transit time of the rf in the delay
lines determines the duration of the compressed pulse. Developmental
SLED-II systems have been tested at SLAC with high-power X-band
klystrons to validate the design and its components~\cite{nan93, vli93,
wan94}. The peak power and efficiency needed for the NLC Test
Accelerator have been demonstrated in a prototype system
(Fig.~\ref{fig:two}). Three SLED-II systems are being manufactured for
the NLCTA. Further validation of the SLED-II design at the higher power
levels and efficiencies needed for the NLC will be performed in
upgrades to the NLCTA. The rf pulse compression systems in the NLC and
NLCTA are designed to accommodate future upgrades in which the peak rf
power is nearly tripled, by replacing each 50-MW klystron with a pair
of 75-MW klystrons.

\begin{figure}[htb]
\leavevmode
\centerline{\epsfbox{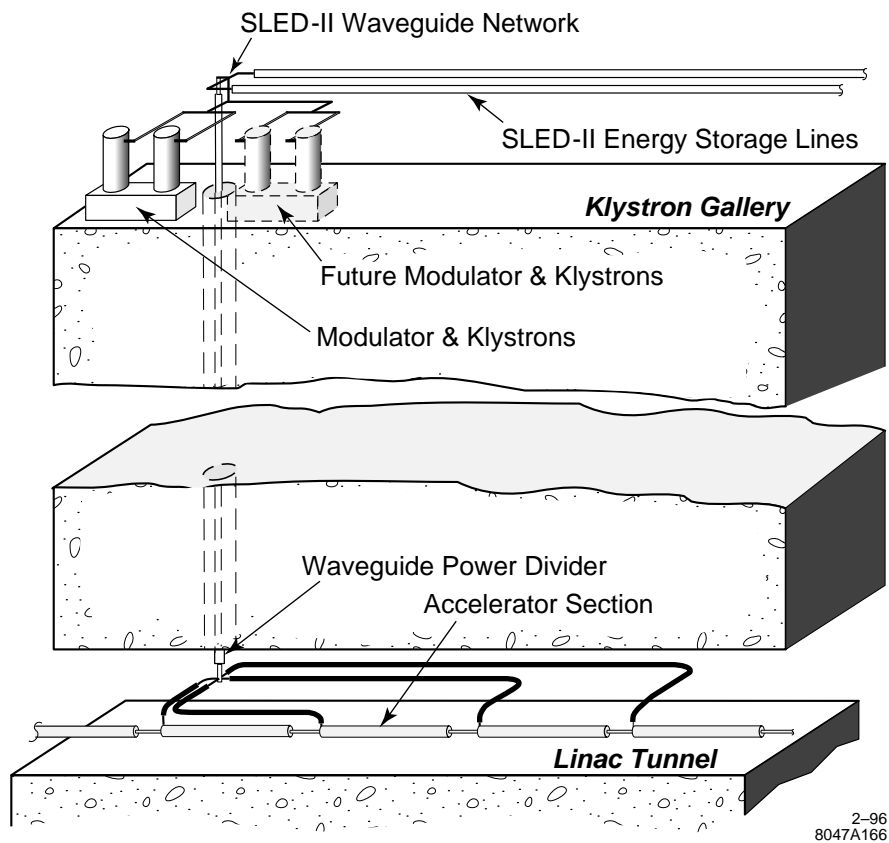}}
\centerline{\parbox{5in}{\caption[*]{Physical layout of the waveguide
network for rf pulse compression and power transmission.}
\label{fig:four}}}
\end{figure} 

Figure~\ref{fig:four} illustrates the physical layout of the waveguide
network for rf pulse compression and power transmission to four
1.8-m-long linac sections in a two-tunnel (SLAC-type) configuration.
The waveguide layout is sufficiently flexible to absorb the absolute
motion of the accelerator sections due to their mechanical movers
($\pm$ 1 mm) and the relative motion between the klystron gallery and
the accelerator tunnel due to settling ($\pm$ 10 mm). The SLED-II
waveguide networks are located in the low-radiation environment of the
klystron gallery so that the rf pulse-compression system may be
modified, station by station, if necessary for upgrade to higher beam
energies, simultaneously with colliding beam operations.

Since the round-trip delay time in the 4.75-in-diameter waveguide (at
the group velocity, 0.964$c$) must equal the rf pulse duration (0.24
$\mu$s), the physical length of the SLED-II delay lines is 35 m. The
physical layout can accommodate this delay-line length by  spatially
overlapping the delay lines from adjacent rf stations. Since the total
length of linac fed by each rf station is approximately eight meters,
each pair of delay lines will overlap parts of five other pairs of
delay lines.

\subsection{Accelerator Structure}
\label{sec:accel}

The design of the X-band accelerator structures for the NLC is based on
theoretical and experimental experience gained by modeling and
building accelerator structures for the NLCTA and operating them at
high gradients.

One of the main challenges is to suppress the deflecting modes that
will otherwise cause severe multibunch emittance growth in the NLC
linacs. Suppression of the transverse wakefield will be achieved
through a combination of precision alignment and by detuning and
damping higher-order modes. Another challenge in the design of the
accelerator structures for the NLC is suppressing field emission at the
high-surface field gradients encountered in these structures. This
suppression, so far, has been achieved through machining, processing,
and handling techniques that minimize surface roughness and eliminate
contamination of the high-gradient surfaces. Other, additional methods
may be adopted later.

The rf accelerator structure is designed to be very nearly a
constant-gradient traveling-wave structure. The design of the structure
has been optimized to suppress the wakefield seen by trailing bunches.
This has been accomplished by tailoring the cell-to-cell frequency
distribution of the dominant deflecting mode to yield an initial
Gaussian-like decay of the wakefield amplitude. On a longer time scale,
the higher-order beam-induced modes of the structure will be damped by
vacuum manifolds to which each cell of the structure is coupled. This
structure is designated by the acronym DDS (damped detuned structure).
The damping manifolds run parallel to the beam channel and are
terminated into matched loads. (The slots that couple the cells to the
manifolds are cut off to the fundamental accelerating mode.) This
damping scheme will reduce the typical quality ($Q$) factors of the
deflecting modes to about 1000. The first prototype 1.8-m-long X-band
accelerator section, which was detuned but not damped, was high-power
tested up to a gradient of 67 MV/m. The effect of the detuning in that
first prototype section was demonstrated experimentally by using
positron and electron bunches from the SLC damping rings as probe and
witness beams, respectively. Another prototype 1.8-m section that is
both damped and detuned is being manufactured and will be used for a
similar test before it is installed in the NLCTA.

\begin{figure}[htb]
\leavevmode
\centerline{\epsfbox{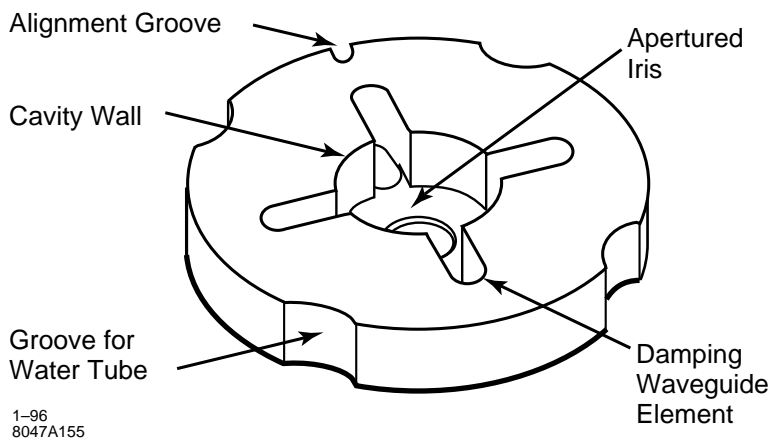}}
{\caption[*]{Basic cell design.}
\label{fig:five}}
\end{figure}

The basic features of the cell are shown in Fig.~\ref{fig:five}. The
central portion is the conventional ``cup'' consisting of a cylindrical
cavity wall and an apertured iris which, when stacked in a row with
other cells, forms the disk-loaded waveguide accelerator structure. The
diameter of the cavity wall, the thickness of the iris, and the
diameter of its aperture vary progressively from cell to cell to
``detune'' the beam-deflecting dipole modes and suppress cumulative
build-up of wakefields, while maintaining the quasi-constant-gradient
characteristics of the fundamental accelerating mode. The central
cavity is slot-coupled to four outer rectangular holes. When the cells
are stacked, the holes form four waveguides which run parallel to the
axis of the structure. When terminated in matched loads, the four
waveguides become ``damping manifolds.'' The slots (which are cut off
for the fundamental accelerating mode) couple power from beam-excited
dipole modes into these damping manifolds, lowering the $Q$ and
suppressing long-range wakefield build-up. Microwave signals from the
damping manifolds can be used to monitor the alignment of the structure
with respect to the beam.

For a single detuned accelerator section, the fractional spacing for
the fundamental dipole-mode frequencies (near the center of the
gaussian distribution of frequencies) is about $3 \times 10^{-4}$.
Machining precisions for conventional machining and diamond-point
machining (obtained at KEK), and alignment tolerances of stacks of
cells, are given in Table~\ref{tab:cell}. Since the cell radius is
about a centimeter, diamond-point machining should produce a random
frequency error, for the fundamental dipole modes,  somewhat less than
the spacing between the frequencies. The systematic error should be
significantly less than this.

Misalignment tolerances, on scales ranging from a few cells within a
structure to several structures, were discussed in the preceding
section. It was found that the tightest tolerances occurred for the
alignment of groups of about 20--40 cells. Thus, great care must be
taken in brazing together the subsections of complete structures after
their initial assembly from individual cups.

The goal is an accelerator section 205 cells long (plus the input
coupler) in which the axes of the individual cavities lie on a straight
line to within 15 $\mu$m (rms). To avoid a cumulative tolerance
build-up which exceeds this goal, many other cell dimensions have to be
held to tolerances in the 2 to 3 $\mu$m range. This applies
particularly to the diameter and coaxiality of the outer surface of the
cell, since this has to be the reference surface used in stacking the
cells prior to bonding. It also applies to the perpendicularity,
flatness and parallelism of the cell faces. Errors here will cause
``bookshelving'' of the cells in the cell stack \cite{see85}. Equally
important is the coaxiality of the iris aperture with respect to the
cavity outer wall. These tight tolerances on cavity dimensions
(including the thickness of the iris and the radius of its edge) are
also necessary to achieve the design fundamental and dipole-mode
frequency characteristics, because provision for tuning after assembly
has been eliminated. Flatness of cell faces is also essential to
achieve diffusion bonding (discussed below) over relatively large
surface areas with a very low expected failure rate.

\bigskip
\begin{table}[ht]
\centering
\caption{Cell machining and assembly tolerances achieved (in microns).}
\label{tab:cell}
\bigskip
\begin{tabular}{|l|ccc|}
\hline\hline
   &  Detuned      & Detuned         & Detuned 1.8-m     \\
   &  1.8-m        & 0.9-m           & Section \#2  \\  
   &  Section \#1  & Sections \#1\&2 & and DDS \#1 \\
\hline
Machining Technique  &  Conventional  &  Conventional  &  Diamond-point  \\  
Alignment Technique  &  Nesting  &  Vee-Block  &  Vee-Block  \\
Diameters  &  $\pm$7  &   $\pm$7  &   $\pm$2  \\   
Concentricity  &  10  &   10      &     1     \\
Thickness  &  $\pm$7  &   $\pm$7  &   $\pm$2  \\   
Parallelism    &  10  &   10      &     0.5     \\
Flatness       &  10  &   10      &     0.5     \\
Surface Finish  &  0.4  & 0.4     &   0.05    \\
\multicolumn{4}{|l|}
{Cell-to-cell Alignment of  Outer Cylindrical  Surfaces:}  \\
\qquad (a) Expected    &  22  &   10  &  3     \\
\multicolumn{4}{|l|}
{\qquad (b) Measured after diffusion bonding or brazing}   \\
    &  10  &   6  &  4  \\
\hline\hline
\end{tabular}
\end{table}
\bigskip

Single-crystal diamond-point machining is to be used on all surfaces of
each cavity, the cell faces and outer periphery. The hardness, high
thermal conductivity, and low thermal expansion of diamond result in a
superior tool which, when used in a vibration-free lathe, yield
mirror-like surfaces on copper which have a roughness of 0.1 $\mu$m or
less. This finish is necessary to obtain good diffusion bonding, high
$Q$ factors, low dark current and high power-handling capability. The
sharpness, mechanical stability, and minimal wear of the tool allow the
dimensional tolerances to be met.

Table~\ref{tab:cell} illustrates what tolerances on dimensions and
alignment have been achieved to date at SLAC, and what can reasonably
be expected with the best technology available today. It can be seen
that the most significant advances have been made by resorting to
diamond-point machining and by using precision granite vee-blocks to
align the stacks of cells prior to brazing or diffusion bonding. Cells
designed to nest into each other have to fit loosely enough to permit
assembly without galling, and this can result in unacceptable
misalignment. Our present experience is that stacks of 38 cells can
have a bow of a few tens of microns after diffusion bonding. This can
be reduced to a few microns by setting the stack horizontally in
cradles on a granite block and applying bending forces while monitoring
movement with a precision coordinate measuring machine.

\begin{figure}[htb]
\leavevmode
\centerline{\epsfbox{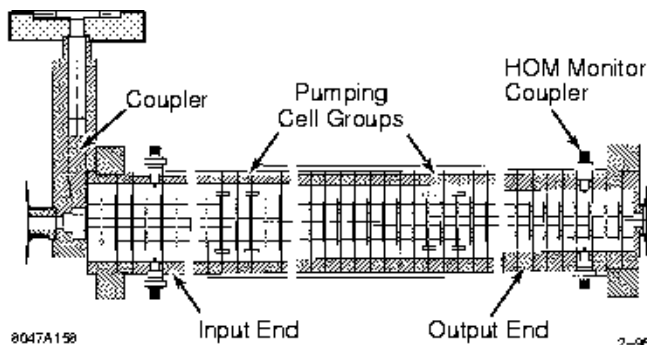}}
\centerline{\parbox{5in}{\caption[*]{Longitudinal cross section of a
1.8-m long X-band accelerator section.}
\label{fig:six}}}
\end{figure}

In addition to direct pumping along the beam-interaction region, the
structure is pumped in parallel by the four damping waveguides. Two
groups of four special pumping cells will be inserted into the
structure, at points one-fourth of the section length from each end. A
longitudinal cross section of the accelerator section is shown in
Fig.~\ref{fig:six}. Each pumping cell group is connected to two pumps
via short transverse manifolds.

Figure~\ref{fig:six} shows no output coupler. The intention is to
provide internal matched terminations for both the fundamental
accelerating mode and the wakefield modes by depositing lossy coatings
on the walls of the cavities and the damping manifolds in the last four
cells. A coaxial loop loosely coupled to the last cavity will be used
to monitor the fundamental-mode power. This signal will pass through a
coaxial rf window in the cell wall, and will be used by the klystron
phasing system. However, if satisfactory internal terminations cannot
be developed, external loads and couplers will have to be used.

\subsubsection{Mass Production of Accelerator Cells}
\label{subsub:quantity}

The two main linacs will contain about 1.8 million cells, which have to
be produced in about three years. Assuming that a specialized plant set
up to fabricate the cells operates 50 weeks per year at two shifts per
day, then, at 80\%\ efficiency, a production rate of about two cells
per minute is required. A close estimate of the time taken to machine a
single cell cannot be made without reference to specific machines and
machining techniques.  (See, for example, the study reports prepared for
the CERN CLIC cells~\cite{read,opti}.)  However, very rough estimates may
be drawn from those reports, noting that the surface area of the X-band
DDS cell is about three times that of the CLIC cell. All the features
of the DDS cell shown in Fig.~\ref{fig:five} could be machined to
conventional tolerances on the end of bar stock (leaving excess
material on the critical surfaces to be diamond-point machined in
subsequent operations), using a multipurpose CNC turning and milling
machine. The cell could thus be cut from the bar using a parting-off
tool, stress-relieved and delivered to a diamond-point lathe for
finishing in two steps. In the first step, the cell could be gripped on
the outside by a precision chuck while the back face is diamond turned.
In the second step, the back face could be  held in a vacuum chuck
while the front face and cavity surfaces are diamond finished. Based on
estimated operation times, the plant will need a minimum of 22
conventional CNC turning and milling machines and 8 diamond-point
lathes, with appropriate numbers of all supporting equipment (automatic
handling and transfer machines, stress-relieving ovens, cleaning, QC,
and packing stations) to maintain the required throughput.
 
\subsubsection{High-power Tests and Dark Current Studies}
\label{sub:dark}
 
The theoretical and experimental studies on the behavior of copper
accelerator structures under extremely high-rf fields have been carried
out at SLAC for several years~\cite{loe88a, wan94a}. We have studied in
considerable detail the problems of rf breakdown and dark current
generated by field emission at high gradient.  The dark current may
absorb rf energy, get captured, and produce undesirable steering
effects and detrimental X-ray radiation. Many experiments have been
done to measure the amplitude and energy spectrum of the dark current
and to study phenomena related to rf breakdown such as outgassing,
radiation, heating, etc. We have concluded that the dark current can be
minimized by improving surface finish and cleanliness, and by rf
processing.
 
High-power tests with an electron beam have demonstrated that the 1.8-m
section is properly tuned to 11.424 GHz and can accelerate a beam at a
gradient of more than 67 MV/m. Dark current from this structure was
measured in a Faraday cup as a function of average accelerating field 
(Fig.~\ref{fig:proc}), and found to be negligible for an accelerating
gradient of 50 MV/m was found to be negligible. At 85 MV/m. The energy
spectrum of the dark current was measured and found to be sufficiently
low that the quadrupoles will overfocus and sweep the field-emission
current away from the primary, high-energy beam.

Experiments performed with S-band accelerators at KEK \cite{tak91,
mat91} have shown that the dark current produced by field emission from
the accelerator disks is reduced by an order of magnitude when
stringent precautions are taken to exclude dust during fabrication,
assembly, processing and testing. These experiments indicate that the
NLC accelerator cells should be cleaned in ultra-pure, dust-free
chemical solutions and rinses, assembled and bonded under clean-room
conditions, and perhaps given a final rinse in ultra-pure water before
vacuum bakeout and installation. We expect application of these
procedures to our X-band structures to yield equally significant
improvements.

\subsection{RF Drive and Phasing Systems}
\label{sec:control}

To achieve a highly monoenergetic multibunch beam pulse for the final
focus, the beam loading induced by the bunch train must be compensated.
The initial transient can be eliminated by pre-loading the sections by
shaping the input rf pulse using an approximately linear rise of the
field amplitude for one filling time of the structure. In this way, the
first electron bunch will see a filled rf structure that appears to be
in the steady state. This requires phase-agile control of the rf before
it is amplified by the klystron.

The design of the rf drive and phasing systems for the NLC will be
based heavily on experience gained from existing systems developed for
the SLAC linac and its 60-GeV upgrade for the SLC, with further
extensions based on the design of the NLCTA. The challenges to be met
for the NLC arise from its greater length, number of components that
must be controlled, required tolerances, maintainability, and
reliability. The requirements imposed on the rf drive and phasing
systems are summarized below:

\begin{itemize}
\item 
The drive system must provide stable, adjustable, and reliable input
power to 4528 klystrons (approximately 1-kW pulsed peak power per
tube). The layout of the system must be such that individual subsystem
failures do not cause the beam energy to decrease to less than 85\%\ of
its operating level. This requirement is dictated by potential
collimator and other damage caused by off-energy beams.
\item 
The amplitude of the drive power must be adjustable so as to
individually saturate the high-power klystrons.
\item 
The phase of the drive power must be adjustable in several ways:
    \begin{itemize}
    \item 
    Slowly, to take care of phase drifts and drifts caused by length
changes in equipment and terrain. Included here are the couplers, coax
lines, intermediate amplifiers, high-power klystrons, SLED-II,
waveguide components, and the changes in time at which the bunch trains
are injected into the X-band linacs by the S-band bunch compressors.
    
    \item 
    Quickly (within each 1.2-$\mu$s klystron rf pulse), to create the
proper resultant amplitude and phase profiles necessary to produce the
desired fields in the accelerator structures and to place the bunches
at the desired phase positions with respect to the X-band waves. These
positions are dependent on bunch number and charge, which can change
from pulse to pulse, and are dictated by BNS phasing, single bunch and
multibunch beam loading, and other considerations.
    
    \item 
     The phasing system design must also include initial adjustments
(upon installation) of the waveguide runs which feed the four
accelerator sections driven by one pair of klystrons in a station. 
    \end{itemize}
\end{itemize}

\begin{figure}[htbp]
\leavevmode
\centerline{\epsfbox{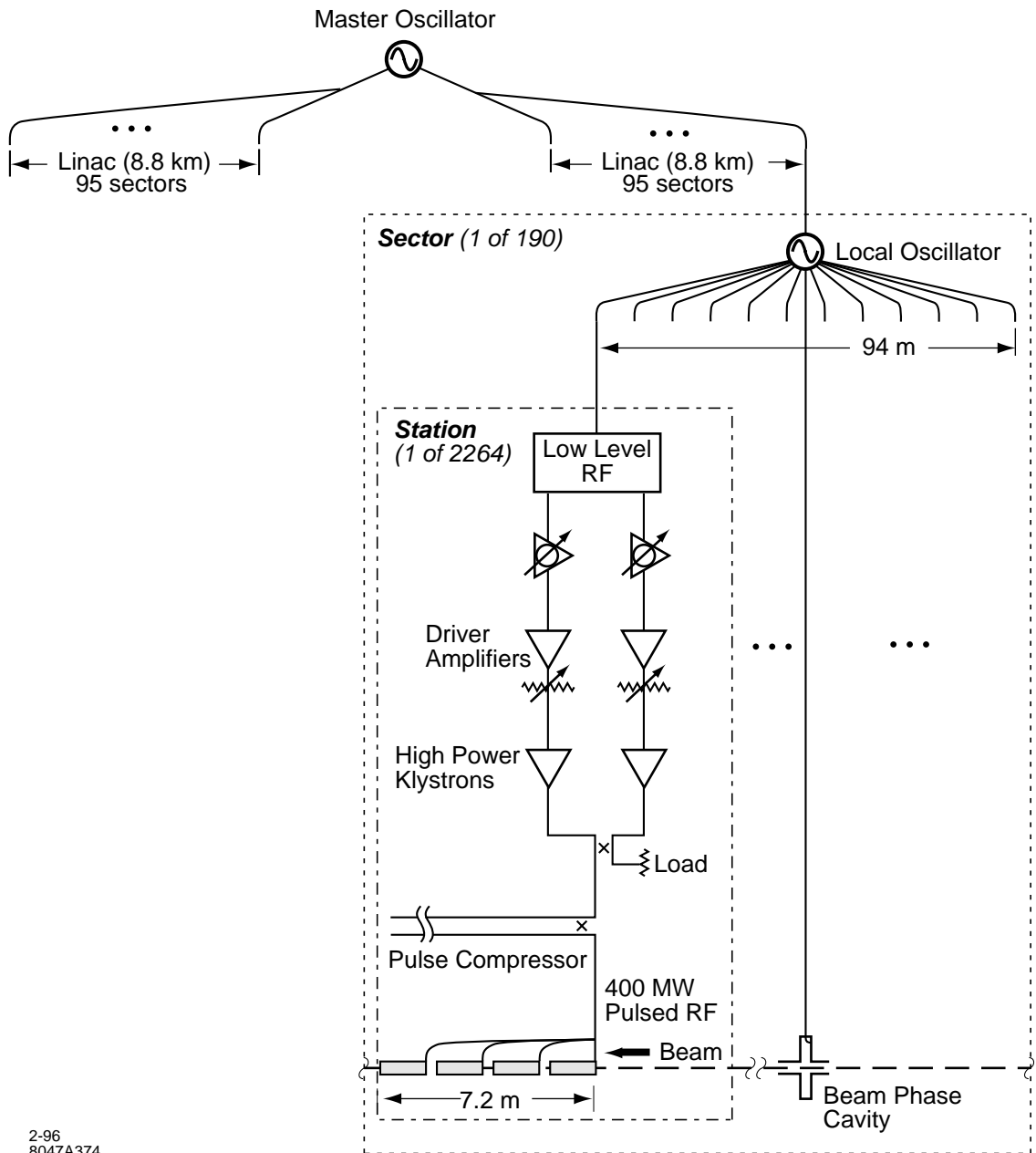}}
\centerline{\parbox{5in}{\caption[*]{Schematic of rf drive system. The
master oscillator provides phase-stable rf to local oscillators in each
sector over a fiber distribution system. Local sources in each sector
provide 11.424 GHz to each rf station in the sector using coaxial
lines. A beam-phase cavity uses the beam to determine the rf phase for
optimum operation. Power levels are low (several milliwatts) until the
final drive for the high-power klystrons.}
\label{fig:seven}}}
\end{figure}

The phases of the high-power rf fields which act on bunch trains
should be adjusted with an accuracy of about 0.5$^\circ$ at 11.424 GHz.
A schematic layout of the rf drive and phasing system for the NLC is
shown in Fig.~\ref{fig:seven}.
    
There are various schemes for providing the approximately 1 kW of
pulsed rf needed to drive each high-power klystron. It is useful to
have individual drivers because the various phase manipulations
involved in the operation of the rf compressors are best done at the
milliwatt level. Power-combining of the two klystrons on the single
modulator is done by phase modulation of the two klystron inputs by
equal amounts in opposite directions.  This has the effect of sending
the desired vectorial sum into the rf pulse compressor and dumping the
vectorial difference into a load.

There exists extensive Traveling Wave Tube (TWT) technology at this
frequency and power level as a product of military applications. The
available TWTs are reliable, but fairly complex to manufacture and
operate. They have already been extensively cost-optimized, and are
still very expensive.

Alternatively, a small X-band driver klystron of fairly narrow bandwidth and
modest gain could be designed.  In construction, it would be
relatively simple compared to a TWT. It is possible that after
the initial engineering models are tested, a modest program of redesign
for automatic manufacturing could significantly reduce the unit cost.

\subsection{Upgrade to 1 TeV}
\label{sub:upgrade}

An essential goal of the NLC is that it be possible to increase the
center-of-mass energy of the collider to a TeV or greater in an
unobstrusive and adiabatic fashion with improvements that can be
expected in the rf system of the main accelerators. The physical layout
that we have chosen for the main linacs and their power sources
(Fig.~\ref{fig:four}) provides access that allows upgrade of the rf system even
as the collider is operating; for example, by taking one or a few
stations off-line at a time. Our choice of X-band rf technology has
been made to assure that accelerating gradients that approach 100 MV/m
can ultimately be used without electrical breakdown or generation of
unacceptable levels of dark current, and there are a number of possible
advances in X-band technology that would lead to the TeV energy.  We
have based our design on what we believe to be improvements that are
near-term and can reasonably be expected to occur in klystron and pulse
compression technologies.

Tests with a prototype X-band klystron have already proven that output
powers of 75 MW can be achieved, and a likely path to the TeV energy
goal is taken by replacing and doubling the number of klystrons used in
the main acelerator (dashed lines in Fig.~3.16). Although the klystron
power rating necessary for the full 1 TeV has been achieved, the power
conversion efficiency of the present tubes (approximately 48\%) is not
sufficiently high to maintain the total power consumed by the linac to
an acceptable level (taken to be about 200 MW).  Klystrons of the
generation now being developed with PPM focusing are expected,
according to computer modeling, to achieve conversion efficiencies of
nearly 60\%.

There are additional improvements that we anticipate will occur in
X-band technologies that will lead to still higher overall power
efficiencies. The finite filling time of the energy storage technique
used in the SLED-II pulse compressor wastes power and results in an
intrinsic 20\% loss that could be avoided with the development of new
microwave switching techniques.  Initial experimental investigations of
an optically triggered silicon switch show promise, and effort is
continuing to develop a usable device based on this idea
\cite{tan95a,tan95b}.   First tests of an alternate compression scheme,
binary pulse compression (BPC), have also been carried out, and have
proven that this technique, though more expensive than SLED-II, works
well with intrinsic efficiency of 100\%.
\cite{far86,lav91}. A nice feature of the BPC scheme is that it reuses
nearly all of the hardware of the SLED-II systems, so that if
conversion to BPC were necessary, then a station-by-station
modification is mechanically reasonable.

Final tests of any of these upgrade paths remain to be done.  It will
be important to verify the power ratings and efficiencies of not only
the power generating components, but also those of the entire system.
This will be one of the primary goals of the experimental program of
the NLCTA.

To obtain the full 1 TeV energy and to maintain the linac power
consumption below 200 MW, the klystron upgrade to 75 MW and
elimination of the intrinsic power loss of the SLED-II system must be
accomplished. From our experience to date, we believe it is reasonable
to expect the net efficiency of the rf system to be increased from the
initial design value of about 30\% to the 40\% required for the
upgrade.  The total length of the main linacs must also be increased slightly,
from 16.3 to 17.7 km, but even if this additional length is not
provided, the upgraded center-of-mass energy will still reach 925 GeV. 
This set of parameters is summarized in Table~\ref{tab:linacs}. Some of
these potential improvements may, in fact, be ready in time to
be included in the initial 500-GeV machine.  This would result in
additional performance margin or cost reductions.  The 500-GeV
parameters listed in Table~\ref{tab:linacs} are conservatively based on
experience with, and measurements on, prototypes which exist at the
present time.

\subsection{Outlook}
\label{sub:outlook}

The design of the rf system for the main linacs of the NLC is supported
by existing and planned developmental prototypes, and by the NLCTA. Key
NLC parameters such as the klystron power, acceleration gradient, and
pulse-compression power gain have been exceeded in prototype systems.
The next steps in the development program are completion of the NLCTA,
the first damped and detuned structure, and the first PPM klystron
prototype. The design of the NLC high-power rf system is mature and is
now progressing toward detailed engineering considerations. Because of
the magnitude of the project, special emphasis is being placed on
designing for manufacturability and for overall system reliability.

\clearpage

\section{Beam Delivery}
\label{chap:beamdel}

A schematic of the beam delivery and removal systems is shown in Fig.
\ref{fig:bdsys}. These systems begin at the end of the linac and
terminate at a post-IP beam dump. The total linac-to-IP length is 5.2
km.  They include a post-linac collimation system, an IP switch and big
bend, a pre-final-focus diagnostic and skew-correction section, a
final-focus system, and a post-IP beam line. The design of all of
these systems has been strongly influenced by experience with similar
systems at the SLC.

\begin{figure}[htb]
\leavevmode
\centerline{\epsfbox{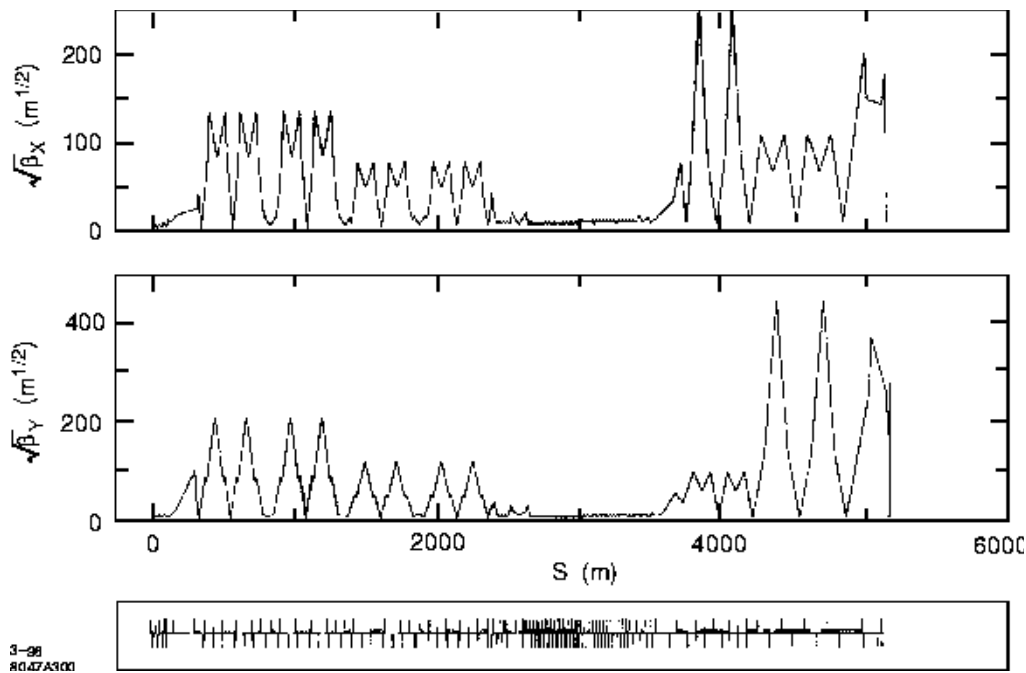}}
\centerline{\parbox{5in}
{\caption{The $\beta$ functions from the linac 
to the IP for the 1-TeV cms beam line.}
\label{fig:11b}}}
\end{figure}

Figure \ref{fig:11b} shows the lattice $\beta$ functions from the
linac to the IP for the 1-TeV cms beam line.  The first 100 meters
contain a post linac diagnostic chicane, which is then followed by a
2.5-km collimation system.  The maximum $\beta$-function points in the
collimation region correspond, for the most part, to the location of
horizontal collimators and chromatic correction sextupoles. The
horizontal $\beta$-function peak at 2.6 km marks the location of the IP
switch.  After that, up to the 3.0-km marker, is a small
$\beta$-function region that contains the big (10 mrad) bend, and
following this, up to the 3.4-km marker, is a small $\beta$-function
region containing a pre-final-focus diagnostic and skew-correction
region.  In this region beam sizes in all phases and planes can be
measured and the presence of coupling detected and corrected. The
region from 3.4 km to 3.8 km contains the $\beta$ match into the final
focus.   Following the $\beta$ match, the first two peaks are the
positions of the horizontal chromaticity compensation sextupoles.  The
last peak, in each plane, at the end of the beam line, is located at
the position of the final-doublet elements.

We have studied beam delivery systems for center-of-mass energies from
350 GeV to 1.5 TeV, for a broad range of assumptions on beam and IP
parameters, and have shown that it is possible to meet system
specifications over this entire range.  Within this energy span, the
final-focus system design we have chosen requires two minor upgrades
consisting of a transverse displacement of magnets by 20 cm or less and
a change of the final doublet.  The tunnel length allotted to
collimation is adequate to collimate energy and the horizontal and
vertical planes at both phases only one time at 1.5 TeV cms energy.
Since the centroid orbit of the collimation system differs only
slightly from a straight line, it would be possible to allocate length
at the end of the linac tunnel for a second IP phase collimation at
this energy.  It is sufficient to collimate the final doublet (FD)
phase only one time.

\subsection{Collimation System}

The design of the post-linac collimation is strongly influenced by
assumptions on incoming beam conditions.  The number of particles
needing to be collimated is difficult to predict precisely, because it
depends on how well upstream systems have been tuned; we have relied
on SLC experience for this estimate.  Downstream, the number of
particles that can be collimated in the final-focus system can be
determined by edge-scattering and muon transport studies. The rough
guideline that evolves from these considerations is that for each bunch
train there may well be a few times 10$^{10}$ particles per bunch train
in the tails at the end of the linac. This number needs to be reduced
to a few times 10$^6$ upon entry into the final-focus system.

Two types of collimation systems have been proposed: linear and
nonlinear.  In the former, beam enlargement at the collimators is
achieved by traditional linear optics methods; in the latter strong
sextupoles are used to blow up the beam.   Since the sextupoles are
exceedingly strong and system lengths are not reduced in the specific
system proposals we have studied, we have opted to look in depth at a
linear collimation scheme.  It is not precluded that a nonlinear (or
combination linear and nonlinear) system could be found that would be
operationally superior and have a lower total cost.  Our primary
objective in the ZDR was to show that at least one collimation system
exists that fulfills all functional requirements.

\begin{figure}[htbp]
\leavevmode
\centerline{\epsfbox{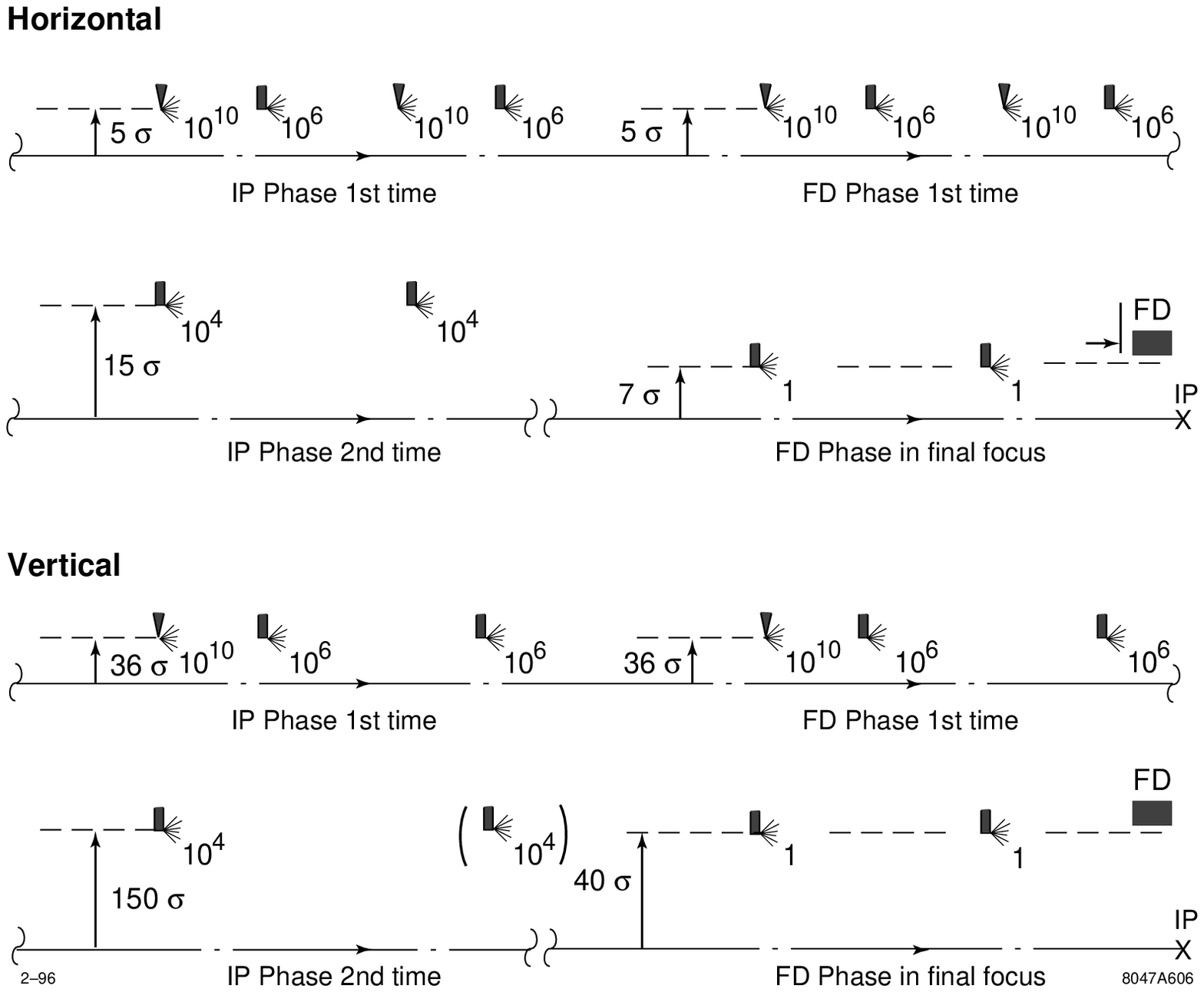}}
\centerline{\parbox{5in}{\caption[*]{A schematic of the collimation
system spoilers and absorbers. The apertures are shown as well as the
estimated number of tail particles rescattered into the beam.}
\label{fig:11c}}}
\end{figure}

Since a small perturbation in upstream conditions could cause a
complete bunch train of 10$^{12}$ particles to be incident on the
collimators of the collimation system, it is necessary in both the
linear and nonlinear systems to rely on a primary collimator that is a
spoiler, followed by a secondary collimator which is the absorber.  The
function of the spoilers is to increase the angular divergence of the
beam, so that when the beam arrives at the absorber it has a much
larger size (2.2 mm radius).  The spoilers must be thermally rugged and
very thin (1/4 radiation length).  The best material we have found is a
titanium alloy plated with titanium nitride for better electrical
conductivity.   The absorber, on the other hand, must be able to
routinely absorb and remove the full energy in the beam tail.  For the
1 TeV cms parameters, 1\%\ of the time-averaged beam power is 84 kW.  
The preferred material for absorbers is copper. Figure \ref{fig:11c} is a
schematic of all spoilers and collimators in the beam delivery system. 
This figure shows the apertures and estimates for the number of tail
particles rescattered into the beam.

The wake fields of the collimators can have a very deleterious effect
on the beam core.  To minimize the wakes, the beam pipe must be tapered
before and after the collimator.  A parallel-plane collimator will have
a quadrupole wake even for an on-axis beam core, which can cause
focusing of the core and alter the trajectory of particles in the
tails. And beams that have been mis-steered close to the wall can
experience very large wake-induced kicks.  Because of the latter effect,
the aperture of the second stage of IP-phase collimation must be quite
large. This in turn places a requirement on the transmission quality of
the beamline from the collimation system to the final-focus system, and
on the dynamic aperture of the final-focus system. These have been
verified with tracking studies.

Because of the large $\beta$-functions and strong focusing that arise
when the beam is blown up with linear optics, there are important
chromatic effects to compensate with sextupole pairs.  And because it
is necessary to collimate each transverse phase at least one time,
there are very large $R_{12}$ and $R_{34}$ functions within the system.
Large $R_{12}$ and $R_{34}$ functions lead to tight tolerances. 
Stability tolerances, that must be held between tunings of the waist
knobs in the final-focus system, are looser than stability tolerances
within the final-focus system.   Though the particle backgrounds in the
collimation system preclude the beam-based stabilization methods
contemplated for the final-focus system, beam changes can be monitored
non-invasively in the skew correction system and aberrations arising
from changes in the collimation system can be continuously compensated
there.

Ground motion studies show that seismic ground motion in the beam
delivery system has a negligible impact on beam collision offsets at
the IP. As a result, vibration tolerances become tolerances between
beam-line elements and the ground beneath them (FFTB quadrupoles
mounted on movers have been shown to follow the ground $\pm$ 1 nm), or
tolerances on ground motion coming from cultural sources.

In the ZDR we have described the system specifications, discussed the
relevant properties of materials, presented spoiler and edge-scattering
distributions, clarified the relevant wake-field of tapered collimators
and defined the optimum choice of collimator shape, quantified the
impact of near-wall wakes for on-axis and mis-steered beam, designed
lattices that implement the required functionality, calculated the
position and strength tolerances of the magnetic elements, clarified
the impact of ground motion, tracked the lattices to confirm their
functionality, tracked spoiler distributions in the collimation system
to determine power deposition, tracked edge-scattered distributions in
the final-focus system to determine the probability of particle impacts
on the final doublet, determined the extent of tail repopulation due to
gas scattering, discussed potential operational problems, and addressed
machine protection issues.  To our knowledge all issues have been
addressed with the satisfactory conclusion that it is possible to build
and operate a collimation system for the proposed beam parameters from
beam energies of 175 GeV to 750 GeV, that will collimate the beam at
the apertures required by the final-focus system.  We believe that we
have demonstrated an existence proof.

\subsection{IP Switch and Big Bend}

Following the collimation system are the IP switch, the big bend, and
the skew correction and diagnostic section, as illustrated in
Fig.~\ref{fig:bdsys}. An NLC design with two IPs will require two IP
switches and four big bends.

\begin{figure}[htb]
\leavevmode
\centerline{\epsfxsize=4in\epsfbox{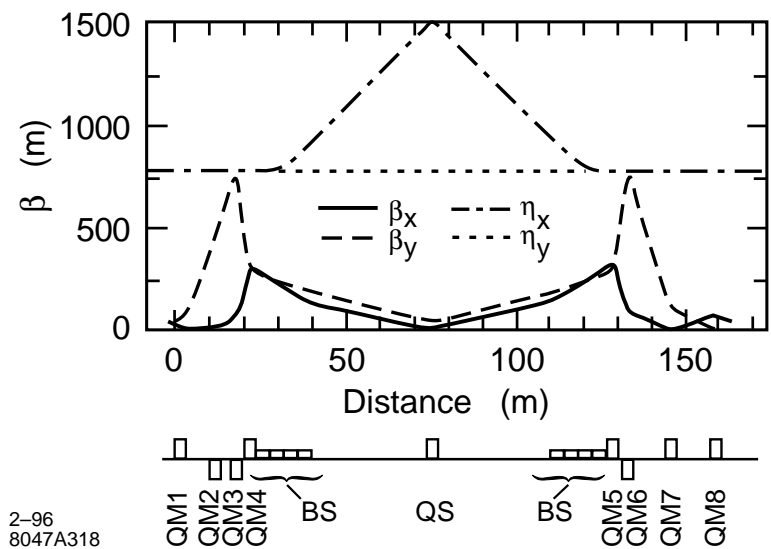}}
\centerline{\parbox{5in}{\caption[*]{IP Switch Optics (500 GeV per beam).}
\label{fig:opt}}}
\end{figure}

The IP switch follows the main linac and collimation section and allows
slow switching between multiple IPs. It bends the beam a total of 1.5
mrad. Figure~\ref{fig:opt} shows the IP switch optics. The QS quadrupole
is horizontally movable in order to switch between IPs. It is displaced
by $\pm$3.25 cm ($\pm$2.6 cm) for the 500-GeV per beam (750-GeV per beam)
configuration.

The big bend is a low-angle arc after the main linac which provides
detector muon protection, IP separation, and an IP crossing angle to
facilitate extraction of the spent beams. The total bend angle
(including 1.5-mrad  IP switch angle) is 10 mrad (20-mrad  IP crossing angle)
which provides approximately 40-m spatial separation between the two IPs
(approximately 700-m transport to an approximately 1600-m-long final focus). At
500--750 GeV per beam, the horizontal emittance growth due to SR sets lower
limits on the system design length. The optical design of the big bend
section was optimized for 500 GeV per beam and 750-GeV per beam electrons (or
positrons). The total emittance growth (in both IP switch and big bend)
for the optimized design, due to chromatic filamentation and due to
synchrotron radiation, is less than 2\% vertically and 3\%
horizontally, for a beam energy of 500 GeV. The big bend rotates the
spin orientation by about 550$^{\circ}$ at 500 GeV. The relative
depolarization is small: 0.06\% for a 0.3\% energy spread.

\begin{figure}[htb]
\leavevmode
\centerline{\epsfbox{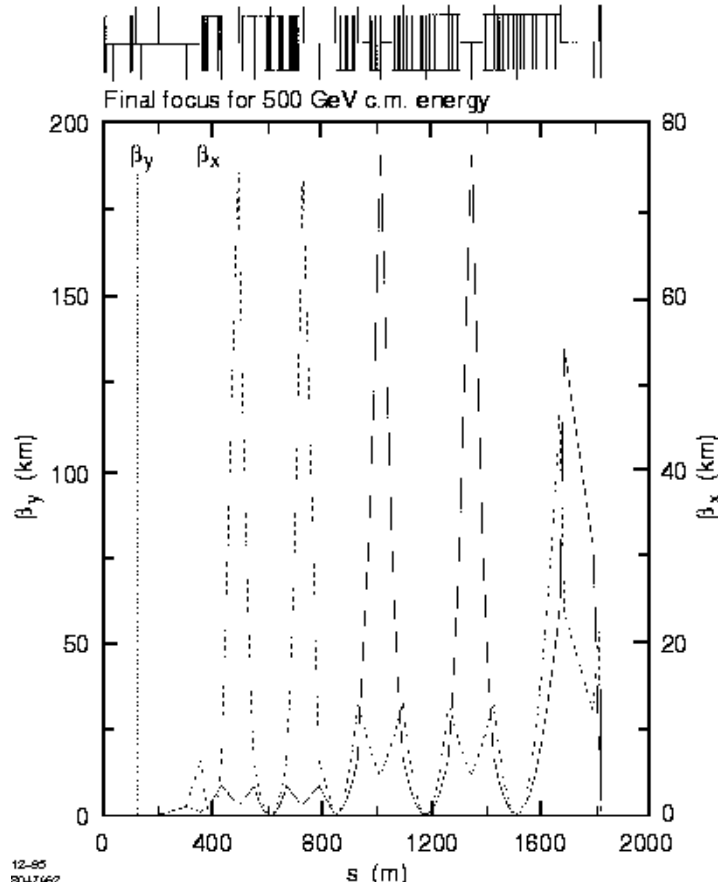}}
\centerline{\parbox{5in}{\caption[*]{Horizontal and vertical beta
functions  for the 500-GeV final focus.}
\label{fig:betxy}}}
\end{figure}

\subsection {Final Focus}

The main purpose of the NLC final-focus system is to transport electron and
positron beams of energy 180 GeV to 750 GeV from the end of the big
bend to the IP, where the demagnified beams are collided with a typical
spot size of 4--8 nm vertically and 200--300 nm horizontally.
One must also
remove the beam remnants cleanly in order to facilitate crucial post-IP
diagnostics.

The basic layout of the final focus is very similar to that of the
Final-Focus Test Beam (FFTB). The final focus is constructed from six
functional modules. These are, in the order of their location:
geometry-adjustment section (GAS), beta- and phase-matching section
(BMS), horizontal chromatic correction section (CCX), beta-exchanger
(BX), vertical chromatic correction section (CCY), and final
transformer (FT). The total distance from the entrance of the GAS to
the IP is about 1820 m. A schematic of the magnet configuration is
depicted in the top part of Figure~\ref{fig:betxy}. The figure also
shows the beta functions corresponding to the largest IP beam
divergences of the NLC parameter plane, for which the system was
designed. This represents a worst case, since the effect of nonlinear
aberrations, the aperture requirements, and the Oide effect
(synchrotron radiation in the last quadrupoles) all become more severe
for larger divergence. Located at the end of the final transformer is
the final doublet, which focuses the beam to the design spot size at
the IP. In the NLC, the final doublet is actually a quartet comprising
four different quadrupoles: 2 conventional magnets, 1 superconducting
and 1 permanent.

The chromaticity (energy-dependent focusing) of the final doublet is
similar to that at the FFTB, and about 5 times larger than in the SLC.
If the chromaticity were not corrected, the vertical IP spot size would
increase by about a factor of 100 from the design value. The chromatic
correction is accomplished in two separate beam-line sections, CCX and
CCY, each of which accommodates a pair of sextupoles separated by an
optical transform ($-I$) that cancels geometric aberrations and second
order dispersion. The momentum bandwidth of the final focus was
optimized with three additional sextupoles and exceeds $\pm 0.6$ \%.

The NLC design calls for a 20-mrad crossing angle of the two beams at
the IP, which requires two separate beam lines. Magnet apertures and
focusing strengths in the two beam lines can be chosen independently,
and are in fact quite different. The crossing angle, furthermore,
requires rotation each beam prior to collision by 10 mrad in the
horizontal/longitudinal plane, by means of a 5-cm-long X-band crab
cavity.

Placed about 150 m in front of the final doublet is a 70-m long soft
bending section with a 12-G field, which serves to deflect the beam
orbit at the final-doublet entrance by about 8 mm with respect to the
hard photons radiated upstream, and, thereby, reduces the detector
background substantially. A similar soft bend has proved invaluable at
the SLC.

\begin{figure}[htb]
\leavevmode
\centerline{\epsfbox{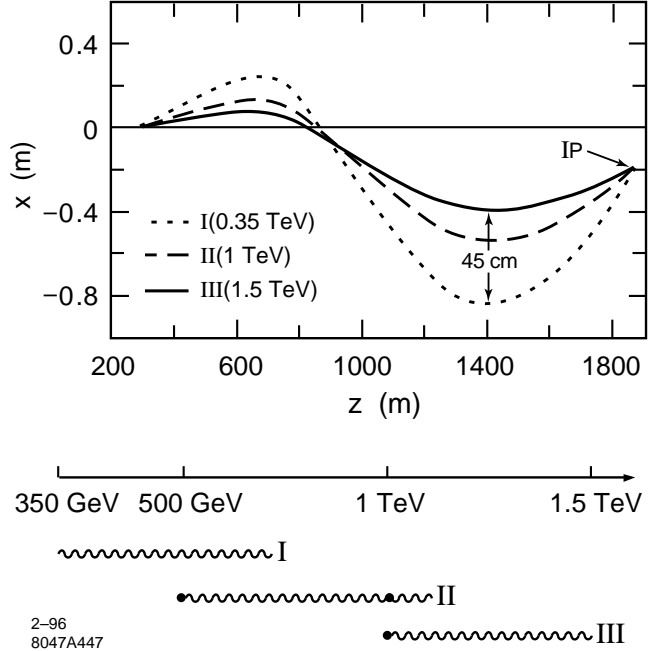}}
\centerline{\parbox{5in}{\caption[*]{Top view of the final focus, and
energy range of the three different geometries. Magnet displacements by
at most 45 cm are necessary during an upgrade from 500 (350) GeV to 1.5
TeV, while the IP-orbit angle changes by about 1.5 mrad.}
\label{fig:erang} }}
\end{figure}

The final focus can operate in the entire cms energy range between 350
GeV and 1.5 TeV. This flexibility is achieved with three slightly
different geometries, illustrated in Fig. \ref{fig:erang}. A dedicated
geometry-adjustment section (GAS) at the entrance to the final focus is
used to adjust the incoming beam-orbit angle for the different
geometries, such as to keep the IP position constant and the transverse
magnet displacements small. Energy adjustability of the final doublet
is provided by the superconducting quadrupole, whose field changes sign
when the beam energy is raised from 250 GeV to 500 GeV. Finally, the
apertures of all magnets are consistent with a minimum vertical beam
stay-clear of 55$\sigma_{y}$ at 500-GeV cms energy.

Higher-order optical aberrations, synchrotron radiation, residual
uncorrected low-order aberrations, timing offsets and crab crossing
errors all reduce the luminosity. A total luminosity loss of about
30\% is included in the NLC design parameters. The low-order
aberrations, such as dispersion, waist shift, skew coupling, {\it
etc.}, which are responsible for almost half of the total loss, need to
be tuned and corrected at regular intervals, {\it e.g.}, every few
hours. An average luminosity loss of 1--2\% was assumed for each
low-order aberration. Actual tuning accuracy in the SLC is about 0.5\%.

The design luminosity is achieved, if, first, the aberrations stay
small between tunings, and, second, the beams collide head-on at the IP
(rms offset smaller than a quarter of half the beam size). The first
condition translates into tolerances on magnet-position drifts over a
1-s time scale and on BPM stability over a few hours. The FFTB magnets
and BPMs fulfill all the NLC tolerances. The second condition is met,
when a) the effect of ground motion is not important and b) the magnet
vibrations with respect to ground are reasonably small.

\begin{figure}[htb]
\leavevmode
\centerline{\epsfbox{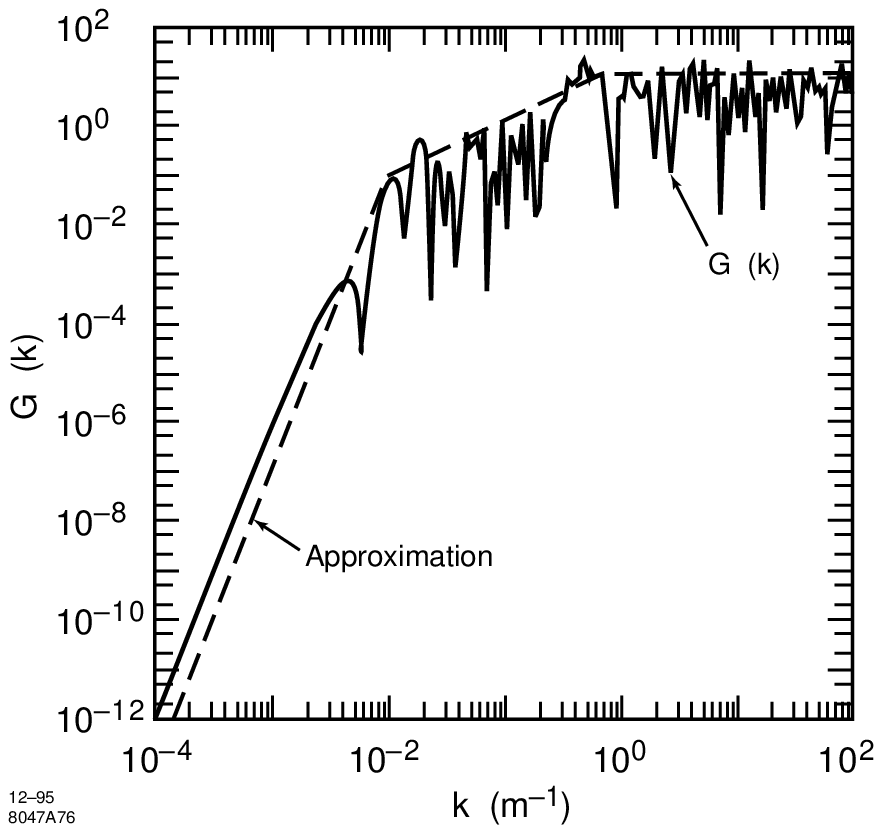}}
\centerline{\parbox{5in}
{\caption[*]{Lattice response function for the NLC final focus
system.}
\label{fig:latresp}}}
\end{figure}

\begin{figure}[htbp]
\leavevmode
\centerline{\epsfbox{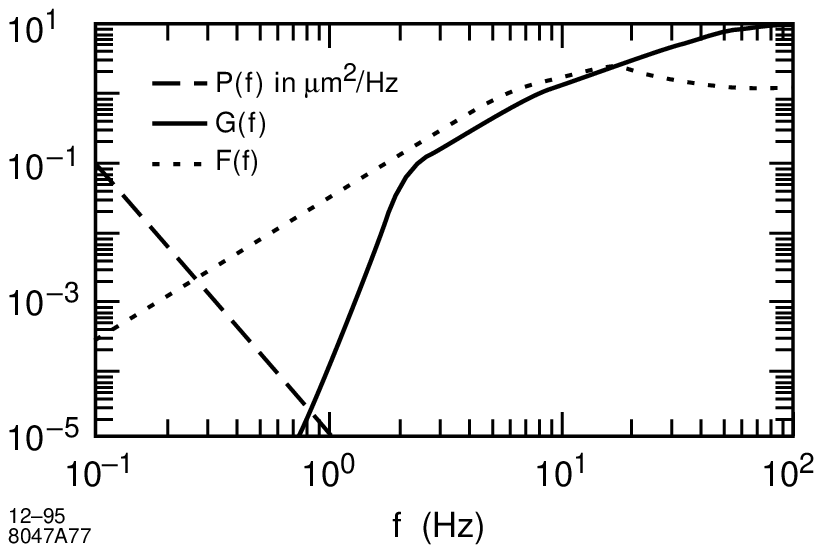}}
\centerline{\parbox{5in}{\caption[*]{Three functions which determine
the rms beam-beam separation due to plane-wave ground motion:
$F(f)$---the feedback response; $P(f)$---the local power density;
$G(f)$---the lattice response function converted into frequency domain.
The integral over the product of these functions gives the square of
the rms beam-beam separation.} 
\label{fig:gmrmssep}}} 
\end{figure}

The ground motion measured above 0.01 Hz can be explained by isotropic
plane ground waves, for which each frequency is associated with a
certain wavelength; longer wavelengths correspond to lower
frequencies. Figure \ref{fig:latresp} shows the response of the NLC
final focus to a harmonic vertical displacement of quadrupoles as a
function of wavelength. Plotted is the average squared ratio of the IP
beam-beam offset and the ground-motion amplitude. The figure
illustrates that the effect of ground motion is heavily suppressed at
long wavelengths, {\it i.e.}\ below about 2 Hz ($k\le 0.01$ m$^{-1}$).
To calculate the expected beam-beam offset at the IP, one has to
convert the curve of Fig.\ \ref{fig:latresp} into frequency domain,
and, after multiplication with the ground-motion frequency spectrum and
a typical SLC feedback response curve (see Fig.\ \ref{fig:gmrmssep}),
to integrate over frequency. The total luminosity loss due to ground
motion is then predicted to be 0.8\% on the SLAC site, and 0.02\% at a
quiet site ({\it e.g.} LEP tunnel). An upper bound on the
luminosity loss due to any additional uncorrelated ground motion is
0.04\%. The conclusion is that one can use the ground (bedrock) as a
reference for stabilization.

Thus magnet supports need to be designed which neither amplify nor damp
the ground motion, but couple the magnet firmly to the ground. At DESY
(SBLC-TF), the rms relative quadrupole-to-ground vibrations above 1 Hz
were measured to be smaller than 1 nm, which would meet all the NLC
tolerances. Quadrupoles at the FFTB were found to vibrate by about 4 nm
with respect to the ground, excited mainly by bad cooling pumps. These
vibration amplitudes would still satisfy the tolerance criteria for all
NLC magnets other than the final doublet. For the latter a special
stabilization system based on an optical anchor and piezo-electric
movers has been devised, which is discussed in the following chapter.

\begin{figure}[htbp]
\leavevmode
\centerline{\epsfbox{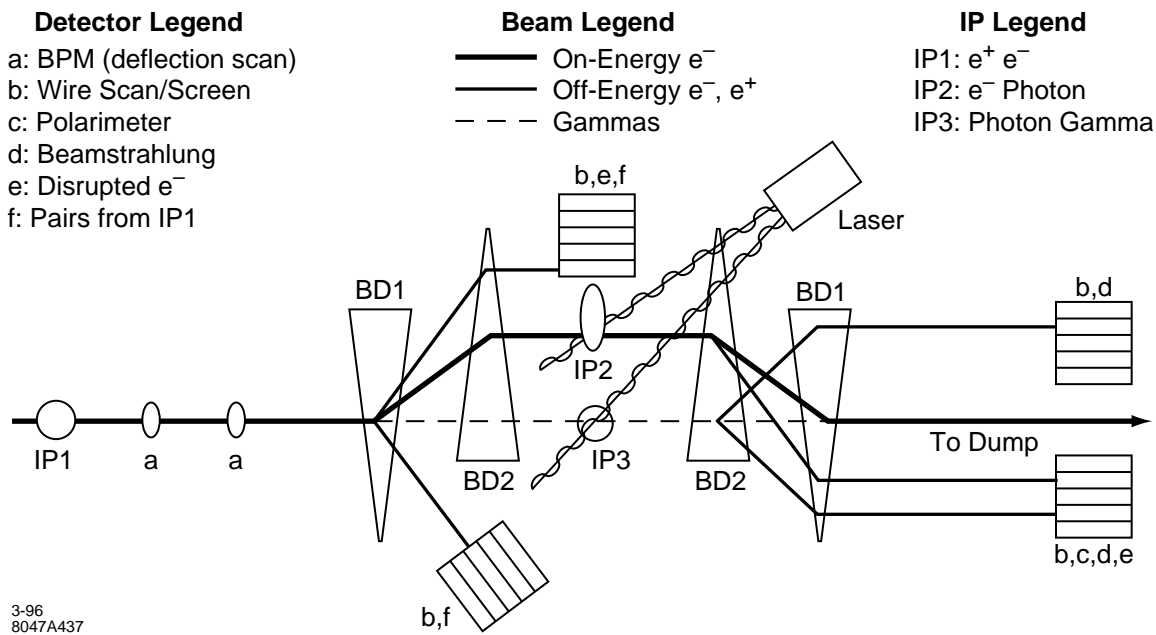}}
\centerline{\parbox{5in}
{\caption[*]{NLC Extraction Line and Diagnostic Layout.}
\label{fig:nlcdiag}}}
\end{figure}

The beam extraction line behind the IP guides the disrupted beam onto a
water-based beam dump, which must dispose of all the beam power
(about 10 MW). Numerous diagnostics in the extraction line allow to
control and stabilize beam orbit and energy, and also to monitor
luminosity and polarization. A schematic is depicted in Fig.\
\ref{fig:nlcdiag}.

In conclusion, the NLC beam delivery system will not only produce small
spot sizes at the IP, but it also promises redundant tunability,
adjustability over a wide energy range, and a tolerable detector
background. Most tolerances on magnet strength or position stability
are not particularly tight, and have, in fact, already been achieved at
the FFTB or at other places. Also natural ground motion is not thought
to be a problem. However care is needed in the mechanical design of
components and supports.

\section{Interaction Region and Detector Backgrounds}
\subsection{Introduction}

The design of the interaction region (IR) of the $e^+e^-$ collider
involves an understanding of beam-beam effects, detector backgrounds,
and the constraints required to maintain very small beam spot sizes at
the interaction point (IP). Luminosity monitors with sufficient spatial
and energy resolution can unfold the energy dependence of the beam-beam
interaction's influence on the luminosity spectrum. With appropriate
shielding around the IP a detector employing a high field solenoid and
sufficiently granular tracking can be designed which is robust against
backgrounds while allowing for excellent tracking and vertexing. Muon
backgrounds in the detector are minimized by distancing the machine
collimation system from the IP, introducing a 10 mrad bend between the
linacs and final focus systems, and installing simple shielding magnets
in the tunnel.

Ground motion from geologic and man-made sources must not move the
final quadrupole doublet by an amount more than the vertical beam spot
size. The rapid fall off of naturally occurring ground motion with
increasing frequency, the use of precision beam position monitor based
feedback loops sampling at the machine cycle frequency (120--180 Hz)
to correct the beam orbits at low frequency, and the coherence arising
from the interplay between the machine's optical lattice and the
wavelength of a ground motion disturbance all conspire to make seismic
motions in bedrock negligibly important to NLC luminosity. Quadrupole
motion above the level of the bedrock, due to laboratory sources
(pumps, etc.) or amplification of the driving term by its mechanical
support within the detector, can be minimized by optically referencing
the doublet to bedrock with a LIGO-style interferometer. Feedback
signals can then drive either piezoelectric actuators or magnetic
beam-corrector elements to reduce motion back down to naturally
occurring seismic levels. The selection of a geologically quiet site,
the design of the physical lab plant to minimize vibration, and the
design of a robust final doublet support will all aid in controlling
the ground motion problem.

\subsection{Layout}

In the schematic layout of Figure~\ref{fig:bdsys} the electron and
positron linac arms are each followed by a 3.2-km-long collimation
section and two final focus tunnels at $\pm$10 mrad. These ``Big
Bends", 2.0 km from the interaction points (IP), protect the detectors
from muons produced by the collimators, provide a 20 mrad crossing
angle between incoming and outgoing beams, and allow for two
experimental halls with IPs separated by 40 m. As trains consisting of
90 bunches of 0.65--1.1 $\times$ 10$^{10}$ particles separated by 1.4
ns will collide at 120--180 Hz, a crossing angle is required so that a
given bunch interacts with only its partner, and not with any other
bunch still traveling to the IP.

\begin{figure}[htb]
\leavevmode
\centerline{\epsfxsize=4in\epsfbox{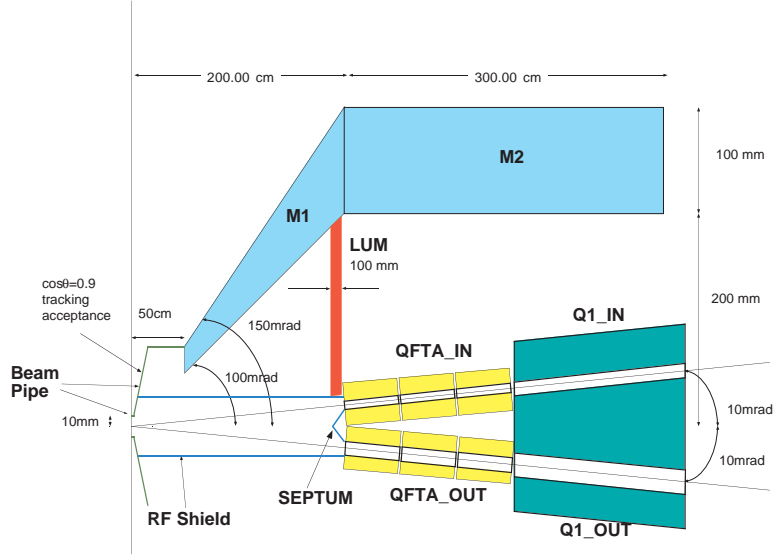}}
\centerline{\parbox{5in}
{\caption{The interaction region masking and magnet layout.} 
\label{fig:ir}}}
\end{figure}

Figure~\ref{fig:ir} shows a schematic of the masking and magnet
layout in the interaction region. In the current final focus lattice
the last quadrupole of the doublet (QFTA) ends 2 m from the IP. Field
strength and quality require that it be constructed of a machinable
permanent magnetic material (samarium cobalt, for example). Its inner aperture
radius is 4.5 mm and outer radius is 10.0 cm. A similar doublet with a
6 mm inner aperture radius to transport the outgoing disrupted beam,
together with any synchrotron radiation and beamstrahlung photons, is
symmetrically positioned horizontally across from QFTA. These two
magnets are followed by a twin bore superconducting magnet to complete
the doublet. Each is surrounded by a superconducting coil which shields
the detector's solenoidal field.

A ``dead cone", within which the vast majority of the low $p_t$, $\ee$
pairs produced by the beam-beam interaction are confined, is defined by
a conical tungsten mask which subtends the angular range from 100 to
150 mrad and 0.5 m$< z <$ 2 m. A cylindrical tungsten skirt, currently
10 cm thick, begins at $z$=2.0 m and $r$=20 cm. Its purpose is to
protect any exterior detector from photons produced when the pair
electrons and positrons strike the from face of the quadrupoles or
luminosity monitor. The beam pipe is assumed at this time to have a
radius of 1.0 cm for the first 2.5 cm from the IP. This will
accommodate
the inner layer of a vertex detector with acceptance out to
$\cos\theta=0.9$. The beam pipe then flares to a radius of 7.2 cm,
switching from beryllium to 1-mm-thick stainless steel once it begins
to follow the contour of the M1 mask. Within the beam pipe
is a thin rf shield and septum which ease the transition from
the narrow apertures of the input and exit quadrupoles to the beam pipe
radius.

\subsection{Backgrounds}

The backgrounds anticipated at the NLC are each described in the
following subsections. The experience of the SLD detector at SLC is
that, when the accelerator is working well, all expected backgrounds
are small. High luminosity correlates strongly with low backgrounds.
While the calculations outlined below indicate that backgrounds will
not be a substantial problem at the NLC, the prudent experimenter will
design a detector which can handle the oft-occurring non-standard
running conditions and {\it unexpected} backgrounds which are in the
nature of linear colliders. 

The background levels that can be supported require that we assume a
detector model. Experience with the SLD's pixel vertex detector, based
on 22 $\mu$m $\times$ 22 $\mu$m CCD pixels, is that hit density is the
figure of merit. Integrating over its readout time of 19 bunch
crossings, the SLD achieves 0.4 hits/mm$^2$ in its innermost layer.
Tracks found in an 80 layer drift chamber are projected back to the
vertex detector and hits lying within a three $\sigma$ extrapolation
error window of roughly 1 mm are added to the track. The traditional
figure of merit for vertex detector backgrounds is thus taken to be 1
hit/mm$^2$ per unit of time. Three to four layer self tracking silicon
vertex detectors, with track extrapolation errors on the order of 10
$\mu$m, should be robust against backgrounds 10 to 20 times worse.

This section assumes that at the NLC the integration time for each
detector is a train of 90 bunches. Depending on the technology actually
used, this assumption may be overly pessimistic; nanosecond level
timing and the ability to isolate background at the bunch level will
certainly be a design goal for NLC detector components.

Raw occupancy is the figure of merit for a gas tracking chamber
comprised of a relatively small number of wires each of which samples a
large volume. The 5120 wire SLD chamber at $r=20$ cm operates with
3--5\% occupancy in good conditions. Simulations indicate that this is
due to between one and ten thousand low energy photons produced from
the secondary interactions of 2-MeV SR photons in the masking system.
We take 1--10 thousand photons per train as the design goal for an NLC
gas tracking chamber.

In SLD muons produced in the upstream collimators traverse the barrel
calorimeter at the rate of one or two per beam crossing. In bad
conditions there can be a factor of ten increase in the muon flux, at
which point the schemes used to keep the muons from contributing to the
trigger rate, and, to a lesser extent, the offline energy
reconstruction offline, break down. The design goal of the muon
protection system is that 1\% of the beam can be scraped off in the
collimation section and produce, on average, one muon per train of
bunches that is transported into the detector.

\subsubsection{Muons}

\begin{figure}[htb]
\leavevmode
\centerline{\epsfysize=3.5in\epsfbox{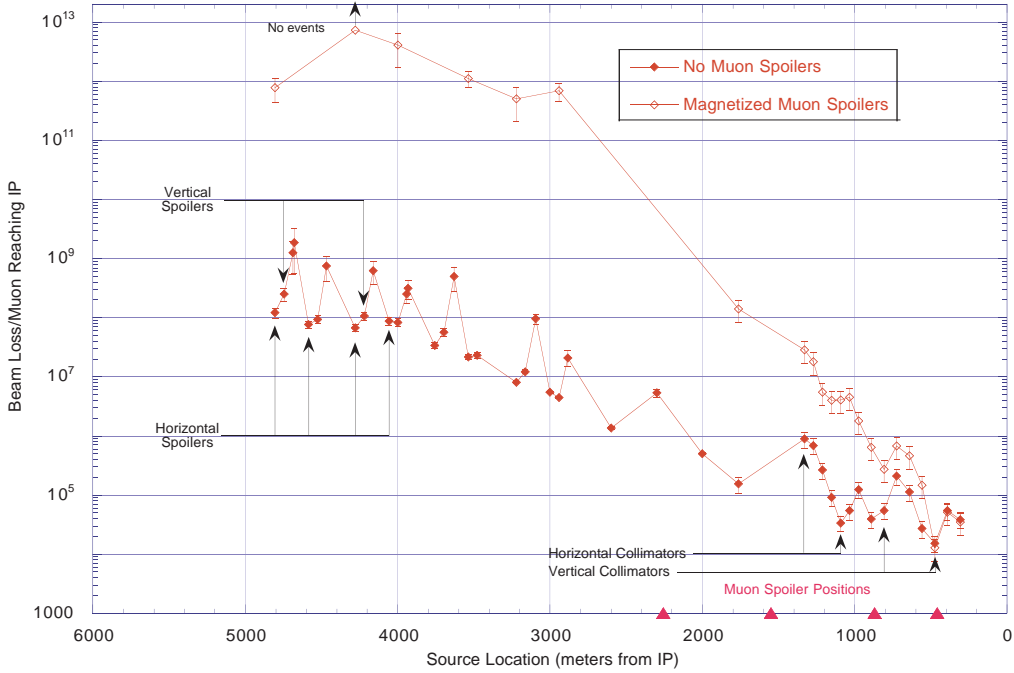}}
\centerline{\parbox{5in}{\caption{Muon backgrounds: the number of beam
particles that can be lost on a given collimeter before producing one
muon that strikes the detector.}
\label{fig:Muon_Result}}}
\end{figure}

Figure~\ref{fig:Muon_Result} shows the results of the muon background
simulation. The black diamonds show the ordinate is the number of muons
that can interact in a given collimator and produce one muon in a 8 m
by 8 m square detector. The abscissa is the location of the collimator
in the tunnel. The points to the left between 4 and 5 km from the IP
correspond to the collimation system; the points between 1 and 2 km
from the IP correspond to locations where it is planned that the clean
up collimators remove the slightest of beam tails. The uppermost of the
two curves corresponds to the calculation when four 9 m long tunnel
filling toroids are added to the final focus to intercept and deflect
errant muons. One sees that this spoiler system added three orders of
magnitude of protection and that we are two orders of magnitude within
the design goal (which would correspond to roughly 10$^{12}$). Without
the spoilers we have a marginal situation. With or without spoilers, we
are much more sensitive to beam particles lost in the FF after the big
bend. Figure~\ref{fig:Muon_Result} shows that 10$^6$ particles lost in
the final focus, which is one part per million per train,  will produce
one muon that reaches the detector.

\subsubsection{Pairs}

\begin{figure}[htb]
\leavevmode
\centerline{\epsfysize=3in\epsfbox{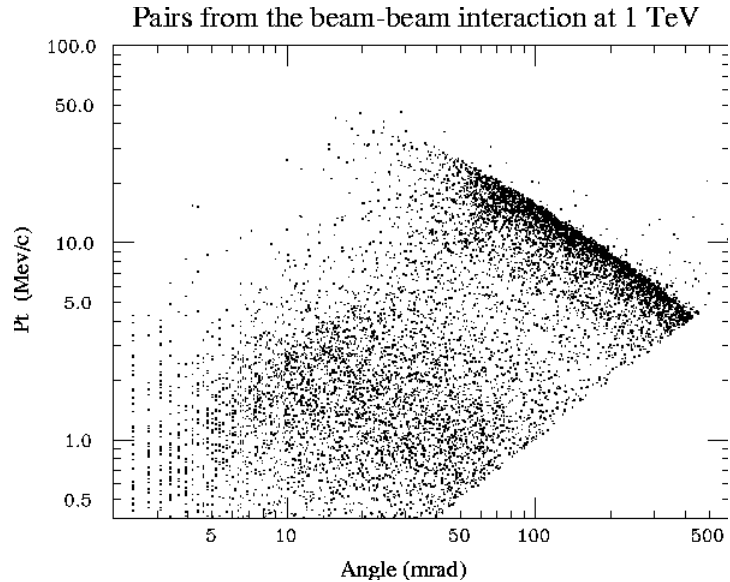}}
{\caption{$p_t$ vs.$\theta$ distribution for pairs.}
\label{fig:pt_theta}}
\end{figure}

Roughly 10$^5$ pairs will be produced by the beam beam interaction each
bunch crossing, predominately through the Bethe-Heitler interaction of
beamstrahlung photons and electrons or positrons. For the most part the
pairs are produced with low intrinsic $p_t$; the same sign partner will
tend to be focused by the opposing beam while the opposite sign partner
will be deflected outside the beam envelope by the magnetic field of
the bunch. The finite beam dimensions then result in a very hard
kinematic edge in the $p_t-\theta$ distribution
(Figure~\ref{fig:pt_theta}).

By introducing a strong solenoidal magnetic field all particles with
$p_t<$ 30 MeV are curled up within the 5 cm minimum radius of the
conical mask of Figure~\ref{fig:ir}. All particles with $\theta$ less
than the dead cone are curled up. The number of particles falling
outside these two cuts is relatively small and manageable.

We have simulated the beam beam interaction at 1 TeV with the ABEL
program. Figure~\ref{fig:abel_hits} in the ``Design of the NLC
Detector'' section of this document 
shows the hit density expected at $r=$ 2 and 3 cm as a function of $z$
for a solenoidal field of 4 Tesla. At $r=$ 2 cm, as long as our VXD
lies within $z=$ 17 cm, the hit density is manageable. Lowering the
field would require that the innermost vertex layer be at
correspondingly larger radius.

\subsubsection{Photons from Pairs}

Pairs will interact with the beam pipe, inner vertex detector layers,
rf shield septum, and front faces of the entrance and exit quads. As
they interact, photons will be produced which will form a secondary
background in the VXD and in any tracking chamber at larger radius.

\begin{figure}[htb]
\leavevmode
\centerline{\epsfysize=3in\epsfbox{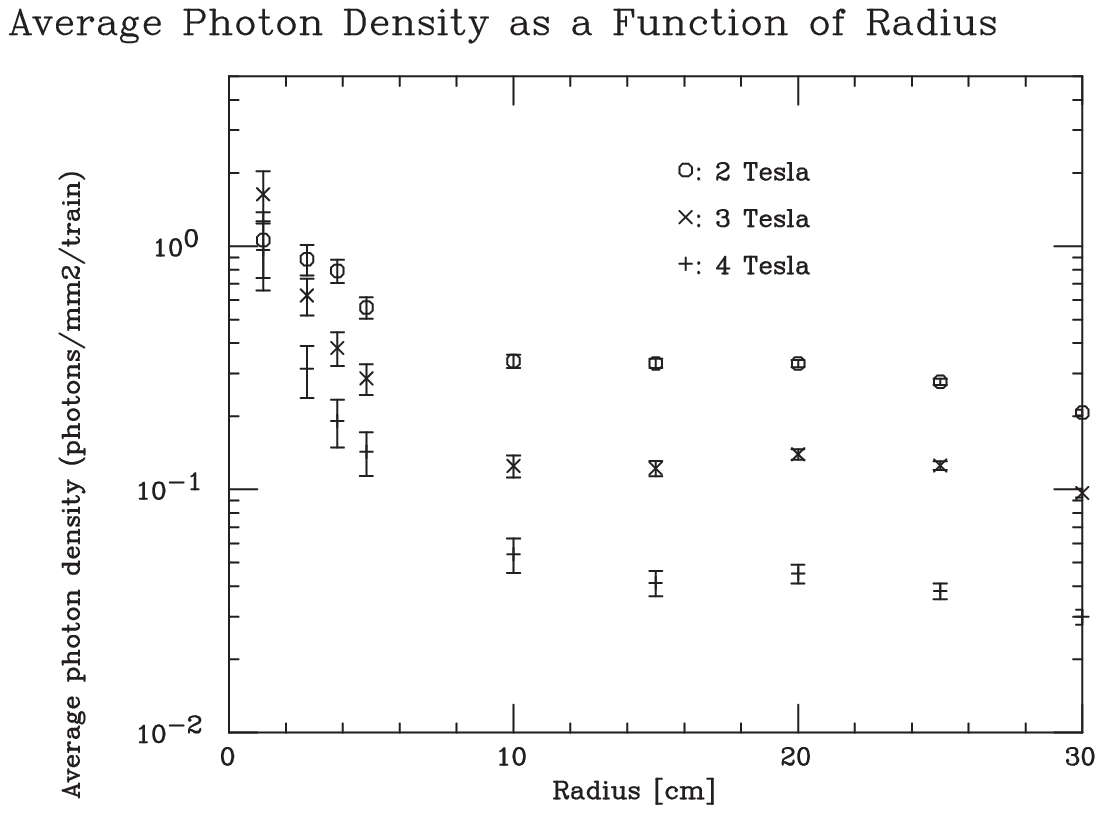}}
{\caption{Photon hit density vs. radius.}
\label{fig:PhotonHitDensity_vs_radius}}
\end{figure}

To study this problem we have simulated the IR of Figure~\ref{fig:ir}
with EGS4 using the 1-TeV ABEL ray files as input. Materials,
thicknesses, and geometries have all been achieved at SLD.
Figure~\ref{fig:PhotonHitDensity_vs_radius} looks at the number of
photons per train intercepting scoring planes with $\cos\theta=0.90$
acceptance. Photon-to-hit conversion efficiencies are expected to be
about the same for silicon tracking or gaseous drift chambers, a couple
of percent. The hit densities are all well below 1/mm$^2$/train, even
before taking into account conversion efficiency. However, the raw
number of photons crossing the r=30 cm plane is rather large,
approximately what we have taken as our figure of merit. The situation,
as indicated in the figure, is made better by going to higher field. It
is also possible that more innovative masking schemes will help. It
does seem however that low granularity tracking devices will be
marginal in the NLC background environment.

\subsubsection{Synchrotron Radiation}

\begin{figure} [htbp]
\leavevmode
\centerline{\epsfig{file=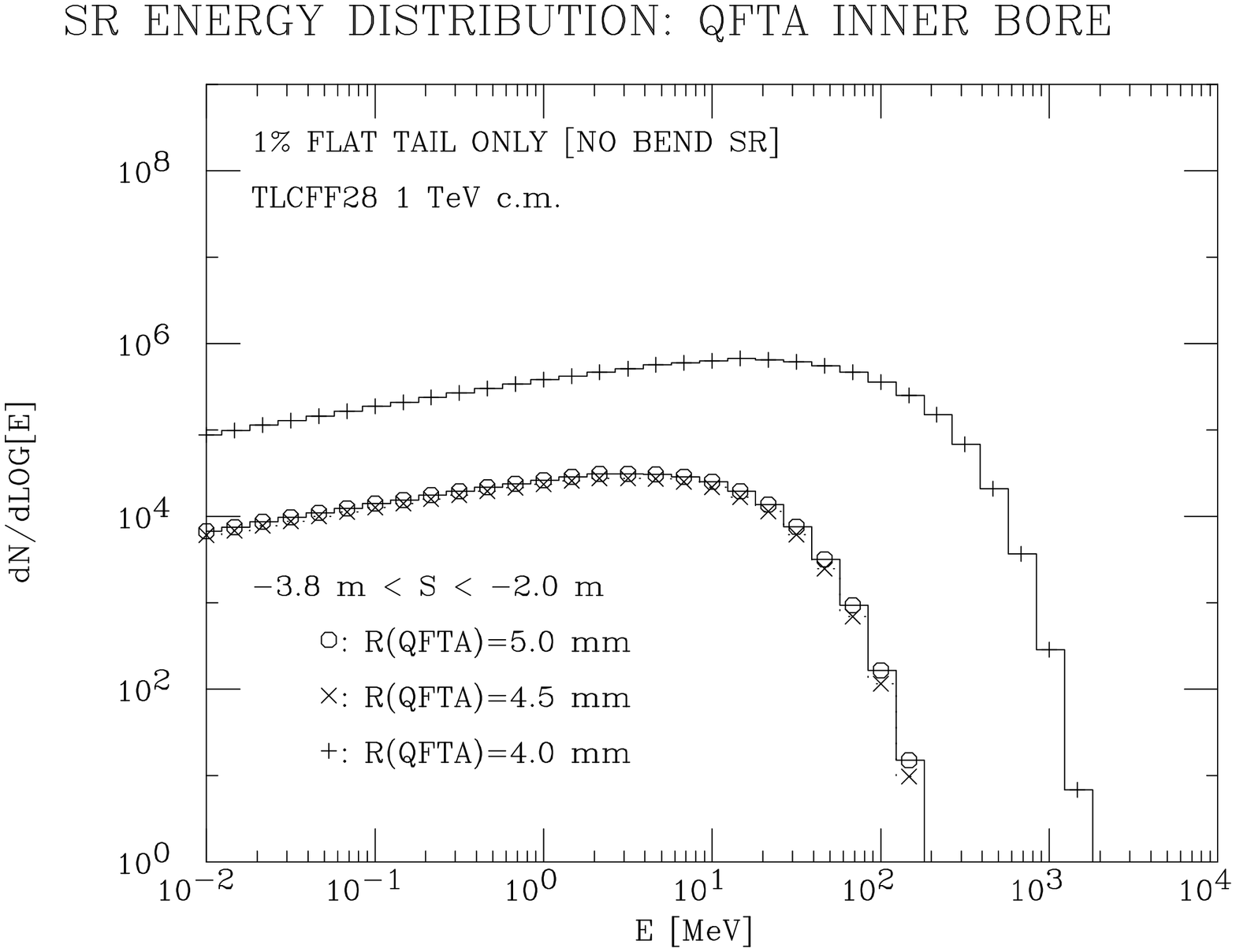,width=3.2 in,angle=90}}
{\caption{Synchrotron radiation energy distribution.}
\label{fig:SR_Energy}}
\bigskip\bigskip
\leavevmode
\centerline{\epsfig{file=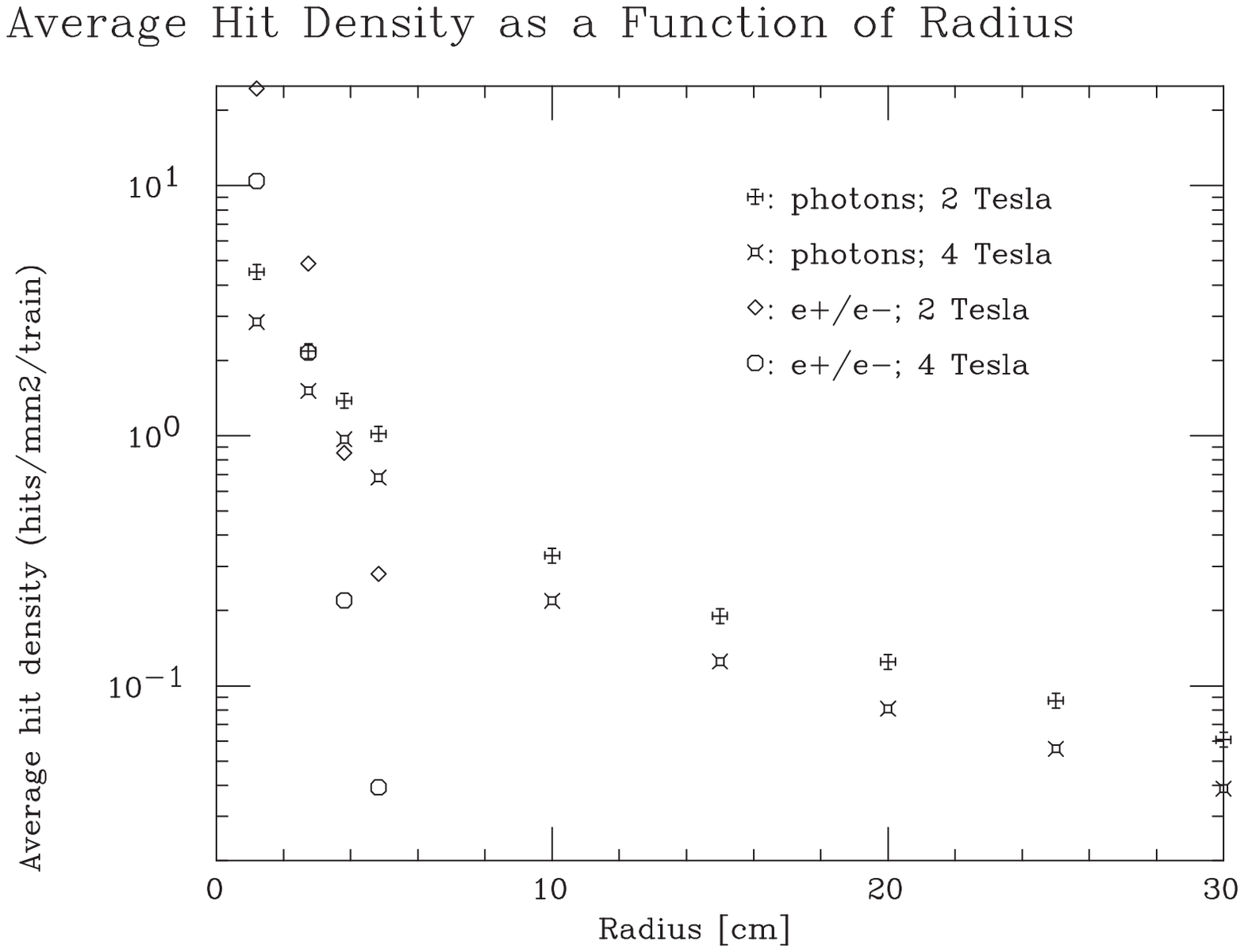,width=3.2 in}}
{\caption{Synchrotron radiation hit distribution.}
\label{fig:SR_Hits}}
\end{figure}

SR produced by the core of the beam in the soft dipole bend just
upstream of the final doublet and SR produced by particles at the edge
of the phase space allowed by the collimation system can cause
backgrounds in the detector. Most of the radiation escapes through the
output quadrupole aperture; photons striking the inside of the incoming
quadrupole aperture near its IP-side tip are the biggest problem. The
soft bend SR from the core can be masked far from the IP.
Figure~\ref{fig:SR_Energy} shows the energy distribution of SR from the
beam tail as calculated for the 1 TeV lattice. As long as the QFTA
aperture is large enough so that radiation produced immediately
upstream in Q1 can escape the situation is not dramatic. If the QFTA
aperture decreases below 4.3 mm radius or the tail collimation is
loosened beyond 7$\sigma_x$ and 35$\sigma_y$ the energy deposited in
the tip of QFTA increases dramatically.

Figure~\ref{fig:SR_Hits} shows the results of the EGS simulation using
the SR energy distribution of Figure~\ref{fig:SR_Energy} as input. The
charged particle hit density is tolerable for $r\geq 2$ cm and $B=4$
Tesla, while the photon hit density is negligible in the region of the
vertex detector. However, the photon density at 30 cm would correspond
to having over 100,000 photons incident on a conventional drift
chamber, a potentially worrisome figure were that technology chosen for
the tracking system. Better masking, especially upstream in the
lattice, and tighter collimation would reduce these numbers. Note also
that the backgrounds would be much less for the 500 GeV lattice.

\subsection{Quad Support and the Optical Anchor}

As outlined in the introduction, the issue of ground motion was once
thought to be a potential ``show stopper" for a machine with nanometer sized
beams. Our current understanding of the relationship between beam
feedback, the ground motion spectrum, and the  coherence resulting from
the machine lattice and the wave nature of the ground motion
is that by minimizing external sources of vibration and by firmly
anchoring the final quadrupoles to bedrock luminosity will be preserved.

\begin{figure}[htb]
\leavevmode
\centerline{\epsfysize=2.8in\epsfbox{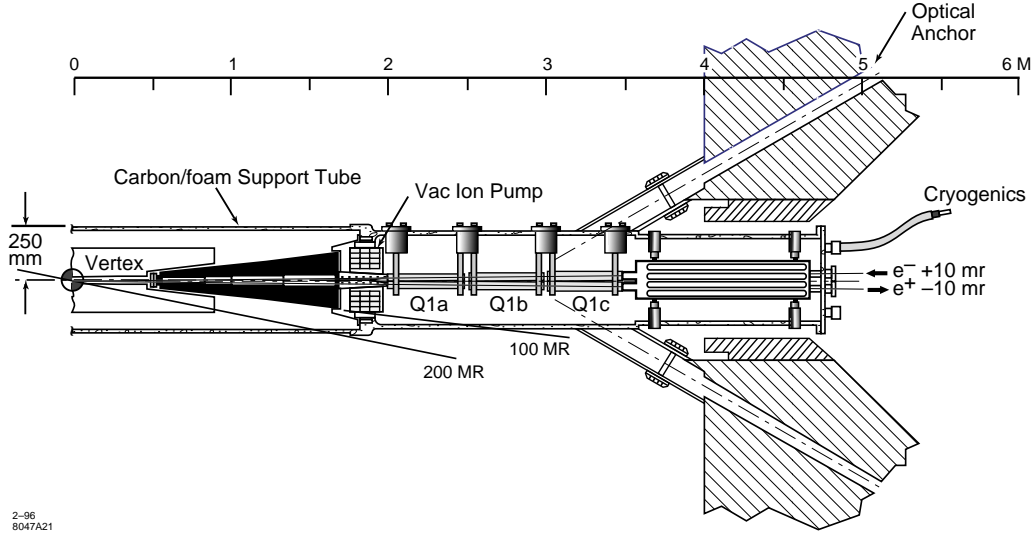}}
\centerline{\parbox{5in}
{\caption{Current final doublet layout with 20-mrad crossing angle.}
\label{fig:current}}}
\end{figure}

However, the final quads will need to be mounted inside a detector.
Engineering the detector and the quad supports to avoid amplifying
natural or cultural jitter sources is mandatory. Developing tools to
remove the jitter that will nonetheless arise is highly prudent.
Figure~\ref{fig:current} shows an engineering drawing of our current
ideas for the final quad layout and support.

In early tests at SLAC and DESY inertial seismometers driving
pizeoelectric movers were used in feedback loops to inertially
stabilize a magnet. A factor of three in amplitude reduction was
achieved at each lab. Fears of using these devices within a particle
detector magnetic field and a desire for larger reduction factors led
to the search for optical solutions to the problem. The LIGO
experiment, geophysicists, and astronomers have all addressed this
problem through laser interferometry. The relative change in the length
of two optical arms, $\ell_1$ and $\ell_2$, is related to an intensity change
at a photodiode,
 $$ \frac{\delta I}{I}=\delta\Phi=2\pi \cdot
\frac{\delta (\ell_1-\ell_2)}{\lambda}$$ 
For $\lambda=633$ nm, and $\delta(\ell_1-\ell_2)=1$ nm, one will have
$\delta\Phi=0.02$ and this will cause a 2$\%$ intensity change on the
photodiode.

While there are many constraints on the laser and optical system in
this interferometer, including tolerances on the laser's intensity
stability, frequency stability and power, as well as tolerances on
pressure and temperature fluctuations in the optical transport arms, it
appears that the solution is viable. Testing its performance and
designing a detector which leaves room for optical lines of sight are
what is required.

\subsection{Conclusion}

While more calculation and testing is required there does not appear to
be any fundamental problem in designing an IR and detector which
maintains the machine luminosity, can handle the backgrounds, and do
the physics.

\newpage
$$ $$
\newpage

\begin{appendix}
\addcontentsline{toc}{chapter}{Appendix: Papers contributed to the NLC
Physics Study}
\centerline{\Large Appendix: Papers Contributed to the NLC Physics Study}

\bigskip
\begin{enumerate}
\item
  T. G. Rizzo, ``Probing Weak Anomalous Couplings with Final State
        Gluons at the NLC.''
                                             
\item

  J. F. Gunion and P. C. Martin, ``Prospects for and Implications
        of Measuring the Higgs to Photon-Photon Branching Ratios
          at the Next Linear $e^+e^-$ Collider.''
                                           
\item

  H. Baer {\it et al.}, ``Determination of Supersymmetric Particle Production
         Cross Sections and Angular Distributions at a High Energy
            Linear Collider.''
                                         
\item

  R. Arnowitt and P. Nath, ``Using Linear Colliders to Probe GUT and
         Post-GUT Physics.''
                                           
\item
  J. Wudka, ``The Meaning of Anomalous Couplings.''
                                           
\item

  M. Gintner, S. Godfrey, and G. Couture, ``Measurement of the
        $WW\gamma$ and $WWZ$ Couplings in the Process $e^+e^- \to
         \ell\nu q q'$.''
                                           
\item

  K. Riles, ``Effects of Detector Resolution on Measurement of
       Anomalous Triple Gauge Boson Couplings.''
                                         
\item

  T. G. Rizzo, ``The Polarization Asymmetry and Triple Gauge Boson
        Couplings in $\gamma e$ Collisions at the NLC.''
                                        
\item

  T. G. Rizzo, ``Below Threshold $Z'$ Mass and Coupling Determinations
          at the NLC.''
                                        
\item
  T. G. Rizzo, ``Anomalous Chromoelectric and Chromomagnetic Moments
         of the Top Quark at the NLC.''
 
 \end{enumerate}                                         
\end{appendix}


\newpage


\newpage

\addcontentsline{toc}{section}{Bibliography for Chapter 1}
\begin{thebibliography}{99}

\bibitem{ahn88}              
    C.~Ahn \etal\ SLAC-Report-329, 1988.

\bibitem{sno88}        
    Proceedings of the 1988 DPF Summer Study: 
    Snowmass '88, 
    High Energy Physics in the 1990s,  
    F. Gilman, ed., Snowmass, Colorado, 1988.

\bibitem{sno90}        
    Proceedings of the 1990 DPF Summer Study on 
    High Energy Physics: Research 
    Directions for the Decade, 
    E. F. Berger, ed., 
    Snowmass, CO, 1990.

\bibitem{lat87}
    Proceedings of the 1987 LaThuile Meeting:
    Results and Perspectives in Particle Physics,
    M.~Greco, ed.,
    Gif-sur-Yvette, France, 1987.

\bibitem{des90}            
    Workshop on Electron-Positron Collisions at 500 GeV: 
    The Physics Potential,   
    DESY, 1990.

\bibitem{jlc89}            
    Proceedings of the First Workshop on Japan Linear Collider 
    (JLC I), S.Kawabata, ed., 
    KEK, 1989.

\bibitem{jlc90}           
    Proceedings of the Second Workshop on Japan Linear Collider 
    (JLC II), 
     KEK, 1990.

\bibitem{fin91}         
    Proceedings of the First International Workshop on 
    Physics and Experiments with Linear Colliders, 
    R. Orava, ed., 
    Saariselka, Finland, 1991.

\bibitem{haw93}          
    Proceedings of the Second International Workshop on 
    Physics and Experiments with Linear Colliders, 
    F. Harris, {\it et al.}, eds., 
    Waikoloa, Hawaii, 1993.

\bibitem{jap95}
Proceedings of the Third International Workshop on
    Physics and Experiments with Linear Colliders, 
   Iwate, Japan, 1995.

\bibitem{newref}
``Zeroth-Order Design Report for the Next Linear Collider,'' SLAC Report
474 (Stanford University May 1996).

\bibitem{ref:neww}
International Linear Collider Technical Review Committee Report,
1995, G. Loew, ed. (Available from the editor.)
\end{thebibliography}

\begin{thebibliography}{99}
\addcontentsline{toc}{section}{Bibliography for Chapter 2}


\bibitem{ZDR}
``Zeroth Order Design Report for the Next 
       Linear Collider,''  SLAC Report 474 (Stanford University, May 1996).


\bibitem{pythia} 
T. Sj\"{o}strand, \journal{Comp. Phys. Comm.}{82}{74}{1994}.


\bibitem{barkmorioka}
T. Barklow, preprint SLAC-PUB-7087,   to appear in the proceedings 
of the Workshop on Physics and Experiments with Linear Collider,
LCWS95.

\bibitem{hpzh}
K.~Hagiwara, K.~Hikasa, R.~D.~Peccei, D.~Zeppenfeld, 
\journal{Nucl. Phys.}{B282}{253}{1987}.

\bibitem{Miyahawaii}  
A. Miyamoto, in {\it Physics and Experiments with Linear $e^+e^-e$
Colliders}, F. A. Harris, S. L. Olsen, S. Pakvasa, and X. Tata, eds.
(World Scientific, Singapore, 1993).


\bibitem{ISAJET} 
F. Paige and S. Protopopescu, in {\it Supercollider Physics} (Snowmass
1986), D. Soper, ed. (World Scientific, Singapore, 1986); H. Baer, F.
Paige, S. Protopopescu, and X. Tata, in {\it Proceedings of the
Workshop on Physics at Current Accelerators and Supercolliders}, J.
Hewett, A. White, and D. Zeppenfeld, eds. (Argonne National Laboratory,
1993).


\bibitem{slac329} 
C. Ahn \etal, {\it Opportunities and requirements for
Experimentation at High Energy e$^{+}$e$^{-}$ Colliders},
SLAC-report-329, May 1988.


\bibitem{CDFDZtop} 
F. Abe, \etal\  (CDF Col\-labo\-ration),
\journal{Phys. Rev. Lett.}{74}{2626}{1995}; 
S. Abachi, \etal\  (D0 Collaboration), 
\journal{Phys. Rev. Lett.}{74}{2632}{1995}.

\bibitem{corwid}
 M. Jezabek, J. H. Kuhn, and T. Teubner, 
\journal{Z. Phys.}{C56}{653}{1992}.

\bibitem{Orrx}
G. Jikia, \journal{Phys. Lett.}{B257}{196}{1991}.

\bibitem{Orr}
V. A. Khoze, L. H. Orr, and W.J. Stirling, 
\journal{Nucl. Phys.}{B378}{413}{1992}.

\bibitem{Fujii}
K. Fujii, T. Matsui, and Y. Sumino, \journal{Phys.
Rev.}{D50}{4341}{1994}.

\bibitem{Peskin}
M. Strassler and M. Peskin, \journal{Phys. Rev.}{D43}{1500}{1991}.

\bibitem{Fadin}
V. Fadin and V. Khoze, \journal{JETP Lett.}{46}{525}{1987}, 
\journal{Sov. J. Nucl. Phys.}{48}{309}{1988}.

\bibitem{topptheory} 
Y. Sumino, K. Fujii, K. Hagiwara, H. Murayama, and C.-K. Ng,
\journal{Phys. Rev.}{D47}{56}{1993};  M. Jezabek, J.
Kuhn, and T. Teubner, \journal{Z. Phys.}{C56}{653}{1992}.

\bibitem{toppphen} 
P. Igo-Kemenes, M. Martinez, R. Miquel, and S. Orteu,
  in 
{\it Physics and Experiments with Linear $\ee$ Colliders},
F. A. Harris, S. L. Olsen, S. Pakvasa, and X. Tata, eds.
(World Scientific, Singapore, 1993).

\bibitem{Summur}
H. Murayama and Y. Sumino, \journal{Phys. Rev.}{D47}{82}{1993}.

\bibitem{Suzuki} 
W. Bernreuther and M. Suzuki, \journal{Rev. Mod.
Phys.}{63}{313}{1991}.

\bibitem{Yuan}  
G. A. Ladinsky and C. P. Yuan, 
\journal{Phys. Rev} {D49}{4415} {1994}; and references therein.

\bibitem{HarlanJK}\
   R. Harlander, M. Jezabek, J. H. Kuhn, and T. Teubner, \journal{Phys. 
        Lett.}{B346}{137}{1995}; R. Harlander, M. Jezabek, J. H. Kuhn,
         and M. Peter, hep-ph/9604328.

\bibitem{Parke} 
S. Parke and Y. Shadmi, preprint FERMILAB-PUB--96/042-T.

\bibitem{Frey} 
R. Frey, preprint SLAC-PUB-7075,  to appear in the proceedings 
      of the Workshop on Physics and Experiments with Linear Colliders,
              LCWS95.

\bibitem{Schmidt} 
C. R. Schmidt, preprint hep-ph/9504434.

\bibitem{Atlas}
ATLAS Collaboration, Technical Proposal.  CERN/LHC/94-43.

\bibitem{Kuhn}
R. Harlander, M. Jezabek, and J. H. K\"uhn, preprint
hep-ph/9506292.

\bibitem{Djouadi}
A. Djouadi, J. Kalinowski and P. M. Zerwas, \journal{Z.
Phys.}{C54}{255}{1992}.

\bibitem{EuroTop}
P. Comas, R. Miquel, M. Martinez, and S. Orteu, preprint CERN-PPE/96-40,
  to appear in the proceedings 
      of the Workshop on Physics and Experiments with Linear Colliders,
              LCWS95.



\bibitem{SSI-Fujii} 
K. Fujii, to appear in the proceedings of the 1995 SLAC Summer Institute.

\bibitem{CPttH}
S. Bar-Shalom, D. Atwood, G. Eilam, R. Mendel, and A. Soni,
\journal{Phys. Rev.}{D53}{1162}{1996}.

\bibitem{Amidei}
D. Amidei and R. Brock, eds., {\it Future ElectroWeak Physics at the
        Fermilab Tevatron}, FERMILAB-PUB-96/046.

\bibitem{PeskinCP}
   C. R. Schmidt and M. E. Peskin, \journal{Phys. Rev. Lett.}{69}{410}{1992}.

\bibitem{YuanCP}
    C.-P. Yuan, \journal{Mod. Phys. Lett.}{A10}{627}{1995}.
  



\bibitem{higgs1}
P. W.~Higgs, \journal{Phys. Lett.}{12}{132}{1964}.

\bibitem{higgsrev}
 J. F.~Gunion, H. E.~Haber, G. L.~Kane and
S.~Dawson, {\sl The Higgs Hunter's Guide}
(Addison-Wesley, Redwood City, 1990).

\bibitem{higgs2}
J. F.~Gunion, A.~Stange and S.~Willenbrock, in    {\it Electroweak
Symmetry Breaking and New Physics at the TeV Scale}, 
  T. Barklow, S. Dawson,
H. Haber, and J. Seigrist, eds.
 (World Scientific, Singapore, 1996).

\bibitem{Nillrev}
H. P. Nilles, \journal{Phys. Repts.}{110}{1}{1984}.

\bibitem{HKrev}
H.~Haber and G.~Kane, \journal{Phys. Repts.}{117}{75}{1985}.

\bibitem{sweep}
T.~Moroi and Y.~Okada, \journal{Phys. Lett.}{B295}{73}{1992};
G. L.~Kane, C.~Kolda, J. D.~Wells,
   \journal{Phys. Rev. Lett.}{70}{2686}{1993}.

\bibitem{LEPlim}
J. F. Grivaz, preprint LAL-95-83 (1995), to appear in the proceedings
of the 
International Europhysics conference on High
Energy Physics, Brussels, 1995.

\bibitem{LEPfit}
 D. Abbaneo, \etal (LEP Electroweak Working Group) and E. Etion, \etal
 (LEP Heavy Flavor Group), preprint LEPEWWG/96-01.

\bibitem{LEP2}
M.~Carena and P.~Zerwas \etal,
{\sl Higgs Physics}, to appear in the proceedings of the
LEP2 Workshop, CERN Yellow Report, eds. G.~Altarelli \etal

\bibitem{wellsco}
  G. L. Kane, G. D. Kribs, S. P. Martin, and J. D. Wells, 
        \journal{Phys. Rev.}{D53}{213}{1996}.

\bibitem{KandMr}
  S. Mrenna and G. L. Kane, preprint hep-ph/9406337.

\bibitem{janot}
P.~Janot,  in 
{\it Physics and Experiments with Linear $\ee$ Colliders},
F. A. Harris, S. L. Olsen, S. Pakvasa, and X. Tata, eds.
(World Scientific, Singapore, 1993).

\bibitem{gunionHH}
 J.~Gunion,  in 
{\it Physics and Experiments with Linear $\ee$ Colliders},
F. A. Harris, S. L. Olsen, S. Pakvasa, and X. Tata, eds.
(World Scientific, Singapore, 1993).

\bibitem{other}
S.~Komamiya, in {\it Physics and Experiments with Linear  Colliders},
R. Orava, P. Eerola, and M.Nordberg, eds. (World Scientific, Singapore,
1992); H.E.~Haber, {\it ibid.}; J.-F.~Grivaz, in  {\it $\ee$ Collisions
at 500 GeV: The Physics Potential}, P. M. Zerwas, ed.,  DESY-92-123.

\bibitem{JLC}
Y.~Fujii, in 
{\it Physics and Experiments with Linear $\ee$ Colliders},
F. A. Harris, S. L. Olsen, S. Pakvasa, and X. Tata, eds.
(World Scientific, Singapore, 1993); {JLC-1
Design Report}, KEK Report 92-16.

\bibitem{janot}
P.~Janot, in {\it Physics and Experiments with Linear Colliders}
F. A. Harris, 
World Scientific, 1993, eds. F.A.~Harris \etal, Waikoloa,
Hawaii.

\bibitem{hildreth}
M. D.~Hildreth, T. L.~Barklow, and D. L.~Burke, 
\journal{Phys. Rev.}{D49}{3441}{1993}.

\bibitem{old}
 P.~Burchat, D.~Burke,
and A.~Petersen, {\it Phys.~Rev.} {\bf D38} (1988) 2735;
J.~Alexander, \etal, in {\it High Energy Physics in the 1990s} 
  (Snowmass 1988), S. Jensen, ed. (World Scientific, Singapore, 1989);
F.~Richard, in {\it  Proceedings of the Workshop on
Physics at Future Accelerators} (La Thuile),  CERN 87-07, 1987.

\bibitem{djouadi}
A.~Djouadi, W.~Kilian, and P.~Ohmann, preprint hep-ph/9512312,
  to appear in the proceedings 
      of the Workshop on Physics and Experiments with Linear Collider,
              LCWS95.

\bibitem{HHHMMM}
  M. Carena, J. R. Espinosa, M. Quiros, and C. E. M. Wagner,
      \journal{Phys. Lett.}{B335}{209}{1995};
     M. Carena, M. Quiros, and C. E. M. Wagner,
    \journal{Nucl. Phys.}{B461}{407}{1996};
   H. E. Haber, R. Hempfling, and A. H. Hoang, preprint
         CERN-TH-95-216.

\bibitem{Rosiek}
A.~Sopczak, {\sl Charged Higgs Boson Discovery Potential at a 500
GeV $e^+e^-$ Linear Collider}, CERN-PPE/93-197;
  A.~Sopczak, in {\it $\ee$ Collisions at 500 GeV: The Physics Potential},
       P. M. Zerwas, ed.,  DESY-93-123C.

\bibitem{sopczak}
J.~Rosiek and A.~Sopczak, \journal{Phys.~Lett.}{B341}{419}{1995}.

\bibitem{kawagoe}
K.~Kawagoe,  in 
{\it Physics and Experiments with Linear $\ee$ Colliders},
F. A. Harris, S. L. Olsen, S. Pakvasa, and X. Tata, eds.
(World Scientific, Singapore, 1993).

\bibitem{CPgunion}
B.~Grz\c{a}dkowski and J. F.~Gunion, \journal{Phys. Lett}{B350}{218}{1995}.

\bibitem{CPgunion2}
J. F.~Gunion and B.~Grz\c{a}dkowski, \journal{Phys.~Lett.}{B294}{361}{1992}.



\bibitem{gamgunion}
J. F.~Gunion and J. G.~Kelly, \journal{Phys. Lett.}{B333}{220}{1994}.

\bibitem{stong}
M. L.~Stong, preprint hep-ph/9504345.

\bibitem{haber}
H.~Haber, preprint hep-ph/9505240.

\bibitem{kamoshita}
J.~Kamoshita, Y.~Okada, and M.~Tanaka, preprint hep-ph/9512307.

\bibitem{nakamura}
I.~Nakamura and K. Kawagoe, preprint hep-ex/9604010; I.~Nakamura,
 to appear in the proceedings 
      of the Workshop on Physics and Experiments with Linear Colliders,
              LCWS95.


\bibitem{HaberTASI}
    H. E. Haber, in {\it Recent Directions in Particle Theory (TASI-92)}, 
          J. Harvey and J. Polchinski, eds. (World Scientific, Singapore,
               1993).

\bibitem{Murarev}
   H. Murayama, in {\it Physics with High Energy Colliders}, 
         S. Yamada and T. Ishii, eds. (World Scientific, Singapore, 1995).

\bibitem{PeskYuk}
   M. E. Peskin, preprint hep-ph/9604339, to appear in the proceedings 
      of the Yukawa International Seminar YKIS'95.

\bibitem{Arno}
   R. Arnowitt and P. Nath, contributed paper in this volume.
 

\bibitem{CSECTANG}  H.~Baer, R.~Dubois, S.~Fahey, 
S.~Manly, R.~Munroe, U.~Nauenberg, X.~Tata, D.~L.~Wagner, 
   contributed paper in this volume.
%
\bibitem{JLC1} T.~Tsukamoto, K.~Fujii, H.~Murayama, 
M.~Yamaguchi, and Y.~Okada, \journal{Phys. Rev.}{D51}{3153}{1995}.
%
\bibitem{HELAS}  H. Murayama, I. Watanabe,and K. Hagiwara,
{\it HELAS: Helicity Amplitude Subroutines for Feynman Diagram
Evaluations},  KEK-91-11 (1992).
%
%

\bibitem{TWOGAMMA}
   R. Becker and C. Vander Velde, in {\it Physics and Experiments with 
        Linear Colliders}, F. A. Harris, S. L. Olsen, S. Pakvasa, and 
        X. Tata, eds. (World Scientific, Singapore, 1993).
%
%
%
\bibitem{JLC2} K.~Fujii, T.~Tsukamoto, and M. M. Nojiri,
 preprint hep-ph/9511338, to appear in the proceedings 
      of the Workshop on Physics and Experiments with Linear Collider,
              LCWS95.

\bibitem{TESTING} J. L. Feng, H. Murayama, M. E. Peskin, and X. Tata, 
          \journal{Phys. Rev.}{D52}{1418}{1995}.

\bibitem{SQUARK}
     J. L. Feng and D. E. Finnell, \journal{Phys. Rev.}{D49}{2369}{1994}.

\bibitem{BCPT} H. Baer, C. H. Chen, F. Paige and X. Tata,
     \journal{Phys. Rev.}{D53}{6241}{1996}.







\bibitem{AGBI}
H.~Aihara \etal,  in    {\it Electroweak
Symmetry Breaking and New Physics at the TeV Scale}, 
  T. Barklow, S. Dawson,
H. Haber, and J. Seigrist, eds.
 (World Scientific, Singapore, 1996).


\bibitem{Einhornsrev}
    M. Einhorn, in   
{\it Physics and Experiments with Linear $\ee$ Colliders},
F. A. Harris, S. L. Olsen, S. Pakvasa, and X. Tata, eds.
(World Scientific, Singapore, 1993).

\bibitem{baggerdaw}
J. Bagger, S. Dawson, and G. Valencia, {\em Nucl. Phys.} {\bf B399},
364 (1993).

\bibitem{aWudka}
J. Wudka, contributed paper in this volume.

\bibitem{TGVSMexpect}
E. N. Argyres, \journal{Nucl. Phys.}{B391}{23}{1993};
J. Papavassiliou and K. Philoppides, \journal{Phys.
Rev.}{D48}{4255}{1993}.

\bibitem{TGVSUSYexpect}
G. Couture {\it et al.,} \journal{Phys. Rev.}{D38}{860}{1988}.

 
\bibitem{HISZ}
H.~Hagiwara, S.~Ishihara, R.~Szalapski, and D.~Zeppenfeld
\plb{283}{353}{1992}

\bibitem{TLBFINLAND} 
T. L.~Barklow, in   in 
{\it Physics and Experiments with Linear  Colliders},
R. Orava, P. Eerola, and M.Nordberg, eds.
(World Scientific, Singapore, 1992).

\bibitem{LEPII}
Z.~Ajaltouni \etal, preprint hep-ph/9601233, to appear in the proceedings
  of the CERN Workshop on LEP II Physics.

\bibitem{TLBUCLA} 
T. L.~Barklow, in {\it Proceedings of the International Symposium on
Vector Boson Self-Interactions},  U.~Baur, S.~Errede, and
T.~M\"uller, eds. (American Inst. Press).
 
\bibitem{KR}
K.~Riles, contributed paper in this volume.

\bibitem{CandCWWZ}
  D. Choudhury and F. Cuypers, \journal{Nucl. Phys.}{B429}{33}{1994}.


\bibitem{Brodsky} 
S. J. Brodsky, T. G. Rizzo, and I. Schmidt, \journal
{Phys. Rev.}{D52}{4929}{1995}.

\bibitem{BDHS} 
T. L.~Barklow, S.~Dawson, H. E.~Haber, and J. L.~Siegrist, in 
{\it Particle Physics: Perspectives and Opportunities}, R. Peccei, 
M. E. Zeller, D. G. Cassel, J. A. Bagger, R. N. Cahn, P. D. Grannis, and
F. J. Sciulli, eds. (World Scientific, Singapore, 1995).


\bibitem{peskinomnes}
M. Peskin, in {\it Physics in Collisions IV},
 A. Seiden, ed. (\'Editions Fronti\`eres, Gif-Sur-Yvette,
1984).
 
\bibitem{bargerhan}
V. Barger, K. Cheung, T. Han, and R.J.N. Phillips,
\journal{Phys. Rev.}{D52}{3815}{1995}.

\bibitem{bargerhanemem}
V. Barger, J. F. Beacom, K. Cheung, and T. Han, 
\journal{Phys. Rev.}{D50}{6704}{1994}.

\bibitem{kauffman}
R. P. Kauffman, \journal{Phys. Rev.}{D41}{3343}{1990}.


\bibitem{rev}
 M. Cvetic and S. Godfrey, in    {\it Electroweak
Symmetry Breaking and New Physics at the TeV Scale}, 
  T. Barklow, S. Dawson,
H. Haber, and J. Seigrist, eds.
 (World Scientific, Singapore, 1996);
 A. Djouadi, J. Ng, and T. G. Rizzo, {\it ibid.}

\bibitem{tgr}
T. G. Rizzo, contributed paper in this volume.


\bibitem{ee95}
C. A. Heusch, ed., {\it Proceedings of the Electron-Electron
          Linear Collider Workshop}, 
\journal{Int. J. Mod. Phys.}{A11}{1523}{1996}. 

\bibitem{ggLBL}
{\it Proceedings of the Workshop on Gamma-Gamma Colliders},
             \journal{NIM}{A355}{1}{1995}. 

\bibitem{GHphoton}
J. F. Gunion and H. E. Haber, \journal{Phys. Rev.}{D48}{5109}{1993}.

\bibitem{oddHiggs} 
J. Gunion and J. Kelly, \journal{Phys. Lett.}{B333}{110}{1994};
S. Moretti, \journal{Phys. Rev.}{D50}{2016}{1994}; M. Kramer
\etal, \journal{Z. Phys.}{C64}{21}{1994}.


\bibitem{Tandean} 
J. Tandean, \journal{Phys. Rev.}{D52}{1398}{1995}.

\bibitem{Anlauf} 
H. Anlauf, W. Bernreuther, and A. Brandenberg, 
           \journal{Phys. Rev.}{D52}{3803}{1995}.              

\bibitem{Borden} 
D. L. Borden, D. A. Bauer, and D. O. Caldwell,
          \journal{Phys. Rev.}{D48}{4018}{1993}.              


\bibitem{Borden2} 
D. L. Borden  V. A. Khoze, W. J. Stirling, and J. Ohnemus, 
          \journal{Phys. Rev.}{D50}{4499}{1994}.              

\bibitem{Ginzburg} 
I. F. Ginzburg and V. G. Serbo, 
          \journal{Phys. Rev.}{D49}{2623}{1994}.              

\bibitem{Jikia} 
G. V. Jikia,      \journal{Phys. Lett.}{B298}{224}{1993},   
   \journal{Nucl. Phys.}{B405}{24}{1993}.              



\bibitem{MCHiggs} 
 O. J. P. Eboli, M. C. Gonzalez-Garcia, F. Halzen, and D. Zeppenfeld,
  \journal{Phys. Rev.}{D48}{1430}{1993};
M. Baillargeon,  G. Belanger, and F. Boudjema,
    \journal{Phys. Rev.}{D51}{4712}{1995}.              

\bibitem{HanEE}
T. Han, \journal{Int. J. Mod. Phys.}{A11}{1541}{1996}.

\bibitem{Hikasa} 
K. Hikasa, \journal{Phys. Lett.}{B164} {385}{1985}.

\bibitem{RizzoEE} 
T. Rizzo,  \journal{Int. J. Mod. Phys.}{A11}{1563}{1996}.

\bibitem{GunionEE} 
J. F. Gunion, \journal{Int. J. Mod. Phys.}{A11}{1551}{1996}. 



\bibitem{Jikia2} 
N. V. Dokholyan and G. V. Jikia, \journal{Nucl. Phys.}{B374}{83}{1992}.

\bibitem{BrodskyHH}
   S. J. Brodsky, in 
{\it Physics and Experiments with Linear $\ee$ Colliders},
F. A. Harris, S. L. Olsen, S. Pakvasa, and X. Tata, eds.
(World Scientific, Singapore, 1993).


\bibitem{Cuypers} 
F. Cuypers  G. J. van Odenborgh, and R. Ruckl,
  \journal{Nucl. Phys.}{B409}{128}{1993}.


\bibitem{ChoudCuy} 
D. Choudhury and F. Cuypers,   \journal{Nucl. Phys.}{B451}{16}{1995}.


\bibitem{Cheung}
   K. Cheung, \journal{Phys. Lett.}{B323}{85}{1994}.



\bibitem{Heusch} 
C. A. Heusch and P. Minkowski, \journal{Nucl. Phys.}{B416}{1}{1994}.

\bibitem{Frampton} 
P. H. Frampton, \journal{Phys. Rev. Lett.}{69}{2889}{1992}. 

\bibitem{Lepore} 
N. Lepor\'e, 
B. Thorndyke, H. Nadeau, and D. London, \journal{Phys. Rev.}{D50}{2031}{1994}.

\bibitem{Nadeau} 
 H. Nadeau and D. London,
 \journal{Phys. Rev.}{D47}{3742}{1993}; G. B\'elanger, D. London, 
and H. Nadeau,
 \journal{Phys. Rev.}{D49}{3140}{1994}.

\bibitem{CLepto}
  F. Cuypers, preprint hep-ph/9508397.

\bibitem{BarklowEE}
  T. Barklow, \journal{Int. J. Mod. Phys.}{A11}{1579}{1996}.           

\bibitem{Choudhury2} 
D. Choudhury, F. Cuypers, and A. Leike, \journal{Phys. Lett.}{B333}{531}{1994}.

\bibitem{Cuypersee} 
F. Cuypers,  \journal{Int. J. Mod. Phys.}{A11}{1571}{1996}, preprint
          hep-ph/9602426. 

\bibitem{BoudjemaEE} 
F. Boudjema, A. Djouadi, and J. L. Kneur, \journal{Z. Phys.}{C57}{425}{1993}.


\bibitem{Telnov} 
V. Telnov, \journal{NIM}{A355}{3}{1995}.

\bibitem{E144} 
J. G. Heinrich, \etal, SLAC Proposal E-144 (1991).


\bibitem{qcd1}  
  P. Langacker and N. Polonsky, \journal{Phys. Rev.}{D47}{4028}{1993},
        \journal{Phys. Rev.}{D52}{3081}{1995}.
 
\bibitem{qcd2} 
S. Bethke, \journal{Nucl. Phys.}{B {\rm Proc. Suppl.} 39C}{198}{1995}.
 
\bibitem{qcd3}
K. Abe \etal\ (SLD Collaboration), \journal{Phys. Rev.}{D51}{962}{1995}.
 
\bibitem{qcd4}
S. Bethke,  
 in {\it Physics and Experiments with Linear $\ee$ Colliders},
F. A. Harris, S. L. Olsen, S. Pakvasa, and X. Tata, eds.
(World Scientific, Singapore, 1993).

\bibitem{qcd5}
B. R. Webber, preprint hep-ph/9411384.
 
\bibitem{qcd7} 
A. Brandenburg, L. Dixon, and Y. Shadmi,
\journal{Phys. Rev.}{D53}{1264}{1996}.
 
\bibitem{qcd8} 
K. Abe \etal\ (SLD Collaboration), \journal{Phys. Rev. Lett.}{74}{4173}{1995}.
 
\bibitem{qcd9}
C. Carone and  H. Murayama, \journal{Phys. Rev. Lett.}{74}{3122}{1995};
D.~Bailey and S.~Davidson, \journal{Phys. Lett.}{B348}{185}{1995}.
 

\bibitem{RizzoQCD}
  T. G. Rizzo, contributed paper in this volume.

\bibitem{KGCP}
J. F. Gunion and P. C. Martin, contributed paper in this volume.
 
\end{thebibliography}

\clearpage
\addcontentsline{toc}{section}{Bibliography for Chapter 3}
\begin{thebibliography}{99}
 
\bibitem{oneref}
``Zeroth-Order Design Report for the Next Linear Collider,'' SLAC
Report 474 (Stanford University, May 1996). 
 
\bibitem{see91}
 J.~Seeman, 
 ``The Stanford Linear Collider'',~
 {\sl Ann. Rev. Nucl. Part. Sci.~}\ 
 (Annual Review, Inc., Palo Alto, California, 1991),
 Vol. 41,
 pp. 389--428.

\bibitem{see93}
 J.~Seeman, 
 ``Accelerator Physics of the Stanford Linear Collider and
 SLC Accelerator Experiments Towards the Next\
 Linear Collider'',~
 {\sl Advances of Acclerator Physics and Technologies}
 (Vol. 12 of the Advanced Series on Directions in
 High Energy Physics),
 edited by Herwig Schopper
 (World Scientific, 1993),
 pp. 219--248.

\bibitem{SLAC93}
 ``Next Linear Collider Test Accelerator Conceptual Design Report'',~
 SLAC Report 411 (Stanford University, August 1993).

 
\bibitem{rut93}
 R. D.~Ruth {\it et al.},
 ``The Next Linear Collider Test Accelerator''~
 (SLAC-PUB-6252), 
 Proceedings of the 1993 IEEE Particle Accelerator Conference, 
 Washington, DC, May 17--30, 1003 (IEEE Catalog No. 93CH3279-7), 
 pp. 543--545.

\bibitem{wil90}
 P. B.~Wilson, Z. D.~Farkas, and R. D.~Ruth,
 ``SLED-II:  A New Method of RF Pulse Compression''~
 (SLAC-PUB-5330), 
 Proceedings of the 1990 Linear Accelerator Conference, 
 Albuquerque, New Mexico, 
 September 10--14, 1990.
 
\bibitem{far74}
 Z. D.~Farkas {\it et al.},\
 ``SLED:  A Method of Doubling SLAC's Energy'',~
 (SLAC-PUB-1453), 
 Proceedings of the 9th International Conference on 
 High Energy Accelerators 
 (Stanford, California, May 2--7, 1974), 
 pp. 576--583.
 
\bibitem{nan93}
 C.~Nantista {\it et al.},\ 
 ``High Power Rf Pulse Compression with SLED-II at SLAC''~
 (SLAC-PUB-6145), 
 Proceedings of the 1993 IEEE Particle Accelerator Conference, 
 Washington, DC, May 17--20, 1993, (IEEE Catalog No. 93CH3279-7), 
 pp. 1196--1198.

\bibitem{vli93}
 A. E.~Vlieks {\it et al.}, 
 ``Accelerator and Rf System Development for NLC''~
 (SLAC-PUB-6148), 
 Proceedings of the 1993 IEEE Particle Accelerator Conference
 Wahington, DC, May 17--20, 1003 (IEEE Catalog No. 93CH3279-7)  
 pp. 620--622.

\bibitem{wan94}
 J. W.~Wang {\it et al.},
 ``High Gradient Tests of SLAC Linear Collider 
 Accelerator Structures''~
 (SLAC-PUB-6617), 
 Proceedings of the 17th International Linear Accelerator Conference
 (LINAC 94),
 Tskuba, Japan, August 21--26, 1994.
 
\bibitem{see85}
 J.~Seeman {\it et al.},\
 ``RF Beam Deflection Measurements and
 Corrections in the SLC Linac'',~
 Proceedings of the 1985 US Particle Accelerator Conference,
 Vancouver, Canada, May 1985
 (IEEE NS32, No. 5),
 pp. 2629--2631.
 
\bibitem{read}
  R.F.J.~Read and W.J.~Wills-Moren, 
 Cranfield Precision Engineering, Ltd., Bedford, England,
 ``Study Report for the Estimation of Unit Cost of 
 Mass Producing Discs for the CLIC Accelerating Sections'',~
 1993.
 
\bibitem{opti}
 ``Technology and Costs for 2 Million Copper Discs 
 of the CLIC Accelerator at CERN'',~
 Optische Werke G. Rodenstock,
 1993.
 
\bibitem{loe88a}
 G. A. Loew and J. W. Wang,
 ``RF Breakdown Studies in Room Temperature 
 Electron Linac Structures'' (SLAC-PUB-4647),~
 Proceedings of the 13th International Symposium on
 Discharges and Electrical Insulation in Vacuum, 
 Paris, France, June 27--30, 1988.

\bibitem{wan94a}
 J. W. Wang {\it et al.},\
 ``High Gradient Tests of SLAC Linear Collider 
 Accelerator Structures''~(SLAC-PUB-6617),
 Proceedings of the 17th International Linear Accelerator 
 Conference (LINAC 94), 
 Tsukuba, Japan, August 21--26, 1994. 

\bibitem{tak91}
 S.~Takeda {\it et al.},\
 ``High Gradient Experiments by the ATF'',~
 1991 IEEE Particle Accelerator Conference,
 San Francisco, California, May 6--9, 1001 (IEEE Catalog No.
 91CH3038-7),
 pp. 2061--2063.

\bibitem{mat91}
 H.~Matsumoto {\it et al.},\
 ``Applications of Hot Isostatic Pressing (HIP)
 for High Gradient Accelerator Structure'',~
 1991 IEEE Particle Accelerator Conference,
 San Francisco, California, May 6--9, 1991 (IEEE Catalog No.
 91CH3038-7),
 pp. 1008--1010.

\bibitem{tan95a}
 S. G. Tantawi, R. D. Ruth, and A. E. Vlieks, ``Active Ratio Frequency
 Pulse Compression using Switched Resonant Delay Lines''
 (SLAC-PUB-95-6748), {\sl Nucl. Instr. Meth.}~ {\bf A370}, 297 (1996).

\bibitem{tan95b} 
S. G. Tantawi, T. G. Lee, R. D. Ruth, A. E. Vlieks, and M. Zolotorev, 
``Design of a Multimegawatt X-Band Solid State Microwave Switch''
(SLAC-PUB-95-6827), Proceedings of the 1995 IEEE Particle Accelerator
Conference, Dallas, Texas.
 
\bibitem{far86}
Z. D. Farkas, ``Binary Peak Power Multiplier and Its Application to
Linear Accelerator Deisgn'',
IEEE Transactions MTT-34 (1986), pp. 1036--1043. 

\bibitem{lav91}
T. L. Lavine {\it et al.},\ ``High-power Radio-frequency Binary Pulse
Compression Experiment at SLAC'' (SLAC-PUB-5451), Proceedins of the
1991 IEEE Particle Accelerator Conference, San Francisco, California,
May 6--9, 1991 (IEEE Catalog 91CH3038-7), pp. 652--654.
 

\end{thebibliography}
\end{document}